\documentclass{article}
\usepackage{authblk}
\usepackage[utf8]{inputenc}
\usepackage{amsmath,amssymb,amsfonts}
\usepackage{physics}
\usepackage{graphicx}
\usepackage{xcolor}
\usepackage[numbers,sort&compress,square]{natbib}
\usepackage{amsthm}
\usepackage{appendix}
\usepackage{subfigure} 
\usepackage{environ}
\usepackage{comment}
\usepackage[margin=1in]{geometry}
\usepackage[expansion=false]{microtype}
\usepackage{enumitem}
\usepackage{booktabs}
\usepackage{tabularx}
\usepackage{mathtools}
\usepackage{tikz}
\usepackage{float}
\usepackage{scalerel}
\usepackage[linktocpage]{hyperref}
\hypersetup{
  colorlinks=true,
  citecolor=blue,
  linkcolor=blue,
  urlcolor=blue,
  breaklinks=true,
  pdftitle={Non-Clifford quantum cellular automata from topological quantum field theories},
  pdfauthor={Meng Sun, Bowen Yang, Zongyuan Wang, Nathanan Tantivasadakarn, and Yu-An Chen}
}
\newcolumntype{Y}{>{\raggedright\arraybackslash}X}

\tikzset{
  op/.style={inner sep=1pt, fill=white},
  every picture/.style={line width=0.5pt}
}
\usetikzlibrary{arrows.meta,calc,positioning,decorations.pathmorphing}

\newcommand{\ZZ}{{\mathbb Z}}

\newcommand{\RR}{{\mathbb R}}

\newcommand{\CC}{{\mathbb C}}

\newcommand{\mP}{\mathcal P}

\newcommand{\bv}{\boldsymbol{v}}
\newcommand{\be}{\boldsymbol e}
\newcommand{\bface}{\boldsymbol{f}}
\newcommand{\bt}{\boldsymbol{t}}
\newcommand{\bc}{\boldsymbol{c}}
\newcommand{\bp}{\boldsymbol{p}}

\newcommand{\bsig}{\boldsymbol{\sigma}}
\newcommand{\btau}{\boldsymbol{\tau}}

\newcommand{\bnu}{\boldsymbol{\nu}}

\newcommand{\Sq}{\mathrm{Sq}}

\newcommand{\br}{\boldsymbol r}

\NewEnviron{eqs}{%
\begin{equation}\begin{split}
    \BODY
\end{split}\end{equation}
}

\title{Non-Clifford quantum cellular automata from invertible topological quantum field theories}

\author[1,*]{\mbox{Meng Sun}}
\author[1,*]{\mbox{Zongyuan Wang}}
\author[2]{\mbox{Bowen Yang}}
\author[3,$\dagger$]{\mbox{Nathanan Tantivasadakarn}}
\author[1,$\ddagger$]{\mbox{Yu-An Chen}}

\affil[1]{International Center for Quantum Materials, School of Physics, Peking University, Beijing 100871, China}
\affil[2]{Center of Mathematical Sciences and Applications, Harvard University, Cambridge, Massachusetts 02138, USA}
\affil[3]{C. N. Yang Institute for Theoretical Physics, Stony Brook University, Stony Brook, NY 11794, USA}

\date{\today}

\begin{document}
\maketitle

\renewcommand*{\thefootnote}{*}
\footnotetext[1]{These authors contributed equally to this work.}
\renewcommand*{\thefootnote}{$\dagger$}
\footnotetext[1]{e-mail: {\color{blue}nathanan.tantivasadakarn@stonybrook.edu}}
\renewcommand*{\thefootnote}{$\ddagger$}
\footnotetext[1]{e-mail: {\color{blue}yuanchen@pku.edu.cn}}
\renewcommand{\thefootnote}{\arabic{footnote}}

\begin{abstract}
Quantum cellular automata (QCAs) describe locality-preserving quantum dynamics and link quantum information, many-body physics, and topological quantum field theory (TQFT). Constructing a QCA from a TQFT, however, remains challenging: although a topological action yields a Hamiltonian with commuting terms that realizes the desired ground state, it does not by itself specify an automorphism of the full operator algebra. We develop a unified algebraic construction that extends such Hamiltonians to complete separator-flipper algebras on the full tensor-product Hilbert space, thereby defining the corresponding QCAs microscopically. In 3 spatial dimensions, our framework unifies all previously known QCA constructions associated with the $\mathbb{Z}_8\times\mathbb{Z}_2$ subgroup of the Witt group, including the $U(1)_2$ and $U(1)_4$ QCAs. The same algebraic structure yields new infinite families of generalized $U(1)_2$ and $U(1)_4$ non-Clifford QCAs in spatial dimensions $d=4k-1$. We also reformulate the 4-dimensional $w_2w_3$ QCA and use it to construct QCAs from TQFTs associated with arbitrary products of Stiefel--Whitney classes. For example, we explicitly construct two infinite families: $w_2^nw_3^m$ QCAs in dimension $d=2n+3m-1$ and $w_2w_{4k-1}$ QCAs in dimension $d=4k$. For the 5-dimensional $w_3^2$ and $w_2^3$ QCAs, our construction gives finite-depth quantum circuits, consistent with the cobordism classification. More generally, we show that our construction yields a finite-depth quantum circuit whenever the corresponding Stiefel-Whitney number vanishes on every closed oriented manifold. Overall, these results convert invertible TQFTs into microscopic QCAs, provide higher-dimensional constructions beyond the Clifford setting, and open a systematic route to characterizing their stable structures and boundary anomalies.
\end{abstract}

\tableofcontents

\section{Introduction}

Quantum cellular automata (QCAs) describe reversible quantum dynamics that preserve locality~\cite{schumacher2004reversibleQCA}. A QCA is an automorphism of the local operator algebra such that both the transformation and its inverse map every local operator to one supported within a uniformly bounded neighborhood. Finite-depth quantum circuits (FDQCs) and lattice translations are the simplest examples, but in spatial dimensions greater than two, a QCA need not be equivalent to any composition of these transformations~\cite{Freedman2020ClassificationQCA, haah_QCA_23}. Index theory and subsequent classification results reveal sharp dimensional distinctions, from the complete index classification in one dimension to intrinsically higher-dimensional stable classes~\cite{Gross2012GNVWindex, Haah2021CliffordQCA}.

Beyond the problem of classifying locality-preserving unitaries, QCAs arise naturally in lattice formulations of quantum field theories~\cite{QFT2}, the study of Floquet phases~\cite{PoChiral, PoRadical, PotterVishwanathFidkowski18, Zhang2021classification, Glorioso21, Zhang2023Floquet}, and tensor-network representations of locality-preserving unitaries~\cite{IgnacioCirac2017,Sahinoglu2018,Piroli2020,Piroli21Fermionic}. They also constrain entanglement and operator growth~\cite{Gutschow2010Clifford,GongNahumPiroli21,GongPiroliCirac21} and provide microscopic realizations of subsystem and higher symmetries and their anomalies~\cite{Stephen2019subsystem,ma2024QCA,jones2024QCA,tu2025anomalies,kapustin2025higher}. QCAs connect together quantum information, many-body physics, and topological quantum field theory. This connection suggests a powerful route to constructing QCAs. An invertible topological quantum field theory (TQFT) encodes the universal response of a gapped phase~\cite{Chen2012cohomology,kapustin2014symmetry}, whereas an appropriate QCA can provide an exact lattice transformation that prepares or disentangles a representative state of that phase~\cite{Shirley2022QCA,fidkowski2024qca,Sun2026Clifford,haah2025topological}. This motivates constructing QCAs directly from topological response actions, thereby turning field-theory data into constructive tools for locality-preserving quantum dynamics.

Several landmark constructions established complementary aspects of this connection. Fidkowski, Haah, and Hastings constructed the first intrinsically nontrivial QCA in three spatial dimensions from the 3-fermion Walker-Wang model~\cite{Walker2012TQFT,haah_QCA_23}. Shirley \textit{et al.} subsequently obtained the 3-dimensional semion QCA by condensing the transparent boson in a pre-modular $\ZZ_4$ Walker-Wang model~\cite{Shirley2022QCA}. The boson-condensation construction separated the physical qubits from disentangled ancillas and extended to the $U(1)_{2^q}$ series. These models provided the first non-Clifford QCAs associated with chiral Abelian surface topological orders and strengthened the conjectured connection between 3-dimensional QCAs and the Witt group~\cite{Drinfeld2010Braided, Davydov2013Witt}.

Parallel progress came from exactly solvable models of beyond-cohomology phases~\cite{Fidkowski2020beyondcohomology}. Chen and Hsin constructed broad classes of such Hamiltonians and, for the $w_2w_3$ TQFT, obtained a locally flippable separator with an anomalous particle-loop boundary theory~\cite{chen2023exactly}. Fidkowski, Haah, and Hastings later developed generalized Walker-Wang QCAs and conjectured a broad correspondence between QCAs and TQFT responses built from Wu or Stiefel-Whitney classes~\cite{fidkowski2024qca}.
Our previous work~\cite{Sun2026Clifford} developed a complementary cup-product approach that related Clifford QCAs directly to TQFTs and produced higher-dimensional families matching the classification predicted by algebraic $L$-theory~\cite{haah2025topological, Yang2026CategorifyingQCA}, with stable nontriviality of the three-dimensional chiral $\mathbb Z_p$ examples established for every odd prime $p$, even allowing non-Clifford circuits~\cite{yang2026threedimensionalqcastablycircuits}. Building on this $L$-theoretic perspective and the ideas of Pedersen and Weibel~\cite{PedersenWeibel}, Ji and Yang subsequently developed a generalized cohomology theory that classifies arbitrary QCAs, including non-Clifford ones~\cite{ji2026quantum, ji2026k}.

Despite this progress, no microscopic QCA-completion procedure was known that unifies all these aforementioned constructions. A topological action determines ground-state information but does not uniquely select a microscopic parent Hamiltonian. Different Hamiltonians can realize the same ground-state phase while differing in their excited-state spectra. A QCA, by contrast, must define an automorphism of the entire local operator algebra. Thus, a generic commuting parent Hamiltonian alone cannot determine the transformation of all microscopic degrees of freedom.

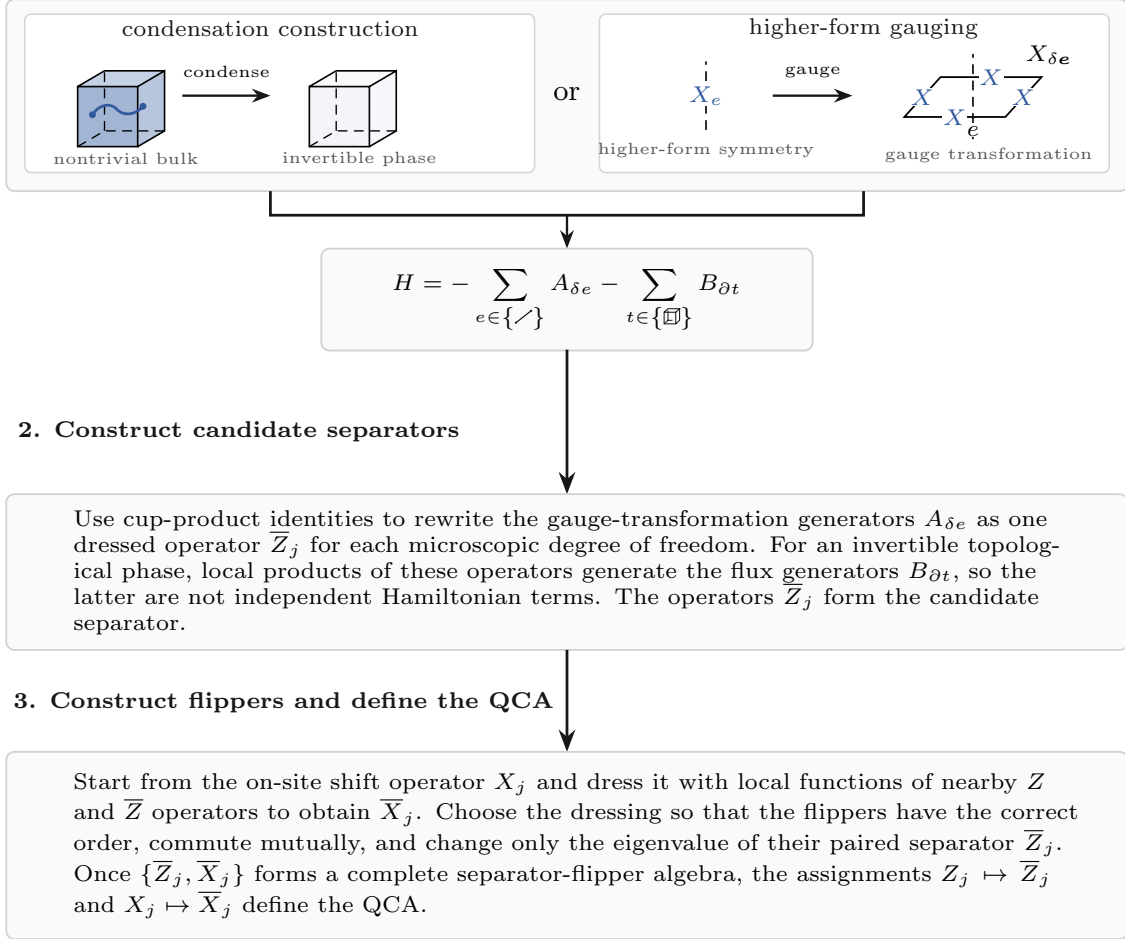
\begin{figure}[ht]
  \centering
  \resizebox{.9\linewidth}{!}{%

\begin{tikzpicture}

\definecolor{workflowInk}{HTML}{171717}
\definecolor{workflowMuted}{HTML}{555555}
\definecolor{workflowBlue}{HTML}{315A9E}

\newcommand{\workflowedgecell}{%
  \vcenter{\hbox{\tikz[baseline=-.38ex,x=.64em,y=.64em]{%
    \draw[workflowInk,line width=.38pt] (0,0)--(.92,.72);
    \fill[workflowInk] (0,0) circle (.34pt);
    \fill[workflowInk] (.92,.72) circle (.34pt);
  }}}%
}
\newcommand{\workflowcubecell}{%
  \vcenter{\hbox{\tikz[baseline=-.42ex,x=.62em,y=.62em]{%
    \draw[workflowInk,line width=.32pt]
      (0,0)--(.72,0)--(.72,.72)--(0,.72)--cycle
      (0,.72)--(.24,.94)--(.96,.94)--(.72,.72)
      (.72,0)--(.96,.22)--(.96,.94);
    \draw[workflowInk,densely dashed,line width=.28pt]
      (0,0)--(.24,.22)--(.96,.22)
      (.24,.22)--(.24,.94);
  }}}%
}

\tikzset{
  font=\small,
  line cap=round,
  line join=round,
  flow/.style={
    -{Stealth[length=2.4mm,width=1.8mm]},
    draw=workflowInk,
    line width=.85pt
  },
  smallflow/.style={
    -{Stealth[length=2.0mm,width=1.5mm]},
    draw=workflowInk,
    line width=.72pt
  },
  stepheading/.style={
    anchor=west,
    font=\bfseries\scriptsize,
    text=workflowInk
  },
  stagebox/.style={
    draw=black!18,
    fill=black!2,
    rounded corners=2.5pt,
    line width=.55pt
  },
  routebox/.style={
    draw=black!16,
    fill=white,
    rounded corners=2pt,
    line width=.50pt
  },
  paneltitle/.style={
    font=\scriptsize,
    text=workflowInk,
    align=center
  }
}

\node[stepheading] at (-6.05,10.66)
  {1. Obtain a Hamiltonian for the quantum phase of interest};

\draw[stagebox] (-6.05,8.22) rectangle (6.05,10.29);
\draw[routebox] (-5.85,8.42) rectangle (-.35,10.13);
\draw[routebox] (.35,8.42) rectangle (5.85,10.13);

\node[paneltitle] at (-3.20,9.98)
  {condensation construction};

\begin{scope}[
  shift={(-5.28,8.70)},
  x={(.54cm,0cm)},
  y={(.29cm,.17cm)},
  z={(0cm,.54cm)}
]
  \fill[workflowBlue!45]
    (0,0,0)--(1.20,0,0)--(1.20,0,1.20)--(0,0,1.20)--cycle;
  \fill[workflowBlue!35]
    (1.20,0,0)--(1.20,1.00,0)--(1.20,1.00,1.20)
    --(1.20,0,1.20)--cycle;
  \fill[workflowBlue!28]
    (0,0,1.20)--(1.20,0,1.20)--(1.20,1.00,1.20)
    --(0,1.00,1.20)--cycle;

  \draw[line width=.50pt]
    (0,0,0)--(1.20,0,0)--(1.20,0,1.20)--(0,0,1.20)--cycle
    (1.20,0,0)--(1.20,1.00,0)--(1.20,1.00,1.20)
    --(1.20,0,1.20)
    (0,0,1.20)--(0,1.00,1.20)--(1.20,1.00,1.20);

  \draw[densely dashed,line width=.42pt]
    (0,0,0)--(0,1.00,0)--(1.20,1.00,0)
    (0,1.00,0)--(0,1.00,1.20);

  \draw[workflowBlue,line width=.85pt]
    (.20,.18,.58)
      .. controls (.32,.18,.78) and (.46,.36,.78) .. (.56,.42,.61)
      .. controls (.68,.48,.44) and (.82,.66,.44) .. (.92,.72,.58);

  \fill[workflowBlue] (.20,.18,.58) circle (1.05pt);
  \fill[workflowBlue] (.92,.72,.58) circle (1.05pt);
\end{scope}

\node[font=\tiny,text=workflowMuted] at (-4.76,8.57)
  {nontrivial bulk};

\draw[smallflow] (-4.15,9.25)--(-3.20,9.25)
  node[midway,above=2pt,font=\tiny] {condense};

\begin{scope}[
  shift={(-2.78,8.70)},
  x={(.54cm,0cm)},
  y={(.29cm,.17cm)},
  z={(0cm,.54cm)}
]
  \fill[workflowBlue!6]
    (0,0,0)--(1.20,0,0)--(1.20,0,1.20)--(0,0,1.20)--cycle;
  \fill[workflowBlue!4]
    (1.20,0,0)--(1.20,1.00,0)--(1.20,1.00,1.20)
    --(1.20,0,1.20)--cycle;
  \fill[workflowBlue!3]
    (0,0,1.20)--(1.20,0,1.20)--(1.20,1.00,1.20)
    --(0,1.00,1.20)--cycle;

  \draw[line width=.50pt]
    (0,0,0)--(1.20,0,0)--(1.20,0,1.20)--(0,0,1.20)--cycle
    (1.20,0,0)--(1.20,1.00,0)--(1.20,1.00,1.20)
    --(1.20,0,1.20)
    (0,0,1.20)--(0,1.00,1.20)--(1.20,1.00,1.20);

  \draw[densely dashed,line width=.42pt]
    (0,0,0)--(0,1.00,0)--(1.20,1.00,0)
    (0,1.00,0)--(0,1.00,1.20);
\end{scope}

\node[font=\tiny,text=workflowMuted,align=center] at (-2.24,8.57)
  {invertible phase};

\node[paneltitle] at (3.20,9.98)
  {higher-form gauging};

\draw[densely dashed,line width=.50pt]
  (1.50,8.88)--(1.50,9.62);

\node[
  fill=white,
  text=workflowBlue,
  font=\scriptsize,
  inner sep=.28pt
] at (1.50,9.25) {$X_e$};

\node[font=\tiny,text=workflowMuted] at (1.50,8.66)
  {higher-form symmetry};

\draw[smallflow] (2.22,9.25)--(3.08,9.25)
  node[midway,above=2pt,font=\tiny] {gauge};

\draw[densely dashed,line width=.50pt]
  (4.38,8.78)--(4.38,9.70);

\node[fill=white,font=\scriptsize,inner sep=.25pt]
  at (4.38,8.87) {$e$};

\draw[line width=.55pt]
  (3.64,9.02)--(4.73,9.02)--(5.12,9.45)--(4.03,9.45)--cycle;

\path (3.64,9.02)--(4.73,9.02)
  node[
    fill=white,
    text=workflowBlue,
    font=\scriptsize,
    inner sep=.25pt,
    pos=.50
  ] {$X$};

\path (4.73,9.02)--(5.12,9.45)
  node[
    fill=white,
    text=workflowBlue,
    font=\scriptsize,
    inner sep=.25pt,
    pos=.50
  ] {$X$};

\path (5.12,9.45)--(4.03,9.45)
  node[
    fill=white,
    text=workflowBlue,
    font=\scriptsize,
    inner sep=.25pt,
    pos=.50
  ] {$X$};

\path (4.03,9.45)--(3.64,9.02)
  node[
    fill=white,
    text=workflowBlue,
    font=\scriptsize,
    inner sep=.25pt,
    pos=.50
  ] {$X$};

\node[font=\scriptsize] at (5.20,9.69)
  {$X_{\delta\boldsymbol e}$};

\node[font=\tiny,text=workflowMuted] at (4.55,8.61)
  {gauge transformation};

\node[font=\small] at (0,9.27) {or};

\draw[draw=workflowInk,line width=.85pt]
  (-3.20,8.22)--(-3.20,7.96)--(3.20,7.96)--(3.20,8.22);

\draw[smallflow] (0,7.96)--(0,7.60);

\draw[stagebox] (-2.65,6.50) rectangle (2.65,7.60);

\node[font=\scriptsize] at (0,7.05) {$\displaystyle
  H=
  -\sum_{e\in\left\{
    \vphantom{\workflowcubecell}\workflowedgecell
  \right\}} A_{\delta e}
  -\sum_{t\in\left\{
    \workflowcubecell
  \right\}} B_{\partial t}
$};

\begin{scope}[yshift=-1.45cm]

\node[stepheading,align=left] at (-6.05,7.08)
  {2. Construct candidate separators};

\draw[flow] (0,7.96)--(0,6.40);

\draw[stagebox] (-6.05,4.72) rectangle (6.05,6.40);

\node[
  anchor=west,
  align=left,
  text width=10.85cm,
  font=\scriptsize
] at (-5.45,5.56) {
Use cup-product identities to rewrite the gauge-transformation generators $A_{\delta e}$ as one dressed operator $\overline Z_j$ for each microscopic degree of freedom. Verify that local products of these operators generate the flux generators $B_{\partial t}$, so that the latter impose no additional constraints. The operators $\overline Z_j$ form the candidate separator. };

\draw[flow] (0,4.72)--(0,3.62);

\node[
  stepheading,
  fill=white,
  inner xsep=2pt
] at (-6.05,4.15)
  {3. Construct flippers and define the QCA};

\draw[stagebox] (-6.05,1.60) rectangle (6.05,3.62);

\node[
  anchor=west,
  align=left,
  text width=10.85cm,
  font=\scriptsize
] at (-5.45,2.61) {
Start from the on-site shift operator $X_j$ and dress it with local functions of nearby $Z$ and $\overline Z$ operators to obtain $\overline X_j$. Choose the dressing so that the flippers have the correct order, commute mutually, and change only the eigenvalue of their paired separator $\overline Z_j$. Once $\{\overline Z_j,\overline X_j\}$ forms a complete separator-flipper algebra, the assignments $Z_j\mapsto\overline Z_j$ and $X_j\mapsto\overline X_j$ define the QCA.
};

\end{scope}
\end{tikzpicture}}
\caption{Workflow for constructing the QCAs in this work. A commuting Hamiltonian is first obtained through condensation or higher-form gauging. Cup-product identities reorganize its generators into one candidate separator per microscopic degree of freedom. Compatible local flippers then complete the separator algebra, whose identification with the on-site clock and shift algebra defines the QCA.}
\label{fig:qca-construction-workflow}
\end{figure}

In this work, we develop such a completion for broad families of invertible topological responses. As summarized in Fig.~\ref{fig:qca-construction-workflow}, we begin with a commuting Hamiltonian realizing the desired phase. Complementary-cell cup products then select a natural separator candidate for each microscopic degree of freedom, while higher-cup identities guide the construction of compatible flippers. Verifying that these operators form a complete separator-flipper algebra allows their one-to-one identification with the standard clock and shift operators, thereby defining the QCA. The same procedure applies to all dimensions and topological response actions considered in this work.

\subsection*{Summary of results}

Our principal advances are as follows:

\begin{enumerate}

\item \emph{A unified framework for known 3-dimensional QCAs.}
We place all previously known QCA constructions associated with the $\mathbb Z_8\times\mathbb Z_2$ subgroup of the Abelian Witt group within a single algebraic framework~\cite{haah_QCA_23, Shirley2022QCA}. This includes the $U(1)_2$ semion and $U(1)_4$ QCAs. In particular, we give an independent topological-action construction of the semion QCA and naturally obtain its complete separator-flipper algebra. Our contribution to the existing literature is a systematic derivation of these constructions and their unification as microscopic automorphisms.

\item \emph{New higher-form generalizations.}
The unified 3-dimensional construction extends directly to new infinite families of generalized $U(1)_2$ and $U(1)_4$ QCAs in spatial dimensions $d=4k-1$, which we construct explicitly. The two families carry distinct higher-form quadratic excitation data, generalizing from their 3-dimensional counterparts~\cite{Drinfeld2010Braided, Davydov2013Witt}. Their 4-dimensional periodicity mirrors the periodic structure predicted by algebraic $L$-theory and explicitly constructed for Clifford QCAs~\cite{Sun2026Clifford,haah2025topological}.

\item \emph{A general construction from products of Stiefel-Whitney classes.}
We reformulate the known 4-dimensional $w_2w_3$ QCA~\cite{chen2023exactly} within the same algebraic framework, revealing it as the simplest member of much broader higher-dimensional families of QCAs. More generally, we construct QCAs for TQFT responses defined by arbitrary products of Stiefel-Whitney classes~\cite{Steenrod1947Products, MilnorStasheff1974, LusztigMilnorPeterson1969}. This construction includes the $w_2^nw_3^m$ QCAs in spatial dimension $d=2n+3m-1$ and the $w_2w_{4k-1}$ QCAs in spatial dimension $d=4k$. These results explicitly realize the Stiefel-Whitney responses anticipated in Ref.~\cite{fidkowski2024qca} and rigorously connect the resulting QCAs to their corresponding cobordism invariants.

\item \emph{Trivial QCAs for vanishing Stiefel-Whitney numbers.}
We construct explicit finite-depth quantum circuits for the $w_3^2$ and $w_2^3$ QCAs in five spatial dimensions, proving that both representatives are trivial, consistent with the vanishing of their oriented-bordism invariants~\cite{kapustin2014symmetry}. 
Appendix~\ref{app:wu-core-reduction} establishes a broader constructive result: for the canonical circuit-compatible representatives covered by its construction, if a mod-two Stiefel-Whitney number vanishes on every closed oriented spacetime manifold, then the associated stable QCA class is trivial, with a stabilized representative explicitly realized by a finite-depth circuit. The five-dimensional circuits provide concrete instances of this general trivialization mechanism.

\end{enumerate}

All together, these results establish a systematic route from invertible TQFT responses to microscopic QCAs. By separating QCA construction from stable nontriviality, our framework provides a foundation for classifying higher-dimensional locality-preserving dynamics and their boundary anomalies.

The paper is organized as follows. In Sec.~\ref{sec:construction-framework}, we introduce our cochain conventions and the definition of separator-flipper underlying our procedure to construct a QCA (Fig.~\ref{fig:qca-construction-workflow}). Sec.~\ref{sec:three-dimensional} constructs the 3-dimensional $U(1)_2$ semion QCA through both condensation and response-action approaches and then develops the $U(1)_4$ QCA. In Sec.~\ref{sec:generalized-u1-families}, we extend these constructions to infinite families of generalized $U(1)_2$ and $U(1)_4$ QCAs in spatial dimensions $d=4k-1$. Sec.~\ref{sec:four-dimensional} reformulates the $w_2w_3$ QCA and establishes its complete separator-flipper algebra. Sec.~\ref{sec:five-dimensional} constructs the $w_2^3$ and $w_3^2$ representatives and gives explicit finite-depth circuits proving that both are stably trivial. Sec.~\ref{sec:generalized-semion-seven-dimensional} specializes the higher-form constructions to seven spatial dimensions and constructs the additional $w_2^4$ QCA. Sec.~\ref{sec:general-family} constructs the general $w_2^nw_3^m$ and Wu-Bockstein families and analyzes their cobordism invariants and non-Clifford properties. Finally, in Sec.~\ref{sec:discussion}, we summarize the framework and discuss open problems concerning QCA classification and boundary anomalies.

Appendices~\ref{app:z4-cup-hamiltonian} and~\ref{sec:z2-semion-construction} provide detailed Hamiltonian derivations for the semion construction. Appendix~\ref{app:sw-number-relations} collects low-dimensional relations among oriented Stiefel-Whitney numbers, while Appendix~\ref{app:w2n-w3m-so-bordism} establishes the oriented-bordism criterion for the $w_2^nw_3^m$ family. Appendix~\ref{app:wu-core-reduction} constructs finite-depth circuit realizations for actions with vanishing Stiefel-Whitney numbers. Appendix~\ref{app:semion-signature-bordism} compares the signature responses and quadratic data of the semion and $U(1)_4$ QCA families.

\section{Construction framework and conventions}
\label{sec:construction-framework}

This section collects the minimal cochain and operator-algebraic background needed to follow our framework for constructing QCAs. The essential point is that the commuting terms obtained from a topological response do not yet define a QCA. They characterize a preferred state or sector, whereas a QCA must act on every local operator and hence on the entire spectrum. Thus, we seek a complete set of local operators, consisting of one separator and one compatible flipper for each microscopic degree of freedom. Further background on our cochain conventions can be found in our earlier work on constructing Clifford QCAs~\cite{Sun2026Clifford}.

\subsection*{Cochains and higher cup products}

Let $M$ be a $d$-dimensional bounded-valence cellulation.  The group $C_p(M,R)$ consists of formal $R$-linear combinations of $p$-cells, where $R$ will usually be $\ZZ$, $\ZZ_2$, or $\ZZ_{2^k}$. A $p$-cochain is an $R$-valued linear function on $p$-chains, and the group of such cochains is denoted by $C^p(M,R)$. For a $p$-cell $\sigma_p$, its indicator cochain $\boldsymbol{\sigma}_p$ is defined as
\begin{equation}
    \boldsymbol{\sigma}_p(\sigma_p') =\delta_{\sigma_p,\sigma_p'}.
    \label{eq:indicator-cochain}
\end{equation}
We use bold symbols such as $\bv$, $\be$, $\bface$, $\bt$, $\bc$ and $\bp$ to denote the indicator cochains of a vertex, edge, face, 3-cell, 4-cell, and 5-cell respectively. Any cochain can be expanded in this basis.

The cellular boundary operator $\partial$ maps a cell to its oriented boundary. Its dual is the coboundary operator
\begin{equation}
 \delta:C^p(M,R)\longrightarrow C^{p+1}(M,R), \qquad (\delta a)(\sigma_{p+1})=a(\partial\sigma_{p+1}), \qquad \delta^2=0.
 \label{eq:coboundary-convention}
\end{equation}
A cochain satisfying $\delta a=0$ is a cocycle and $a=\delta\lambda$ is a coboundary. For a top-degree cochain, we use the shorthand
\begin{equation}
    \int_M a:=\langle a,[M]\rangle,
    \label{eq:cochain-integration-convention}
\end{equation}
where $[M]$ is the fundamental chain.

The ordinary cup product combines a $p$-cochain and a $q$-cochain into a $(p+q)$-cochain. Its higher versions is given by
\begin{equation}
    \cup_i:C^p(M,R)\times C^q(M,R) \longrightarrow C^{p+q-i}(M,R).
\label{eq:higher-cup-degree}
\end{equation}
We use the local cellular realization of higher cup products on the hypercubic cellulation developed in Ref.~\cite{Chen2023HigherCup}. The identity used repeatedly in this work is
\begin{equation}
    \begin{aligned}
        \delta(a\cup_i b)
        ={}&\delta a\cup_i b+(-1)^p a\cup_i\delta b
        +(-1)^{p+q-i}a\cup_{i-1}b
        +(-1)^{pq+p+q}b\cup_{i-1}a .
    \end{aligned}
    \label{eq:higher-cup-recursion}
\end{equation}
Here, $\cup_0=\cup$ is the ordinary cup product, and any term containing $\cup_i$ with $i<0$ is evaluated to 0 \cite{Steenrod1947Products,Chen2023HigherCup}. For $i=0$, Eq.~\eqref{eq:higher-cup-recursion} reduces to the Leibniz rule for the ordinary cup product. For $\ZZ_2$ cochains all signs disappear as $-1 = 1$ in $\ZZ_2$. The expressions retain the signs otherwise.

The cochain formalism plays two roles. First, it expresses the topological action and its Hamiltonian descendants in a common language. Second, its cellular products have bounded support, ensuring locality. On the hypercubic cellulation used in this paper, complementary cells pair locally and nondegenerately~\cite{Chen2023HigherCup}. For the constructions of QCAs, we use this pairing to select an operator $\overline{Z}_j$ for each independent generator of the Hamiltonian. This is our separator candidate. We then check that the remaining constraints in the Hamiltonian are generated by local products of these separators and that no additional relations remain.

\subsection*{Separators, flippers, and QCAs}

A QCA is an invertible automorphism of the local operator algebra that preserves locality in both directions. That is, it maps every operator supported in a finite-sized region to an operator supported within a uniformly bounded neighborhood of that region, and the same condition holds for the inverse transformation~\cite{schumacher2004reversibleQCA,Gross2012GNVWindex,haah_QCA_23}. Automorphisms corresponding to conjugation by finite-depth quantum circuits and shifts form a special class of QCAs, but a general QCA need not admit such a circuit realization.

We consider an infinite hypercubic lattice. All operators written down in this paper have bounded supports. Each microscopic degree of freedom is labeled by $j$ and has a finite $N$-dimensional basis $\{|s\rangle:s\in\mathbb Z_N\}$. And, its matrix algebra $\text{Mat}_N(\CC)$ is generated by the generalized Pauli operators
\begin{equation}
    Z|s\rangle=\omega^s|s\rangle, \qquad X|s\rangle=|s+1\rangle, \qquad ZX=\omega XZ, \qquad \omega=e^{2\pi i/N},
    \label{eq:generalized-pauli-algebra}
\end{equation}
where addition in $s+1$ is computed modulo $N$. On the tensor-product space formed by multiple microscopic degrees of freedom, we then have the generalized Pauli algebra of
\begin{equation}
    Z_k X_j =\omega^{\delta_{jk}} X_j Z_k, \qquad [ X_j,X_k]=[Z_j,Z_k]=0, \qquad X_j^N=Z_j^N=1,
    \label{eq:generalized_Pauli_algebra_tensor_product_space}
\end{equation}
where the subscripts $j,k$ label the sites.

A complete local separator is a collection of mutually commuting and independent local unitary operators $\{\overline Z_j\}$ obtained from the Hamiltonian of the quantum phase of interest, with one operator of order $N$ for each site \cite{haah_QCA_23}. A local unitary $\overline X_j$ is a flipper paired with $\overline Z_j$ if it only changes the eigenvalue of $\overline Z_j$. A separator admitting one such flipper for the site $j$ is called locally flippable \cite{haah_QCA_23}. In our construction of the QCAs, we further choose the flippers such that they commute with one another and have order $N$. The complete separator-flipper algebra is
\begin{equation}
    \overline Z_k\overline X_j =\omega^{\delta_{jk}}\overline X_j\overline Z_k, \qquad [\overline X_j,\overline X_k]=[\overline Z_j,\overline Z_k]=0, \qquad \overline X_j^N=\overline Z_j^N=1,
    \label{eq:separator-flipper-algebra}
\end{equation}
where beyond the definition in Ref.~\cite{haah_QCA_23}, we further demand that the flippers $\overline{X}_j$ commute with each other. Notice that this is of the same form as the algebra for each degree of freedom as in Eq.~\eqref{eq:generalized_Pauli_algebra_tensor_product_space}. The corresponding QCA can then be defined by the one-to-one assignment
\begin{equation}
    Z_j\longmapsto \overline{Z}_j, \qquad X_j\longmapsto \overline{X}_j .
    \label{eq:qca-from-separator}
\end{equation}
Since the separators and flippers have bounded support and obey the same algebra as the on-site operators, this assignment then defines a reversible locality-preserving transformation i.e. a QCA~\cite{haah_QCA_23}.

\section{Non-Clifford QCAs in three spatial dimensions}
\label{sec:three-dimensional}

We begin with the 1-semion QCA, whose boundary anyon theory is that of the chiral semion, which is also known as the $U(1)_2$ Chern-Simons theory. We do this in Sec.~\ref{sec:semion-qca}. In Sec.~\ref{sec:u14-qca}, we move on to construct a QCA for the order-four $U(1)_4$ extension. Both QCAs have been constructed in Ref.~\cite{Shirley2022QCA}, here, we reproduce them using our TQFT and cup-product framework.

\subsection{Semion QCA}
\label{sec:semion-qca}

The semion QCA is among the simplest examples of a non-Clifford QCA. The cubic lattice construction of the 1-semion, or $\{1,s\}$, Walker-Wang QCA was first given in Ref.~\cite{Shirley2022QCA}. Here, we follow the same physical route. However, instead of using the tools of category theory, we use our framework of passing the topological response action to a lattice Hamiltonian and explicitly write every operator using the higher-cup-product method. We explicitly construct the separator and furthermore flippers that are mutually commuting, which were absent in Ref.~\cite{Shirley2022QCA}. Our formulation allows us to directly generalize to higher dimensions, in particular, a QCA in seven spatial dimensions (Sec.~\ref{sec:7d-generalized-semion-qca}). In Appendix~\ref{sec:z2-semion-construction}, we take an alternative route to obtain the same separator and flipper expressions via gauging a $\mathbb Z_2$ 1-form SPT.

The strategy to reach the semion QCA is as follows. We start from a pre-modular theory, which, in the category language, is the Walker-Wang model whose input category corresponds to the $\mathbb Z_4^{(1)}$ topological order \cite{Walker2012TQFT,Keyserlingk2013Threedimensional}. This $\mathbb Z_4^{(1)}$ topological order is the subtheory generated by $em$ in the $\ZZ_4$ toric code \cite{Bonderson2007}. The Hamiltonian of the $\mathbb Z_4^{(1)}$ Walker-Wang can therefore be written down using $\ZZ_4$ qudits. The $\mathbb Z_4^{(1)}$ Walker-Wang has an intrinsic bulk $\mathbb Z_2$ topological order and an order-two point-like bosonic excitation \cite{Keyserlingk2013Threedimensional}. Condensing this boson trivializes the bulk but leaves a surface semion topological order, which one can verify by opening up a boundary \cite{Shirley2022QCA}. The Hamiltonian obtained after the condensation is the desired semion Walker-Wang model. Then, following the procedure in Fig.~\ref{fig:qca-construction-workflow} together with a reduction of $\ZZ_4$ qudits to two qubits and a Clifford disentangler to keep only one qubit per face, we find the desired separator and flipper algebra. Next, we demonstrate in detail how our formalism reproduces the semion QCA.

Let us consider a cubic lattice and place one $\ZZ_4$ qudit on every face of the cubic lattice, with generalized Pauli operators
\begin{equation}
    X_f:=\sum_{j\in\mathbb Z_4}\ket{j+1}\bra{j},
    \qquad
    Z_f:=\sum_{j\in\mathbb Z_4}\omega^j\ket{j}\bra{j},
\end{equation}
where addition on the basis labels is defined modulo four and
\begin{equation}
    \omega:=\exp\!\left(\frac{2\pi i}{4}\right)=i.
\end{equation}
These operators satisfy
\begin{equation}
    X_f^4=1, \qquad Z_f^4=1,
\end{equation}
and, for a pair of faces $f$ and $f'$, obey
\begin{equation}
    Z_fX_{f'}=
    \begin{cases}
        \omega X_{f'}Z_f, & f=f',\\
        X_{f'}Z_f, & f\neq f'.
    \end{cases}
\end{equation}

We begin with the topological action for the $\ZZ^{(1)}_4$ Walker-Wang, which is
\begin{equation}
    \frac14 B_2\cup B_2\in H^4(B^2\mathbb Z_4,\mathbb R/\mathbb Z).
    \label{eq:z4-semion-action}
\end{equation}
The corresponding Hamiltonian can be constructed from the procedure described in Refs.~\cite{Chen2012cohomology,Chen2023HigherCup} via the cup-product. We save the details to Appendix~\ref{app:z4-cup-hamiltonian}. The result is
\begin{equation}
    \begin{aligned}
        H^{\mathbb Z_4^{(1)}}={}&
        -\sum_e\left[X_{\delta\be}\prod_{f'}Z_{f'}^{\int(\bface'\cup\be+\be\cup\bface')}+\mathrm{h.c.}\right]
        -\sum_c\left(Z_{\partial c}+Z_{\partial c}^{\dagger}\right) .
    \end{aligned}
    \label{eq:z4-initial-hamiltonian}
\end{equation}
To facilitate later constructions, let us massage the above Hamiltonian using higher cup products. Focusing on the gauge transformation term, we employ the higher-cup recursion formula of $\int(\bface'\cup\be+\be\cup\bface') =\int(\delta\be\cup_1\bface'-\be\cup_1\delta\bface' +2\be\cup\bface')$ to obtain
\begin{equation}
    X_{\delta\be}
    \prod_{f'}Z_{f'}^{\int(
    \delta\be\cup_1\bface'+2\be\cup\bface')}
    =G_e\prod_{f'}Z_{f'}^{2\int\be\cup\bface'},
    \label{eq:z4-edge-term-rewrite}
\end{equation}
where
\begin{equation}
    G_e:=X_{\delta\be}
    \prod_{f'}Z_{f'}^{\int\delta\be\cup_1\bface'}
    \label{eq:z4-Ge}
\end{equation}
and the $-\be\cup_1\delta\bface'$ contribution has been removed as it is generated by the 3-cell flux terms $Z_{\partial c}$. Diagrammatically, the $G_e$ terms on all three edge orientations are 
\begin{equation}
    \begin{tikzpicture}[
  x={(1.23cm,0cm)},
  y={(.78cm,.47cm)},
  z={(0cm,1.26cm)},
  line cap=round,
  line join=round
]
  \begin{scope}[xshift=0cm]
    \draw[densely dashed,line width=.50pt]
      (-.6,.6,.6)--(.6,.6,.6);

    \draw[line width=.58pt]
      (0,0,0)--(0,0,1.2)
      node[pos=0.5,fill=white,inner sep=.35pt,text=blue,font=\small]{$X^\dagger Z$};
    \draw[line width=.58pt]
      (0,0,1.2)--(0,1.2,1.2)
      node[pos=0.5,fill=white,inner sep=.35pt,text=blue,font=\small] {$X$};
    \draw[line width=.58pt]
      (0,1.2,0)--(0,1.2,1.2)
      node[pos=0.5,fill=white,inner sep=.35pt,text=blue,font=\small] {$X$};
    \draw[line width=.58pt]
      (0,0,0)--(0,1.2,0)
      node[pos=0.6,fill=white,inner sep=.35pt,text=blue,font=\small]
      {$X^\dagger Z^\dagger$};

    \draw[line width=.58pt]
      (0,0,-1.2)--(0,0,0)
      node[pos=0.5,fill=white,inner sep=.35pt,text=blue,font=\small]
      {$Z^\dagger$};
    \draw[line width=.58pt]
      (0,-1.2,0)--(0,0,0)
      node[pos=0.5,fill=white,inner sep=.35pt,text=blue,font=\small] {$Z$};
    \draw[line width=.58pt]
      (0,0,0)--(1,0,0)
      node[pos=0.5,fill=white,inner sep=.01pt,text=blue,font=\small] {$Z$};
    \draw[line width=.58pt]
      (0,1.2,1.2)--(1,1.2,1.2)
      node[pos=0.5,fill=white,inner sep=.35pt,text=blue,font=\small] {$Z^\dagger$};

    \node[font=\small,fill=white,inner sep=.6pt] at (0,.6,.6) {$e$};
    \node[font=\normalsize] at (.50,0,-1.52) {$G_{e_x}$};
  \end{scope}

  \begin{scope}[xshift=5.15cm]
    \draw[densely dashed,line width=.50pt]
      (.6,-0.62,.6)--(.6,.62,.6);

    \draw[line width=.58pt]
      (0,0,0)--(0,0,1.2)
      node[pos=0.5,fill=white,inner sep=.35pt,text=blue,font=\small] {$X$};
    \draw[line width=.58pt]
      (0,0,1.2)--(1.2,0,1.2)
      node[pos=0.5,fill=white,inner sep=.01pt,text=blue,font=\small] {$X$};
    \draw[line width=.58pt]
      (1.2,0,0)--(1.2,0,1.2)
      node[pos=0.5,fill=white,inner sep=.35pt,text=blue,font=\small] {$X^\dagger Z$};
    \draw[line width=.58pt]
      (0,0,0)--(1.2,0,0)
      node[pos=0.45,fill=white,inner sep=.35pt,text=blue,font=\small] {$X^\dagger Z^\dagger$};

    \draw[line width=.58pt]
      (0,-1.2,0)--(0,0,0)
      node[pos=0.5,fill=white,inner sep=.35pt,text=blue,font=\small] {$Z^\dagger$};
    \draw[line width=.58pt]
      (1.2,-1.2,0)--(1.2,0,0)
      node[pos=0.5,fill=white,inner sep=.35pt,text=blue,font=\small] {$Z$};
    \draw[line width=.58pt]
      (1.2,0,0)--(1.2,1.2,0)
      node[pos=0.5,fill=white,inner sep=.35pt,text=blue,font=\small] {$Z^\dagger$};
    \draw[line width=.58pt]
      (1.2,0,1.2)--(1.2,1.2,1.2)
      node[pos=0.5,fill=white,inner sep=.01pt,text=blue,font=\small] {$Z$};
    \draw[line width=.58pt]
      (1.2,0,-1.2)--(1.2,0,0)
      node[pos=0.5,fill=white,inner sep=.35pt,text=blue,font=\small] {$Z^\dagger$};
    \draw[line width=.58pt]
      (1.2,0,0)--(2.4,0,0)
      node[pos=0.5,fill=white,inner sep=.35pt,text=blue,font=\small] {$Z$};

    \node[font=\small,fill=white,inner sep=.6pt] at (.6,0,.6) {$e$};
    \node[font=\normalsize] at (.70,0,-1.52) {$G_{e_y}$};
  \end{scope}

  \begin{scope}[xshift=10.30cm]
    \draw[densely dashed,line width=.50pt]
      (.6,.6,-.6)--(.6,.6,.6);

    \draw[line width=.58pt]
      (0,0,0)--(0,1.2,0)
      node[pos=0.5,fill=white,inner sep=.35pt,text=blue,font=\small] {$X$};
    \draw[line width=.58pt]
      (0,1.2,0)--(1.2,1.2,0)
      node[pos=0.5,yshift=1pt,fill=white,inner sep=.35pt,
      text=blue,font=\small] {$X^\dagger Z^\dagger$};
    \draw[line width=.58pt]
      (1.2,0,0)--(1.2,1.2,0)
      node[pos=0.5,yshift=-1pt,fill=white,inner sep=.35pt,text=blue,font=\small]
      {$X^\dagger Z$};
    \draw[line width=.58pt]
      (0,0,0)--(1.2,0,0)
      node[pos=0.5,fill=white,inner sep=.35pt,text=blue,font=\small] {$X$};

    \draw[line width=.58pt]
      (0,0,-1.2)--(0,0,0)
      node[pos=0.5,fill=white,inner sep=.35pt,text=blue,font=\small] {$Z$};
    \draw[line width=.58pt]
      (1.2,1.2,-1.2)--(1.2,1.2,0)
      node[pos=0.5,fill=white,inner sep=.35pt,text=blue,font=\small]
      {$Z^\dagger$};
    \draw[line width=.58pt]
      (1.2,1.2,0)--(1.2,2.4,0)
      node[pos=0.5,fill=white,inner sep=.35pt,text=blue,font=\small] {$Z^\dagger$};
    \draw[line width=.58pt]
      (1.2,1.2,0)--(2.4,1.2,0)
      node[pos=0.5,fill=white,inner sep=.35pt,text=blue,font=\small] {$Z$};

    \node[font=\small,fill=white,inner sep=.6pt] at (.6,.6,0) {$e$};
    \node[font=\normalsize] at (1.30,0,-1.52) {$G_{e_z}$};
  \end{scope}
\end{tikzpicture}
\label{eq:z4-Ge-diagram}
\end{equation}
where the solid lines denote the dual lattice and the dashed line represents the primal edge $e$.

The bulk of the pre-modular $\ZZ_4^{(1)}$ Walker-Wang model is nontrivial and exhibits the 3D $\ZZ_2$ topological order \cite{Walker2012TQFT,Keyserlingk2013Threedimensional,Shirley2022QCA}. To obtain the semion Walker-Wang model, we choose to condense the order-two boson, which is a point-like excitation, in the bulk of the $\ZZ_4^{(1)}$ Walker-Wang. A segment of the boson's string-operator should commute with every $G_e$ and violate only the 3-cell flux terms at its endpoints. The naive choice of $X_f^2$ does not commute with the $G_e$ terms as
\begin{equation}
    X_f^2G_e =(-1)^{\int\delta\be\cup_1\bface}G_eX_f^2.
    \label{eq:z4-bare-boson-edge-commutator}
\end{equation}
We can, however, decorate the naive $X_f^2$ by products of Pauli-$Z^2$'s. First, note that over $\ZZ_2$, we have the following cup-product relation at our disposal
\begin{equation}
    \sum_{f'}\delta\be(f')
    \int\left(\bface'\cup_1\bface
    +\bface\cup_2\delta\bface'\right) =\int\left(\delta\be\cup_1\bface
    +\bface\cup_2\delta^2\be\right)
    =\int\delta\be\cup_1\bface 
    \pmod 2.
    \label{eq:z4-condensate-dressing-identity}
\end{equation}
The left-hand side is precisely the exponent needed to cancel the unwanted phase factor in Eq.~\eqref{eq:z4-bare-boson-edge-commutator}. The short string-operator for the boson can then be defined as
\begin{equation}
    C_f:=X_f^2\prod_{f'}Z_{f'}^{2\int(
    \bface'\cup_1\bface+\bface\cup_2\delta\bface')}.
    \label{eq:z4-condensate}
\end{equation}
or, diagrammatically,
\begin{equation}
    \begin{tikzpicture}[
  x={(1.18cm,0cm)},
  y={(.72cm,.43cm)},
  z={(0cm,1.18cm)},
  scale=1.12,
  transform shape,
  line cap=round,
  line join=round
]
  \begin{scope}[xshift=0cm]
    \draw[line width=.58pt]
      (0,0,0)--(1,0,0)
      node[pos=.45,yshift=1pt,fill=white,inner sep=.01pt,
      text=blue,font=\small] {$X_f^2Z_f^2$};
    \draw[line width=.58pt]
      (0,0,0)--(0,0,1)
      node[pos=.5,fill=white,inner sep=.01pt,
      text=blue,font=\small] {$Z^2$};
    \draw[line width=.58pt]
      (1,0,-1)--(1,0,0)
      node[pos=.5,fill=white,inner sep=.01pt,
      text=blue,font=\small] {$Z^2$};
    \draw[line width=.58pt]
      (1,0,0)--(1,0,1)
      node[pos=.5,fill=white,inner sep=.01pt,
      text=blue,font=\small] {$Z^2$};
    \draw[line width=.58pt]
      (0,-1,0)--(0,0,0)
      node[pos=.5,fill=white,inner sep=.01pt,
      text=blue,font=\small] {$Z^2$};
    \draw[line width=.58pt]
      (1,-1.2,0)--(1,0,0)
      node[pos=.35,fill=white,inner sep=.01pt,
      text=blue,font=\small] {$Z^2$};
    \draw[line width=.58pt]
      (1,0,0)--(1,1,0)
      node[pos=.5,fill=white,inner sep=.01pt,
      text=blue,font=\small] {$Z^2$};
    \draw[line width=.58pt]
      (1,0,0)--(2,0,0)
      node[pos=.6,fill=white,inner sep=.01pt,
      text=blue,font=\small] {$Z^2$};
    \node[font=\normalsize] at (1.00,0,-1.48) {$C_{f_x}$};
  \end{scope}

  \begin{scope}[xshift=5.55cm]
    \draw[line width=.58pt]
      (0,0,0)--(0,1.5,0)
      node[pos=.6,fill=white,inner sep=.01pt,
      text=blue,font=\small] {$X_f^2Z_f^2$};
    \draw[line width=.58pt]
      (0,0,-1)--(0,0,0)
      node[pos=.5,fill=white,inner sep=.01pt,
      text=blue,font=\small] {$Z^2$};
    \draw[line width=.58pt]
      (0,0,0)--(0,0,1)
      node[pos=.5,fill=white,inner sep=.01pt,
      text=blue,font=\small] {$Z^2$};
    \draw[line width=.58pt]
      (0,1.5,0)--(0,1.5,1)
      node[pos=.5,fill=white,inner sep=.01pt,
      text=blue,font=\small] {$Z^2$};
    \draw[line width=.58pt]
      (0,-1.2,0)--(0,0,0)
      node[pos=.5,fill=white,inner sep=.01pt,
      text=blue,font=\small] {$Z^2$};
    \draw[line width=.58pt]
      (0,0,0)--(1,0,0)
      node[pos=.5,fill=white,inner sep=.01pt,
      text=blue,font=\small] {$Z^2$};
    \node[font=\normalsize] at (.48,0,-1.48) {$C_{f_y}$};
  \end{scope}

  \begin{scope}[xshift=10.30cm]
    \begin{scope}[shift={(0,0,-1)}]
      \draw[line width=.58pt]
        (0,0,0)--(0,0,1)
        node[pos=.5,fill=white,inner sep=.01pt,
        text=blue,font=\small] {$X_f^2Z_f^2$};
      \draw[line width=.58pt]
        (0,0,1)--(0,0,2)
        node[pos=.5,,fill=white,inner sep=.35pt,
        text=blue,font=\small] {$Z^2$};
      \draw[line width=.58pt]
        (0,-1.2,1)--(0,0,1)
        node[pos=.5,fill=white,inner sep=.35pt,
        text=blue,font=\small] {$Z^2$};
      \draw[line width=.58pt]
        (0,0,1)--(1,0,1)
        node[pos=.5,fill=white,inner sep=.35pt,
        text=blue,font=\small] {$Z^2$};
    \end{scope}
    \node[font=\normalsize] at (.50,0,-1.48) {$C_{f_z}$};
  \end{scope}
\end{tikzpicture}
\label{eq:z4-condensate-diagram}
\end{equation}
which satisfies $[C_f,C_{f'}]=0$ for any faces $f,f'$. Moreover, $C_f$ is order-two as
\begin{equation}
    C_f^2=X_f^4\prod_{f'}Z_{f'}^{4\int(\bface'\cup_1\bface+\bface\cup_2\delta\bface')}=1.
\label{eq:z4-condensate-algebra}
\end{equation}
It is straightforward to check that $C_f$ excites exactly a pair of adjacent 3-cell stabilizers
\begin{equation}
    Z_{\partial c}C_f
    =(-1)^{\delta\bface(c)}C_fZ_{\partial c}.
    \label{eq:z4-condensate-flux-commutator}
\end{equation}
Here, $\delta\bface(c)$ is the incidence of $f$ in $\partial c$ and is nonzero only for exactly two 3-cells adjacent to $f$. $C_f$ therefore creates exactly two flux excitations next to the face $f$. And, along a path on the dual lattice, the string-operator creates excitations only at its two ends. Thus, $C_f$ is the desired short string-operator for the point-like boson excitation.

To condense this excitation, we impose $C_f=1$ on every face. The stabilizers that survive this condensation are those that commute with all the $C_f$'s. The kinetic terms remain in the condensed Hamiltonian since they commute with $C_f$. $Z_{\partial c}$, on the other hand, does not commute with $C_f$ and is thus not in the resultant Hamiltonian, but $Z^2_{\partial c}$ commutes with $C_f$ and survives the condensation. The resulting semion Walker-Wang Hamiltonian written in terms of $\ZZ_4$ qudits is then
\begin{equation}
    \begin{aligned}
        H^{\{1,s\}}={}&-\sum_fC_f-\sum_e\left[G_e\prod_{f'}Z_{f'}^{2\int\be\cup\bface'}+\mathrm{h.c.}\right]-\sum_c Z_{\partial c}^2 \, .
    \end{aligned}
    \label{eq:z4-condensed-semion-hamiltonian}
\end{equation}
In the condensed space of $C_f =1$, we notice that the kinetic terms involving $G_e$ already generate the cubic terms. To make this explicit, let us write the kinetic terms by noting that, on the cubic lattice, an edge $e$ and a face $f$ are in one-to-one correspondence, so the kinetic terms may be relabeled by the faces
\begin{equation}
    (\text{kinetic term})=Z_f^2
    \prod_eG_e^{\int\be\cup\bface} = Z_f^2\prod_e\left[
    X_{\delta\be}\prod_{f'}Z_{f'}^{\int\delta\be\cup_1\bface'}
    \right]^{\int\be\cup\bface} =:\widetilde Z_f.
    \label{eq:z4-face-separator}
\end{equation}
This $\widetilde{Z}_f$ is our separator candidate. For illustration, we draw the kinetic terms $\widetilde{Z}_f$ on the dual cubic lattice and explicitly label the corresponding $f$
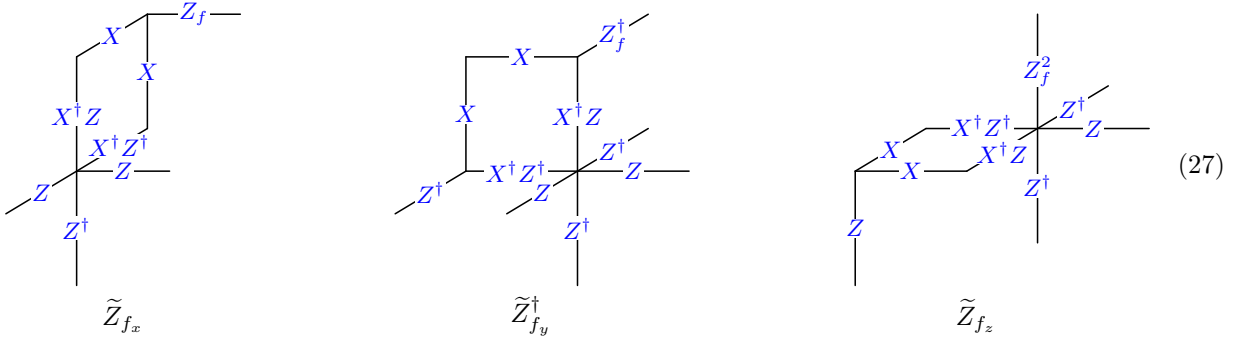
\begin{equation}
\begin{tikzpicture}[
  baseline=(current bounding box.center),
  x={(1.23cm,0cm)},
  y={(.78cm,.47cm)},
  z={(0cm,1.26cm)},
  line cap=round,
  line join=round,
  dual/.style={draw=black,line width=.58pt},
  op/.style={fill=white,inner sep=.35pt,text=blue,font=\small}
]

\begin{scope}[xshift=0cm]
  \draw[dual] (0,0,0)--(0,0,1.2)
    node[pos=.5,op] {$X^\dagger Z$};
  \draw[dual] (0,0,1.2)--(0,1.2,1.2)
    node[pos=.5,op] {$X$};
  \draw[dual] (0,1.2,0)--(0,1.2,1.2)
    node[pos=.5,op] {$X$};
  \draw[dual] (0,0,0)--(0,1.2,0)
    node[pos=.6,op] {$X^\dagger Z^\dagger$};

  \draw[dual] (0,0,-1.2)--(0,0,0)
    node[pos=.5,op] {$Z^\dagger$};
  \draw[dual] (0,-1.2,0)--(0,0,0)
    node[pos=.5,op] {$Z$};
  \draw[dual] (0,0,0)--(1,0,0)
    node[pos=.5,op] {$Z$};
  \draw[dual] (0,1.2,1.2)--(1,1.2,1.2)
    node[pos=.5,op] {$Z_f$};

  \node[font=\normalsize] at (.50,0,-1.52)
    {$\widetilde Z_{f_x}$};
\end{scope}

\begin{scope}[xshift=5.15cm]
  \draw[dual] (0,0,0)--(0,0,1.2)
    node[pos=.5,op] {$X$};
  \draw[dual] (0,0,1.2)--(1.2,0,1.2)
    node[pos=.5,op] {$X$};
  \draw[dual] (1.2,0,0)--(1.2,0,1.2)
    node[pos=.5,op] {$X^\dagger Z$};
  \draw[dual] (0,0,0)--(1.2,0,0)
    node[pos=.45,op] {$X^\dagger Z^\dagger$};

  \draw[dual] (0,-1.2,0)--(0,0,0)
    node[pos=.5,op] {$Z^\dagger$};
  \draw[dual] (1.2,-1.2,0)--(1.2,0,0)
    node[pos=.5,op] {$Z$};
  \draw[dual] (1.2,0,0)--(1.2,1.2,0)
    node[pos=.5,op] {$Z^\dagger$};
  \draw[dual] (1.2,0,1.2)--(1.2,1.2,1.2)
    node[pos=.5,op] {$Z_f^\dagger$};
  \draw[dual] (1.2,0,-1.2)--(1.2,0,0)
    node[pos=.5,op] {$Z^\dagger$};
  \draw[dual] (1.2,0,0)--(2.4,0,0)
    node[pos=.5,op] {$Z$};

  \node[font=\normalsize] at (.70,0,-1.52)
    {$\widetilde Z_{f_y}^{\dagger}$};
\end{scope}

\begin{scope}[xshift=10.30cm]
  \draw[dual] (0,0,0)--(0,1.2,0)
    node[pos=.5,op] {$X$};
  \draw[dual] (0,1.2,0)--(1.2,1.2,0)
    node[pos=.5,yshift=1pt,op] {$X^\dagger Z^\dagger$};
  \draw[dual] (1.2,0,0)--(1.2,1.2,0)
    node[pos=.5,yshift=-1pt,op] {$X^\dagger Z$};
  \draw[dual] (0,0,0)--(1.2,0,0)
    node[pos=.5,op] {$X$};

  \draw[dual] (0,0,-1.2)--(0,0,0)
    node[pos=.5,op] {$Z$};
  \draw[dual] (1.2,1.2,-1.2)--(1.2,1.2,0)
    node[pos=.5,op] {$Z^\dagger$};
  \draw[dual] (1.2,1.2,0)--(1.2,1.2,1.2)
    node[pos=.5,op] {$Z_f^2$};
  \draw[dual] (1.2,1.2,0)--(1.2,2.4,0)
    node[pos=.5,op] {$Z^\dagger$};
  \draw[dual] (1.2,1.2,0)--(2.4,1.2,0)
    node[pos=.5,op] {$Z$};

  \node[font=\normalsize] at (1.30,0,-1.52)
    {$\widetilde Z_{f_z}$};
\end{scope}

\end{tikzpicture}
\label{eq:z4-face-separator-diagram}
\end{equation}
where the above only differs from the $G_e$ terms in Eq.~\eqref{eq:z4-Ge-diagram} by a relabeling of $e \to f$ and an additional $Z^2_f$ operator. Let us now show that the cubic term in Eq.~\eqref{eq:z4-condensed-semion-hamiltonian} is already generated by $\widetilde{Z}_f$'s. This follows from the fact that taking the product of $\widetilde{Z}_f$ around a cube $c$
\begin{equation}
    \prod_f \left(\widetilde Z_f\right)^{\delta\bface(c)} =Z_{\partial c}^2 \, ,
    \label{eq:z4-face-product-flux}
\end{equation}
where the contribution from the product of $G_e$'s evaluates to 1 due to $\delta^2=0$. Thus, $Z_{\partial c}^2$ in the Hamiltonian can be suppressed, and in the condensed space the only independent stabilizer generators of the Hamiltonian are given by the kinetic $\widetilde{Z}_f$ terms.

For $\widetilde{Z}_f$ to define a candidate separator for the semion QCA, it must also be an order-two operator in the condensed space of $C_f = 1$.
This can be verified by noting that
\begin{equation}
    G_e^2
    =X_{\delta\be}^2
    \prod_{f'}Z_{f'}^{2\int\delta\be\cup_1\bface'}
    =\prod_{f'}C_{f'}^{\delta\be(f')},
    \label{eq:z4-Ge-square}
\end{equation}
where we have used $\int\delta\be\cup_1\delta\be = 0$ the second equality follows from the higher-cup recursion, over $\ZZ_2$ coefficients, $\int\delta\be\cup_1\bface' = \int\left(\bface'\cup_1\delta\be+\delta\be\cup_2\delta\bface'\right)$.

Having found the candidate separator, we now move on to construct a candidate flipper. We look for a local operator that flips exactly one separator while commuting with every $C_f$. Again following the strategy in Fig.~\ref{fig:qca-construction-workflow}, let us consider the operator
\begin{equation}
    V_f:=X_f\prod_{f'}Z_{f'}^{\int(
    \bface'\cup_1\bface+\delta\bface'\cup_2\bface)},
    \label{eq:z4-bare-flipper}
\end{equation}
pictorially,
\begin{equation}
\begin{tikzpicture}[
  x={(1.18cm,0cm)},
  y={(.72cm,.43cm)},
  z={(0cm,1.18cm)},
  scale=1.12,
  transform shape,
  line cap=round,
  line join=round
]
  \begin{scope}[xshift=0cm]
    \draw[line width=.58pt]
      (0,0,0)--(1.2,0,0)
      node[pos=.50,yshift=-1pt,fill=white,inner sep=.01pt,text=blue,font=\small] {$X_fZ_f$};

    \draw[line width=.58pt]
      (-1,0,0)--(0,0,0)
      node[pos=.5,fill=white,inner sep=.01pt,
      text=blue,font=\small] {$Z^\dagger$};
    \draw[line width=.58pt]
      (0,0,0)--(0,1.5,0)
      node[pos=.6,fill=white,inner sep=.01pt,
      text=blue,font=\small] {$Z^\dagger$};
    \draw[line width=.58pt]
      (0,0,-1)--(0,0,0)
      node[pos=.5,fill=white,inner sep=.35pt,
      text=blue,font=\small] {$Z^\dagger$};

    \node[font=\normalsize] at (.30,0,-1.55) {$V_{f_x}$};
  \end{scope}

  \begin{scope}[xshift=5.55cm]
    \draw[line width=.58pt]
      (0,-1.5,0)--(0,0,0)
      node[pos=.5,fill=white,inner sep=.35pt,text=blue,font=\small] {$X_fZ_f$};

    \draw[line width=.58pt]
      (-0.9,-1.5,0)--(0,-1.5,0)
      node[pos=.5,fill=white,inner sep=.01pt,
      text=blue,font=\small] {$Z$};
    \draw[line width=.58pt]
      (-1,0,0)--(0,0,0)
      node[pos=.5,yshift=2pt,fill=white,inner sep=.01pt,
      text=blue,font=\small] {$Z^\dagger$};
    \draw[line width=.58pt]
      (0,0,0)--(1,0,0)
      node[pos=.5,fill=white,inner sep=.01pt,
      text=blue,font=\small] {$Z$};
    \draw[line width=.58pt]
      (0,0,0)--(0,1.3,0)
      node[pos=.5,fill=white,inner sep=.01pt,
      text=blue,font=\small] {$Z^\dagger$};
    \draw[line width=.58pt]
      (0,0,-1)--(0,0,0)
      node[pos=.5,fill=white,inner sep=.01pt,
      text=blue,font=\small] {$Z^\dagger$};

    \node[font=\normalsize] at (-.28,0,-1.55) {$V_{f_y}$};
  \end{scope}

  \begin{scope}[xshift=10.55cm]
    \draw[line width=.58pt]
      (0,0,0)--(0,0,1)
      node[pos=.5,fill=white,inner sep=.01pt,
      text=blue,font=\small] {$X_fZ_f$};

    \draw[line width=.58pt]
      (0,0,-1)--(0,0,0)
      node[pos=.5,fill=white,inner sep=.01pt,
      text=blue,font=\small] {$Z^\dagger$};
    \draw[line width=.58pt]
      (0,-1.2,0)--(0,0,0)
      node[pos=.4,fill=white,inner sep=.01pt,
      text=blue,font=\small] {$Z$};
    \draw[line width=.58pt]
      (0,0,0)--(0,1.2,0)
      node[pos=.6,fill=white,inner sep=.01pt,
      text=blue,font=\small] {$Z^\dagger$};
    \draw[line width=.58pt]
      (-1,0,0)--(0,0,0)
      node[pos=.5,fill=white,inner sep=.01pt,
      text=blue,font=\small] {$Z^\dagger$};
    \draw[line width=.58pt]
      (0,0,0)--(1,0,0)
      node[pos=.6,fill=white,inner sep=.01pt,
      text=blue,font=\small] {$Z$};

    \draw[line width=.58pt]
      (0,0,1)--(0,1,1)
      node[pos=.5,fill=white,inner sep=.01pt,
      text=blue,font=\small] {$Z$};
    \draw[line width=.58pt]
      (-1,0,1)--(0,0,1)
      node[pos=.5,fill=white,inner sep=.01pt,
      text=blue,font=\small] {$Z$};

    \node[font=\normalsize] at (.12,0,-1.55) {$V_{f_z}$};
  \end{scope}
\end{tikzpicture}
\label{eq:z4-Vf-dual-lattice}
\end{equation}
$V_{f}$ commutes with $C_f$. Indeed, commuting $V_{f_2}$ through $C_{f_1}$ produces the phase
\begin{equation}
    \begin{aligned}
        &(-1)^{\int(\bface_2\cup_1\bface_1
        +\bface_1\cup_2\delta\bface_2
        +\bface_1\cup_1\bface_2
        +\delta\bface_1\cup_2\bface_2)}
        =(-1)^{\int\delta(\bface_1\cup_2\bface_2)}=1.
    \end{aligned}
    \label{eq:z4-C-V-commutator}
\end{equation}
Moreover, $V_f$ commutes with every $G_e$
\begin{equation*}
\begin{aligned}
V_fG_e
={}&
i^{
\sum_{f'}\delta\be(f')
\int\left(
\bface'\cup_1\bface
+\delta\bface'\cup_2\bface
\right)
-\int\delta\be\cup_1\bface
}
G_eV_f
=
i^{
\int\delta\be\cup_1\bface
+\int\delta(\delta\be)\cup_2\bface
-\int\delta\be\cup_1\bface
}
G_eV_f
=G_eV_f,
\end{aligned}
\end{equation*}
where, in the second equality, we have used $\sum_{f'}\delta\be(f')\,\bface'=\delta\be$ and $\sum_{f'}\delta\be(f')\,\delta\bface'=\delta(\delta\be)=0$. The only nontrivial contribution to the separator-flipper commutation comes from $Z_{f'}^2$, for which $Z_{f'}^2X_f=(-1)^{\delta_{f,f'}}X_fZ_{f'}^2$. Therefore, $V_f$ has the desired property of
\begin{equation}
    \widetilde Z_{f'}V_f = (-1)^{\delta_{f,f'}} V_f\widetilde Z_{f'}.
\end{equation}
However, it does not yet satisfy other properties of a flipper defined in this paper. First,
\begin{equation}
    V_{f_2}V_{f_1}
    =i^{\int(
    \bface_1\cup_1\bface_2+\delta\bface_1\cup_2\bface_2
    -\bface_2\cup_1\bface_1-\delta\bface_2\cup_2\bface_1)}
    V_{f_1}V_{f_2},
    \label{eq:z4-bare-flipper-commutator}
\end{equation}
so $V_f$ do not all commute with each other. Second, $V_f$ does not square to one in the condensed space of $C_f=1$ because
\begin{equation}
    V_f^2=iC_f\prod_{f'}
    \left(\widetilde Z_{f'}\right)^{
    \int\delta\bface'\cup_3\delta\bface}.
    \label{eq:z4-bare-flipper-square}
\end{equation}
To resolve this, we need to add ``square roots" of $\widetilde Z_{f'}$. Since $\widetilde Z_{f'}$ square to one, we can define a squareroot as $\tilde S_{f'} := i^{(1-\widetilde Z_{f'})/2}$. Moreover, we will need to define its corresponding $CZ$, which we define as follows. For two operators $a$ and $b$ which commute and square to one, define
\begin{equation}
    CZ(a,b):=(-1)^{\frac{1-a}{2}\frac{1-b}{2}}.
    \label{eq:z4-spectral-cz}
\end{equation}
We claim that the flipper is given by
\begin{equation}
\begin{aligned}
V'_f
:={}&i\,V_f\prod_{f'}\left[
(\widetilde S_{f'}^\dagger)^{
\int\left(\bface'\cup_1\bface+\delta\bface'\cup_2\bface\right)}
CZ\!\left(\widetilde Z_f,\widetilde Z_{f'}\right)^{
\int\left(\bface\cup_1\bface'+\delta\bface\cup_2\bface'\right)}
\right]
\\
={}&i\,X_f
\left[
\prod_{f'}Z_{f'}^{
\int\left(\bface'\cup_1\bface+\delta\bface'\cup_2\bface\right)}
\right]
\left[
\prod_{f'}
(\widetilde S_{f'}^\dagger)^{
\int\left(\bface'\cup_1\bface+\delta\bface'\cup_2\bface\right)}
CZ\!\left(\widetilde Z_f,\widetilde Z_{f'}\right)^{
\int\left(\bface\cup_1\bface'+\delta\bface\cup_2\bface'\right)}
\right],
\end{aligned}
\label{eq:z4-corrected-flipper}
\end{equation}
where the $Z_{f'}$ in the second line is the microscopic $\mathbb Z_4$ clock operator, whereas $\widetilde Z_{f'}$ is the separator in Eq.~\eqref{eq:z4-face-separator}. The two products in the second line are therefore kept in the indicated order. The prefactor $i$ ensures that the dressed operator squares exactly to $C_f$.

We now verify the square and commutation properties of $V'_f$. We use
\begin{equation}
    (\widetilde S_f^\dagger)^2=\widetilde Z_f,
    \qquad
    CZ(\widetilde Z_f,\widetilde Z_f)=\widetilde Z_f,
    \qquad
    \widetilde S_f^\dagger\widetilde Z_f=\widetilde S_f,
    \qquad
    V_f\widetilde S_f=i\widetilde S_f^\dagger V_f.
\end{equation}
In our oriented cubic lattice, the local self-incidences are
\begin{equation}
    \int\bface\cup_1\bface=0,
    \qquad
    \int\delta\bface\cup_2\bface=1.
\end{equation}
Thus, the operator in Eq.~\eqref{eq:z4-corrected-flipper} involving the self-incidences is
\begin{equation*}
    \widetilde S_f^\dagger
    CZ(\widetilde Z_f,\widetilde Z_f)
    =\widetilde S_f^\dagger\widetilde Z_f
    =\widetilde S_f,
\end{equation*}
and hence $V_f\widetilde S_fV_f\widetilde S_f=iV_f^2$. For $f'\neq f$, the two square-root factors contribute $\widetilde Z_{f'}^{\int(\bface'\cup_1\bface+ \delta\bface'\cup_2\bface)}$, while the two controlled-phase factors contribute $\widetilde Z_{f'}^{\int(\bface\cup_1\bface'+ \delta\bface\cup_2\bface')}$. So,
\begin{eqs}
{V'_f}^{\,2}
={}&
-\,V_f
\prod_{f'}\left[
(\widetilde S_{f'}^\dagger)^{
\int\left(
\bface'\cup_1\bface
+\delta\bface'\cup_2\bface
\right)}
CZ(\widetilde Z_f,\widetilde Z_{f'})^{
\int\left(
\bface\cup_1\bface'
+\delta\bface\cup_2\bface'
\right)}
\right]
\\
&\times
V_f
\prod_{f'}\left[
(\widetilde S_{f'}^\dagger)^{
\int\left(
\bface'\cup_1\bface
+\delta\bface'\cup_2\bface
\right)}
CZ(\widetilde Z_f,\widetilde Z_{f'})^{
\int\left(
\bface\cup_1\bface'
+\delta\bface\cup_2\bface'
\right)}
\right]
\\
={}&
-iV_f^2
\prod_{f'\neq f}
\widetilde Z_{f'}^{
\int\left(
\bface'\cup_1\bface
+\delta\bface'\cup_2\bface
+\bface\cup_1\bface'
+\delta\bface\cup_2\bface'
\right)}
\\
={}&
C_f
\prod_{f'}
\widetilde Z_{f'}^{
\int\left(
\delta\bface'\cup_3\delta\bface
+\bface'\cup_1\bface
+\delta\bface'\cup_2\bface
+\bface\cup_1\bface'
+\delta\bface\cup_2\bface'
\right)}
=C_f,
\end{eqs}
which yields $1$ on the condensed space. We briefly explain the computation above. The leading minus sign comes from squaring the prefactor $i$, while the additional factor $i$ in the second equality is the self-pair contribution discussed above. In the third equality, we have used Eq.~\eqref{eq:z4-bare-flipper-square} and extended the product from $f'\neq f$ to all $f'$ as for $f'=f$, the additional exponent is $1+1=2$, which simply involves a multiplication by the operator $\widetilde Z_f^2=1$ to the product. Finally, the higher-cup recursion gives the $\mathbb Z_2$ cochain identity
\begin{equation}
    \begin{aligned}
        &\int\left(
        \delta\bface'\cup_3\delta\bface
        +\bface'\cup_1\bface
        +\delta\bface'\cup_2\bface
        +\bface\cup_1\bface'
        +\delta\bface\cup_2\bface'
        \right)
        =
        \int\delta\left(
        \bface'\cup_2\bface
        +\bface'\cup_3\delta\bface
        \right)
        =0
        \pmod 2,
    \end{aligned}
\end{equation}
where the last equality uses the fact that we are on a closed manifold.

We now check the dressed flippers $V'_f$ commute. For two distinct faces $f_1$ and $f_2$, only the $f'_1=f_2$ factor in the first dressing and the $f'_2=f_1$ factor in the second dressing fail to commute with the opposite bare flipper. We find
\begin{eqs}
V'_{f_1}V'_{f_2}
={}&
-\,V_{f_1}
\prod_{f'_1}
\left[
(\widetilde S_{f'_1}^\dagger)^{
\int\left(
\bface'_1\cup_1\bface_1
+\delta\bface'_1\cup_2\bface_1
\right)}
CZ(\widetilde Z_{f_1},\widetilde Z_{f'_1})^{
\int\left(
\bface_1\cup_1\bface'_1
+\delta\bface_1\cup_2\bface'_1
\right)}
\right]
\\
&\times
V_{f_2}
\prod_{f'_2}
\left[
(\widetilde S_{f'_2}^\dagger)^{
\int\left(
\bface'_2\cup_1\bface_2
+\delta\bface'_2\cup_2\bface_2
\right)}
CZ(\widetilde Z_{f_2},\widetilde Z_{f'_2})^{
\int\left(
\bface_2\cup_1\bface'_2
+\delta\bface_2\cup_2\bface'_2
\right)}
\right]
\\
={}&
-\,i^{-\int\left(
\bface_1\cup_1\bface_2
+\delta\bface_1\cup_2\bface_2
-\bface_2\cup_1\bface_1
-\delta\bface_2\cup_2\bface_1
\right)}
V_{f_2}V_{f_1}
\\
&\times
i^{-\int\left(
\bface_2\cup_1\bface_1
+\delta\bface_2\cup_2\bface_1
\right)}
\widetilde Z_{f_2}^{
\int\left(
\bface_2\cup_1\bface_1
+\delta\bface_2\cup_2\bface_1
\right)}
\widetilde Z_{f_1}^{
\int\left(
\bface_1\cup_1\bface_2
+\delta\bface_1\cup_2\bface_2
\right)}
\\
&\times
\prod_{f'_1}
\left[
(\widetilde S_{f'_1}^\dagger)^{
\int\left(
\bface'_1\cup_1\bface_1
+\delta\bface'_1\cup_2\bface_1
\right)}
CZ(\widetilde Z_{f_1},\widetilde Z_{f'_1})^{
\int\left(
\bface_1\cup_1\bface'_1
+\delta\bface_1\cup_2\bface'_1
\right)}
\right]
\\
&\times
\prod_{f'_2}
\left[
(\widetilde S_{f'_2}^\dagger)^{
\int\left(
\bface'_2\cup_1\bface_2
+\delta\bface'_2\cup_2\bface_2
\right)}
CZ(\widetilde Z_{f_2},\widetilde Z_{f'_2})^{
\int\left(
\bface_2\cup_1\bface'_2
+\delta\bface_2\cup_2\bface'_2
\right)}
\right]
\\
={}&
-\,i^{-\int\left(
\bface_1\cup_1\bface_2
+\delta\bface_1\cup_2\bface_2
-\bface_2\cup_1\bface_1
-\delta\bface_2\cup_2\bface_1
\right)}
V_{f_2}
\\
&\times
i^{-\int\left(
\bface_2\cup_1\bface_1
+\delta\bface_2\cup_2\bface_1
\right)}
\widetilde Z_{f_2}^{
\int\left(
\bface_2\cup_1\bface_1
+\delta\bface_2\cup_2\bface_1
\right)}
(-\widetilde Z_{f_1})^{
\int\left(
\bface_1\cup_1\bface_2
+\delta\bface_1\cup_2\bface_2
\right)}
\\
&\times
i^{-\int\left(
\bface_1\cup_1\bface_2
+\delta\bface_1\cup_2\bface_2
\right)}
\widetilde Z_{f_1}^{
\int\left(
\bface_1\cup_1\bface_2
+\delta\bface_1\cup_2\bface_2
\right)}
\widetilde Z_{f_2}^{
\int\left(
\bface_2\cup_1\bface_1
+\delta\bface_2\cup_2\bface_1
\right)}
\\
&\times
\prod_{f'_2}
\left[
(\widetilde S_{f'_2}^\dagger)^{
\int\left(
\bface'_2\cup_1\bface_2
+\delta\bface'_2\cup_2\bface_2
\right)}
CZ(\widetilde Z_{f_2},\widetilde Z_{f'_2})^{
\int\left(
\bface_2\cup_1\bface'_2
+\delta\bface_2\cup_2\bface'_2
\right)}
\right]
\\
&\times
V_{f_1}
\prod_{f'_1}
\left[
(\widetilde S_{f'_1}^\dagger)^{
\int\left(
\bface'_1\cup_1\bface_1
+\delta\bface'_1\cup_2\bface_1
\right)}
CZ(\widetilde Z_{f_1},\widetilde Z_{f'_1})^{
\int\left(
\bface_1\cup_1\bface'_1
+\delta\bface_1\cup_2\bface'_1
\right)}
\right]
\\
={}&
-\,V_{f_2}
\prod_{f'_2}
\left[
(\widetilde S_{f'_2}^\dagger)^{
\int\left(
\bface'_2\cup_1\bface_2
+\delta\bface'_2\cup_2\bface_2
\right)}
CZ(\widetilde Z_{f_2},\widetilde Z_{f'_2})^{
\int\left(
\bface_2\cup_1\bface'_2
+\delta\bface_2\cup_2\bface'_2
\right)}
\right]
\\
&\times
V_{f_1}
\prod_{f'_1}
\left[
(\widetilde S_{f'_1}^\dagger)^{
\int\left(
\bface'_1\cup_1\bface_1
+\delta\bface'_1\cup_2\bface_1
\right)}
CZ(\widetilde Z_{f_1},\widetilde Z_{f'_1})^{
\int\left(
\bface_1\cup_1\bface'_1
+\delta\bface_1\cup_2\bface'_1
\right)}
\right]
\\
={}&V'_{f_2}V'_{f_1}.
\end{eqs}
In the second to last equality, the separator powers reduce to
\begin{equation*}
    (-1)^{\int\left(
    \bface_1\cup_1\bface_2+
    \delta\bface_1\cup_2\bface_2
    \right)},
\end{equation*}
whereas the remaining powers of $i$ reduce to
\begin{equation*}
    i^{-2\int\left(
    \bface_1\cup_1\bface_2+
    \delta\bface_1\cup_2\bface_2
    \right)},
\end{equation*}
so they cancel. Therefore,
\begin{equation}
    (V'_f)^2=1,
    \qquad
    [V'_{f_1},V'_{f_2}]=0
    \qquad
    \text{in the sector $C_f=1$ for every face $f$.}
\end{equation}
Thus, in the subspace $C_f=1$, the operators $\widetilde Z_f$ and $V'_f$ form an order-two operator algebra satisfying the separator and flipper algebra for the qubit. That is,
\begin{equation}
    \widetilde{Z}_k V'_j =(-1)^{\delta_{jk}} V'_j \widetilde{Z}_k, \qquad [V'_j,V'_k]=[\widetilde{Z}_j,\widetilde{Z}_k]=0, \qquad {V'_j}^2={\widetilde{Z}_j}^2=1 \Big|_{C_f = 1},
    \label{eq:semion-z4-candidate-separator-flipper-algebra}
\end{equation}
All that remains to do to obtain a qubit QCA for the semion Walker-Wang at this point is to translate our local degree of freedom from a $\ZZ_4$ qudit to two qubits and find the FDQC that maps the $C_f$ operator to a single qubit operator and disentangles it from our system.

Consider the following $\ZZ_4$ qudit to qubits mapping
\begin{equation}
    X_f = X^A_f CZ^{AB}_f,\qquad
    Z_f = X^B_f S^A_f,\qquad
    X^2_f = Z^B_f,\qquad
    Z^2_f = Z^A_f,
    \label{eq:z4-two-qubit-map}
\end{equation}
where the Pauli operators with $A,B$ superscripts are the qubit operators. Under this mapping, the condensate operator $C_f$ can be written as
\begin{equation}
    C_f =Z_f^B\prod_{f'}(Z_{f'}^A)^{\int(
    \bface'\cup_1\bface+\bface\cup_2\delta\bface')}.
    \label{eq:z4-encoded-condensate}
\end{equation}
The desired FDQC mapping to the qubit semion Walker-Wang is the Clifford circuit of
\begin{equation}
    U_C:=\prod_{f,f'}CX^{AB}(f',f)^{\int(\bface'\cup_1\bface+\bface\cup_2\delta\bface')}.
    \label{eq:z4-condensate-circuit}
\end{equation}
Here, $CX^{AB}(f',f)$ has control qubit $A$ on face $f'$ and target qubit $B$ on face $f$. Its action on the single-qubit Pauli operators that will come in handy in later computations is
\begin{equation}
    \begin{aligned}
        U_C Z_f^B U_C^\dagger
        &=Z_f^B\prod_{f'}(Z_{f'}^A)^{\int(
        \bface'\cup_1\bface+\bface\cup_2\delta\bface')},\\
        U_C X_f^A U_C^\dagger
        &=X_f^A\prod_{f'}(X_{f'}^B)^{\int(
        \bface\cup_1\bface'+\bface'\cup_2\delta\bface)},\\
        U_C Z_f^A U_C^\dagger &= Z_f^A, \\
        U_C X_f^B U_C^\dagger &= X_f^B .
    \end{aligned}
    \label{eq:z4-disentangler-actionqubit}
\end{equation}
It then follows that
\begin{equation}
    U_C C_f U_C^\dagger=Z_f^B.
    \label{eq:z4-disentangled-condensate}
\end{equation}
The $B$ qubits are now disentangled from the $A$ qubit system. Since $C_f = 1$, we have $Z_f^B = 1$ with the semion Walker-Wang defined on only the $A$ qubits.

At this point, we have found the semion QCA in terms of qubits since the conjugation by a unitary operator does not change the operator algebra formed by $\widetilde{Z}_f$ and $V'_{f'}$ in Eq.~\eqref{eq:semion-z4-candidate-separator-flipper-algebra}. For completeness, let us write out the qubit separators and flippers explicitly by acting the above transformations on $\widetilde{Z}_f$ and $V'_{f'}$. We start with $\widetilde{Z}_f$. Note that $X_{\delta\be}$ contains not just $X$ but also $X^\dagger$. The faces where $X^\dagger$ appears are precisely given by $\be\cup_1\delta\be$ for which we will have an extra factor of $X^2 = Z^B$. Therefore, the separator expressed as qubits is
\begin{equation}
\begin{aligned}
\widetilde Z_f = Z_f^A\prod_e\Bigg[&X_{\delta\be}^A
\prod_{f'\in\delta \be}
CZ(Z_{f'}^A,Z_{f'}^B) \times 
Z_{\be\cup_1\delta\be}^B \times 
\prod_{f'}(X_{f'}^BS_{f'}^A)^{\int\delta\be\cup_1\bface'}
\Bigg]^{\int\be\cup\bface}.
\end{aligned}
\label{eq:z4-encoded-separator}
\end{equation}
Before acting by the Clifford circuit $U_C$ in Eq.~\eqref{eq:z4-condensate-circuit}, it is helpful to first move $X^B_{f'}$ to the front of the expression in the bracket since $U_C \left(X_{\delta\be}^A \prod_{f'}(X_{f'}^B)^{\int\delta\be\cup_1\bface'}\right) U_C^\dagger = X_{\delta\be}^A$. When moving $X^B_{f'}$ through $Z^B_{f''}$, we get a minus sign whenever $f' = f''$. Therefore, we have
\begin{equation}
\begin{aligned}
&\widetilde Z_f \\
=& Z_f^A\prod_e\Bigg[\left(X_{\delta\be}^A \prod_{f'}(X_{f'}^B)^{\int\delta\be\cup_1\bface'}\right)
\prod_{f'\in\delta \be}
CZ(Z_{f'}^A,(-1)^{\int\delta\be\cup_1\bface'}Z_{f'}^B)
\times 
(-1)^{\int\delta\be\cup_1(\be\cup_1\delta\be)}Z_{\be\cup_1\delta\be}^B\times 
\prod_{f'}(S_{f'}^A)^{\int\delta\be\cup_1\bface'}
\Bigg]^{\int\be\cup\bface}\\
=& Z_f^A\prod_e\Bigg[\left(X_{\delta\be}^A \prod_{f'}(X_{f'}^B)^{\int\delta\be\cup_1\bface'}\right)
\prod_{f'\in\delta \be}
(Z_{f'}^A)^{\int\delta\be\cup_1\bface'}CZ(Z_{f'}^A,Z_{f'}^B)
\times Z_{\be\cup_1\delta\be}^B\times 
\prod_{f'}(S_{f'}^A)^{\int\delta\be\cup_1\bface'}
\Bigg]^{\int\be\cup\bface},
\end{aligned}
\label{eq:z4-encoded-separator-transformed}
\end{equation}
where we have used the fact that $CZ(Z^A_f, -1) = Z^A_f$ and $\int\delta\be\cup_1(\be\cup_1\delta\be)=0\pmod 2$ on the cubic lattice. Acting on $\widetilde Z_f$ by $U_C$, we obtain
\begin{equation}
\begin{aligned}
U_C\widetilde Z_fU_C^\dagger
=&Z_f^A\prod_e\Bigg[
X_{\delta\be}^A
\prod_{f'\in\delta\be}
\Bigg\{
\left(Z_{f'}^A\right)^{
\int\delta\be\cup_1\bface'}
CZ\!\left(
Z_{f'}^A,
Z_{f'}^B
\prod_{f''}\left(Z_{f''}^A\right)^{
\int\left(
\bface''\cup_1\bface'
+\bface'\cup_2\delta\bface''
\right)}
\right)
\Bigg\}
\\
&\hspace{10mm}\times
\Bigg\{
Z_{\be\cup_1\delta\be}^B
\prod_{f'}\left(Z_{f'}^A\right)^{
\int\left(
\bface'\cup_1(\be\cup_1\delta\be)
+(\be\cup_1\delta\be)\cup_2\delta\bface'
\right)}
\Bigg\} \times
\prod_{f'}\left(S_{f'}^A\right)^{
\int\delta\be\cup_1\bface'}
\Bigg]^{\int\be\cup\bface}.
\end{aligned}
\label{eq. Z after conjugation}
\end{equation}
Observe that the above expression contains only Pauli-$Z$ operators on the $B$ qubits. To obtain the qubit separators acting only on the $A$ qubits, we simply set $Z^B_f = +1$, which is exactly the condensation constraint we needed to impose in the first place c.f. Eq.~\eqref{eq:z4-disentangled-condensate}. Using $CZ\!\left(A,\prod_j B_j\right)= \prod_j CZ(A,B_j)$, we have
\begin{equation}
\boxed{
\begin{aligned}
\overline Z_f
=&Z_f^A\prod_e\Bigg[
X_{\delta\be}^A
\prod_{f'\in\delta\be} \Bigg\{
\left(Z_{f'}^A\right)^{
\int\delta\be\cup_1\bface'}
\prod_{f''}
CZ\!\left(Z_{f'}^A,Z_{f''}^A\right)^{
\int\left(
\bface''\cup_1\bface'
+\bface'\cup_2\delta\bface''
\right)} \Bigg\}
\\
&\quad\times
\prod_{f'}\left(Z_{f'}^A\right)^{
\int
\bface'\cup_1(\be\cup_1\delta\be)
+(\be\cup_1\delta\be)\cup_2\delta\bface'
}
\left(S_{f'}^A\right)^{
\int\delta\be\cup_1\bface'}
\Bigg]^{\int\be\cup\bface}.
\end{aligned}}
\label{eq:z4-reduced-separator}
\end{equation}

For reference, we explicitly draw the above separators on a cubic-lattice support as shown below.
\begin{equation}
\begin{tikzpicture}[
  baseline=(current bounding box.center),
  x={(.410cm,0cm)},
  y={(.260cm,.1567cm)},
  z={(0cm,.420cm)},
  dual/.style={draw=black,line width=.58pt},
  dualbg/.style={draw=black!32,line width=.52pt,densely dotted},
  op/.style={fill=white,inner sep=.35pt,text=blue,font=\small},
  cz/.style={draw=black,line width=.55pt},
  czdot/.style={circle,fill=black,inner sep=0pt,minimum size=3pt}
]

\def\FXSupport#1{%
  \draw[#1] (0,0,-3)--(0,0,6);
  \draw[#1] (0,-3,0)--(0,3,0);
  \draw[#1] (0,-3,3)--(0,3,3);
  \draw[#1] (0,3,0)--(0,3,6);
  \draw[#1] (0,0,0)--(3,0,0);
  \draw[#1] (0,0,3)--(3,0,3);
  \draw[#1] (0,3,3)--(3,3,3);
}

\def\FYSupport#1{%
  \draw[#1] (0,0,-3)--(0,0,6);
  \draw[#1] (0,-3,0)--(0,0,0);
  \draw[#1] (0,-3,3)--(0,0,3);
  \draw[#1] (0,0,0)--(6,0,0);
  \draw[#1] (0,0,3)--(6,0,3);
  \draw[#1] (3,0,-3)--(3,0,6);
  \draw[#1] (3,-3,0)--(3,3,0);
  \draw[#1] (3,-3,3)--(3,3,3);
}

\def\FZSupport#1{%
  \draw[#1] (0,0,-3)--(0,0,3);
  \draw[#1] (0,-3,0)--(0,3,0);
  \draw[#1] (0,0,0)--(6,0,0);
  \draw[#1] (3,-3,0)--(3,6,0);
  \draw[#1] (0,3,0)--(6,3,0);
  \draw[#1] (3,0,-3)--(3,0,3);
  \draw[#1] (0,3,0)--(0,3,3);
  \draw[#1] (3,3,-3)--(3,3,3);
}

\def\FXOneQ{%
  \path (0,0,0)--(0,0,3) node[pos=.5,op] {$XS$};
  \path (0,0,0)--(0,3,0) node[pos=.5,op] {$XS$};
  \path (0,3,0)--(0,3,3) node[pos=.4,op] {$X$};
  \path (0,0,3)--(0,3,3) node[pos=.5,op] {$XZ$};
  \path (0,0,-3)--(0,0,0) node[pos=.5,op] {$S$};
  \path (0,-3,0)--(0,0,0) node[pos=.5,op] {$S^\dagger$};
  \path (0,0,0)--(3,0,0) node[pos=.5,op] {$S^\dagger$};
  \path (0,0,3)--(3,0,3) node[pos=.45,op] {$Z$};
  \path (0,-3,3)--(0,0,3) node[pos=.5,op] {$Z$};
  \path (0,0,3)--(0,0,6) node[pos=.5,op] {$Z$};
  \path (0,3,3)--(3,3,3) node[pos=.5,op] {$S_f$};
}

\def\FYDaggerOneQ{%
  \path (0,0,0)--(0,0,3) node[pos=.5,op] {$X$};
  \path (0,0,0)--(3,0,0) node[pos=.5,op] {$XS$};
  \path (3,0,0)--(3,0,3) node[pos=.5,op] {$XS$};
  \path (0,0,3)--(3,0,3) node[pos=.5,op] {$XZ$};
  \path (0,-3,0)--(0,0,0) node[pos=.5,op] {$S$};
  \path (3,0,-3)--(3,0,0) node[pos=.5,op] {$S$};
  \path (3,-3,0)--(3,0,0) node[pos=.45,op] {$S^\dagger$};
  \path (3,0,0)--(3,3,0) node[pos=.5,op] {$S$};
  \path (3,0,0)--(6,0,0) node[pos=.5,op] {$S^\dagger$};
  \path (3,0,3)--(6,0,3) node[pos=.5,op] {$Z$};
  \path (3,-3,3)--(3,0,3) node[pos=.5,op] {$Z$};
  \path (3,0,3)--(3,0,6) node[pos=.5,op] {$Z$};
  \path (3,0,3)--(3,3,3) node[pos=.5,op] {$S_f^\dagger$};
}

\def\FZOneQ{%
  \path (0,0,0)--(0,3,0) node[pos=.5,op] {$X$};
  \path (0,0,0)--(3,0,0) node[pos=.5,op] {$XZ$};
  \path (3,0,0)--(3,3,0) node[pos=.45,op] {$XS$};
  \path (0,3,0)--(3,3,0) node[pos=.6,op] {$XS$};
  \path (0,0,-3)--(0,0,0) node[pos=.5,op] {$S$};
  \path (3,3,-3)--(3,3,0) node[pos=.5,op] {$S$};
  \path (3,3,0)--(3,6,0) node[pos=.5,op] {$S$};
  \path (3,3,0)--(6,3,0) node[pos=.5,op] {$S^\dagger$};
  \path (3,0,0)--(6,0,0) node[pos=.5,op] {$Z$};
  \path (3,-3,0)--(3,0,0) node[pos=.4,op] {$Z$};
  \path (0,3,0)--(0,3,3) node[pos=.5,op] {$Z$};
  \path (3,0,-3)--(3,0,0) node[pos=.5,op] {$Z$};
  \path (3,0,0)--(3,0,3) node[pos=.7,op] {$Z$};
  \path (3,3,0)--(3,3,3) node[pos=.5,op] {$Z_f$};
}

\def\Arc#1#2#3{\draw[cz] #1 to[#2] #3;}

\def\FXCZ#1{%
  #1{(0,1.5,0)}{bend right=55}{(0,-1.5,0)}
  #1{(0,1.5,0)}{bend left=40}{(0,0,-1.5)}
  #1{(0,1.5,0)}{bend right=40}{(0,0,1.5)}
  #1{(0,1.5,0)}{bend left=40}{(1.5,0,0)}
  #1{(0,0,1.5)}{bend left=50}{(0,0,4.5)}
  #1{(0,0,1.5)}{bend left=40}{(0,-1.5,3)}
  #1{(0,0,1.5)}{bend right=40}{(1.5,0,3)}
  #1{(0,1.5,3)}{bend right=55}{(0,-1.5,3)}
  #1{(0,1.5,3)}{bend left=40}{(0,0,1.5)}
  #1{(0,1.5,3)}{bend right=40}{(0,0,4.5)}
  #1{(0,1.5,3)}{bend left=40}{(1.5,0,3)}
  #1{(0,1.5,0)}{bend left=40}{(0,3,1.5)}
  #1{(0,3,1.5)}{bend left=50}{(0,3,4.5)}
  #1{(0,3,1.5)}{bend left=40}{(0,1.5,3)}
  #1{(0,3,1.5)}{bend right=40}{(1.5,3,3)}
  #1{(0,1.5,3)}{bend left=40}{(0,3,4.5)}
}
\def\FXCZNodes{%
  \foreach \p in {(0,1.5,0),(0,-1.5,0),(0,0,-1.5),(0,0,1.5),
    (1.5,0,0),(0,0,4.5),(0,-1.5,3),(1.5,0,3),
    (0,1.5,3),(0,3,1.5),(0,3,4.5),(1.5,3,3)}
    \node[czdot] at \p {};
}

\def\FYCZ#1{%
  #1{(1.5,0,0)}{bend left=40}{(0,-1.5,0)}
  #1{(1.5,0,0)}{bend right=40}{(0,0,1.5)}
  #1{(0,0,1.5)}{bend left=50}{(0,0,4.5)}
  #1{(0,0,1.5)}{bend left=40}{(0,-1.5,3)}
  #1{(0,0,1.5)}{bend right=40}{(1.5,0,3)}
  #1{(1.5,0,3)}{bend left=40}{(0,-1.5,3)}
  #1{(1.5,0,3)}{bend right=40}{(0,0,4.5)}
  #1{(1.5,0,0)}{bend left=50}{(4.5,0,0)}
  #1{(1.5,0,0)}{bend right=40}{(3,-1.5,0)}
  #1{(1.5,0,0)}{bend left=40}{(3,1.5,0)}
  #1{(1.5,0,0)}{bend right=40}{(3,0,-1.5)}
  #1{(1.5,0,0)}{bend left=40}{(3,0,1.5)}
  #1{(3,0,1.5)}{bend left=50}{(3,0,4.5)}
  #1{(3,0,1.5)}{bend left=40}{(3,-1.5,3)}
  #1{(3,0,1.5)}{bend right=40}{(4.5,0,3)}
  #1{(1.5,0,3)}{bend left=50}{(4.5,0,3)}
  #1{(1.5,0,3)}{bend right=40}{(3,-1.5,3)}
  #1{(1.5,0,3)}{bend left=40}{(3,1.5,3)}
  #1{(1.5,0,3)}{bend right=40}{(3,0,1.5)}
  #1{(1.5,0,3)}{bend left=40}{(3,0,4.5)}
}
\def\FYCZNodes{%
  \foreach \p in {(1.5,0,0),(0,-1.5,0),(0,0,1.5),(0,0,4.5),
    (0,-1.5,3),(1.5,0,3),(4.5,0,0),(3,-1.5,0),
    (3,1.5,0),(3,0,-1.5),(3,0,1.5),(3,0,4.5),
    (3,-1.5,3),(4.5,0,3),(3,1.5,3)}
    \node[czdot] at \p {};
}

\def\FZCZ#1{%
  #1{(0,1.5,0)}{bend right=55}{(0,-1.5,0)}
  #1{(0,1.5,0)}{bend left=40}{(0,0,-1.5)}
  #1{(0,1.5,0)}{bend right=40}{(0,0,1.5)}
  #1{(0,1.5,0)}{bend left=40}{(1.5,0,0)}
  #1{(1.5,0,0)}{bend left=40}{(0,-1.5,0)}
  #1{(1.5,0,0)}{bend right=40}{(0,0,1.5)}
  #1{(0,1.5,0)}{bend left=40}{(0,3,1.5)}
  #1{(1.5,3,0)}{bend left=40}{(0,1.5,0)}
  #1{(1.5,3,0)}{bend right=40}{(0,3,1.5)}
  #1{(1.5,0,0)}{bend left=50}{(4.5,0,0)}
  #1{(1.5,0,0)}{bend right=40}{(3,-1.5,0)}
  #1{(1.5,0,0)}{bend left=40}{(3,1.5,0)}
  #1{(1.5,0,0)}{bend right=40}{(3,0,-1.5)}
  #1{(1.5,0,0)}{bend left=40}{(3,0,1.5)}
  #1{(3,1.5,0)}{bend right=55}{(3,-1.5,0)}
  #1{(3,1.5,0)}{bend left=40}{(3,0,-1.5)}
  #1{(3,1.5,0)}{bend right=40}{(3,0,1.5)}
  #1{(3,1.5,0)}{bend left=40}{(4.5,0,0)}
  #1{(1.5,3,0)}{bend left=40}{(3,4.5,0)}
  #1{(1.5,3,0)}{bend right=40}{(3,3,-1.5)}
  #1{(3,1.5,0)}{bend left=40}{(3,3,1.5)}
  #1{(1.5,3,0)}{bend left=50}{(4.5,3,0)}
  #1{(1.5,3,0)}{bend right=40}{(3,1.5,0)}
  #1{(1.5,3,0)}{bend left=40}{(3,3,1.5)}
}
\def\FZCZNodes{%
  \foreach \p in {(0,1.5,0),(0,-1.5,0),(0,0,-1.5),(0,0,1.5),
    (1.5,0,0),(0,3,1.5),(1.5,3,0),(4.5,0,0),
    (3,-1.5,0),(3,1.5,0),(3,0,-1.5),(3,0,1.5),
    (3,4.5,0),(3,3,-1.5),(3,3,1.5),(4.5,3,0)}
    \node[czdot] at \p {};
}

\begin{scope}[yshift=5.10cm]
  \begin{scope}[xshift=1.55cm]
    \node[anchor=east,font=\normalsize] at (0,0,1.5)
      {$\overline Z_{f_x}=$};
  \end{scope}
  \begin{scope}[xshift=2.80cm,scale=1.15]
    \FXSupport{dual}\FXOneQ
  \end{scope}
  \begin{scope}[xshift=6.20cm]
    \node[font=\large] at (0,0,1.5) {$\times$};
  \end{scope}
  \begin{scope}[xshift=7.50cm,scale=1.15]
    \FXSupport{dualbg}\FXCZ{\Arc}\FXCZNodes
  \end{scope}
\end{scope}

\begin{scope}
  \begin{scope}[xshift=1.55cm]
    \node[anchor=east,font=\normalsize] at (0,0,1.5)
      {$\overline Z_{f_y}^{\dagger}=$};
  \end{scope}
  \begin{scope}[xshift=2.80cm,scale=1.15]
    \FYSupport{dual}\FYDaggerOneQ
  \end{scope}
  \begin{scope}[xshift=6.20cm]
    \node[font=\large] at (0,0,1.5) {$\times$};
  \end{scope}
  \begin{scope}[xshift=7.50cm,scale=1.15]
    \FYSupport{dualbg}\FYCZ{\Arc}\FYCZNodes
  \end{scope}
\end{scope}

\begin{scope}[yshift=-5.10cm]
  \begin{scope}[xshift=1.55cm]
    \node[anchor=east,font=\normalsize] at (0,1.5,0)
      {$\overline Z_{f_z}=$};
  \end{scope}
  \begin{scope}[xshift=2.80cm,scale=1.15]
    \FZSupport{dual}\FZOneQ
  \end{scope}
  \begin{scope}[xshift=6.20cm]
    \node[font=\large] at (0,1.5,0) {$\times$};
  \end{scope}
  \begin{scope}[xshift=7.50cm,scale=1.15]
    \FZSupport{dualbg}\FZCZ{\Arc}\FZCZNodes
  \end{scope}
\end{scope}

\end{tikzpicture}
\label{eq:semion-separator-pictorial}
\end{equation}
where the black arcs represent the $CZ$ gates.

We now move on to the commuting qubit flippers by repeating the above process on $V'_f$. Consider first acting the above transformations on the bare operator $V_f$. We find
\begin{equation}
    \begin{aligned}
        V_f
        &=X_f\prod_{f'}Z_{f'}^{\int
        \left(\bface'\cup_1\bface+\delta\bface'\cup_2\bface\right)}
        \\
        &=X_f^A CZ_f^{AB}
        \prod_{f'}\left(X_{f'}^B S_{f'}^A\right)^{\int
        \left(\bface'\cup_1\bface+\delta\bface'\cup_2\bface\right)}
        \\
        &=X_f^A
        \prod_{f'}\left(X_{f'}^B\right)^{\int
        \left(\bface'\cup_1\bface+\delta\bface'\cup_2\bface\right)}
        \left(Z_f^A\right)^{\int
        \left(\bface\cup_1\bface+\delta\bface\cup_2\bface\right)}
        CZ_f^{AB}
        \prod_{f'}\left(S_{f'}^A\right)^{\int
        \left(\bface'\cup_1\bface+\delta\bface'\cup_2\bface\right)}.
    \end{aligned}
    \label{eq:z4-Vf-UC}
\end{equation}
In the last line, we have used $CZ_f^{AB}X_f^B=X_f^BZ_f^ACZ_f^{AB}$ to move the $X_{f'}^B$ factors to the left. An additional $Z^A_f$ operator appears only when $f'=f$ as a result of the action by the $CZ_f$ gate. The $X^B$ string generated by conjugating $X_f^A$ with $U_C$ cancels the one in the last equality of Eq.~\eqref{eq:z4-Vf-UC}, as a result of the higher-cup recursion
\begin{equation*}
    \int \left(\bface'\cup_1\bface+\delta\bface'\cup_2\bface+\bface\cup_1\bface'+\bface'\cup_2\delta\bface\right)=\int \delta\left(\bface'\cup_2\bface\right)=0 \pmod 2.
\end{equation*}
Thus,
\begin{equation}
    \begin{aligned}
    U_CV_fU_C^\dagger
    ={}&X_f^A
    \left(Z_f^A\right)^{\int
    \left(\bface\cup_1\bface+\delta\bface\cup_2\bface\right)}
    CZ\!\left(
    Z_f^A,
    Z_f^B\prod_{f'}\left(Z_{f'}^A\right)^{\int
    \left(\bface'\cup_1\bface+\bface\cup_2\delta\bface'\right)}
    \right)
    \prod_{f'}\left(S_{f'}^A\right)^{\int
    \left(\bface'\cup_1\bface+\delta\bface'\cup_2\bface\right)}.
    \end{aligned}
    \label{eq:z4-Vf-conjugated}
\end{equation}
Substituting the decomposition of $V_f'$ from Eq.~\eqref{eq:z4-corrected-flipper}, we have
\begin{equation}
    \begin{aligned}
        \overline X_f
        =&i\left(U_CV_fU_C^\dagger\right) \\
        &\times
        \left.\prod_{f'}\Bigg[
        \left(U_C\widetilde S_{f'}^\dagger U_C^\dagger\right)^{\int_{M_3}
        \left(\bface'\cup_1\bface+\delta\bface'\cup_2\bface\right)}
        CZ\!\left(
        U_C\widetilde Z_fU_C^\dagger,
        U_C\widetilde Z_{f'}U_C^\dagger
        \right)^{\int_{M_3}
        \left(\bface\cup_1\bface'+\delta\bface\cup_2\bface'\right)}
        \Bigg]\right|_{Z_{f'}^B=1\,;\,\forall f'}.
    \end{aligned}
    \label{eq:X_bar_Z4_to_Z2}
\end{equation}
Using $U_C\widetilde Z_f U_C^\dagger = \overline Z_f$ and $U_C\widetilde S_fU_C^\dagger=i^{(1-\overline Z_f)/2}=:\overline S_f$, we finally arrive at the qubit flipper
\begin{equation}
    \boxed{\begin{aligned}
    \overline X_f
    ={}&iX_f^A
    \left(Z_f^A\right)^{\int
    \left(
    \bface\cup_1\bface
    +\delta\bface\cup_2\bface
    \right)}
    \\
    &\times
    \prod_{f'}\Bigg[
    \left(S_{f'}^A\right)^{\int
    \left(
    \bface'\cup_1\bface
    +\delta\bface'\cup_2\bface
    \right)}
    CZ\!\left(Z_f^A,Z_{f'}^A\right)^{\int
    \left(
    \bface'\cup_1\bface
    +\bface\cup_2\delta\bface'
    \right)}
    \Bigg]
    \\
    &\times
    \prod_{f'}\Bigg[
    \left(\overline S_{f'}^\dagger\right)^{\int
    \left(
    \bface'\cup_1\bface
    +\delta\bface'\cup_2\bface
    \right)}
    CZ\!\left(\overline Z_f,\overline Z_{f'}\right)^{\int
    \left(
    \bface\cup_1\bface'
    +\delta\bface\cup_2\bface'
    \right)}
    \Bigg].
    \end{aligned}}
    \label{eq:z4-reduced-flipper-direct}
\end{equation}
We omit a drawing of $\overline{X}_f$ since it is similar to that of $\overline{Z}_f$.

Finally, since conjugation by unitaries preserves the operator algebra of $\widetilde{Z}_f$ and $V'_f$ in Eq.\eqref{eq:semion-z4-candidate-separator-flipper-algebra}, the qubit separators and flippers obey the same algebra
\begin{equation}
        \overline Z_f^2=\overline X_f^2=1,\qquad 
        [\overline Z_f,\overline Z_{f'}] =[\overline X_f,\overline X_{f'}]=0, \qquad
        \overline Z_{f'}\overline X_f =(-1)^{\delta_{ff'}}\overline X_f\overline Z_{f'}.
    \label{eq:z4-final-separator-flipper-algebra}
\end{equation}
Hence, the mapping
\begin{equation}
    Z_f^A \longmapsto \overline{Z}_f, \qquad X_f^A \longmapsto \overline{X}_f
    \label{eq:z4-semion-qca-map}
\end{equation}
defines an automorphism of the local operator algebra and therefore defines the semion QCA.

\subsection{\texorpdfstring{$U(1)_4$}{U(1)4} QCA}
\label{sec:u14-qca}


The $U(1)_4$ construction follows the same sequence of steps as the semion in Sec.~\ref{sec:semion-qca}. We begin with the pre-modular $\mathbb Z_8^{(1)}$ Walker-Wang model built from $\ZZ_8$ qudits, and then condense the order-two boson in the bulk. The surviving separator is order-four. After correcting the flippers, we use a $\mathbb Z_8\to\mathbb Z_4\times\mathbb Z_2$ algebra isomorphism to map every $\ZZ_8$ qudit to a $\ZZ_4$ qudit and a qubit, and an FDQC to disentangle the qubits from the system. This leaves one $\ZZ_4$ qudit on every face and produces the $\ZZ_4$-qudit $U(1)_4$ QCA introduced in Ref.~\cite{Shirley2022QCA}.

Condensing the bulk boson turns the pre-modular $\mathbb Z_8^{(1)}$ theory into the pointed modular category $U(1)_4$. The corresponding quadratic form is
\begin{equation}
    \mathcal A_4=\mathbb Z_4, \qquad q_4(j)=\frac{j^2}{8}, \qquad b_4(j,k)=\frac{jk}{4} \quad\text{in }\mathbb R/\mathbb Z.
\label{eq:u14-quadratic-module}
\end{equation}
Indeed, $q_8(j)=j^2/8$ obeys
$q_8(j+4)-q_8(j)=j+2\in\mathbb Z$, so it descends through $\mathbb Z_8/\langle4\rangle\cong\mathbb Z_4$ to the displayed function $q_4$. The normalized Gauss-Milgram sum is $e^{2\pi i/8}$, so the associated invertibly normalized Crane-Yetter partition function has the same $\sigma(X^4)/8=\int_{X^4}p_1/24$ signature phase as the semion theory, see Appendix~\ref{app:semion-signature-bordism} for further details. This phase is not a complete Witt or QCA invariant. Indeed, semion $U(1)_2$ has the different quadratic module $(\mathbb Z_2,q_2(1)=1/4)$, and $[U(1)_4]\neq[U(1)_2]$ in the pointed Witt group \cite{Drinfeld2010Braided,Davydov2013Witt}.  Thus, the common signature character does not identify the two boundary theories or their QCAs. The microscopic $U(1)_4$ QCA exists by the factorization recalled below, whereas its precise stable order and its relation to the pointed Witt class remain conjectural.

The $\ZZ_8^{(1)}$ Walker--Wang model~\cite{Walker2012TQFT} has surface anyon theory $\mathcal{A}=\{1,a,a^2,\ldots,a^7\}$, with $\theta(a)=\exp(2\pi i/8)$ and $a^8=1$.  Its four-dimensional response is
\begin{equation}
    \frac{1}{8} B_2 \cup B_2
    \;\;\in\;\; H^4(B^2 \ZZ_8, \RR/\ZZ),
\label{eq:z8-action}
\end{equation}
where $B_2$ is a dynamical $\mathbb Z_8$-valued two-cocycle. With a $\mathbb Z_8$ qudit on each face, define
\begin{equation}
 X_f:=\sum_{j\in\mathbb Z_8}\ket{j+1}_f\!\bra j,
 \qquad
 Z_f:=\sum_{j\in\mathbb Z_8}\omega_8^j\ket j_f\!\bra j,
 \qquad
 \omega_8:=e^{2\pi i/8}.
 \label{eq:z8-clock-shift}
\end{equation}
Thus, $X_f^8=Z_f^8=1$ and $Z_fX_f=\omega_8X_fZ_f$. The corresponding commuting Hamiltonian is
\begin{equation}
\begin{aligned}
H^{\ZZ_8^{(1)}}={}&-\frac12\sum_c
\left(Z_{\partial c}+Z_{\partial c}^{\dagger}\right)
-\frac12\sum_e\left[
X_{\delta \be}\prod_{f'}Z_{f'}^{\int(
\bface'\cup\be+\be\cup\bface')}+\mathrm{h.c.}\right],
\end{aligned}
\label{eq:z8-initial-hamiltonian}
\end{equation}
Exactly as in Eq.~\eqref{eq:z4-edge-term-rewrite}, we use the identity
\begin{equation}
 \int(\bface'\cup\be+\be\cup\bface')
 =\int\left(
 \delta\be\cup_1\bface'
 -\be\cup_1\delta\bface'
 +2\be\cup\bface'
 \right)
 \quad(\mathrm{mod}\ 8).
 \label{eq:z8-higher-cup-rewrite}
\end{equation}
The middle term is a product of $Z$ stabilizers and therefore can be dropped. Thus,
\begin{equation}
X_{\delta\be}\prod_{f'}Z_{f'}^{\int(
\delta\be\cup_1\bface'+2\be\cup\bface')}
=G_e\prod_{f'}Z_{f'}^{2\int\be\cup\bface'},
\label{eq:z8-edge-term-rewrite}
\end{equation}
where
\begin{equation}
    G_e:=X_{\delta\be}\prod_{f'}Z_{f'}^{\int
\delta\be\cup_1\bface'}.
\end{equation}
Note that diagrammatically this $G_e$ looks identical to that in Eq.~\eqref{eq:z4-Ge-diagram} since it has the same cup-product expression albeit the generalized Pauli operators now are $\ZZ_8$.

After the relabeling from edge to faces on the cubic lattice, an equivalent generating set is therefore
\begin{equation}
H'^{\ZZ_8^{(1)}}
=-\sum_c
\left(Z_{\partial c}+Z_{\partial c}^{\dagger}\right)
-\sum_f\left[
Z_f^2\prod_eG_e^{\int\be\cup\bface}+\mathrm{h.c.}\right].
\label{eq:z8-rewritten-hamiltonian}
\end{equation}
The input theory is pre-modular with a transparent boson. In the Walker-Wang construction, this transparent sector gives rise to a deconfined order-two particle in the 3-dimensional bulk. Condensing this particle removes the intrinsic bulk topological order, leaving an invertible phase. As in the semion construction Sec.~\ref{sec:semion-qca}, the bare operator $X_f^4$ violates the $Z$ stabilizers, but must be properly dressed so that it commutes with the $G_e$ terms along its interior. Similarly to the semion case, the dressed short boson string is
\begin{equation}
 C_f:=X_f^4\prod_{f'}Z_{f'}^{4\int(
 \bface'\cup_1\bface+\bface\cup_2\delta\bface')}.
 \label{eq:z8-condensate}
\end{equation}
A graphical representation of $C_f$ is identical to that of Eq.~\eqref{eq:z4-condensate-diagram} with the substitution of $2 \to 4$ for all exponents.

Its elementary algebra is
\begin{equation}
    \begin{gathered}
     C_f^2=1,\quad
     [C_f,C_{f'}]=[C_f,G_e]=0,\quad
     Z_{\partial c}C_f
     =(-1)^{[\delta\bface]_2(c)}C_fZ_{\partial c}.
    \end{gathered}
 \label{eq:z8-condensate-algebra}
\end{equation}
Thus, $Z_{\partial c}$ may anticommute with a condensate, whereas $Z_{\partial c}^2$ commutes with every $C_f$.  Imposing $C_f=1$ therefore removes $Z_{\partial c}$ and only retains $Z_{\partial c}^2$.

After the complementary edge--face relabeling, define
\begin{equation}
 \widetilde Z_f:=Z_f^2\prod_eG_e^{\int\be\cup\bface}.
 \label{eq:z8-separator}
\end{equation}
The condensed Hamiltonian may first be written as
\begin{equation}
 H_{\mathrm{cond}}^{U(1)_4}
 =-\sum_fC_f
 -\sum_c\left[
 Z_{\partial c}^2+(Z_{\partial c}^{\dagger})^2\right]
 -\sum_f\left(\widetilde Z_f+\widetilde Z_f^\dagger\right).
 \label{eq:z8-condensed-hamiltonian-with-flux}
\end{equation}
Now, $Z_{\partial c}^{2}$ is a redundant stabilizer since
\begin{equation}
    \prod_f\left(\widetilde Z_f\right)^{\delta\bface(c)}=Z_{\partial c}^{2}.
    \label{eq:z8-face-product-flux}
\end{equation}
and therefore can be removed. After the condensation, we are left with
\begin{equation}
 H^{U(1)_4}
 =-\sum_fC_f
 -\sum_f\left(\widetilde Z_f+\widetilde Z_f^\dagger\right).
 \label{eq:z8-condensed-hamiltonian}
\end{equation}
The two terms mutually commute.  Their local independence and completeness will follow from the paired flippers constructed next.

The separator becomes order four once the condensation constraints are imposed. Indeed, using the higher cup-product recursion relation gives
\begin{equation}
\begin{aligned}
G_e^4
&=X_{\delta\be}^4
\prod_{f'}Z_{f'}^{4\int\delta\be\cup_1\bface'}
=X_{\delta\be}^4
\prod_{f'}Z_{f'}^{4\int(
\bface'\cup_1\delta\be+\delta\be\cup_2\delta\bface')}
=\prod_fC_f^{[\delta\be]_2(f)} .
\end{aligned}
\label{eq:z8-G-fourth-power}
\end{equation}
Hence, $G_e^4=1$ when all $C_f=1$, and Eq.~\eqref{eq:z8-separator} implies $\widetilde Z_f^4=1$ in the same sector.

The natural conjugate to paired with $\widetilde Z_f$ is
\begin{equation}
    V_f:=X_f\prod_{f'}Z_{f'}^{\int(
    \bface'\cup_1\bface+\delta\bface'\cup_2\bface)}
    \label{eq:z8-bare-flipper}
\end{equation}
Its commutation with $C_f$ follows from
\begin{equation}
    \begin{aligned}
        &(-1)^{\int(
        \bface_2\cup_1\bface_1+\bface_1\cup_2\delta\bface_2
        +\bface_1\cup_1\bface_2+\delta\bface_1\cup_2\bface_2)}
        =(-1)^{\int\delta(\bface_1\cup_2\bface_2)}=1.
    \end{aligned}
    \label{eq:z8-bare-flipper-condensate-commutator}
\end{equation}
After relabeling complementary cells, one obtains
\begin{equation}
    \widetilde Z_{f'}V_f
    =i^{\delta_{f,f'}}V_f\widetilde Z_{f'}.
    \label{eq:z8-bare-flipper-Weyl-relation}
\end{equation}
Thus, the paired separator really has exact order four. However, the bare $V_f$ do not commute among themselves
\begin{align}
    V_{f_2}V_{f_1}
    ={}&\exp\!\left\{\frac{2\pi i}{8}\int\left(
    \bface_1\cup_1\bface_2+\delta\bface_1\cup_2\bface_2
    -\bface_2\cup_1\bface_1-\delta\bface_2\cup_2\bface_1
    \right)\right\}V_{f_1}V_{f_2},
    \label{eq:z8-bare-flipper-commutator}\\
    V_f^4
    ={}&-iC_f\prod_{f'}{\widetilde{Z}_{f'}}^{
    2\int\delta\bface'\cup_3\delta\bface}.
    \label{eq:z8-bare-flipper-fourth-power}
\end{align}
In order to make them commute, we again need to introduce phase factors coming from $\widetilde Z$. To do so, we will need the Clifford gates that generalize the $S$ and $CZ$ gates to qudits.

First, the $\ZZ_4$ qudit $CZ$ can be defined for any two order-four commuting operators $A$ and $B$ as
\begin{equation}
    CZ(A,B)\ket{a,b}=i^{ab}\ket{a,b},
    \qquad
    A\ket{a,b}=i^a\ket{a,b},\quad
    B\ket{a,b}=i^b\ket{a,b}.
    \label{eq:z8-CZ4}
\end{equation}
Note that this definition includes the case where $A=B$, in which case the phase is $i^{a^2}$ on the $i^a$ eigenspace. We can then define the $S$ gate as $\sqrt{CZ(Z,Z)}$. It acts on the computational basis as
 \begin{align}
     S \ket{a} = e^{\pi i a^2/4}\ket{a},
 \end{align}
 and satisfies
 \begin{align}
     SXS^\dagger = e^{\pi i/4} XZ.
 \end{align}
Note that unlike the qubit case, $S^2 = CZ(Z,Z) \ne Z$ (indeed, $\sqrt{Z}$ is not a Clifford gate). Instead, it satisfies $S^4 = Z^2$.

We now claim that the commuting flipper is given by
\begin{equation}
    \begin{aligned}
        V_f':=e^{\pi i/4}V_f\prod_{f'}\Big[
        &({\widetilde S}_{f'}^\dagger)^{\int(
        \bface'\cup_1\bface+\delta\bface'\cup_2\bface)}
         CZ^\dagger(\widetilde Z_f,\widetilde Z_{f'})^{
        \int(\bface\cup_1\bface'+\delta\bface\cup_2\bface')}
        \Big].
    \end{aligned}
    \label{eq:z8-corrected-flipper}
\end{equation}
where $\tilde S = \sqrt{CZ(\tilde Z,\tilde Z)}$. This is nearly identical to the semion version in Eq.~\eqref{eq:z4-corrected-flipper}, where the only thing we had to keep track of was the dagger for $CZ$. One can similarly show that they have the right commutation relations. That is,
\begin{equation}
    V_f'^4 = C_f, \qquad [V'_f, V'_{f'}] =0.
\end{equation}

Next, we perform a basis transformation to reduce to the subspace where $C_f=1$. First, we need to map the $\ZZ_8$ qudit into a $\ZZ_4\times \ZZ_2$ qudit. We follow the general factorization $\mathbb Z_{2N}\to\mathbb Z_N\times\mathbb Z_2$ of Ref.~\cite{Shirley2022QCA}.  On every face, write label $A$ for the $\ZZ_4$ qudit and $B$ for the qubit. Let $X^A,Z^A$ obey $Z^AX^A=iX^AZ^A$, and define
\begin{equation}
    R^A :=  \sqrt{Z^A} =\operatorname{diag}
    \left(1,e^{\pi i/4},i,e^{3\pi i/4}\right).
    \label{eq:z8-on-site-square-root}
\end{equation}
Note again that in the qudit case, this is distinct from the $S$-gate defined earlier. Moreover, we will define the following carry-controlled phase gate between $A$ and $B$ as
\begin{align}
     {CZ}^{AB} = (-1)^{\frac{1 + i Z^A - (Z^A)^2 - i (Z^A)^3}{4} \frac{1-Z^B}{2}} ,
\end{align}
which acts in the computational basis as
\begin{equation}
    {CZ}^{AB}\ket{a,b}
    =(-1)^{\delta_{a,3} b}\ket{a,b},
    \qquad a\in\mathbb Z_4,\quad b\in\mathbb Z_2.
    \label{eq:z8-carry-CZ}
\end{equation}
which defines the gate using projectors, and is a different way from defining $CZ$ than earlier. The on-site algebra isomorphism is
\begin{equation}
    \begin{aligned}
        X_f &=  X^A_f CZ_f^{AB}, & Z_f &= R^A_f X^B_f,\\
        X^4_f &= Z_f^B, & Z_f^2&= Z_f^A.
    \end{aligned}
    \label{eq:z8-ququart-qubit-map}
\end{equation}

Under this equivalence, the condensation term can be written as
\begin{equation}
    C_f =Z_f^B\prod_{f'}(Z_{f'}^A)^{2\int(
    \bface'\cup_1\bface+\bface\cup_2\delta\bface')}.
    \label{eq:z8-encoded-condensate}
\end{equation}
To perform the basis transformation, define the controlled-gate
\begin{equation}
    CX^{AB}\ket{a,b}
    =\ket{a,b+[a]_2},
    \label{eq:z8-parity-controlled-X}
\end{equation}
where $[]_2$ is the value modulo $2$. Its relevant conjugation rules are
\begin{equation}
    \begin{aligned}
    Z_f^B&\longmapsto (Z_{f'}^A)^2Z_f^B,
    &Z_{f'}^A&\longmapsto Z_{f'}^A,\\
    X_f^B&\longmapsto X_f^B,
    &X_{f'}^A&\longmapsto X_{f'}^AX_f^B.
    \end{aligned}
    \label{eq:z8-parity-controlled-X-algebra}
\end{equation}
Define the basis transformation as
\begin{equation}
    U_C
    :=\prod_{f,f'}CX^{AB}(f',f)^{
    \int(\bface'\cup_1\bface+\bface\cup_2\delta\bface')}.
    \label{eq:z8-condensate-circuit}
\end{equation}
It acts on the generalized Pauli operators as
\begin{equation}
    \begin{aligned}
    U_CZ_f^BU_C^\dagger
    &=Z_f^B\prod_{f'}(Z_{f'}^A)^{2\int(
    \bface'\cup_1\bface+\bface\cup_2\delta\bface')},\\
    U_CX_f^AU_C^\dagger
    &=X_f^A\prod_{f'}(X_{f'}^B)^{\int(
    \bface\cup_1\bface'+\bface'\cup_2\delta\bface)},
    \end{aligned}
    \label{eq:z4-disentangler-action}
\end{equation}
while $Z_f^A$ and $X_f^B$ are unchanged. It then follows that
\begin{equation}
    U_C C_f U_C^\dagger=Z_f^B.
    \label{eq:z8-disentangled-condensate}
\end{equation}

Since $\widetilde Z_f$ and $V_f'$ commute with every condensate, their conjugated images commute with each $Z_f^B$. They therefore preserve the joint $Z^B=+1$ sector, so the following restrictions are well defined. Similarly, we define the separator and the commuting flippers via 
\begin{equation}
    \boxed{\begin{aligned}
        \overline Z_f^A
        &:=\left. U_C \widetilde{Z}_f U_C\dagger\right|_{Z^B=1},\\
        \overline X_f^A
        &:=\left. U_C V_f' U_C\dagger\right|_{Z^B=1} .
    \end{aligned}}
\end{equation}
The final explicit expressions are very complicated, but we know that they satisfy
\begin{equation}
    \begin{gathered}
        (\overline Z_f^A)^4=(\overline X_f^A)^4=1,\qquad
        [\overline Z_f^A,\overline Z_{f'}^A]
        =[\overline X_f^A,\overline X_{f'}^A]=0,\\
        \overline Z_{f'}^A\overline X_f^A
        =i^{\delta_{f,f'}}\overline X_f^A\overline Z_{f'}^A.
    \end{gathered}
\label{eq:z8-reduced-Weyl-algebra}
\end{equation}
Thus, this generates the $\ZZ_4$ clock and shift algebra for every face. We can then define the QCA for $U(1)_4$ as
\begin{equation}
    Z_f^A\longmapsto\overline Z_f^A,\qquad
    X_f^A\longmapsto\overline X_f^A
\label{eq:z8-u14-qca-map}
\end{equation}


\section{Generalized $U(1)_2$ and $U(1)_4$ QCAs in $4k-1$ dimensions}
\label{sec:generalized-u1-families}

The classification of Clifford QCAs from L-theory exhibits a period-four structure in spatial dimension, with nontrivial classes occurring in dimensions $d=4k-1$~\cite{haah2025topological}. In our previous work, we constructed explicit lattice representatives of these classes~\cite{Sun2026Clifford}. Here we show that an analogous period-four structure occurs for certain families beyond the Clifford setting\footnote{This statement does not imply a period-four structure for general non-Clifford QCAs. In particular, QCAs associated with more general cobordism invariants need not exhibit this periodicity.}. Specifically, we construct generalized QCAs associated with the chiral $U(1)_2$ semion and $U(1)_4$ theories in every spatial dimension $d=4k-1$ with $k > 1$.

As a particular example, we discuss the first higher-dimensional instance in this series in 7 spatial dimensions. The invertible topological phase underlying the generalized $U(1)_2$ QCA supports semionic membrane excitations on its 6-dimensional boundary~\cite{feng2026paulistatistics}; we therefore refer to the corresponding QCA as the semionic-membrane QCA. The cup-product formulation makes the extension to the entire period-four family transparent: the construction retains the same algebraic form, with the cochain degrees shifting appropriately as the spatial dimension increases by four.

\subsection{Generalized semion QCA}
\label{sec:7d-generalized-semion-qca}

In this section, we generalize the 3D semion QCA to $(4k-1)$ spatial dimensions, following the same condensation procedure as in the three-dimensional construction. 

The higher-form analogue of the
premodular $\mathbb Z_4$ Walker--Wang action is
\begin{equation}
\frac14\int_{M_{4k}}B_{2k}\cup B_{2k},
\qquad
 B_{2k}\in Z^{2k}(M_{4k},\mathbb Z_4).
\label{eq:7d-gs-z4-action}
\end{equation}

On the $4k-1$ dimensional hypercubic lattice, place a $\mathbb Z_4$ qudit on every $2k$-cell. We will denote the $2k$-cochains $\bsig$ and $2k-1$ cochains $\btau$. Integrals in the following expressions are implicitly over this hypercubic lattice.

The Walker-Wang Hamiltonian is
\begin{equation}
H_{\mathbb Z_4}
=-\frac12\sum_\nu
\left(Z_{\partial\nu}+Z_{\partial\nu}^{\dagger}\right)
-\frac12\sum_\tau
\left(Q_\tau+Q_\tau^{\dagger}\right).
\label{eq:7d-gs-z4-uncondensed}
\end{equation}
where $\nu$ ranges over elementary $2k+1$ cells and
\begin{equation}
\begin{aligned}
Q_\tau
&:=X_{\delta\btau}
\prod_{\sigma'}Z_{\sigma'}^{
\int(\delta\btau\cup_1\bsig'+2\btau\cup\bsig')}
=:G_\tau
\prod_{\sigma'}Z_{\sigma'}^{2\int\btau\cup\bsig'},
\qquad
G_\tau:=X_{\delta\btau}
\prod_{\sigma'}Z_{\sigma'}^{\int\delta\btau\cup_1\bsig'} .
\end{aligned}
\label{eq:7d-gs-z4-move}
\end{equation}
The model contains a transparent order-two excitation generated by the local operator
\begin{equation}
C_\sigma
:=X_\sigma^2
\prod_{\sigma'}Z_{\sigma'}^{
2\int(\bsig'\cup_1\bsig+\bsig\cup_2\delta\bsig')}.
\label{eq:7d-gs-condensate}
\end{equation}
The two terms in the $Z^2$ dressing cancel the sign obtained by commuting
$X_\sigma^2$ through $G_\tau$.  Explicitly,
\begin{equation}
\begin{aligned}
&\sum_{\sigma'}[\delta\btau]_2(\sigma')
\int\left(\bsig'\cup_1\bsig
+\bsig\cup_2\delta\bsig'\right)
=\int\delta\btau\cup_1\bsig
\quad (\mathrm{mod}\ 2).
\end{aligned}
\label{eq:7d-gs-condensate-commutation}
\end{equation}
which satisfies
\begin{equation} 
C_\sigma^2=1,\qquad
[C_\sigma,C_{\sigma'}]=[C_\sigma,Q_\tau]=0,\qquad
Z_{\partial\nu}C_\sigma
=(-1)^{[\delta\bsig]_2(\nu)}C_\sigma Z_{\partial\nu}.
\label{eq:7d-gs-condensate-algebra}
\end{equation}

After condensing $C_{\sigma}$, the commuting Hamiltonian becomes
\begin{equation}
H_{\mathrm{GS}}
=-\sum_\sigma C_\sigma
-\sum_\nu Z_{\partial\nu}^2
-\frac12\sum_\tau
\left(Q_\tau+Q_\tau^{\dagger}\right).
\label{eq:7d-gs-z4-condensed}
\end{equation}
Using the pairing between $\sigma$ and $\tau$ on the hypercubic lattice, define
\begin{equation}
\widetilde{Z}_\sigma
:=Z_\sigma^2
\prod_{\tau}G_\tau^{\int\btau\cup\bsig}
=\prod_\tau
Q_\tau^{\int\btau\cup\bsig}.
\label{eq:7d-gs-separator}
\end{equation}
We again find the $Z^2_{\partial \nu}$ is a redundant stabilizer generated by the product of $\widetilde{Z}_\sigma$'s and therefore can be removed from the Hamiltonian. Then
\begin{equation}
H_{\mathrm{GS,min}}
=-\sum_\sigma C_\sigma
-\frac12\sum_\sigma
\left(\widetilde Z_\sigma+\widetilde Z_\sigma^\dagger\right).
\label{eq:7d-gs-minimal-z4-Hamiltonian}
\end{equation}
Now we construct the flippers. 
\begin{equation}
V_\sigma
:=X_\sigma
\prod_{\sigma'}Z_{\sigma'}^{
\int(\bsig'\cup_1\bsig+\delta\bsig'\cup_2\bsig)}.
\label{eq:7d-gs-bare-flipper}
\end{equation}
Its commutation relation with the separator is
\begin{equation}
\widetilde{Z}_{\sigma'}V_\sigma
=(-1)^{\delta_{\sigma\sigma'}}
V_\sigma \widetilde{Z}_{\sigma'}.
\label{eq:7d-gs-bare-flipper-commutation}
\end{equation}
but it is not yet a commuting flipper because
\begin{equation}
\begin{aligned}
V_{\sigma_2}V_{\sigma_1}
=&i^{\int
\bsig_1\cup_1\bsig_2+\delta\bsig_1\cup_2\bsig_2
-\bsig_2\cup_1\bsig_1-\delta\bsig_2\cup_2\bsig_1}
V_{\sigma_1}V_{\sigma_2},
\end{aligned}
\label{eq:7d-gs-bare-commutator}
\end{equation}

We now decorate $V_{\sigma}$ with $\widetilde Z_\sigma$ so that the resulting flippers commute. Consider
\begin{equation}
\begin{aligned}
V_\sigma'
:=V_\sigma\prod_{\sigma'}\Bigg[
&(\widetilde S_{\sigma'}^\dagger)^{
\int(\bsig'\cup_1\bsig+\delta\bsig'\cup_2\bsig)}
CZ(\widetilde Z_\sigma,\widetilde Z_{\sigma'})^{
\int(\bsig\cup_1\bsig'+\delta\bsig\cup_2\bsig')}
\Bigg].
\end{aligned}
\label{eq:7d-gs-corrected-flipper}
\end{equation}
To restrict to the $C_\sigma=1$ we decompose each $\mathbb Z_4$ qudit into two qubits, denoted $A$ and $B$. Using the Clifford circuit
\begin{equation}
U_C
=\prod_{\sigma,\sigma'}
CX^{AB}(\sigma',\sigma)^{
\int(\bsig'\cup_1\bsig+\bsig\cup_2\delta\bsig')}
\label{eq:7d-gs-disentangler}
\end{equation}
and restricting to the subspace $Z^B=1$ we obtain
\begin{equation}
\begin{aligned}
\overline Z_\sigma^A
:=\left.
U_C \widetilde Z_\sigma U_C^\dagger
\right|_{Z^B=1}
=&Z_\sigma^A\prod_\tau \Bigg[
X_{\delta\btau}^A
\prod_{\sigma'\in\delta\btau} \Bigg\{
\left(Z_{\sigma'}^A\right)^{
\int\delta\btau\cup_1\bsig'}
\prod_{\sigma''}
CZ\!\left(Z_{\sigma'}^A,Z_{\sigma''}^A\right)^{
\int\left(
\bsig''\cup_1\bsig'
+\bsig'\cup_2\delta\bsig''
\right)} \Bigg\}
\\
&\quad\times
\prod_{\sigma'}\left(Z_{\sigma'}^A\right)^{
\int
\bsig'\cup_1(\btau\cup_{2k-1}\delta\btau)
+(\btau\cup_{2k-1}\delta\btau)\cup_2\delta\bsig'
}
\left(S_{\sigma'}^A\right)^{
\int\delta\btau\cup_1\bsig'}
\Bigg]^{\int\btau\cup\bsig},\\
\overline X_\sigma^A
:=\left.
U_CV'_\sigma U_C^\dagger
\right|_{Z^B=1} =& X_\sigma^A
\left(Z_\sigma^A\right)^{\int
\left(
\bsig\cup_1\bsig
+\delta\bsig\cup_2\bsig
\right)}
\\
&\times
\prod_{\sigma'}\Bigg[
\left(S_{\sigma'}^A\right)^{\int
\left(
\bsig'\cup_1\bsig
+\delta\bsig'\cup_2\bsig
\right)}
CZ\!\left(Z_\sigma^A,Z_{\sigma'}^A\right)^{\int
\left(
\bsig'\cup_1\bsig
+\bsig\cup_2\delta\bsig'
\right)}
\Bigg]
\\
&\times
\prod_{\sigma'}\Bigg[
\left(\overline S_{\sigma'}^\dagger\right)^{\int
\left(
\bsig'\cup_1\bsig
+\delta\bsig'\cup_2\bsig
\right)}
CZ\!\left(\overline Z_\sigma,\overline Z_{\sigma'}\right)^{\int
\left(
\bsig\cup_1\bsig'
+\delta\bsig\cup_2\bsig'
\right)}
\Bigg].
\label{eq:7d-gs-reduced-flipper}
\end{aligned}
\end{equation}
The QCA maps
\begin{equation}
    Z_\sigma^A\longmapsto\overline Z_\sigma^A, \qquad X_\sigma^A\longmapsto\overline X_\sigma^A.
    \label{eq:7d-gs-reduced-qca-map}
\end{equation}

\subsection{Generalized \texorpdfstring{$U(1)_4$}{U(1)4} QCA}
\label{sec:7d-u14-qca}

We now construct the corresponding higher-$U(1)_4$
QCA. We obtain an explicit parent Hamiltonian by shifting the degree of the
$\mathbb Z_8^{(1)}$ construction.  Place one $\ZZ_8$ qudit on every $2k$-cell
The bulk parent action is
\begin{equation}
\frac18\int_{M_{4k}}
 B_{2k}\cup B_{2k},
\qquad
 B_{2k}\in Z^{2k}(M_{4k},\mathbb Z_8).
\label{eq:7d-u14-z8-action}
\end{equation}
The corresponding Hamiltonian is again
\begin{equation}
\begin{aligned}
H_{\mathbb Z_8}
={}&-\frac12\sum_\nu
\left(Z_{\partial\nu}
+Z_{\partial\nu}^{\dagger}\right)
-\frac12\sum_\tau\left[
G_\tau
\prod_{\sigma'}(Z_{\sigma'})^{
2\int\btau\cup\bsig'}
+\mathrm{h.c.}\right],
\end{aligned}
\label{eq:7d-u14-z8-parent-Hamiltonian}
\end{equation}
where
\begin{equation}
G_\tau
:=X_{\delta\btau}
\prod_{\sigma'}(Z_{\sigma'})^{
\int\delta\btau\cup_1\bsig'}.
\label{eq:7d-u14-Gtau}
\end{equation}
The condensate and flipper have the same form as in the three-dimensional
construction, with all cochain degrees shifted by multiples of two.  Their dressings
therefore contain the same $\cup_1$ and $\cup_2$ products.

The transparent order-two excitation is condensed by
\begin{equation}
C_\sigma
:=(X_\sigma)^4
\prod_{\sigma'}(Z_{\sigma'})^{
4\int(\bsig'\cup_1\bsig+\bsig\cup_2\delta\bsig')}.
\label{eq:7d-u14-condensate}
\end{equation}

Define
\begin{equation}
\widetilde Z_\sigma
:=(Z_\sigma)^2
\prod_\tau(G_\tau)^{\int\btau\cup\bsig}.
\label{eq:7d-u14-separator}
\end{equation}
Multiplying the separators around a $2k+1$ cell $\nu$ gives
\begin{equation}
    \prod_\sigma \widetilde Z_\sigma^{(\delta\bsig)(\nu)} =(Z_{\partial\nu})^2.
    \label{eq:7d-u14-flux-from-separators}
\end{equation}
The minimal condensed Hamiltonian is therefore
\begin{equation}
H_{U(1)_4}^{(4k-1)}
=-\sum_\sigma C_\sigma
-\frac12\sum_\sigma
\left(\widetilde{Z}_\sigma
+\widetilde{Z}_\sigma^{\dagger}\right),
\label{eq:7d-u14-minimal-Hamiltonian}
\end{equation}
with no independent $2k+1$-cell flux term.

The non-commuting flippers are
\begin{equation}
V_\sigma
:=X_\sigma
\prod_{\sigma'}(Z_{\sigma'})^{
\int(\bsig'\cup_1\bsig+\delta\bsig'\cup_2\bsig)}.
\label{eq:7d-u14-bare-flipper}
\end{equation}
which satisfy
\begin{align}
V_{\sigma_2}V_{\sigma_1}
={}&
\exp\!\left\{\frac{2\pi i}{8}\int\left(
\bsig_1\cup_1\bsig_2+\delta\bsig_1\cup_2\bsig_2
-\bsig_2\cup_1\bsig_1-\delta\bsig_2\cup_2\bsig_1
\right)\right\}
V_{\sigma_1}V_{\sigma_2},
\label{eq:7d-u14-bare-flipper-commutator}\\
(V_\sigma)^4
={}&C_\sigma
\prod_{\sigma'}(\widetilde{Z}_{\sigma'})^{
2\int\delta\bsig'\cup_3\delta\bsig}.
\label{eq:7d-u14-bare-flipper-fourth-power}
\end{align}
We will decorate this with functions created from separators. Define
\begin{equation}
\widetilde S_\sigma
:=\sqrt{CZ(\widetilde Z_\sigma,\widetilde Z_\sigma)},
\label{eq:7d-u14-S-gate}
\end{equation}
where the order-four controlled phase $CZ$ is defined in Eq.~\eqref{eq:z8-CZ4}.  The corrected flipper is
\begin{equation}
\begin{aligned}
 V_\sigma'
:=V_\sigma
\prod_{\sigma'}\Big[
&(\widetilde S_{\sigma'}^\dagger)^{\int(
\bsig'\cup_1\bsig+\delta\bsig'\cup_2\bsig)}
 CZ^\dagger(
\widetilde Z_\sigma,\widetilde Z_{\sigma'})^{
\int(\bsig\cup_1\bsig'+\delta\bsig\cup_2\bsig')}
\Big].
\end{aligned}
\label{eq:7d-u14-corrected-flipper}
\end{equation}
The checks of the fourth power and mutual commutativity are the degree-shifted versions of the three-dimensional calculations, but are identical.

We use the same on-site
$\mathbb Z_8\to\mathbb Z_4\times\mathbb Z_2$ operator map as in Eq.~\eqref{eq:z8-ququart-qubit-map}, now on every $2k$-cell. Under this identification, the condensate becomes
\begin{equation}
C_\sigma =Z_\sigma^B
\prod_{\sigma'}(Z_{\sigma'}^A)^{
2\int(\bsig'\cup_1\bsig+\bsig\cup_2\delta\bsig')}.
\label{eq:7d-u14-encoded-condensate}
\end{equation}
Using the controlled gate in Eq.~\eqref{eq:z8-parity-controlled-X}, define
\begin{equation}
U_{C}
:=\prod_{\sigma,\sigma'}
{CX}^{AB}(\sigma',\sigma)^{
\int(\bsig'\cup_1\bsig+\bsig\cup_2\delta\bsig')}.
\label{eq:7d-u14-condensate-circuit}
\end{equation}
Conjugating $Z_\sigma^B$ by this circuit produces exactly the $Z^A$ factors in Eq.~\eqref{eq:7d-u14-encoded-condensate}.  Hence
\begin{equation}
U_{C} C_\sigma
U_{C}\dagger=Z_\sigma^B.
\label{eq:7d-u14-disentangled-condensate}
\end{equation}
Setting $Z_\sigma^B=1$ removes the $B$ qubits as product-state ancillas and leaves one physical $\ZZ_4$ qudit $A_\sigma$ on every $2k$-cell.

Both $\widetilde Z_\sigma$ and $V_\sigma'$ commute with the condensates. Their conjugated images therefore preserve the sector $Z_\sigma^B=1$. Using the on-site map above, we define
\begin{align}
\overline Z_\sigma^A
&:=\left.
U_{C}
(\widetilde Z_\sigma)
U_{C}^{\dagger}\right|_{Z^B=1},
\label{eq:7d-u14-reduced-separator}\\
\overline X_\sigma^A
&:=\left.
U_{C}
 (V_\sigma')
U_{C}^{\dagger}\right|_{Z^B=1}.
\label{eq:7d-u14-reduced-flipper}
\end{align}
They satisfy
\begin{equation}
\begin{gathered}
(\overline Z_\sigma^A)^4=(\overline X_\sigma^A)^4=1,\qquad
[\overline Z_\sigma^A,\overline Z_{\sigma'}^A]
=[\overline X_\sigma^A,\overline X_{\sigma'}^A]=0,\\
\overline Z_{\sigma'}^A\overline X_\sigma^A
=i^{\delta_{\sigma,\sigma'}}
\overline X_\sigma^A\overline Z_{\sigma'}^A.
\end{gathered}
\label{eq:7d-u14-reduced-Pauli-algebra}
\end{equation}
Thus
\begin{equation}
Z_\sigma^A\longmapsto\overline Z_\sigma^A,\qquad
X_\sigma^A\longmapsto\overline X_\sigma^A
\label{eq:7d-u14-qca-map}
\end{equation}
defines the $U(1)_4$ QCA on $\ZZ_4$ qudits.

\section{$w_2w_3$ QCA in four spatial dimensions}
\label{sec:four-dimensional}

This section revisits the four-dimensional QCA associated with the $(4{+}1)$-dimensional $w_2w_3$ phase, originally constructed in Ref.~\cite{chen2023exactly}. Although that work established the QCA, several structural features of the construction were left implicit. In particular, it was not explained why the flux constraints are generated by the remaining commuting terms and are therefore redundant, nor how those terms furnish a complete separator--flipper algebra. Here, we make these points explicit and develop a formulation that will serve as the foundation for the higher-dimensional constructions considered below.

Starting from a cocycle representative of the action, we derive and gauge
the operators generated by arbitrary cochain shifts.  We show that every
flux constraint is a product of these commuting operators, so the flux
terms may be omitted without changing the ground space or the generated
commuting algebra.  The
remaining operators provide a natural set of separators.  We construct
their flippers and verify the required order and commutation relations,
completing the separator--flipper description of the $w_2w_3$ QCA.

We consider a $\mathbb{Z}_2$ one-form and $\mathbb{Z}_2$ two-form gauge
theory in $(4{+}1)$ dimensions with the topological action
\cite{chen2023exactly}
\begin{equation}
    S_{w_2w_3}(A_2,B_3):=\frac{1}{2}\int
    \bigl[B_3\cup_1 B_3+A_2\cup B_3
    +A_2\cup(A_2\cup_1 A_2)\bigr]~,
\label{eq:w2w3-action}
\end{equation}
where $A_2$ and $B_3$ are dynamical $\mathbb{Z}_2$ cocycles on a closed oriented
five-manifold.  In the integrated cohomology class, the Wu formula gives
\(
\int B_3\cup_1 B_3=\int \operatorname{Sq}^2B_3=\int w_2\cup B_3
\) modulo two.  The first two terms in Eq.~\eqref{eq:w2w3-action} may therefore be combined as $(w_2+A_2)\cup B_3$ under the integral.  Summing over $B_3$ imposes $A_2=w_2$ in cohomology.  On an oriented manifold, $\operatorname{Sq}^1w_2=w_3$; hence the final term, with $w_2\cup_1w_2$ representing $\operatorname{Sq}^1w_2$, reduces the action to the topological invariant
\begin{equation}
    \frac{1}{2} \int w_2\cup w_3~,
\label{eq:w2w3-oriented-response}
\end{equation}
which classifies the invertible topological phase.

We now derive the parent higher-form SPT Hamiltonian, which gauges to this invertible phase.
Before gauging, place one
qubit on every edge and one qubit on every face, with computational-basis
configurations $\alpha\in C^1(M_4,\mathbb Z_2)$ and
$\beta\in C^2(M_4,\mathbb Z_2)$.  Define
\begin{equation}
    \phi_4(\alpha,\beta)
    :=\beta\cup\beta+\beta\cup_1\delta\beta
    +\alpha\cup\bigl(\delta\alpha\cup_1\delta\alpha
    +\delta\beta\bigr).
\label{eq:w2w3-parent-phase}
\end{equation}
If $\omega_5(A_2,B_3)$ denotes the bracketed integrand in
Eq.~\eqref{eq:w2w3-action}, then the cochain-level transgression identity is
\begin{equation}
    \delta\phi_4(\alpha,\beta)
    =\omega_5(\delta\alpha,\delta\beta)
    \qquad (\mathrm{mod}\ 2).
\label{eq:w2w3-transgression}
\end{equation}
Start from the trivial paramagnet
\begin{equation}
    H_0=-\sum_eX_e-\sum_fX_f.
\label{eq:w2w3-trivial-parent}
\end{equation}
Edges and faces carry the $\alpha$ and $\beta$ degrees of freedom,
respectively.
The SPT entangler is the unitary
$U\lvert\alpha,\beta\rangle
=(-1)^{\int\phi_4(\alpha,\beta)}
\lvert\alpha,\beta\rangle$. 
For later use, we determine its conjugation action not only on a single
Pauli $X$, but also on an arbitrary product of $X$ operators. This will
allow us to show that the stabilizers imposing the flatness conditions of
higher-form symmetries are redundant and may be removed.

For arbitrary cochains
$\epsilon_1\in C^1(M_4,\mathbb Z_2)$ and
$\varphi_2\in C^2(M_4,\mathbb Z_2)$, we define
\begin{equation}
\begin{aligned}
    X_{\epsilon_1}
    &:=\prod_e(X_e)^{\epsilon_1(e)},&
    T_{\epsilon_1}
    &:=UX_{\epsilon_1}U^\dagger,\\
    X_{\varphi_2}
    &:=\prod_f(X_f)^{\varphi_2(f)},&
    T_{\varphi_2}
    &:=UX_{\varphi_2}U^\dagger.
\end{aligned}
\label{eq:w2w3-parent-general-generator}
\end{equation}
First consider $T_{\epsilon_1}$. Since $X_{\epsilon_1}$ maps basis elements as
$\lvert\alpha,\beta\rangle\mapsto
\lvert\alpha+\epsilon_1,\beta\rangle$, the conjugated operator acts as
\begin{align}
    T_{\epsilon_1} \ket{\alpha,\beta} = (-1)^{\int\phi_4(\alpha +\epsilon_1,\beta) - \phi_4(\alpha ,\beta)}  \ket{\alpha+\epsilon_1,\beta}
\end{align}
So the relevant phase change is the cochain
\begin{equation}
\begin{aligned}
\Delta_{\epsilon_1}\phi_4(\alpha,\beta)
&:=\phi_4(\alpha+\epsilon_1,\beta)-\phi_4(\alpha,\beta)
=\epsilon_1\cup(a\cup_1a+b)
+(\alpha+\epsilon_1)\cup
\bigl(a\cup_1\delta\epsilon_1
+\delta\epsilon_1\cup_1a
+\delta\epsilon_1\cup_1\delta\epsilon_1\bigr).
\end{aligned}
\label{eq:w2w3-parent-general-difference}
\end{equation}
Here $a:=\delta\alpha$ and $b:=\delta\beta$.  Since $\delta a=0$ and $\delta^2\epsilon_1=0$, the higher-cup recursion gives
\begin{equation}
\begin{aligned}
\delta\bigl[
a\cup_2\delta\epsilon_1
+\epsilon_1\cup_1\delta\epsilon_1
+\epsilon_1\cup\epsilon_1
\bigr]
={}&a\cup_1\delta\epsilon_1
+\delta\epsilon_1\cup_1a
+\delta\epsilon_1\cup_1\delta\epsilon_1.
\end{aligned}
\label{eq:w2w3-general-epsilon-one-primitive}
\end{equation}
Therefore,
\begin{equation}
\begin{aligned}
\int \Delta_{\epsilon_1}\phi_4(\alpha,\beta)
=\int \Bigl[&\epsilon_1\cup(a\cup_1a+b)
+(a+\delta\epsilon_1)\cup
\bigl(a\cup_2\delta\epsilon_1
+\epsilon_1\cup_1\delta\epsilon_1
+\epsilon_1\cup\epsilon_1\bigr)\Bigr].
\end{aligned}
\label{eq:w2w3-general-epsilon-one-phase}
\end{equation}

Similarly, shifting $\beta$ by the general $2$-cochain $\varphi_2$ changes the phase according to
\begin{equation}
    \int \Delta_{\varphi_2}\phi_4(\alpha,\beta)
    =\int \bigl[
    a\cup\varphi_2+b\cup_2\delta \varphi_2
    +\varphi_2\cup\varphi_2+\varphi_2\cup_1\delta \varphi_2
    \bigr].
\label{eq:w2w3-general-phi-two-phase}
\end{equation}
The last two terms are essential when $\varphi_2$ contains more than one face.

Thus, 
\begin{equation}
\begin{aligned}
T_{\epsilon_1} &= (-1)^{\int\Delta_{\epsilon_1}\phi_4} X_{\epsilon_1},\\
T_{\varphi_2} &= (-1)^{\int\Delta_{\varphi_2}\phi_4} X_{\varphi_2}
\end{aligned}
\label{eq:w2w3-parent-general-actions}
\end{equation}

For the basis cochains labeled by edges $\be$ and faces $\bface$, the corresponding phases are
\begin{equation}
\begin{aligned}
\int\Delta_{\be}\phi_4
&=\int\Bigl[
a\cup\bigl(\be\cup_1\delta\be+a\cup_2\delta\be\bigr)
+\be\cup(a\cup_1a+b)\Bigr],\\
\int\Delta_{\bface}\phi_4
&=\int\bigl[b\cup_2\delta\bface+a\cup\bface\bigr].
\end{aligned}
\label{eq:w2w3-parent-elementary-phases}
\end{equation}
In the second line we used
$\bface\cup\bface+\bface\cup_1\delta\bface=0$ for a single face
indicator.  The canonical local parent Hamiltonian is therefore
\begin{equation}
    H_{\mathrm{SPT}}
    :=UH_0U^\dagger
    =-\sum_eT_{\be}-\sum_fT_{\bface}.
\label{eq:w2w3-parent-spt-hamiltonian}
\end{equation}
Its terms are commuting involutions.  On a finite closed lattice, the
unique ground state is
\begin{equation}
\lvert\Psi_{\mathrm{SPT}}\rangle
=U\left(\bigotimes_e\lvert+\rangle_e^\alpha\right)
\left(\bigotimes_f\lvert+\rangle_f^\beta\right).
\label{eq:w2w3-parent-ground-state}
\end{equation}

Now we act with the gauging map. The Hilbert space after gauging consists of faces (which we label $A$) and 3-cells (which we label $B$). The gauging map acts as 
\begin{equation}
\begin{aligned}
    \lvert\alpha,\beta\rangle
    &\longmapsto\lvert\delta\alpha,\delta\beta\rangle,\\
    X_{\epsilon_1}
    &\longmapsto
    \prod_e\bigl(X^A_{\delta\be}\bigr)^{\epsilon_1(e)}
    =X^A_{\delta\epsilon_1},\\
    X_{\varphi_2}
    &\longmapsto
    \prod_f\bigl(X^B_{\delta\bface}\bigr)^{\varphi_2(f)}
    =X^B_{\delta\varphi_2}.
\end{aligned}
\label{eq:w2w3-general-cochain-gauging-map}
\end{equation}
On the image of the gauging map, $a=\delta\alpha$ and $b=\delta\beta$ are
exact.  In the enlarged gauged Hilbert space, however, we use the same
symbols for arbitrary configurations
$a\in C^2(M_4,\mathbb Z_2)$ and $b\in C^3(M_4,\mathbb Z_2)$.  
We denote the corresponding gauged stabilizers with tildes. The gauged
$2$-cochain generator is
\begin{equation}
\begin{aligned}
\widetilde T_{\varphi_2}
:={}&(-1)^{\int[
\varphi_2\cup\varphi_2+\varphi_2\cup_1\delta \varphi_2]}
X^B_{\delta \varphi_2}
\prod_{t'}(Z_{t'}^B)^{\int\bt'\cup_2\delta \varphi_2}
\prod_{f'}(Z_{f'}^A)^{\int\bface'\cup\varphi_2}.
\end{aligned}
\label{eq:w2w3-general-phi-two-gauged-generator}
\end{equation}
For an elementary face $\varphi_2=\bface$, the $\varphi_2\cup\varphi_2+\varphi_2\cup_1\delta \varphi_2$ vanishes. Thus,
\begin{equation}
    \widetilde T_{\bface}
    =G_f^B\prod_{f'}(Z_{f'}^A)^{\int\bface'\cup\bface},
    \qquad
    G_f^B:=X^B_{\delta\bface}
    \prod_{t'}(Z_{t'}^B)^{\int\bt'\cup_2\delta\bface}.
\label{eq:w2w3-Gf}
\end{equation}

Expanding the phase in Eq.~\eqref{eq:w2w3-general-epsilon-one-phase} gives
the gauged $1$-cochain generator.
\begin{equation}
\begin{aligned}
\widetilde T_{\epsilon_1}
:={}&(-1)^{\int\delta\epsilon_1\cup
(\epsilon_1\cup_1\delta\epsilon_1+\epsilon_1\cup\epsilon_1)}
X^A_{\delta\epsilon_1}
\prod_f(Z_f^A)^{\int\left[
\bface\cup(\epsilon_1\cup_1\delta\epsilon_1
+\epsilon_1\cup\epsilon_1)
+\delta\epsilon_1\cup(\bface\cup_2\delta\epsilon_1)\right]}\\
&\times\prod_t(Z_t^B)^{\int\epsilon_1\cup\bt}
\prod_{f_1,f_2}CZ(Z_{f_1}^A,Z_{f_2}^A)^{
\int\left[\epsilon_1\cup(\bface_1\cup_1\bface_2)
+\bface_1\cup(\bface_2\cup_2\delta\epsilon_1)\right]}.
\end{aligned}
\label{eq:w2w3-general-epsilon-one-gauged-generator}
\end{equation}
Diagonal pairs in the last product are interpreted as
$CZ(Z_f^A,Z_f^A)=Z_f^A$.   For the special case of an
elementary edge cochain $\epsilon_1=\be$, integration by parts and the
elementary-cell identities reduce this expression, on a closed $M_4$
modulo a total coboundary, to
\begin{equation}
\begin{aligned}
\widetilde T_{\be}
={}&X^A_{\delta\be}
\prod_f(Z_f^A)^{\int\bface\cup(\be\cup_1\delta\be)}
\prod_t(Z_t^B)^{\int\be\cup\bt}
\prod_{f_1,f_2}CZ(Z_{f_1}^A,Z_{f_2}^A)^{
\int\left[\be\cup(\bface_1\cup_1\bface_2)
+\bface_1\cup(\bface_2\cup_2\delta\be)\right]}.
\end{aligned}
\label{eq:w2w3-elementary-edge-generator}
\end{equation} 
This is the other term in the gauged Hamiltonian.

Combining the two gauged kinetic families with the two flux families gives
\begin{equation}
\begin{aligned}
H_{w_2w_3}
:={}&
-\sum_f \widetilde T_{\bface}
-\sum_e \widetilde T_{\be}
-\sum_tZ_{\partial t}^A-\sum_{c_4}Z_{\partial c_4}^B.
\end{aligned}
\label{eq:w2w3-hamiltonian}
\end{equation}
This is Eq.~(3.22) of Ref.~\cite{chen2023exactly}.  We now show that its
two explicit flux sums are redundant.  On the hypercubic lattice, define
the edge cochain complementary to a 3-cell $t$ and retain only the two
kinetic families:
\begin{equation}
\begin{aligned}
H_{w_2w_3}^{\mathrm{kin}}
:={}&-\sum_{f} \widetilde T_{\bface}
-\sum_t \widetilde T_{\epsilon_{1,t}}, \qquad
\epsilon_{1,t}:=\sum_e
\left(\int_{M_4}\be\cup\bt\right)\be.
\end{aligned}
\label{eq:w2w3-two-family-hamiltonian}
\end{equation}
Hypercubic complementarity makes every $\epsilon_{1,t}$ a single
elementary edge and gives
$\int_{M_4}\epsilon_{1,t}\cup\bt'=\delta_{tt'}$.  In particular,
the term $\epsilon_{1,t}\cup b$ in
Eq.~\eqref{eq:w2w3-general-epsilon-one-phase} supplies the second family with
one undressed $Z_t^B$.  Thus both sums in
Eq.~\eqref{eq:w2w3-two-family-hamiltonian} have separator form.

We first use only the first sum.  Let $e(t)$ be the edge paired with $t$
under the reversed cup pairing, $\int_{M_4}\bt\cup\be(t)=1$.  The
complementary-cell incidence identity gives
\cite{Chen2023HigherCup}
\begin{equation}
\begin{aligned}
&\prod_{f'}
\left[
Z_{f'}^A
\prod_f(G_f^B)^{\int_{M_4}\bface'\cup\bface}
\right]^{(\delta\bface')(t)}
=Z_{\partial t}^A
\prod_f(G_f^B)^{
\sum_{f'}(\delta\bface')(t)
\int_{M_4}\bface'\cup\bface}
=Z_{\partial t}^A
\prod_f(G_f^B)^{(\delta\be(t))(f)}
=Z_{\partial t}^A.
\end{aligned}
\label{eq:w2w3-A-flux-from-first-family}
\end{equation}
The last step follows from
\begin{equation*}
\prod_f
(G_f^B)^{(\delta\be(t))(f)}
=
(-1)^{\int_{M_4}\delta\be(t)\cup\delta\be(t)}
=1.
\end{equation*}

On a basis state,
\begin{equation}
Z_{\partial t}^A\lvert a,b\rangle
=(-1)^{(\delta a)(t)}\lvert a,b\rangle.
\label{eq:w2w3-A-flatness-from-first-family}
\end{equation}
Consequently, the common $+1$ constraints from the first sum of
Eq.~\eqref{eq:w2w3-two-family-hamiltonian} imply $\delta a=0$.

We may now use $\delta a=0$ when analyzing the second sum.  For a
four-cell $c_4$, let $v=v(c_4)$ be its complementary vertex,
$\int_{M_4}\bv\cup\bc_4=1$.  Complementary-cell incidence gives the
cochain identity
\begin{equation}
\sum_t(\delta\bt)(c_4)\epsilon_{1,t}=\delta\bv.
\label{eq:w2w3-boundary-pairing}
\end{equation}
The exact endpoint product law therefore yields
\begin{equation}
\prod_t
\left(\widetilde T_{\epsilon_{1,t}}\right)^{(\delta\bt)(c_4)}
=\widetilde T_{\delta\bv}
\qquad\text{on }~\delta a=0.
\label{eq:w2w3-boundary-general-generator}
\end{equation}
Because $\delta^2\bv=0$, this composite does not shift $a$.  Substituting
$\epsilon_1=\delta\bv$ into
Eq.~\eqref{eq:w2w3-general-epsilon-one-gauged-generator}, its action is
\begin{equation}
\begin{aligned}
\left.\widetilde T_{\delta\bv}\lvert a,b\rangle\right|_{\delta a=0}
&=(-1)^{\int_{M_4}\delta\bv\cup(a\cup_1a+b)}
\lvert a,b\rangle
=(-1)^{\int_{M_4}\bv\cup(
\delta a\cup_1a+a\cup_1\delta a+\delta b)}
\lvert a,b\rangle\\
&=(-1)^{\int_{M_4}\bv\cup\delta b}\lvert a,b\rangle
=(-1)^{(\delta b)(c_4)}\lvert a,b\rangle
=Z_{\partial c_4}^B\lvert a,b\rangle.
\end{aligned}
\label{eq:w2w3-B-flux-from-second-family}
\end{equation}
The equalities follow from integration by parts and the higher-cup recursion.
Hence, the common $+1$ constraints from the second sum, together with the already-derived condition $\delta a=0$, imply $\delta b=0$.  
The two kinetic families in Eq.~\eqref{eq:w2w3-two-family-hamiltonian} therefore generate both flux families and have the same common $+1$ eigenspace as Eq.~\eqref{eq:w2w3-hamiltonian}.

Within its common $+1$ sector, the first family in
Eq.~\eqref{eq:w2w3-two-family-hamiltonian} gives the constraint
\begin{equation}
Z_f^A
\prod_{f'}(G_{f'}^B)^{\int_{M_4}\bface\cup\bface'}=1
\qquad\Longrightarrow\qquad
Z_f^A
=\prod_{f'}(G_{f'}^B)^{\int_{M_4}\bface\cup\bface'}.
\label{eq:w2w3-first-family-substitution}
\end{equation}
We can therefore obtain the second separator family directly by replacing
every $Z_f^A$ in the second term of
Eq.~\eqref{eq:w2w3-two-family-hamiltonian} by the product on the
right-hand side of Eq.~\eqref{eq:w2w3-first-family-substitution}.  The
products of $G_f^B$ that occur in this substitution are commuting
involutions, so their controlled phases are well defined.

To define a full-space completion, we first move the bare $Z^B_t$ in
Eq.~\eqref{eq:w2w3-elementary-edge-generator} to the left, before making the replacement. We then temporarily factor out this $Z^B_t$ and replace
each $Z_f^A$ in the remaining edge dressing by the corresponding product of $G^B$ operators. This defines
\begin{equation}
\begin{aligned}
\widehat T_e^A
:={}&X_{\delta\be}^A
\prod_f
\left[
\prod_{f'}(G_{f'}^B)^{
\int\bface\cup\bface'}
\right]^{
\int\bface\cup(\be\cup_1\delta\be)}
\\
&\times
\prod_{f_1,f_2}
CZ\!\left(
\prod_{f'_1}(G_{f'_1}^B)^{
\int\bface_1\cup\bface'_1},
\prod_{f'_2}(G_{f'_2}^B)^{
\int\bface_2\cup\bface'_2}
\right)^{
\int
\be\cup(\bface_1\cup_1\bface_2)
+\bface_1\cup(\bface_2\cup_2\delta\be)}.
\end{aligned}
\label{eq:w2w3-completed-edge-move}
\end{equation}
By Eq.~\eqref{eq:w2w3-two-family-hamiltonian}, hypercubic complementarity gives
\begin{equation*}
\prod_{t'}
(Z_{t'}^B)^{
\int_{M_4}\epsilon_{1,t}\cup\bt'}
=
Z_t^B.
\end{equation*}
Thus $Z_t^B$ is the unique undressed $Z^B$ factor in
$\widetilde T_{\epsilon_{1,t}}$, while the remaining dressing is obtained from
the corresponding product of the operators
$\widehat T_e^A$.

We therefore define the two separator families by
\begin{equation}
\boxed{
\begin{aligned}
\overline Z_f^A
&:=
Z_f^A
\prod_{f'}(G_{f'}^B)^{
\int_{M_4}\bface\cup\bface'},
\\
\overline Z_t^B
&:=
Z_t^B
\prod_e(\widehat T_e^A)^{
\int_{M_4}\be\cup\bt}.
\end{aligned}}
\label{eq:w2w3-completed-separators}
\end{equation}
All incidence numbers and operator exponents below are understood modulo
$2$. On the oriented hypercubic cellulation, the pairing between edges
and three-cells is perfect. Thus, for every three-cell $t$, there is a
unique edge $e$ such that
\begin{equation*}
\int_{M_4}\be\cup\bt=1
\pmod 2
\end{equation*}
Consequently, $\overline Z_t^B=Z_t^B\widehat T_e^A$.
On the simultaneous $\overline Z_f^A=1$ sector, the first line of
Eq.~\eqref{eq:w2w3-completed-separators} gives
\begin{equation*}
Z_f^A
=
\prod_{f'}(G_{f'}^B)^{
\int_{M_4}\bface\cup\bface'}.
\end{equation*}
The second line therefore agrees with the original kinetic operator
$\widetilde T_{\epsilon_{1,t}}$ on this sector. Away from this sector,
Eq.~\eqref{eq:w2w3-completed-separators} specifies a local completion on
the full tensor-product Hilbert space.

We now verify the separator algebra on the full Hilbert space. The
operators $G_f^B$ are commuting involutions:
\begin{equation*}
(G_f^B)^2=1,
\qquad
[G_f^B,G_{f'}^B]=0.
\end{equation*}
Moreover, the paired $G^B$ products satisfy the same local flatness
relation as the original $A$ variables. For every 3-cell $s$, let
$e(s)$ be its unique complementary edge under the reversed cup pairing,
characterized by
\begin{equation*}
\int_{M_4}\boldsymbol s\cup\be'
=
\delta_{e',e(s)}
\pmod 2
\end{equation*}
for every elementary edge $e'$. Complementary-cell incidence then gives
\begin{equation}
\begin{aligned}
\prod_f
\left[
\prod_{f'}(G_{f'}^B)^{
\int_{M_4}\bface\cup\bface'}
\right]^{(\delta\bface)(s)}
&=
\prod_{f'}(G_{f'}^B)^{
\sum_f(\delta\bface)(s)
\int_{M_4}\bface\cup\bface'}
=
\prod_{f'}(G_{f'}^B)^{(\delta\be(s))(f')}
=1.
\end{aligned}
\label{eq:w2w3-paired-G-flatness}
\end{equation}

Next, fix a face $f$ and a complementary pair $(e,t)$ under the cup pairing
in the order $\be\cup\bt$. This pairing should be distinguished from the
reversed pairing $\boldsymbol s\cup\be(s)$ used above. Directly from the
definition of $G_{f'}^B$ and complementary-cell incidence,
\begin{equation}
\begin{aligned}
Z_t^B
\left[
\prod_{f'}(G_{f'}^B)^{
\int_{M_4}\bface\cup\bface'}
\right]
Z_t^B
&=
(-1)^{
\sum_{f'}
\left(\int_{M_4}\bface\cup\bface'\right)
(\delta\bface')(t)}
\prod_{f'}(G_{f'}^B)^{
\int_{M_4}\bface\cup\bface'}
=
(-1)^{(\delta\be)(f)}
\prod_{f'}(G_{f'}^B)^{
\int_{M_4}\bface\cup\bface'}.
\end{aligned}
\label{eq:w2w3-complementary-G-conjugation}
\end{equation}
Thus conjugation by $Z_t^B$ shifts each substituted face variable by
$(-1)^{(\delta\be)(f)}$. This is precisely the endpoint shift required
when two parent edge moves are composed.

Indeed, recall that
\begin{equation*}
\Delta^\alpha_{\be}\phi_4(\alpha,\beta)
=
\phi_4(\alpha+\be,\beta)
-\phi_4(\alpha,\beta)
\pmod 2.
\end{equation*}
Applying the same move twice gives
\begin{equation}
\begin{aligned}
&\Delta^\alpha_{\be}\phi_4(\alpha+\be,\beta)
+\Delta^\alpha_{\be}\phi_4(\alpha,\beta)
=
\phi_4(\alpha+2\be,\beta)
-\phi_4(\alpha,\beta)
=0,
\end{aligned}
\label{eq:w2w3-parent-square-telescoping}
\end{equation}
because $2\be=0$ over $\mathbb Z_2$. Similarly, two distinct moves satisfy
\begin{equation}
\begin{aligned}
&\Delta^\alpha_{\be}\phi_4(\alpha+\be',\beta)
+\Delta^\alpha_{\be'}\phi_4(\alpha,\beta)
=
\phi_4(\alpha+\be+\be',\beta)
-\phi_4(\alpha,\beta)
=
\Delta^\alpha_{\be'}\phi_4(\alpha+\be,\beta)
+\Delta^\alpha_{\be}\phi_4(\alpha,\beta).
\end{aligned}
\label{eq:w2w3-parent-commutation-telescoping}
\end{equation}
The ordered composite $Z_t^B\widehat T_e^A$ realizes the full substituted
parent move: $Z_t^B$ supplies the bare $Z^B$ factor omitted from
$\widehat T_e^A$, while
Eq.~\eqref{eq:w2w3-complementary-G-conjugation} implements the required
endpoint shift in its remaining factors. Therefore
Eq.~\eqref{eq:w2w3-parent-square-telescoping} gives
\begin{equation}
\left(
Z_t^B\widehat T_e^AZ_t^B
\right)
\widehat T_e^A
=1,
\label{eq:w2w3-completed-edge-square-identity}
\end{equation}
and, for complementary pairs $(e,t)$ and $(e',s)$,
Eq.~\eqref{eq:w2w3-parent-commutation-telescoping} gives
\begin{equation}
\left(
Z_s^B\widehat T_e^AZ_s^B
\right)
\widehat T_{e'}^A
=
\left(
Z_t^B\widehat T_{e'}^AZ_t^B
\right)
\widehat T_e^A.
\label{eq:w2w3-completed-edge-commutation-identity}
\end{equation}
The bare $X^A$ shifts introduce no additional phase, since
\begin{equation*}
(X_{\delta\be}^A)^2=1,
\qquad
X_{\delta\be}^AX_{\delta\be'}^A
=
X_{\delta\be'}^AX_{\delta\be}^A.
\end{equation*}
It now follows directly that
\begin{equation*}
\left(\overline Z_t^B\right)^2
=
Z_t^B\widehat T_e^AZ_t^B\widehat T_e^A
=1.
\end{equation*}
For complementary pairs $(e,t)$ and $(e',t')$,
\begin{equation*}
\begin{aligned}
\overline Z_t^B\overline Z_{t'}^B
&=
Z_t^BZ_{t'}^B
\left(
Z_{t'}^B\widehat T_e^AZ_{t'}^B
\right)
\widehat T_{e'}^A
=
Z_{t'}^BZ_t^B
\left(
Z_t^B\widehat T_{e'}^AZ_t^B
\right)
\widehat T_e^A
=
\overline Z_{t'}^B\overline Z_t^B,
\end{aligned}
\end{equation*}
where the second equality uses
Eq.~\eqref{eq:w2w3-completed-edge-commutation-identity}.

The first separator family also consists of commuting involutions, since
the $Z_f^A$ operators commute with all $G_{f'}^B$. It remains to verify
the mixed commutator. The operator $\widehat T_e^A$ commutes with every
paired $G^B$ product because it is built from $X^A$ operators and
commuting functions of the $G_{f'}^B$. Using
Eq.~\eqref{eq:w2w3-complementary-G-conjugation}, we obtain
\begin{equation*}
\begin{aligned}
\overline Z_t^B\overline Z_f^A
&=
Z_t^B\widehat T_e^A
Z_f^A
\prod_{f'}(G_{f'}^B)^{
\int_{M_4}\bface\cup\bface'}
=
(-1)^{(\delta\be)(f)}
Z_f^AZ_t^B
\left[
\prod_{f'}(G_{f'}^B)^{
\int_{M_4}\bface\cup\bface'}
\right]
\widehat T_e^A
\\
&=
(-1)^{2(\delta\be)(f)}
Z_f^A
\left[
\prod_{f'}(G_{f'}^B)^{
\int_{M_4}\bface\cup\bface'}
\right]
Z_t^B\widehat T_e^A
=
\overline Z_f^A\overline Z_t^B.
\end{aligned}
\end{equation*}
The first sign arises when $Z_f^A$ is commuted through
$X_{\delta\be}^A$, while the second identical sign arises when the paired
$G^B$ product is commuted through $Z_t^B$. The two signs cancel.

We have therefore established the full separator algebra:
\begin{equation}
\begin{gathered}
\left(\overline Z_f^A\right)^2
=
\left(\overline Z_t^B\right)^2
=1,
\quad
\left[\overline Z_f^A,\overline Z_{f'}^A\right]
=
\left[\overline Z_f^A,\overline Z_t^B\right]
=
\left[\overline Z_t^B,\overline Z_{t'}^B\right]
=0.
\end{gathered}
\label{eq:w2w3-completed-separator-algebra}
\end{equation}
To prove independence and completeness, we construct one bounded-range
flipper for each separator:
\begin{equation}
\boxed{
\begin{aligned}
\overline X_f^A
&:=
X_f^A,
\\
\overline X_t^B
&:=
\left[
X_t^B
\prod_{t'}(Z_{t'}^B)^{
\int_{M_4}\bt\cup_2\bt'}
\right]
\left[
\prod_{t'}(\overline Z_{t'}^B)^{
\int_{M_4}\bt\cup_2\bt'}
\right].
\end{aligned}}
\label{eq:w2w3-flippers}
\end{equation}
Consider
the first bracket in $\overline X_t^B$. Its $X_t^B$ factor overlaps with
the $Z^B$ dressing of $G_f^B$ with exponent
$\int_{M_4}\bt\cup_2\delta\bface$,
while its $Z^B$ dressing overlaps with $X_{\delta\bface}^B$ with exponent
\begin{equation*}
\sum_{t'}(\delta\bface)(t')
\int_{M_4}\bt\cup_2\bt'
=
\int_{M_4}\bt\cup_2\delta\bface.
\end{equation*}
The two overlap phases therefore cancel modulo two, so the first bracket
commutes with every $G_f^B$. Moreover,
Eq.~\eqref{eq:w2w3-completed-edge-move} shows that
$\widehat T_e^A$ consists of the shift $X_{\delta\be}^A$ multiplied by
products and controlled phases of the commuting $G_f^B$ operators, with
no remaining bare $Z^B$ factor. The first bracket therefore commutes with
every $\widehat T_e^A$ and, among the $B$ separators, flips only the
explicit $Z_t^B$ in $\overline Z_t^B$.

The second bracket in $\overline X_t^B$ is a product of mutually commuting
separators. It therefore commutes with every separator and does not alter
this pairing. Similarly, $\overline X_f^A=X_f^A$ anticommutes only with
the explicit $Z_f^A$ in $\overline Z_f^A$. It commutes with every
$B$ separator because $\widehat T_e^A$ contains only $X^A$ shifts and
functions of the $G_f^B$, with no $Z^A$ operators. Consequently,
\begin{equation}
\overline Z_{f'}^A\overline X_f^A
=
(-1)^{\delta_{ff'}}
\overline X_f^A\overline Z_{f'}^A,
\qquad
\overline Z_{t'}^B\overline X_t^B
=
(-1)^{\delta_{tt'}}
\overline X_t^B\overline Z_{t'}^B,
\label{eq:w2w3-separator-flipper-pairing}
\end{equation}
and all cross-species separator--flipper commutators vanish.

It remains to verify the algebra among the flippers. The first bracket in
$\overline X_t^B$ has square phase
$(-1)^{\int_{M_4}\bt\cup_2\bt}$. Since it flips only
$\overline Z_t^B$, commuting it through the second bracket produces the
same phase. Hence
\begin{equation*}
\left(\overline X_t^B\right)^2
=
(-1)^{\int_{M_4}\bt\cup_2\bt}
(-1)^{\int_{M_4}\bt\cup_2\bt}
=1.
\end{equation*}
For distinct three-cells $t$ and $t'$, the mutual commutator of their
first brackets is
\begin{equation*}
(-1)^{
\int_{M_4}
\left(
\bt\cup_2\bt'
+
\bt'\cup_2\bt
\right)}.
\end{equation*}
Commuting the first bracket in $\overline X_t^B$ through the separator
bracket in $\overline X_{t'}^B$ contributes
$(-1)^{\int_{M_4}\bt'\cup_2\bt}$, while the opposite cross-commutation
contributes $(-1)^{\int_{M_4}\bt\cup_2\bt'}$. Their product reproduces
the same mutual-commutator phase. Therefore,
\begin{equation*}
\begin{aligned}
\overline X_t^B\overline X_{t'}^B
&=
(-1)^{
\int_{M_4}
\left(
\bt\cup_2\bt'
+
\bt'\cup_2\bt
\right)}
(-1)^{
\int_{M_4}
\left(
\bt\cup_2\bt'
+
\bt'\cup_2\bt
\right)}
\overline X_{t'}^B\overline X_t^B
=
\overline X_{t'}^B\overline X_t^B.
\end{aligned}
\end{equation*}
The $A$ flippers commute among themselves. They also commute with the
$B$ flippers: $X_f^A$ commutes with the first bracket by species
separation and with the second bracket because it commutes with every
$\overline Z_{t'}^B$.

Combining these identities with the separator algebra gives the complete
operator algebra
\begin{equation}
\left(\overline X_i\right)^2
=
\left(\overline Z_i\right)^2
=1,
\qquad
[\overline X_i,\overline X_j]
=
[\overline Z_i,\overline Z_j]
=0,
\qquad
\overline Z_i\overline X_j
=
(-1)^{\delta_{ij}}
\overline X_j\overline Z_i,
\label{eq:w2w3-full-Weyl-algebra}
\end{equation}
where $i,j$ range over both qubit species.

To prove independence, suppose that a finite-support product of
separators is scalar:
\begin{equation*}
\prod_f
\left(\overline Z_f^A\right)^{n_f}
\prod_t
\left(\overline Z_t^B\right)^{m_t}
=
\kappa\,1,
\qquad
n_f,m_t\in\mathbb Z_2,
\qquad
\kappa\in\{+1,-1\}.
\end{equation*}
Conjugating this relation by $\overline X_f^A$ gives
$(-1)^{n_f}=1$, and hence $n_f=0$ for every face. Conjugation by
$\overline X_t^B$ similarly gives $m_t=0$ for every three-cell. The
left-hand side then reduces to the identity, so $\kappa=1$. Thus the
separators obey no nontrivial finite-support relation.

The cup-$2$ incidences in Eq.~\eqref{eq:w2w3-flippers} are nonzero only
within a bounded neighborhood of the corresponding cell, so both flipper
families have uniformly bounded support. Since there is exactly one
independent separator--flipper pair for each physical qubit, the
separator family is independent and complete.

The corresponding separator Hamiltonian is
\begin{equation}
H^{\mathrm{QCA}}_{w_2w_3}
:=
-\sum_f\overline Z_f^A
-\sum_t\overline Z_t^B.
\label{eq:w2w3-QCA-hamiltonian}
\end{equation}
It is a commuting Hamiltonian of involutions. Equivalently, up to an
additive constant and an overall rescaling,
\begin{equation*}
\sum_f\frac{1-\overline Z_f^A}{2}
+
\sum_t\frac{1-\overline Z_t^B}{2}
\end{equation*}
is a full-space commuting-projector Hamiltonian. The bounded-range
assignments
\begin{equation*}
Z_i\longmapsto\overline Z_i,
\qquad
X_i\longmapsto\overline X_i
\end{equation*}
therefore define a QCA.

The boundary anomaly supplies an independent obstruction to a
finite-depth realization. The framing description identifies a fermionic
particle and a fermionic loop with $\pi$ mutual statistics at the
boundary of the $w_2w_3$ Hamiltonian~\cite{chen2023exactly}. This is one
of the anomalous $(3{+}1)$-dimensional $\mathbb Z_2$ topological orders
discussed in Ref.~\cite{johnson2020topological}. Under the
boundary-realizability assumption used in
Refs.~\cite{chen2023exactly,haah_QCA_23}---namely, that this anomalous
order cannot be realized by a strictly $(3{+}1)$-dimensional local
commuting-projector Hamiltonian---the QCA cannot be a finite-depth
circuit. Otherwise, terminating the circuit at a boundary would produce
precisely such a realization.

\section{QCAs in five spatial dimensions as finite-depth quantum circuits}
\label{sec:five-dimensional}

In $(5{+}1)$D, a cochain representative that appears non-Pauli need not define a nontrivial QCA. The oriented bordism group of dimension 6 is trivial. So, degree-6 Stiefel-Whitney numbers, including $w_2^3$ and $w_3^2$, vanish on every closed oriented 6-manifold \cite{MilnorStasheff1974}. The relevant relations among these characteristic numbers are reviewed in Appendix \ref{app:sw-number-relations}.

Nevertheless, the cochain constructions developed in the preceding sections can still be applied to topological actions whose integrated characteristic-class representatives are $w_2^3$ or $w_3^2$.  These actions give rise to candidate five-dimensional QCAs. In this section, we construct both candidates explicitly on the lattice using cup products. We show that they are trivial as QCAs. Each automorphism is implemented by conjugation with a finite-depth quantum circuit.

We begin with the complete transgression and gauging construction for the $w_2^3$ representative. We then relate it to the $w_3^2$ representative through a generally non-Clifford finite-depth field-redefinition circuit. The term $\beta(B_2)\cup\beta(B_2)$ is a coboundary and contributes only a local boundary circuit. After removing this contribution, we construct an explicit Clifford finite-depth circuit for the remaining  quadratic part of the action. This proves that both complete $5$-dimensional representatives define finite-depth, and hence trivial, QCAs.

\subsection{The $w_2^3$ QCA construction}
\label{sec:w2-cubed}

Consider the six-dimensional topological action
\begin{equation}
    \Phi_{w_2^3} =\frac{1}{2}\int_{M_6}\left(A_4\cup_2 A_4 + B_2\cup A_4+B_2\cup B_2\cup B_2\right),
    \label{eq:w2-cubed-action}
\end{equation}
where $A_4$ and $B_2$ are closed $\ZZ_2$ cochains. By the Wu formula, on a closed oriented 6-manifold,
\begin{equation*}
    \int_{M_6}A_4\cup_2 A_4 = \int_{M_6}\Sq^2 A_4 = \int_{M_6}w_2\cup A_4.
\end{equation*}
The terms involving $A_4$ can be combined into $\frac{1}{2}\int_{M_6}(w_2+B_2)\cup A_4$. Summing over $A_4$ imposes $[B_2]=w_2$. Substituting this constraint into the remaining cubic term yields
\begin{equation*}
    \frac{1}{2}\int_{M_6}B_2\cup B_2\cup B_2 = \frac{1}{2}\int_{M_6}w_2^3,
\end{equation*}
which yields the $w_2^3$ response.

In the trivial cohomology sector, i.e. locally on a contractible patch, we can write $A_4=\delta a$ and $B_2=\delta b$, with $a\in C^3(N,\mathbb Z_2)$ and $b\in C^1(N,\mathbb Z_2)$. The transgression at the cochain level is
\begin{equation}
    \begin{aligned}
        \Phi_{w_2^3}
        &=\frac12\int_{M_6}\delta\phi_5(a,b)
        =\frac12\int_{N}\phi_5(a,b),\\
        \phi_5(a,b)
        &:=a\cup_2 \delta a + a\cup_1a + b\cup \delta a
        +b\cup \delta b \cup \delta b,
    \end{aligned}
    \label{eq:w2-cubed-transgression}
\end{equation}
where $N=\partial M_6$. Following the endpoint construction of Sec.~\ref{sec:four-dimensional}, define the diagonal entangler by $U_5\lvert a,b\rangle=(-1)^{\int_N\phi_5(a,b)}\lvert a,b\rangle$. For cochains $r\in C^3(N,\mathbb Z_2)$ and $\lambda\in C^1(N,\mathbb Z_2)$, let us write $X_r^a:=\prod_tX_t^{r(t)}$, $X_\lambda^b:=\prod_eX_e^{\lambda(e)}$, and
\begin{equation*}
    T_r^a:=U_5X_r^aU_5^\dagger,
    \qquad
    T_\lambda^b:=U_5X_\lambda^bU_5^\dagger.
\end{equation*}

For a general three-cochain $r\in C^3(N,\mathbb Z_2)$, let us define
\begin{equation*}
    \Delta_r^a\phi_5(a,b) := \phi_5(a+r,b)-\phi_5(a,b).
\end{equation*}
Its integrated finite difference is
\begin{equation}
    \begin{aligned}
        \int_N\Delta_r^a\phi_5(a,b)
        &=
        \int_N\left[
        \delta r\cup_3\delta a
        +\delta b\cup r
        +r\cup_2\delta r
        +r\cup_1r
        \right]
        \pmod 2.
    \end{aligned}
    \label{eq:w2-cubed-general-a-shift}
\end{equation}

For an elementary edge indicator $\be$, define
\begin{equation*}
    \Delta_{\be}^b\phi_5(a,b) := \phi_5(a,b+\be)-\phi_5(a,b).
\end{equation*}
Direct expansion, followed by integration by parts and the local ordered-cup identities, gives
\begin{equation}
    \begin{aligned}
        \int_N\Delta_{\be}^b\phi_5(a,b)
        &=
        \int_N\left[
        \be\cup\delta a
        +\be\cup\delta b\cup\delta b
        +\delta b\cup\be\cup\delta b
        +\delta b\cup\delta b\cup\be
        \right]
        \pmod 2.
    \end{aligned}
    \label{eq:w2-cubed-b-shift}
\end{equation}
Eqs.~\eqref{eq:w2-cubed-general-a-shift} and \eqref{eq:w2-cubed-b-shift} completely specify the two types of Hamiltonian terms. Thus, no separate expansion into elementary moves is required.

We now gauge the two higher-form symmetries. After gauging, the resulting qubits live on faces and four-cells. After separating from each gauged move the factor that is linear in the other field, let us define
\begin{equation}
    \begin{aligned}
        G_t &:=
        \prod_{c\supset t}X_c \prod_{c'}Z_{c'}^{\int_N\delta\bt\cup_3\bc'},
        \\
        K_e &:=
        \prod_{f\supset e}X_f \prod_{f_1,f_2} CZ(Z_{f_1},Z_{f_2})^{\int_N\left[\be\cup\bface_1\cup\bface_2+\bface_1\cup\be\cup\bface_2+\bface_1\cup\bface_2\cup\be\right]}.
    \end{aligned}
    \label{eq:w2-cubed-Gt}
\end{equation}
All exponents are modulo $2$. The product over $(f_1,f_2)$ includes diagonal pairs, which are evaluated using $CZ(Z_f,Z_f)=Z_f$. For an elementary three-cell indicator $\bt$, the local hypercubic cup-product rules give
\begin{equation*}
    \bt\cup_2\delta\bt=0, \qquad \bt\cup_1\bt=0.
\end{equation*}
Hence, those contributions vanish and no additional sign is required in the definition of $G_t$. 

Given that $Z_{\partial s}:=\prod_{f\subset s}Z_f$ and $Z_{\partial p}:=\prod_{c\subset p}Z_c$, where $s$ and $p$ label
three- and five-cells, respectively, the gauged Hamiltonian is
\begin{equation}
    \begin{aligned}
        H_{\mathrm{gauged}}
        :={}&
        -\sum_tG_t\prod_fZ_f^{\int_N\bface\cup\bt}
        -\sum_eK_e\prod_cZ_c^{\int_N\be\cup\bc}
        -\sum_sZ_{\partial s}
        -\sum_pZ_{\partial p}.
    \end{aligned}
    \label{eq:w2-cubed-gauged-hamiltonian}
\end{equation}
The cup products $\int_N\bface\cup\bt$ and $\int_N\be\cup\bc$ give the same local perfect pairings used in Sec.~\ref{sec:four-dimensional}. Relabeling the two kinetic families by these pairings yields
\begin{equation}
    H_{\mathrm{kin}}
    :=
    -\sum_fZ_f\prod_tG_t^{\int_N\bface\cup\bt}
    -\sum_cZ_c\prod_eK_e^{\int_N\be\cup\bc}.
    \label{eq:w2-cubed-reduced-hamiltonian}
\end{equation}
Next, we show that the two omitted flux families impose no additional constraints on the ground space.

The operators $G_t$ are commuting involutions. Although the self-phase vanishes for each elementary indicator $\bt$, it need not vanish for a composite three-cochain. The exact endpoint product law gives, for every $r\in C^3(N,\mathbb Z_2)$,
\begin{equation}
    \prod_tG_t^{r(t)}
    =
    (-1)^{\int_N(r\cup_2\delta r+r\cup_1r)}
    \prod_cX_c^{(\delta r)(c)}
    \prod_{c'}Z_{c'}^{\int_N\delta r\cup_3\bc'}.
    \label{eq:w2-cubed-G-product-law}
\end{equation}
Thus, the phase on the right-hand side is generated by multiplying $G_t$.

For every three-cell $s$, let $f(s)$ be its complementary face under the reversed cup pairing, characterized by
\begin{equation*}
    \int_N\boldsymbol s\cup\bface'
    =
    \delta_{f',f(s)}
    \pmod 2
\end{equation*}
for every face $f'$. Complementary-cell incidence gives
\begin{equation*}
    \sum_f(\delta\bface)(s)
    \int_N\bface\cup\bt
    =
    (\delta\boldsymbol{f(s)})(t)
    \pmod 2.
\end{equation*}
It follows from Eq.~\eqref{eq:w2-cubed-G-product-law} that
\begin{equation}
    \begin{aligned}
        \prod_f
        \left[
        Z_f\prod_tG_t^{\int_N\bface\cup\bt}
        \right]^{(\delta\bface)(s)}
        &=
        Z_{\partial s}
        \prod_tG_t^{(\delta\boldsymbol{f(s)})(t)}
        =
        Z_{\partial s}
        (-1)^{\int_N
        \delta\boldsymbol{f(s)}\cup_1
        \delta\boldsymbol{f(s)}}
        =
        Z_{\partial s},
    \end{aligned}
    \label{eq:w2-cubed-face-flux-generated}
\end{equation}
where $Z_{\partial s}:=\prod_{f\subset s}Z_f$. Thus, the first kinetic family generates every face-flux operator and imposes $\delta B_2=0$.

For every five-cell $p$, let $v(p)$ be the unique vertex characterized by
\begin{equation*}
    \int_N\bv'\cup\boldsymbol p
    =
    \delta_{v',v(p)}
    \pmod 2
\end{equation*}
for every vertex $v'$. Complementary-cell incidence gives
\begin{equation*}
    \sum_c(\delta\bc)(p)
    \sum_e\left(\int_N\be\cup\bc\right)\be
    =
\delta\bv(p).
\end{equation*}
Hence, the product of the second-family generators around $p$ is the exact endpoint operator for the composite shift
$b\mapsto b+\delta\bv(p)$
\begin{equation}
    \begin{aligned}
        \prod_c
        \left[
        Z_c\prod_eK_e^{\int_N\be\cup\bc}
        \right]^{(\delta\bc)(p)}
        &=
        Z_{\partial p}
        \prod_eK_e^{(\delta\bv(p))(e)}
        =
        Z_{\partial p}
        (-1)^{\int_N\delta\bv(p)\cup B_2\cup B_2},
    \end{aligned}
    \label{eq:w2-cubed-four-cell-flux-generated}
\end{equation}
where $Z_{\partial p}:=\prod_{c\subset p}Z_c$. On the sector $\delta B_2=0$ imposed by the first kinetic family,
\begin{equation*}
    \int_N\delta\bv(p)\cup B_2\cup B_2
    =
    \int_N\delta\left[
    \bv(p)\cup B_2\cup B_2
    \right]
    =
    0,
\end{equation*}
because $N$ is closed. Therefore,
Eq.~\eqref{eq:w2-cubed-four-cell-flux-generated} reduces to $Z_{\partial p}$. And, the second kinetic family generates every four-cell flux once the first family has imposed $\delta B_2=0$. The two kinetic families in Eq.~\eqref{eq:w2-cubed-reduced-hamiltonian} have the same common $+1$ eigenspace as the four-family Hamiltonian Eq.~\eqref{eq:w2-cubed-gauged-hamiltonian}.

It remains to define a completion on the full tensor-product Hilbert space. Define
\begin{equation*}
    \widehat K_e
    :=
    \left.K_e\right|_{Z_f\,\mapsto\,\prod_tG_t^{\int_N\bface\cup\bt}},
\end{equation*}
where the replacement is made in every controlled phase appearing in
$K_e$. We then define
\begin{equation}
    \boxed{
    \begin{aligned}
    \overline Z_f
    &:=
    Z_f\prod_tG_t^{\int_N\bface\cup\bt},
    \\
    \overline Z_c
    &:=
    Z_c\prod_e(\widehat K_e)^{\int_N\be\cup\bc}.
    \end{aligned}}
    \label{eq:w2-cubed-completed-separators}
\end{equation}
On the simultaneous $\overline Z_f=1$ sector, the first line gives
\begin{equation*}
    Z_f
    =
    \prod_tG_t^{\int_N\bface\cup\bt}.
\end{equation*}
The second line of Eq.~\eqref{eq:w2-cubed-completed-separators} agrees with the original second kinetic family on this sector and defines a particular local extension away from it.

Eq.~\eqref{eq:w2-cubed-face-flux-generated} also implies
\begin{equation*}
    \prod_{f\subset s} \prod_tG_t^{\int_N\bface\cup\bt} =1
\end{equation*}
for every three-cell $s$. Thus, the substituted face operators satisfy the flatness relation required of a $\ZZ_2$ face configuration.

We finally verify the algebra on the full Hilbert space. For every four-cell $c$, let $e(c)$ be its complementary edge under the cup pairing, characterized by
\begin{equation*}
    \int_N\be'\cup\bc = \delta_{e',e(c)} \pmod 2
\end{equation*}
for every edge $e'$. The perfect pairing implies
\begin{equation*}
    \overline Z_c=Z_c\widehat K_{e(c)}.
\end{equation*}
For every face $f$, complementary-cell incidence gives
\begin{equation*}
    \begin{aligned}
        Z_c
        \left[
        \prod_tG_t^{\int_N\bface\cup\bt}
        \right]
        Z_c
        &=
        (-1)^{
        \sum_t
        \left(\int_N\bface\cup\bt\right)
        (\delta\bt)(c)}
        \prod_tG_t^{\int_N\bface\cup\bt}
        =
        (-1)^{(\delta\boldsymbol{e(c)})(f)}
        \prod_tG_t^{\int_N\bface\cup\bt}.
    \end{aligned}
\end{equation*}
Thus, conjugation by $Z_c$ implements the endpoint shift $\delta\boldsymbol{e(c)}$ in every substituted face operator. Consequently, the ordered composite $Z_c\widehat K_{e(c)}$ realizes the full substituted endpoint move. The telescoping identities Eqs.~\eqref{eq:w2w3-parent-square-telescoping} and \eqref{eq:w2w3-parent-commutation-telescoping} then show that
\begin{equation*}
    (\overline Z_c)^2=1,
    \qquad
    [\overline Z_c,\overline Z_{c'}]=0.
\end{equation*}

The mixed commutator is also trivial. Indeed,
\begin{equation*}
    \begin{aligned}
    \overline Z_c\overline Z_f
    &=
    Z_c\widehat K_{e(c)}Z_f
    \prod_tG_t^{\int_N\bface\cup\bt}
    \\
    &=
    (-1)^{(\delta\boldsymbol{e(c)})(f)}
    Z_fZ_c
    \left[
    \prod_tG_t^{\int_N\bface\cup\bt}
    \right]
    \widehat K_{e(c)}
    \\
    &=
    (-1)^{2(\delta\boldsymbol{e(c)})(f)}
    Z_f
    \left[
    \prod_tG_t^{\int_N\bface\cup\bt}
    \right]
    Z_c\widehat K_{e(c)}
    \\
    &=
    \overline Z_f\overline Z_c.
    \end{aligned}
\end{equation*}
The first sign arises from commuting $Z_f$ through the face shift in $\widehat K_{e(c)}$, while the second arises from commuting $Z_c$ through the paired $G_t$ product. The two signs cancel. Therefore,
\begin{equation}
    (\overline Z_i)^2=1, \qquad [\overline Z_i,\overline Z_j]=0,
    \label{eq:w2-cubed-completed-separator-algebra}
\end{equation}
where $i$ and $j$ range over faces and four-cells. These identities show that the completed operators are commuting involutions. Their independence and completeness are established by the local flippers constructed below.

We now construct the flippers. The natural four-cell operator is
\begin{equation}
U_c
:=
X_c
\prod_{c'}Z_{c'}^{\int_N\bc'\cup_3\bc}.
\label{eq:w2-cubed-natural-flippers}
\end{equation}
The order $\bc'\cup_3\bc$ is essential.  Indeed,
\begin{equation*}
    U_cG_t = (-1)^{\int_N\delta\bt\cup_3\bc+\sum_{c'}(\delta\bt)(c')\int_N\bc'\cup_3\bc}G_tU_c =G_tU_c.
\end{equation*}
Thus, $U_c$ commutes with every $G_t$ and every $\widehat K_e$. It
anticommutes only with the explicit $Z_c$ in $\overline Z_c$. $X_f$ anticommutes only with the explicit $Z_f$ in $\overline Z_f$. However, the $U_c$ themselves obey
\begin{equation}
    \begin{aligned}
        U_c^2
        &=
        (-1)^{\int_N\bc\cup_3\bc},
        \\
        U_cU_{c'}
        &=
        (-1)^{\int_N(
        \bc\cup_3\bc'
        +\bc'\cup_3\bc)}
        U_{c'}U_c.
    \end{aligned}
    \label{eq:w2-cubed-provisional-flipper-algebra}
\end{equation}
As in the flipper construction following Eq.~\eqref{eq:w2w3-completed-separator-algebra}, both defects are removed by an ordered separator dressing
\begin{equation}
    \boxed{
    \begin{aligned}
    \overline X_f
    &:=X_f,
    \\
    \overline X_c
    &:=
    \left[
    X_c
    \prod_{c'}Z_{c'}^{\int_N\bc'\cup_3\bc}
    \right]
    \left[
    \prod_{c'}(\overline Z_{c'})^{\int_N\bc'\cup_3\bc}
    \right].
    \end{aligned}}
    \label{eq:w2-cubed-dressed-flippers}
\end{equation}
For the square, the separator bracket contributes a second factor $(-1)^{\int_N\bc\cup_3\bc}$. For the commutator of the $c$ and $c'$ flippers, the two cross-commutations with the separator brackets contribute a second factor $(-1)^{\int_N(\bc\cup_3\bc'+\bc'\cup_3\bc)}$. These factors cancel the two phase factors in Eq.~\eqref{eq:w2-cubed-provisional-flipper-algebra}. Moreover, $X_f$ commutes with every $\overline Z_c$ and every $\overline X_c$. We then obtain the full local Pauli algebra
\begin{equation}
    \begin{gathered}
        (\overline X_i)^2=(\overline Z_i)^2=1,
        \qquad
        [\overline X_i,\overline X_j]
        =[\overline Z_i,\overline Z_j]=0,
        \\
        \overline Z_i\overline X_j
        =(-1)^{\delta_{ij}}
        \overline X_j\overline Z_i,
    \end{gathered}
    \label{eq:w2-cubed-full-Pauli-algebra}
\end{equation}
where $i,j$ again range over faces and four-cells.

The flipper argument following the $w_2w_3$ case now applies verbatim: conjugating a finite-support product of separators by its paired flippers shows that every exponent vanishes. Hence, the separators are independent. There is one separator-flipper pair for each physical qubit, so the family is complete. Moreover, all cup-product incidences and complementary-cell pairings have uniformly bounded diameter. That is, all operators above have bounded range. The assignments of $Z_i\mapsto\overline Z_i$ and $X_i\mapsto\overline X_i$ define a QCA, with separator Hamiltonian
\begin{equation}
    H_{w_2^3}^{\mathrm{QCA}} := -\sum_f\overline Z_f -\sum_c\overline Z_c.
    \label{eq:w2-cubed-QCA-hamiltonian}
\end{equation}

The preceding construction establishes the complete QCA independently of any circuit realization. We next determine whether this QCA is a finite-depth circuit. The argument below separately accounts for the cochain-coboundary circuit, the field-redefinition circuit, and the circuit implementing the quadratic part of the action. Their composition gives an FDQC realization of the complete $w_2^3$ representative.

\subsection{The \texorpdfstring{$w_3^2$}{w3 squared} representative}
\label{sec:w3-squared}

Let $\widetilde B_2$ denote the canonical integer $0$--$1$ lift of $B_2$. We can write down the representative associated with $w_3^2$ as
\begin{equation}
     \Phi_{w_3^2}
     =\frac{1}{2}\int_{M_6}\left[
     A_4\cup_2A_4+B_2\cup A_4
     +\frac{\delta \widetilde B_2}{2}\cup\frac{\delta \widetilde B_2}{2}\right].
     \label{eq:w3-squared-action}
\end{equation}
Here, $A_4$ and $B_2$ are closed modulo two. Eq.~\eqref{eq:w3-squared-action} specifies the response, but does not by itself construct the corresponding QCA.  The last term is a cochain coboundary of
\begin{equation}
     \frac{1}{2} \frac{\delta \widetilde B_2}{2}\cup\frac{\delta \widetilde B_2}{2}
     =\delta\left(\frac18\widetilde B_2\cup\delta\widetilde B_2\right).
     \label{eq:bockstein-square-coboundary}
\end{equation}
On a spatial boundary $M_5$, it contributes only the local diagonal circuit
\begin{equation}
     \mathcal W_{\beta}\ket B
     :=\exp\!\left(
     \frac{2\pi i}{8}\int_{M_5}\widetilde B\cup\delta\widetilde B
     \right)\ket B.
     \label{eq:bockstein-square-boundary-circuit}
\end{equation}
The integrand has bounded range, so its local phase gates can be colored into a system-size-independent number of layers.  Thus, $\mathcal W_{\beta}$ is an FDQC.  After this coboundary circuit has been disentangled, we are only left with the quadratic part of the action $A_4\cup_2A_4+B_2\cup A_4$.  In Sec.~\ref{sec:A4-cup2-A4-A4-cup-B2-FDQC} we construct an explicit Clifford FDQC for that contribution. Their composition gives a finite-depth realization of the full $w_3^2$ QCA.

This representative is also equivalent to the $w_2^3$ QCA. As shown in the next subsection, the local relabeling $A_4\mapsto A_4+B_2\cup B_2$ maps the two complete responses into one another up to cochain coboundaries. The relabeling is implemented by a finite-depth circuit built from controlled-controlled phase gates and is generally non-Clifford. Hence, the $w_2^3$ and $w_3^2$ QCA representatives differ by a generally non-Clifford FDQC, rather than by a nontrivial stable QCA class.

\subsection{Equivalence of the \texorpdfstring{$w_2^3$}{w2 cubed} and
\texorpdfstring{$w_3^2$}{w3 squared} descriptions}
\label{sec:w2-w3-five-dimensional-equivalence}

The last terms in Eqs.~\eqref{eq:w2-cubed-action} and
\eqref{eq:w3-squared-action} look different:
\begin{equation}
 \frac{1}{2}B_2\cup B_2\cup B_2,
 \qquad
 \frac{1}{2}\beta(B_2)\cup\beta(B_2).
\end{equation}
The first represents the nonzero class in
$H^6(K(\ZZ_2,2),\RR/\ZZ)$, whereas the second is a coboundary at the cochain
level by Eq.~\eqref{eq:bockstein-square-coboundary}.
Once the $A_4$ sector is included, however, the two complete actions are
related by the local field redefinition
\begin{equation}
 A_4\longmapsto A_4':=A_4+B_2\cup B_2.
 \label{eq:five-dimensional-field-redefinition}
\end{equation}
Because the path integral sums over all $A_4$ cocycles, this is a bijective
change of variables.

\paragraph{Finite-depth implementation of the field redefinition.}
The change of variables in
Eq.~\eqref{eq:five-dimensional-field-redefinition} is itself implemented by
a finite-depth quantum circuit.  On a spatial slice $M_5$, put an $A$-qubit
on every four-cell $c$ and a $B$-qubit on every face $f$, and set
\begin{equation}
 \mu_{c;ff'}:=\bigl(\boldsymbol f\cup\boldsymbol f'\bigr)(c)
 \in\mathbb F_2.
 \label{eq:field-redefinition-incidence}
\end{equation}
For three mutually commuting involutions, define the gate
\begin{equation}
CCZ(P,Q,R)
 :=\exp\!\left[\frac{i\pi}{8}(1-P)(1-Q)(1-R)\right].
 \label{eq:spectral-CCZ-field-redefinition}
\end{equation}
The required circuit is
\begin{equation}
 \mathcal V_{\mathrm{fr}}
 :=\prod_{c,f,f'}
CCZ\!\left(X_c^A,Z_f^B,Z_{f'}^B\right)^{
 \mu_{c;ff'}} .
 \label{eq:field-redefinition-circuit}
\end{equation}
Indeed,
$\operatorname{CCZ}(X_c^A,Z_f^B,Z_{f'}^B)
=H_c^A\operatorname{CCZ}(Z_c^A,Z_f^B,Z_{f'}^B)H_c^A$,
so for $f\ne f'$ it is a Toffoli gate with the two $B$-qubits as controls
and the $A$-qubit as target.  The product is over \emph{ordered} pairs
$(f,f')$, as required by the cochain-level cup-product. A term with
$f=f'$ must also be retained; the spectral definition then reduces that factor to $\operatorname{CNOT}(B_f\mathbin{\to}A_c)$.  Consequently,
\begin{equation}
 \mathcal V_{\mathrm{fr}}\ket{A,B}
 =\ket{A+B\cup B,B},
 \qquad
 \mathcal V_{\mathrm{fr}}^2=1.
 \label{eq:field-redefinition-basis-action}
\end{equation}
An equivalent operator-level check is
\begin{equation}
 \mathcal V_{\mathrm{fr}}Z_c^A\mathcal V_{\mathrm{fr}}^\dagger
 =Z_c^A\prod_{f,f'}
CZ(Z_f^B,Z_{f'}^B)^{\mu_{c;ff'}},
 \label{eq:field-redefinition-Z-conjugation}
\end{equation}
whose extra factor has eigenvalue
$(-1)^{(B\cup B)(c)}$ on a $B$-configuration.  The circuit fixes every
$X_c^A$ and $Z_f^B$.  Writing
$X_B(\lambda):=\prod_f(X_f^B)^{\lambda(f)}$, the conjugation of a $B$-shift
acts on basis states as
\begin{equation}
 \mathcal V_{\mathrm{fr}}X_B(\lambda)
 \mathcal V_{\mathrm{fr}}^\dagger\ket{A,B}
 =\ket{A+B\cup\lambda+\lambda\cup B+\lambda\cup\lambda,
 B+\lambda},
 \label{eq:field-redefinition-XB-conjugation}
\end{equation}
which is precisely the finite difference of the quadratic shift $B\cup B$.
All factors in Eq.~\eqref{eq:field-redefinition-circuit} commute.  Every
nonzero coefficient in
Eq.~\eqref{eq:field-redefinition-incidence} involves cells in a uniformly
bounded neighborhood.  On a bounded-valence cellulation, the resulting
local interaction hypergraph has bounded degree and can be colored with a
system-size-independent number of colors.  Thus
$\mathcal V_{\mathrm{fr}}$ has finite depth, even though it is generally a
non-Clifford circuit.

Starting from Eq.~\eqref{eq:w2-cubed-action}, one finds
\begin{eqs}
 \Phi_{w_2^3}
 &=\frac12\int_{M_6}\bigl[
 \Sq^2(A_4')+B_2\cup A_4'+B_2\cup B_2\cup B_2\bigr] \\
 &=\frac12\int_{M_6}\bigl[
 \Sq^2(A_4+\Sq^2B_2)+B_2\cup A_4\bigr] \\
 &=\frac12\int_{M_6}\bigl[
 \Sq^2(A_4)+\Sq^2\Sq^2B_2
 +\delta(A_4\cup_3\Sq^2B_2)+B_2\cup A_4\bigr] \\
 &=\frac12\int_{M_6}\bigl[
 \Sq^2(A_4)+\Sq^3\Sq^1B_2+B_2\cup A_4\bigr] \\
 &=\frac12\int_{M_6}\bigl[
 A_4\cup_2A_4+B_2\cup A_4
 +\beta(B_2)\cup\beta(B_2)\bigr],
 \label{eq:w2-w3-response-equivalence}
\end{eqs}
where the fourth line uses the Adem relation
\begin{equation}
 \Sq^2\Sq^2\alpha=\Sq^3\Sq^1\alpha+\delta x(\alpha).
\end{equation}
Thus the two gauged responses agree after a local change of variables and
cochain coboundaries.  More precisely, for a local
$\mathbb R/\mathbb Z$-valued five-cochain $\Lambda_{\mathrm{cb}}$ collecting
the cross-term, Adem, and lift/Bockstein cochain homotopies above, one may
write
\begin{equation}
 \Phi_{w_2^3}[A+B\cup B,B]
 =\Phi_{w_3^2}[A,B]
 +\int_{M_6}\delta\Lambda_{\mathrm{cb}}(A,B).
 \label{eq:field-redefinition-with-coboundary}
\end{equation}
On $M_5=\partial M_6$, this exact term is implemented by the local diagonal
boundary circuit
\begin{equation}
 \mathcal W_{\mathrm{cb}}\ket{A,B}
 :=\exp\!\left(2\pi i\int_{M_5}\Lambda_{\mathrm{cb}}(A,B)\right)
 \ket{A,B}.
 \label{eq:field-redefinition-coboundary-circuit}
\end{equation}
It is an FDQC by the same bounded-range coloring argument.  For parent
Hamiltonians equipped with the same matched off-flat completion, the exact
comparison is therefore
\begin{equation}
 H_{w_3^2}
 =\mathcal U_{\mathrm{eq}}H_{w_2^3}\mathcal U_{\mathrm{eq}}^\dagger,
 \qquad
 \mathcal U_{\mathrm{eq}}
 :=\mathcal W_{\mathrm{cb}}^\dagger\mathcal V_{\mathrm{fr}}.
 \label{eq:five-dimensional-Hamiltonian-equivalence}
\end{equation}
Eq.~\eqref{eq:field-redefinition-circuit} therefore implements the
nonlinear field redefinition exactly, while
$\mathcal W_{\mathrm{cb}}$ implements the separate cochain coboundary.  The
CCZ circuit alone maps the complete Hamiltonians only if that boundary
phase has already been absorbed into their transgression convention.  An
independently chosen off-flat extension need agree only in the flat sector
until the above matching is imposed.  This field-redefinition circuit,
which is generally non-Clifford, should also be distinguished from the
Clifford circuit for the quadratic part of the action constructed in the next
subsection.

\subsection{Explicit finite-depth circuit for the quadratic term in the action}
\label{sec:A4-cup2-A4-A4-cup-B2-FDQC}
In this section we give an explicit circuit trivialization of the QCA
associated with
\begin{equation}
    S[A,B]
    =\frac{1}{2}\int_{M_6}
    \bigl(A_4\cup_2A_4+A_4\cup B_2\bigr).
    \label{eq:AAB-action}
\end{equation}
For closed cocycles, reversing the order of the mixed cup-product changes
the cochain only by a local coboundary,
\begin{equation}
    A_4\cup B_2+B_2\cup A_4
      =\delta(A_4\cup_1B_2).
\end{equation}
The corresponding boundary circuit is
\begin{equation}
    \mathcal W_{\mathrm{ord}}\lvert A,B\rangle
      :=(-1)^{\int_N A\cup_1B}\lvert A,B\rangle;
    \label{eq:order-change-circuit}
\end{equation}
it is a bounded-depth product of controlled-$Z$ gates between $A$- and
$B$-qubits.  We prove the circuit statement below in the $B_2\cup A_4$
convention used by the displayed lattice Hamiltonian.  The circuit for the
ordering in Eq.~\eqref{eq:AAB-action} is then
$\mathcal W_{\mathrm{ord}}\mathcal U$.

The proof applies to the hypercubic cellulation used above, and more
generally whenever the two complementary-cell cup pairings introduced
below are local and nondegenerate.  No ancillas or shifts are required; the
circuit will be a Clifford FDQC.

We work over $\mathbb F_2$ on a closed spatial five-manifold $M_5$.  The
same calculation applies to finitely supported operators in an infinite
bulk, while a physical boundary requires the corresponding boundary
completion.  Put an $A$-qubit on every four-cell $c$ and a $B$-qubit on
every face $f$.  Plain letters $t,e,f,c$ label three-, one-, two-, and
four-cells, while bold letters denote their indicator cochains.  In the
$B_2\cup A_4$ convention, the two independent kinetic families are
\begin{align}
&X^A_{\delta\bt}
  \prod_c(Z_c^A)^{\int_N\delta\bt\cup_3\bc}
  \prod_f(Z_f^B)^{\int_N\bface\cup\bt},
\label{eq:five-dimensional-A-kinetic}\\
&X^B_{\delta\be}
  \prod_c(Z_c^A)^{\int_N\be\cup\bc}.
\label{eq:five-dimensional-B-kinetic}
\end{align}
The usual $A$- and $B$-flux operators are products of these kinetic
generators on the hypercubic lattice and are therefore redundant.

Depending on the transgression convention, the first line may carry the
configuration-independent sign
$(-1)^{\int_N(\bt\cup_2\delta\bt+\bt\cup_1\bt)}$.  Once the
complementary pairing below is chosen, all such signs are removed by the
depth-one product of the paired $X_f^B$ operators.  We therefore suppress
them without assuming that either self-cup vanishes.

For every face $f$, let $t(f)$ be its complementary 3-cell, and for
every four-cell $c$, let $e(c)$ be its complementary edge, characterized by
\begin{equation}
\int_N\boldsymbol f'\cup\boldsymbol{t(f)}=\delta_{f'f},
\qquad
\int_N\boldsymbol{e(c)}\cup\boldsymbol c'=\delta_{cc'}.
\label{eq:five-dimensional-complementary-pairings}
\end{equation}
Relabeling Eqs.~\eqref{eq:five-dimensional-A-kinetic} and
\eqref{eq:five-dimensional-B-kinetic} gives one separator for every
physical qubit:
\begin{equation}
\boxed{
\begin{aligned}
\overline Z_f^B
&:=Z_f^B X^A_{\delta\boldsymbol{t(f)}}
\prod_c(Z_c^A)^{
\int_N\delta\boldsymbol{t(f)}\cup_3\bc},\\
\overline Z_c^A
&:=Z_c^A X^B_{\delta\boldsymbol{e(c)}}.
\end{aligned}}
\label{eq:five-dimensional-direct-separators}
\end{equation}
Cellular Stokes' formula gives the incidence compatibility
\begin{equation}
\begin{aligned}
(\delta\boldsymbol{t(f)})(c)
&=\int_N\boldsymbol{e(c)}\cup\delta\boldsymbol{t(f)}
=\int_N\delta\boldsymbol{e(c)}\cup\boldsymbol{t(f)}
=(\delta\boldsymbol{e(c)})(f).
\end{aligned}
\label{eq:five-dimensional-complementary-incidence}
\end{equation}
This identity is also the cancellation of the two signs in the mixed
commutator of the separators.

Two further higher-cup identities will be used in the circuit.  For any
faces $f,f'$, the relevant incidence is symmetric and has zero diagonal:
\begin{align}
&\int_N\left[
\delta\boldsymbol{t(f)}\cup_3\delta\boldsymbol{t(f')}
+\delta\boldsymbol{t(f')}\cup_3\delta\boldsymbol{t(f)}
\right]
=\int_N\delta\left(
\delta\boldsymbol{t(f)}\cup_4\delta\boldsymbol{t(f')}
\right)=0,
\label{eq:five-dimensional-cup3-symmetry}\\
&\int_N\delta\boldsymbol{t(f)}\cup_3\delta\boldsymbol{t(f)}
=\int_N\left[
\boldsymbol{t(f)}\cup_2\delta\boldsymbol{t(f)}
+\delta\boldsymbol{t(f)}\cup_2\boldsymbol{t(f)}
\right]
=\int_N\delta\left(
\boldsymbol{t(f)}\cup_2\boldsymbol{t(f)}
\right)=0.
\label{eq:five-dimensional-cup3-diagonal}
\end{align}
The second identity also shows that the overlapping $X^A$ and $Z^A$
factors in $\overline Z_f^B$ commute, so no additional factor of $i$ is
required.

We now construct the circuit directly in cup-product notation.  Let
\begin{equation}
CZ_X(i,j):=(H_iH_j)CZ(i,j)(H_iH_j)
\label{eq:X-basis-CZ}
\end{equation}
denote controlled-$Z$ in the $X$ basis.  It fixes $X_i$ and $X_j$, while
$Z_i\mapsto Z_iX_j$ and $Z_j\mapsto X_iZ_j$.  Consider the three-layer
Clifford circuit
\begin{equation}
\begin{aligned}
\mathcal U
={}&
\prod_{c,f}
CZ_X(A_c,B_f)^{
 \int_N\boldsymbol{e(c)}\cup\delta\boldsymbol{t(f)}}
\prod_{c,f}
\operatorname{CNOT}(A_c\rightarrow B_f)^{
 \int_N\delta\boldsymbol{t(f)}\cup_3\bc}
\prod_{f<f'}
CZ_X(B_f,B_{f'})^{
 \int_N\delta\boldsymbol{t(f)}
 \cup_3\delta\boldsymbol{t(f')}}.
\end{aligned}
\label{eq:explicit-FDQC}
\end{equation}
The rightmost layer acts first under conjugation.  The last product contains
each unordered pair of distinct faces once.  This is well defined because
Eq.~\eqref{eq:five-dimensional-cup3-symmetry} makes its exponent symmetric,
while Eq.~\eqref{eq:five-dimensional-cup3-diagonal} shows that no
single-qubit phase gate is required.

The action on an $A$-type bare separator is immediate.  Only the leftmost
layer acts nontrivially, and hence
\begin{equation}
\begin{aligned}
\mathcal U Z_c^A\mathcal U^\dagger
&=Z_c^A\prod_f(X_f^B)^{
\int_N\boldsymbol{e(c)}\cup\delta\boldsymbol{t(f)}}
=Z_c^A X^B_{\delta\boldsymbol{e(c)}}
=\overline Z_c^A,
\end{aligned}
\label{eq:five-dimensional-circuit-on-ZA}
\end{equation}
where the second equality follows from
Eq.~\eqref{eq:five-dimensional-complementary-incidence}.

For $Z_f^B$, the rightmost layer first contributes
\begin{equation}
\prod_{f'}(X_{f'}^B)^{
\int_N\delta\boldsymbol{t(f)}
\cup_3\delta\boldsymbol{t(f')}}.
\label{eq:five-dimensional-first-XB-dressing}
\end{equation}
The middle layer produces the $Z^A$ dressing in
Eq.~\eqref{eq:five-dimensional-direct-separators}.  When the leftmost
layer acts on that $Z^A$ dressing, it produces a second copy of
Eq.~\eqref{eq:five-dimensional-first-XB-dressing}.  The two copies cancel
because every exponent is evaluated modulo two.  Explicitly,
\begin{align}
\mathcal U Z_f^B\mathcal U^\dagger
={}&
Z_f^B X^A_{\delta\boldsymbol{t(f)}}
\prod_c(Z_c^A)^{
\int_N\delta\boldsymbol{t(f)}\cup_3\bc}
\left[
\prod_{f'}(X_{f'}^B)^{
\int_N\delta\boldsymbol{t(f)}
\cup_3\delta\boldsymbol{t(f')}}
\right]^2
=\overline Z_f^B.
\label{eq:five-dimensional-circuit-on-ZB}
\end{align}

The same circuit sends the bare $X$ operators to
\begin{align}
\overline X_c^A
&:=\mathcal U X_c^A\mathcal U^\dagger
=X_c^A\prod_f(X_f^B)^{
\int_N\delta\boldsymbol{t(f)}\cup_3\bc},
\label{eq:five-dimensional-A-flipper}\\
\overline X_f^B
&:=\mathcal U X_f^B\mathcal U^\dagger
=X_f^B.
\label{eq:five-dimensional-B-flipper}
\end{align}
Eqs.~\eqref{eq:five-dimensional-circuit-on-ZA}--
\eqref{eq:five-dimensional-B-flipper} are the images of a complete set of
onsite Pauli generators.  Thus the dressed separators and flippers form a
complete local Pauli algebra, and Eq.~\eqref{eq:explicit-FDQC} implements
the QCA exactly.  Eq.~\eqref{eq:five-dimensional-cup3-diagonal}
ensures that the displayed separators are Hermitian involutions.  If a
different transgression convention supplies fixed signs to the kinetic
generators, those signs are removed by a depth-one product of the paired
$X_f^B$ operators described above.

Returning to the complete $w_3^2$ representative in Eq.~\eqref{eq:w3-squared-action}, the remaining Bockstein-square term is implemented by the boundary circuit $\mathcal W_\beta$ in Eq.~\eqref{eq:bockstein-square-boundary-circuit}. Thus, in the $B_2\cup A_4$ ordering used for that representative, the full QCA is the finite-depth composition of $\mathcal U$ and $\mathcal W_\beta$. The Hamiltonian equivalence in Eq.~\eqref{eq:five-dimensional-Hamiltonian-equivalence} then transports this circuit to the $w_2^3$ representative. The transport circuit $\mathcal U_{\mathrm{eq}}$ is generally non-Clifford because its field-redefinition factor $\mathcal V_{\mathrm{fr}}$ contains CCZ gates. Consequently, both five-dimensional QCAs are FDQCs, although the FDQC relating their two cochain representatives is generally non-Clifford.

As a cohomological check, for a closed oriented 6-manifold and a closed $A_4$ one has
\begin{equation}
    \int_{M_6}A_4\cup_2A_4
      =\int_{M_6}\Sq^2(A_4)
      =\int_{M_6}w_2\cup A_4\pmod 2
\end{equation}
by the Wu formula.  The change of variables
$B_2\mapsto B_2+w_2$ therefore reduces Eq.~\eqref{eq:AAB-action} to
the hyperbolic $A_4\cup B_2$ pairing.  This explains conceptually why the
quadratic form is metabolic.  The explicit circuit
Eq.~\eqref{eq:explicit-FDQC} is the lattice-level statement that upgrades
this cohomological observation to an actual finite-depth trivialization.
The same quadratic pairing is the basic building block of the general
reduction in Appendix~\ref{app:wu-core-reduction}. There, a Wu certificate
for a vanishing Stiefel-Whitney number is converted into field shifts of
quadratic and BF cores, giving a finite-depth realization of the
corresponding canonical circuit-compatible QCA representative.

\section{$w_2^4$ QCA in seven spatial dimensions}
\label{sec:generalized-semion-seven-dimensional}

The eight-dimensional oriented bordism group is
$\Omega_8^{\mathrm{SO}}\cong\ZZ\oplus\ZZ$, with generators represented by
$\CC P^4$ and $\CC P^2\times\CC P^2$. As shown in
Appendix~\ref{app:semion-signature-bordism}, the seven-dimensional
generalized-semion (semionic-membrane) and higher-\(U(1)_4\) QCAs have the
same invertibly normalized response,
\[
S(X)=\frac{\sigma(X)}{8}\pmod{1}.
\]
This defines an order-eight bordism character. At the level of responses,
the fourth power of the semionic-membrane QCA satisfies
\[
4S(X)=\frac{\sigma(X)}{2}
     =\frac{1}{2}\bigl\langle w_8(TX),[X]\bigr\rangle
     \pmod{1}.
\]
This agreement motivates our conjecture that it is stably equivalent to
the seven-dimensional generalization of the 3-fermion Clifford QCA constructed in
Ref.~\cite{fidkowski2024qca,Sun2026Clifford}, with response $w_8$.

In this section, we instead consider the independent order-two bordism
character
\[
[X]\longmapsto
(-1)^{\langle w_2(TX)^4,[X]\rangle}
\]
and construct the corresponding $w_2^4$ QCA.\footnote{On any closed oriented 8-manifold $X$, the Stiefel-Whitney numbers satisfy
$\langle w_8,[X]\rangle=\langle w_4^2+w_2^4,[X]\rangle$. Thus, once the $w_8$ response is included, it suffices to consider $w_2^4$ as the remaining independent invariant.}
The construction is based on a coupled $A_6$--$B_2$ action.  Let
$A_6\in Z^6(M_8,\mathbb Z_2)$ and
$B_2\in Z^2(M_8,\mathbb Z_2)$ be dynamical gauge fields on a closed oriented 8-manifold, and set
\begin{equation}
 S
 :=\frac12\int_{M_8}\left[
 A_6\cup_4A_6+B_2\cup A_6+B_2\cup B_2 \cup B_2 \cup B_2 \right].
 \label{eq:7d-w2-fourth-action}
\end{equation}
All cochain exponents in this subsection are understood modulo two.  The Wu relation gives
\begin{equation}
 \int_{M_8}A_6\cup_4A_6
 =\int_{M_8}\Sq^2A_6
 =\int_{M_8}w_2\cup A_6.
 \label{eq:7d-w2-fourth-Wu-pairing}
\end{equation}
Summing over $A_6$ imposes $B_2=w_2$.  The resulting
gravitational response is
\begin{equation}
 S
 =\frac12\int_{M_8}w_2^4.
 \label{eq:7d-w2-fourth-response}
\end{equation}

\paragraph{Boundary action and SPT wavefunction.}
Let $M_7=\partial M_8$ and, in the trivial flat sector, write
\begin{equation}
 A_6=\delta a,\qquad B_2=\delta b=:c,\qquad
 a\in C^5(M_8,\mathbb Z_2),\quad
 b\in C^1(M_8,\mathbb Z_2).
\end{equation}
Using
\begin{align}
 \delta\left(a\cup_4\delta a+a\cup_3a\right)
 &=\delta a\cup_4\delta a,
 &
 \delta(b\cup\delta a)&=\delta b\cup\delta a,
 &
 \delta(b\cup c^{\cup3})&=c^{\cup4}
 \label{eq:7d-w2-fourth-descent-identities}
\end{align}
gives the boundary phase
\begin{equation}
 \phi_{w_2^4}(a,b)
 :=\int_{M_7}\left[
 a\cup_4\delta a+a\cup_3a+b\cup\delta a
 +b\cup c\cup c\cup c\right].
 \label{eq:7d-w2-fourth-boundary-phase}
\end{equation}
The term $a\cup_3a$ is required by the higher-cup recursion and must be
retained.  The fixed-point boundary state is therefore
\begin{equation}
 \ket{\psi_{w_2^4}}
 =\sum_{a,b}(-1)^{\phi_{w_2^4}(a,b)}\ket{a,b}.
 \label{eq:7d-w2-fourth-boundary-state}
\end{equation}

Let $\nu$ be a five-cell and $e$ an edge, with indicator cochains $\bnu$ and
$\be$, and define $u_e:=\delta\be$.  Shifting $a$ by $\bnu$ or $b$ by
$\be$ changes the phase by
\begin{align}
 \Delta_\nu\phi_{w_2^4}
 &=\int_{M_7}\left[
 \delta\bnu\cup_5\delta a+c\cup\bnu
 +\bnu\cup_4\delta\bnu+\bnu\cup_3\bnu\right],
 \label{eq:7d-w2-fourth-a-variation}\\
 \Delta_e\phi_{w_2^4}
 &=\int_{M_7}\left[
 \be\cup\delta a+\mathcal F_e^{\mathrm{ex}}(c)\right],
 \qquad
 \mathcal F_e^{\mathrm{ex}}(c)
 :=\sum_{i=0}^{3}c^{\cup i}\cup\be\cup
 (c+u_e)^{\cup(3-i)}.
 \label{eq:7d-w2-fourth-b-variation}
\end{align}
The factors containing $c+u_e$ are required because this is a finite edge
flip.  The identity
\begin{equation}
 (c+u_e)^{\cup3}+c^{\cup3}
 =\sum_{i=0}^{2}c^{\cup i}\cup u_e\cup
 (c+u_e)^{\cup(2-i)}
 \label{eq:7d-w2-fourth-telescoping}
\end{equation}
together with integration by parts gives
Eq.~\eqref{eq:7d-w2-fourth-b-variation}.  Thus the phase contains all terms
produced by the shift, including the terms independent of $c$.

For commuting operators $P,Q,R$ that square to one, define
\begin{equation}
CZ(P,Q)
 :=(-1)^{\frac{1-P}{2}\frac{1-Q}{2}},\qquad
CCZ(P,Q,R)
 :=(-1)^{\frac{1-P}{2}\frac{1-Q}{2}\frac{1-R}{2}}.
 \label{eq:7d-w2-fourth-diagonal-gates}
\end{equation}
For any commuting family $W_f$, the corresponding diagonal phase operator
is
\begin{equation}
 \begin{aligned}
 \mathcal V_e(W)
 :={}&(-1)^{\int_{M_7}\be\cup u_e\cup u_e\cup u_e}\\
 &\times\prod_fW_f^{\int_{M_7}
 \be\cup\bface\cup u_e\cup u_e
 +\be\cup u_e\cup\bface\cup u_e
 +\be\cup u_e\cup u_e\cup\bface
 +\bface\cup\be\cup u_e\cup u_e}\\
 &\times
 \prod_{f_1,f_2}\!\!
CZ(W_{f_1},W_{f_2})^{
 \int_{M_7}
 \be\cup\bface_1\cup\bface_2\cup u_e
 +\be\cup\bface_1\cup u_e\cup\bface_2
 +\be\cup u_e\cup\bface_1\cup\bface_2
 +\bface_1\cup\be\cup\bface_2\cup u_e
 +\bface_1\cup\be\cup u_e\cup\bface_2
 +\bface_1\cup\bface_2\cup\be\cup u_e}\\
 &\times
 \prod_{f_1,f_2,f_3}\!\!
CCZ(W_{f_1},W_{f_2},W_{f_3})^{
 \int_{M_7}
 \be\cup\bface_1\cup\bface_2\cup\bface_3
 +\bface_1\cup\be\cup\bface_2\cup\bface_3
 +\bface_1\cup\bface_2\cup\be\cup\bface_3
 +\bface_1\cup\bface_2\cup\bface_3\cup\be
 } .
 \end{aligned}
 \label{eq:7d-w2-fourth-diagonal-phase}
\end{equation}
The products run over ordered tuples, including tuples with repeated face
labels.  If $W_f\ket c=(-1)^{c(f)}\ket c$, then
Eq.~\eqref{eq:7d-w2-fourth-diagonal-phase} acts by
$(-1)^{\int\mathcal F_e^{\mathrm{ex}}(c)}$.  Note that repeated labels are included;
for example, a $CCZ$ with repeated arguments reduces to the appropriate
lower-body phase gate.

\paragraph{Gauging and the flat-sector Hamiltonian.}
After gauging, place a $B$ qubit on every face and an $A$ qubit on every
six-cell.  We denote their configurations by
$B\in C^2(M_7,\mathbb Z_2)$ and $A\in C^6(M_7,\mathbb Z_2)$.  The
corresponding shifts are
\begin{equation}
 X^B_{\delta\be}:=\prod_f(X_f^B)^{(\delta\be)(f)},\qquad
 X^A_{\delta\bnu}:=\prod_r(X_r^A)^{(\delta\bnu)(r)}.
\end{equation}
Eqs.~\eqref{eq:7d-w2-fourth-a-variation} and
\eqref{eq:7d-w2-fourth-b-variation} give the two elementary kinetic
operators
\begin{align}
 T_e^B
 &:=X^B_{\delta\be}\,\mathcal V_e(Z^B),
 \label{eq:7d-w2-fourth-B-move}\\
 T_\nu^A
 &:=(-1)^{\int_{M_7}(\bnu\cup_4\delta\bnu+\bnu\cup_3\bnu)}
 X^A_{\delta\bnu}
 \prod_r(Z_r^A)^{\int_{M_7}\delta\bnu\cup_5\br}.
 \label{eq:7d-w2-fourth-A-move}
\end{align}
The constant sign in Eq.~\eqref{eq:7d-w2-fourth-A-move} is included in the
definition of the elementary move.  With this choice, the $T_\nu^A$ are
commuting Hermitian involutions.

For each face $f$, let $\nu(f)$ be its complementary five-cell.  For each
six-cell $r$, let $e(r)$ be its complementary edge:
\begin{equation}
 \int_{M_7}\bface'\cup\boldsymbol{\nu(f)}=\delta_{ff'},
 \qquad
 \int_{M_7}\boldsymbol{e(r)}\cup\boldsymbol r'=\delta_{rr'}.
 \label{eq:7d-w2-fourth-complementary-pairing}
\end{equation}
If $\tau$ is a three-cell and $q$ a seven-cell, write
$Z^B_{\partial\tau}:=\prod_{f\subset\partial\tau}Z_f^B$ and
$Z^A_{\partial q}:=\prod_{r\subset\partial q}Z_r^A$.  The kinetic terms
above are obtained for flat configurations.  Adding flux terms that impose
these constraints gives
\begin{equation}
 \begin{aligned}
 H_{w_2^4,\mathrm{flat}}
 ={}&-\sum_eT_e^B
       \prod_r(Z_r^A)^{\int_{M_7}\be\cup\br}
     -\sum_\nu T_\nu^A
       \prod_f(Z_f^B)^{\int_{M_7}\bface\cup\bnu}\\
 &-\sum_\tau Z^B_{\partial\tau}-\sum_qZ^A_{\partial q}.
 \end{aligned}
 \label{eq:7d-w2-fourth-flat-Hamiltonian}
\end{equation}
The operator $\mathcal V_e(Z^B)$ is fixed by the exact ratio of SPT
amplitudes when $\delta B=0$.  The last two sums impose the flatness
constraints.  The derivation above does not yet specify a commuting operator
for configurations with $\delta B\ne0$.

\paragraph{Separators on the full Hilbert space.}
The complementary cells satisfy
\begin{equation}
 (\delta\boldsymbol{e(r)})(f)
 =(\delta\boldsymbol{\nu(f)})(r).
 \label{eq:7d-w2-fourth-pairing-compatibility}
\end{equation}
The product of the complementary $A$ moves around a three-cell is the
identity:
\begin{equation}
 \prod_{f\subset\partial\tau}T_{\nu(f)}^A=1
 \qquad\text{for every three-cell }\tau.
 \label{eq:7d-w2-fourth-boundary-product}
\end{equation}
The total $X^A$ support vanishes because it is a second coboundary, and the
$Z^A$ factors cancel by the higher-cup recursion.  Since these operators
commute and have eigenvalues $\pm1$, we may substitute them for the
variables $W_f$ in Eq.~\eqref{eq:7d-w2-fourth-diagonal-phase}.  Define
\begin{equation}
 \widehat T_e^B
 :=X^B_{\delta\be}
 \left[\mathcal V_e(Z^B)\right]_{Z_f^B\mapsto T_{\nu(f)}^A}.
 \label{eq:7d-w2-fourth-substituted-B-move}
\end{equation}
The separators are
\begin{equation}
 \boxed{
 \overline Z_f^B:=Z_f^B T_{\nu(f)}^A,
 \qquad
 \overline Z_r^A:=Z_r^A\widehat T_{e(r)}^B.}
 \label{eq:7d-w2-fourth-separators}
\end{equation}
In the common $\overline Z_f^B=1$ eigenspace,
$Z_f^B=T_{\nu(f)}^A$.  Therefore these separators have the same common
$+1$ eigenspace as the first two terms of
Eq.~\eqref{eq:7d-w2-fourth-flat-Hamiltonian}.  The substitution above
defines the operator when $\delta B\ne0$; it is not an identity between the
unsubstituted operators.

Conjugating $T_{\nu(f)}^A$ by $Z_r^A$ changes its sign by
$(-1)^{(\delta\boldsymbol{e(r)})(f)}$.  The shift
$X^B_{\delta\boldsymbol{e(r)}}$ produces the same sign, so the two signs
cancel by Eq.~\eqref{eq:7d-w2-fourth-pairing-compatibility}.  Performing a
move twice shows that $(\overline Z_r^A)^2=1$, while performing two moves in
opposite orders shows that the second family commutes.  Thus all separators
are commuting Hermitian involutions.

The flux operators are already products of these separators.  First,
\begin{equation}
 \prod_{f\subset\partial\tau}\overline Z_f^B=Z^B_{\partial\tau},
 \label{eq:7d-w2-fourth-B-flux-generated}
\end{equation}
because $\prod_{f\subset\partial\tau}T_{\nu(f)}^A=1$.  Hence the first
separator family imposes $\delta B=0$.  Next, multiplying
$\overline Z_r^A$ over $r\subset\partial q$ gives the shift
$b\mapsto b+\delta v$, where $v$ is the complementary vertex.  When
$\delta B=0$, the quartic $B$ phase is unchanged, and the remaining phase is
\begin{equation}
 \left.\prod_{r\subset\partial q}\overline Z_r^A\right|_{\delta B=0}
 =(-1)^{\int_{M_7}\delta\boldsymbol v\cup A}
 =(-1)^{\int_{M_7}\boldsymbol v\cup\delta A}
 =Z^A_{\partial q}.
 \label{eq:7d-w2-fourth-A-flux-generated}
\end{equation}
Thus the second separator family imposes $\delta A=0$.  No separate flux
terms are needed in the QCA Hamiltonian.

\paragraph{Flippers, completeness, and the QCA.}
Let $s$ denote a six-cell and $\boldsymbol s$ its indicator cochain.  The
flippers are
\begin{equation}
\boxed{
\begin{aligned}
\overline X_f^B&:=X_f^B,\\
\overline X_r^A
&:=\left[
X_r^A\prod_s(Z_s^A)^{
\int_{M_7}\boldsymbol s\cup_5\br}
\right]
\left[
\prod_s(\overline Z_s^A)^{
\int_{M_7}\boldsymbol s\cup_5\br}
\right].
\end{aligned}}
 \label{eq:7d-w2-fourth-flippers}
\end{equation}
The two commutation signs between the first bracket and $T_\nu^A$ cancel:
\begin{equation}
 \int_{M_7}\delta\bnu\cup_5\br
 +\sum_s(\delta\bnu)(s)
 \int_{M_7}\boldsymbol s\cup_5\br
 =2\int_{M_7}\delta\bnu\cup_5\br=0\pmod2.
 \label{eq:7d-w2-fourth-U-L-commutator}
\end{equation}
Hence the first bracket commutes with every $T_{\nu(f)}^A$, with every
diagonal function of these operators, and with every $\widehat T_e^B$.
Therefore $\overline X_r^A$ flips only $\overline Z_r^A$, while $X_f^B$
flips only $\overline Z_f^B$.  The square of the first bracket is
$(-1)^{\int\br\cup_5\br}$, and commuting two such brackets gives
$(-1)^{\int(\br\cup_5\boldsymbol s+
\boldsymbol s\cup_5\br)}$.  The separator factors in the second bracket
produce the same signs, so they cancel.  The complete pairs therefore
satisfy
\begin{equation}
 \begin{gathered}
 (\overline X_j)^2=(\overline Z_j)^2=1,
 \qquad
 [\overline X_j,\overline X_k]
 =[\overline Z_j,\overline Z_k]=0,\\
 \overline Z_k\overline X_j
 =(-1)^{\delta_{jk}}\overline X_j\overline Z_k,
 \end{gathered}
 \label{eq:7d-w2-fourth-Pauli-algebra}
\end{equation}
and can be used to define the QCA.

\section{Higher-dimensional Stiefel-Whitney QCA families}
\label{sec:general-family}

In this section, we will use the same technique to derive four higher-dimensional classes of QCAs. The first realizes the polynomial responses $w_2^n w_3^m$.  The second shifts the cochain degrees in the $w_2w_3$ construction and produces the Wu--Bockstein response $\nu_p\Sq^1\nu_p$, with $p=2k$ and $k\geq1$. The final construction applies to an arbitrary product of Wu classes.

For each family, we follow the same recipe.  We identify the gravitational
response, transgress the action to a spatial wave-function phase, gauge its
higher-form symmetries, and use the complementary-cell pairing on the hypercubic lattice to choose a
full-space representative with local flippers.  This proves the existence of
the corresponding QCAs.  Determining their stable classes is a separate
classification problem.

\subsection{The \texorpdfstring{$w_2^nw_3^m$}{w2n w3m} family}
\label{subsec:w2n-w3m-family}

Let $n,m\geq0$, set
\begin{equation}
 d+1=2n+3m,
\end{equation}
and assume $d\geq3$.  Note that any term
containing a negative higher-cup index is omitted.  For cocycles
$A_{d-1}\in Z^{d-1}(M_{d+1};\mathbb Z_2)$ and
$B_2\in Z^2(M_{d+1};\mathbb Z_2)$, consider the action
\begin{equation}
 \Phi_{n,m}[A,B]
 =\frac12\int_{M_{d+1}}\left[
 A_{d-1}\cup_{d-3}A_{d-1}+B_2\cup A_{d-1}
 +B_2^{\cup n}\cup(B_2\cup_1B_2)^{\cup m}\right].
 \label{eq:general-family-action}
\end{equation}
Here $x^{\cup j}$ denotes the $j$-fold ordinary cup product, with
$x^{\cup0}=1$.

On a closed oriented manifold, the Wu formula gives
\begin{equation}
 \int_{M_{d+1}} A_{d-1}\cup_{d-3}A_{d-1}
 =\int_{M_{d+1}}\Sq^2A_{d-1}
 =\int_{M_{d+1}}w_2\cup A_{d-1}.
 \label{eq:general-family-response-reduction}
\end{equation}
Summing over $A_{d-1}$ therefore imposes $[B_2]=w_2$. Moreover, $B_2\cup_1B_2$ represents $\Sq^1B_2$, and $\Sq^1w_2=w_3$ on an oriented manifold. The resulting gravitational
response is consequently
\begin{equation}
    \frac12\int_{M_{2n+3m}} w_2^nw_3^m.
\end{equation}

The corresponding oriented characteristic number is not always nonzero. As proved in Appendix~\ref{app:w2n-w3m-so-bordism}, its exact detection criterion is
\begin{equation}
\boxed{
\begin{aligned}
&\exists\ \text{a closed oriented }M_{d+1}\text{ with }
 \left\langle w_2^nw_3^m,[M_{d+1}]\right\rangle=1
\Longleftrightarrow\quad
\left\{
\begin{array}{l}
m=0\ \text{and }n\text{ is even},\quad\text{or}\\
n\text{ and }m\text{ are both odd}.
\end{array}
\right.
\end{aligned}}
\label{eq:general-family-so-bordism-criterion}
\end{equation}
In particular, a mixed response with $n,m>0$ is detected precisely when both exponents are odd. This criterion concerns only the induced $SO$-bordism invariant and therefore does not, by itself, determine the stable class of the microscopic QCA. Motivated by the explicit finite-depth realizations of the $w_2^3$ and $w_3^2$ representatives in Sec.~\ref{sec:five-dimensional}, we conjecture that a member whose characteristic number vanishes on every closed oriented manifold is implementable, after stabilization, by a finite-depth circuit. Appendix~\ref{app:wu-core-reduction} proves this finite-depth direction for the canonical circuit-compatible representatives considered there. The completeness of the Wu relations supplies a certificate for a vanishing number, and shifts of the associated quadratic and BF cores convert that certificate into an FDQC. The appendix also works out the $w_7$ and $w_2w_8$ examples explicitly.

Working in the trivial cohomology sector, we can substitute
\begin{equation}
 A_{d-1}=\delta a,
 \qquad B_2=\delta b,
 \qquad a\in C^{d-2}(M_{d+1};\mathbb Z_2),
 \quad b\in C^1(M_{d+1};\mathbb Z_2).
\end{equation}
For $n>0$, the spacetime action transgresses onto the spatial manifold as
\begin{eqs}
 \Phi_{n,m}
 &=\frac12\int_{M_{d+1}}\delta\Bigl[
 a\cup_{d-3}\delta a+a\cup_{d-4}a+b\cup\delta a
 +b\cup(\delta b)^{\cup(n-1)}
       \cup(\delta b\cup_1\delta b)^{\cup m}\Bigr]
 \nonumber\\
 &=\frac12\int_{M_d}\Bigl[
 a\cup_{d-3}\delta a+a\cup_{d-4}a+b\cup\delta a
 +b\cup(\delta b)^{\cup(n-1)}
       \cup(\delta b\cup_1\delta b)^{\cup m}\Bigr].
 \label{eq:general-family-transgression}
\end{eqs}
Henceforth, unless specified, all integrals will be over $M_d$, which we occasionally write as just $M$.

When $n=0$, the final term in the boundary integrand is instead
\begin{equation}
 (b\cup_1\delta b+b\cup b)
 \cup(\delta b\cup_1\delta b)^{\cup(m-1)}.
 \label{eq:general-family-nzero-boundary-term}
\end{equation}
Indeed,
$\delta(b\cup_1\delta b+b\cup b)=\delta b\cup_1\delta b$.

The corresponding higher-form SPT states are
\begin{align}
 \ket{\psi_{n,m}}
 &=\sum_{a,b}(-1)^{\int_M\phi_{n,m}(a,b)}\ket{a,b},
 \label{eq:general-family-spt-state}\\
 \phi_{n,m}(a,b)
 &:={}
 a\cup_{d-3}\delta a+a\cup_{d-4}a+b\cup\delta a
 +b\cup(\delta b)^{\cup(n-1)}
       \cup(\delta b\cup_1\delta b)^{\cup m},
 \qquad n>0,
\end{align}
and
\begin{align}
 \ket{\psi_{0,m}}
 &=\sum_{a,b}(-1)^{\int_M\phi_{0,m}(a,b)}\ket{a,b},
 \label{eq:general-family-spt-state-nzero}\\
 \phi_{0,m}(a,b)
 &:={}
 a\cup_{d-3}\delta a+a\cup_{d-4}a+b\cup\delta a
 +(b\cup_1\delta b+b\cup b)
       \cup(\delta b\cup_1\delta b)^{\cup(m-1)}.
\end{align}

Let $t$ be a $(d-2)$-cell and let $\bt$ denote its indicator cochain. Applying the higher-cup recursion and integrating by parts on the closed spatial manifold gives
\begin{equation}
 \kappa_t
 :=\int_M\left[
 \bt\cup_{d-3}\delta\bt+\bt\cup_{d-4}\bt\right],
\end{equation}
and
\begin{equation}
 \frac{\braket{a+\bt,b}{\psi_{n,m}}}
      {\braket{a,b}{\psi_{n,m}}}
 =(-1)^{\kappa_t+
 \int_M\left[
 \delta\bt\cup_{d-2}\delta a+\delta b\cup\bt\right]}.
 \label{eq:general-family-a-shift}
\end{equation}
The same formula holds for the $n=0$ branch.  The self-phase depends only on
the chosen elementary cell; we retain it explicitly in the definition of
$L_t^A$ below.

The finite change under $b\mapsto b+\be$ contains the nonlinear part of the
construction.  To keep the ordering of all cup products explicit, define
\begin{equation}
 c:=\delta b,
 \qquad u:=\delta\be,
 \label{eq:general-family-shifted-fields}
\end{equation}
and
\begin{equation}
    \begin{aligned}
        \Theta(c)&:=c\cup_1c,\nonumber\\
 \Lambda(c;\be)
 &:=c\cup_1\be+\be\cup_1c
   +\be\cup_1u+\be\cup\be. 
    \end{aligned}
     \label{eq:general-family-theta-kernel}
\end{equation}
The higher-cup recursion gives the exact identity
\begin{equation}
 \Theta(c+u)+\Theta(c)=\delta\Lambda(c;\be).
 \label{eq:general-family-theta-difference}
\end{equation}

For $n>0$, introduce the ordered telescoping cochains
\begin{align}
 \mathcal P_n(c;\be)
 &:=\sum_{i=0}^{n-2}
 c^{\cup i}\cup\be\cup(c+u)^{\cup(n-2-i)},\nonumber\\
 \mathcal Q_m(c;\be)
 &:=\sum_{j=0}^{m-1}
 \Theta(c)^{\cup j}\cup\Lambda(c;\be)
 \cup\Theta(c+u)^{\cup(m-1-j)}.
 \label{eq:general-family-telescoping-polynomials}
\end{align}
We use $\mathcal P_1=0$ and $\mathcal Q_0=0$.  Since both $c$ and $c+u$
are closed, these cochains satisfy
\begin{align}
 (c+u)^{\cup(n-1)}+c^{\cup(n-1)}
 &=\delta\mathcal P_n(c;\be),\nonumber\\
 \Theta(c+u)^{\cup m}+\Theta(c)^{\cup m}
 &=\delta\mathcal Q_m(c;\be).
 \label{eq:general-family-ordered-telescoping}
\end{align}
All dependence on the undifferentiated cochain $b$ can therefore be removed
by integration by parts.  The remaining local endpoint polynomial is
\begin{align}
 \mathcal V_{n,m}(c;\be)
 :={}&\be\cup(c+u)^{\cup(n-1)}
       \cup\Theta(c+u)^{\cup m}
 +c\cup\mathcal P_n(c;\be)
       \cup\Theta(c+u)^{\cup m}
 +c^{\cup n}\cup\mathcal Q_m(c;\be),
 \quad n>0.
 \label{eq:general-family-exact-b-polynomial}
\end{align}
Thus
\begin{equation}
 \frac{\braket{a,b+\be}{\psi_{n,m}}}
      {\braket{a,b}{\psi_{n,m}}}
 =(-1)^{\int_M\left[
 \be\cup\delta a+\mathcal V_{n,m}(\delta b;\be)\right]}.
 \label{eq:general-family-b-amplitude-ratio}
\end{equation}
For $(n,m)=(4,0)$, this reproduces the ordered four-term expression in
Eq.~\eqref{eq:7d-w2-fourth-b-variation}, including the lower-degree terms
generated by the finite shift.

The pure-$w_3$ branch has an especially simple form.  If
\begin{equation}
 \mathcal R(b,c):=b\cup_1c+b\cup b,
\end{equation}
then
\begin{align}
 \delta\mathcal R(b,c)&=\Theta(c),\nonumber\\
 \mathcal R(b+\be,c+u)+\mathcal R(b,c)
 &=\delta(b\cup_1\be)+\Lambda(c;\be).
 \label{eq:general-family-nzero-identities}
\end{align}
Using Eq.~\eqref{eq:general-family-ordered-telescoping} with the
$(m-1)$st power and integrating by parts shows that the complete endpoint
polynomial is simply
\begin{equation}
    \mathcal V_{0,m}(c;\be):=\mathcal Q_m(c;\be).
    \label{eq:general-family-nzero-variation}
\end{equation}
Consequently,
\begin{equation}
 \frac{\braket{a,b+\be}{\psi_{0,m}}}
      {\braket{a,b}{\psi_{0,m}}}
 =(-1)^{\int_M\left[
 \be\cup\delta a+\mathcal V_{0,m}(\delta b;\be)\right]}.
 \label{eq:general-family-nzero-b-amplitude-ratio}
\end{equation}
This expression is both shorter and fully ordered. No commutation of cochain factors is required.

For either branch, define the diagonal endpoint operator on a closed face configuration $c$ by
\begin{equation}
    \mathcal V_e(Z^B)\ket c :=(-1)^{\int_M\mathcal V_{n,m}(c;\be)}\ket c, \qquad Z_f^B\ket c=(-1)^{c(f)}\ket c.
    \label{eq:general-family-spectral-phase}
\end{equation}
Since $\mathcal{V}_{n,m}$ is a finite local polynomial, $\mathcal{V}_e(Z^B)$ is a bounded product of diagonal multi-qubit phase gates.

Gauging sends $\ket{a,b}\mapsto\ket{A=\delta a,B=\delta b}$. The $A$ qubits live on $(d-1)$-cells $r$, and the $B$ qubits live on faces $f$. For an edge $e$ and a $(d-2)$-cell $t$, define
\begin{align}
 L_t^A
 &:=(-1)^{\kappa_t}X^A_{\delta\bt}
 \prod_r(Z_r^A)^{\int_M\delta\bt\cup_{d-2}\br},
 \label{eq:general-family-Lt}\\
 K_e^B
 &:=X^B_{\delta\be}\,\mathcal V_e(Z^B).
 \label{eq:general-family-Ke}
\end{align}
The explicit factor $(-1)^{\kappa_t}$ implements the elementary-cell phase convention fixed after Eq.~\eqref{eq:general-family-a-shift} and makes every $L_t^A$ an involution. On closed cochain configurations, the gauged commuting Hamiltonian is
\begin{align}
 H_{n,m}^{\mathrm{flat}}
 ={}&-\sum_eK_e^B
 \prod_r(Z_r^A)^{\int_M\be\cup\br}
 -\sum_tL_t^A
 \prod_f(Z_f^B)^{\int_M\bface\cup\bt}
 -\sum_{p\in M_3}Z^B_{\partial p}
 -\sum_{q\in M_d}Z^A_{\partial q}.
 \label{eq:general-family-gauged-hamiltonian}
\end{align}
The endpoint phase in $K_e^B$ is initially fixed only on the closed $B$ configurations.  The next step gives its canonical commuting extension to the full Hilbert space.

For every face $f$, let $t(f)$ be its complementary $(d-2)$-cell, and for every $(d-1)$-cell $r$, let $e(r)$ be its complementary edge. They are characterized by
\begin{equation}
 \int_M\bface'\cup\boldsymbol{t(f)}=\delta_{ff'},
 \qquad
 \int_M\boldsymbol{e(r)}\cup\br'=\delta_{rr'}.
 \label{eq:general-family-complementary-pairing}
\end{equation}
The commuting involutions $L_t^A$ obey the same local closure relation as a closed face configuration,
\begin{equation}
 \prod_{f\subset\partial p}L_{t(f)}^A=1
 \qquad\text{for every 3-cell }p.
 \label{eq:general-family-paired-closure}
\end{equation}
The $X^A$ support in this product is a second coboundary, while the $Z^A$ dressing cancels by the higher-cup recursion. It is therefore consistent to evaluate the endpoint polynomial on the joint spectrum of the $L_{t(f)}^A$.

Replace every $Z_f^B$ in the diagonal part of $K_e^B$ by $L_{t(f)}^A$ and write
\begin{equation}
 \widehat K_e^B
 :=X^B_{\delta\be}
 \left[\mathcal V_e(Z^B)\right]_{Z_f^B\mapsto L_{t(f)}^A}.
 \label{eq:general-family-substituted-Ke}
\end{equation}
The two separator families are
\begin{equation}
\boxed{
\begin{aligned}
 \overline Z_f^B&:=Z_f^B L_{t(f)}^A,\\
 \overline Z_r^A&:=Z_r^A\widehat K_{e(r)}^B.
\end{aligned}}
\label{eq:general-family-final-separators}
\end{equation}
On the common $\overline Z_f^B=1$ sector, the relation $Z_f^B=L_{t(f)}^A$ makes the second family identical to the original kinetic term in Eq.~\eqref{eq:general-family-gauged-hamiltonian}. Away from that sector, the substitution specifies the full-space representative. It is not an operator identity.

The two complementary-cell pairings satisfy
\begin{equation}
 (\delta\boldsymbol{t(f)})(r)
 =(\delta\boldsymbol{e(r)})(f),
 \qquad
 Z_r^A L_{t(f)}^A Z_r^A
 =(-1)^{(\delta\boldsymbol{e(r)})(f)}L_{t(f)}^A.
 \label{eq:general-family-pairing-compatibility}
\end{equation}
Conjugating by $Z_r^A$ therefore changes the substituted spectrum by the same finite shift $\delta\boldsymbol{e(r)}$ that appears in the endpoint phase. Performing one finite shift twice gives unity, while two shifts in opposite orders give the same phase. These two facts prove $(\overline Z_r^A)^2=1$ and $[\overline Z_r^A,\overline Z_s^A]=0$. In the mixed commutator, the two signs cancel by Eq.~\eqref{eq:general-family-pairing-compatibility}. Hence, all separators in Eq.~\eqref{eq:general-family-final-separators} are commuting involutions.

The explicit flux terms in Eq.~\eqref{eq:general-family-gauged-hamiltonian} are generated by products of the separators. Multiplying $\overline Z_f^B$ over $f\subset\partial p$ gives $Z^B_{\partial p}$ because of Eq.~\eqref{eq:general-family-paired-closure}. After imposing $\delta B=0$, multiplying $\overline Z_r^A$ over the boundary of a $d$-cell $q$ produces the exact $B$ move with parameter $\delta\boldsymbol{v(q)}$.  Here $v(q)$ is the complementary vertex, defined by
\begin{equation}
    \int_M\boldsymbol{v(q)}\cup\boldsymbol q'=\delta_{qq'}.
\end{equation}
The polynomial response is unchanged, and
\begin{equation}
    (-1)^{\int_M\delta\boldsymbol{v(q)}\cup A}=(-1)^{\int_M\boldsymbol{v(q)}\cup\delta A}=Z^A_{\partial q}.
    \label{eq:general-family-flux-generation}
\end{equation}
Thus, the separator constraints impose both $\delta B=0$ and $\delta A=0$, and the full-space separator Hamiltonian is
\begin{equation}
 H_{n,m}^{\mathrm{QCA}}
 :=-\sum_f\overline Z_f^B-\sum_r\overline Z_r^A.
 \label{eq:general-family-final-hamiltonian}
\end{equation}

The paired flippers are
\begin{equation}
\boxed{
\begin{aligned}
 \overline X_f^B&:=X_f^B,\\
 \overline X_r^A
 &:={}
 \left[X_r^A\prod_s(Z_s^A)^{
 \int_M\boldsymbol s\cup_{d-2}\br}\right]
 \left[\prod_s(\overline Z_s^A)^{
 \int_M\boldsymbol s\cup_{d-2}\br}\right].
\end{aligned}}
\label{eq:general-family-flippers}
\end{equation}
The displayed order is part of the definition. The first bracket commutes with every $L_t^A$: its two overlap phases are equal and cancel. Its remaining square phase and mutual-commutator phase are reproduced once more when it is moved through the separator dressing in the second bracket, and
therefore also cancel. It follows that
\begin{equation}
 \overline Z_k\overline X_j
 =(-1)^{\delta_{jk}}\overline X_j\overline Z_k,
 \qquad
 [\overline X_j,\overline X_k]
 = [\overline Z_j,\overline Z_k]=0,
 \label{eq:general-family-full-weyl-algebra}
\end{equation}
where $j,k$ range over both face and $(d-1)$-cell qubits. The paired flippers rule out any nontrivial local relation among the separators. Since there is one separator-flipper pair per qubit and every operator has bounded support, the assignments $Z_j\mapsto\overline Z_j$ and $X_j\mapsto\overline X_j$ then define a QCA for every admissible $(n,m)$.

This construction simultaneously contains several earlier examples. The
case $(n,m,d)=(1,1,4)$ reproduces the $w_2w_3$ QCA of
Sec.~\ref{sec:four-dimensional} after exchanging the $A$- and $B$-field
labels; $(3,0,5)$ gives the complete $w_2^3$
representative; and the separate $n=0$ calculation includes the pure
$w_3^m$ family.

\subsection{Uniform degree shifts: the Wu--Bockstein family
\texorpdfstring{$w_2w_{4k-1}$}{w2 w(4k-1)}}
\label{subsec:degree-shifted-wu-family}

We next generalize the cochain degrees in the $w_2w_3$ construction. Fix
$k\geq1$ and let $p=2k$. The spacetime dimension is $2p+1=4k+1$, and the spatial dimension is $d=2p=4k$. Let
\begin{equation}
 A_p\in Z^p(M_{d+1};\mathbb Z_2),
 \qquad
 B_{p+1}\in Z^{p+1}(M_{d+1};\mathbb Z_2).
 \label{eq:shifted-family-degrees}
\end{equation}
 The case $k=1$ reproduces the $w_2w_3$ QCA of Sec.~\ref{sec:four-dimensional}. We will show that the same construction works for every $k$.

The action is
\begin{equation}
 S_k[A,B]
 =\frac12\int_{M_{d+1}}\left[
 B_{p+1}\cup_1B_{p+1}+A_p\cup B_{p+1}
 +A_p\cup\bigl(A_p\cup_{p-1}A_p\bigr)
 \right].
 \label{eq:shifted-family-action}
\end{equation}
For closed cochains,
\begin{equation}
 B_{p+1}\cup_1B_{p+1}=\Sq^pB_{p+1},
 \qquad
 A_p\cup_{p-1}A_p=\Sq^1A_p
 \quad\text{in cohomology}.
 \label{eq:shifted-family-steenrod-identification}
\end{equation}
The Wu formula converts the first term under the integral to $\nu_p\cup B_{p+1}$. Summing over $B_{p+1}$ therefore imposes $[A_p]=\nu_p$, leaving the pure gravitational response
\begin{equation}
 S_k[M_{d+1}]
 =\frac12\int_{M_{d+1}}\nu_p\cup\Sq^1\nu_p.
 \label{eq:shifted-family-wu-response}
\end{equation}
The Bockstein formulation makes the lift convention precise. For an integer lift $\widetilde\nu_p$ of $\nu_p$, define
\begin{equation}
 \beta_{\mathbb Z}(\nu_p)
 :=\left[\frac{\delta\widetilde\nu_p}{2}\right]
 \in H^{p+1}(M_{d+1};\mathbb Z),
 \qquad
 \rho_2\beta_{\mathbb Z}(\nu_p)=\Sq^1\nu_p.
 \label{eq:shifted-family-bockstein}
\end{equation}
In general, this class differs from the integral Stiefel-Whitney class $W_{p+1}=\beta_{\mathbb Z}(w_p)$, and its mod-2 reduction need not equal $w_{p+1}$. The two agree for $p=2$ and $p=4$ on oriented manifolds, but already differ at $p=6$.

The Wu polynomials follow recursively from $w=\Sq(\nu)$. On an oriented manifold,
\begin{equation}
 \nu_j=w_j+
 \sum_{i=1}^{\lfloor j/2\rfloor}\Sq^i\nu_{j-i},
 \qquad w_1=0.
 \label{eq:shifted-family-wu-recursion}
\end{equation}
The first even-degree cases are
\begin{align}
 \nu_2&=w_2,
 &\Sq^1\nu_2&=w_3,
 \nonumber\\
 \nu_4&=w_4+w_2^2,
 &\Sq^1\nu_4&=w_5,
 \nonumber\\
 \nu_6&=w_2w_4+w_3^2,
 &\Sq^1\nu_6&=w_2w_5+w_3w_4,
 \label{eq:shifted-family-low-wu}\\
 \nu_8&=w_8+w_2w_6+w_3w_5+w_4^2+w_2^4,
 &\Sq^1\nu_8&=w_9+w_2w_7+w_3w_6.
 \nonumber
\end{align}

For even $p$ and a closed oriented $(2p+1)$-manifold, one has~\cite{LusztigMilnorPeterson1969}
\begin{equation}
\nu_p\Sq^1\nu_p
 = w_2w_{2p-1}.
 \label{eq:shifted-family-lmp}
\end{equation}
Hence, the closed oriented TQFT defined by Eq.~\eqref{eq:shifted-family-action} has response
\begin{equation}
 S_k^{\mathrm{grav}}[X^{4k+1}]
 =\frac12\int_{X^{4k+1}} w_2w_{4k-1}.
 \label{eq:shifted-family-sw-response}
\end{equation}
This response can be detected on the manifold $W^5\times\mathbb {CP}^{p-2}$  where $W^5=SU(3)/SO(3)$ is the Wu manifold. For $p\geq 4$, let $h$ denote the pullback of the mod-two hyperplane class of $\mathbb{CP}^{p-2}$. The Whitney product formula then gives 
\begin{equation}
 w_{2p-1}
 =h^{p-2}w_3(W)
 +\binom{p-1}{2}h^{p-3}w_5(W).
 \label{eq:shifted-family-detector-expansion}
\end{equation}
Multiplication by $w_2=h+w_2(W)$ shows that the coefficient of $h^{p-2}$ is $w_2(W)w_3(W)+\binom{p-1}{2}w_5(W)$. The second term integrates to zero because $w_5(W)$ is the mod-2 Euler class of a closed odd-dimensional
manifold. Therefore,
\begin{equation}
    \int_{W^5\times\mathbb {CP}^{p-2}}w_2w_{2p-1}=1,
    \label{eq:shifted-family-detector}
\end{equation}
where for $p = 2$, the detecting manifold is $W^5$ itself, and $ \int_{W^5} w_2w_3 = 1$.

Let $M^{2p}$ be a closed spatial manifold and work first in the locally trivial sector
\begin{equation}
 A=\delta a,
 \qquad B=\delta b,
 \qquad a\in C^{p-1}(M;\mathbb Z_2),
 \quad b\in C^p(M;\mathbb Z_2).
\end{equation}
The spatial cochain
\begin{equation}
 \phi_{2p}(a,b)
 :=b\cup b+b\cup_1\delta b
 +a\cup\left(\delta a\cup_{p-1}\delta a+\delta b\right)
 \label{eq:shifted-family-spatial-cocycle}
\end{equation}
obeys the exact transgression identity
\begin{equation}
 \delta\phi_{2p}
 =B\cup_1B+A\cup B+A\cup(A\cup_{p-1}A).
 \label{eq:shifted-family-transgression}
\end{equation}

The finite endpoint changes are most transparent before restricting to single-cell moves. For $s\in C^p(M;\mathbb Z_2)$ and $r\in C^{p-1}(M;\mathbb Z_2)$, the higher-cup recursion gives
\begin{align}
 \int_M\!\left[\phi(a,b+s)+\phi(a,b)\right]
 ={}&\int_M\left[
 B\cup_2\delta s+A\cup s+s\cup s+s\cup_1\delta s
 \right],
 \label{eq:shifted-family-b-variation}\\
 \int_M\!\left[\phi(a+r,b)+\phi(a,b)\right]
 ={}&\int_M\Bigl\{
 r\cup\left(A\cup_{p-1}A+B\right)
 +(A+\delta r)\cup\left[
 A\cup_p\delta r+r\cup_{p-1}\delta r+r\cup_{p-2}r
 \right]\Bigr\}.
 \label{eq:shifted-family-a-variation}
\end{align}

For the indicator $\boldsymbol\sigma$ of one $p$-cell and the indicator $\boldsymbol r$ of one $(p-1)$-cell, the configuration-dependent parts reduce to
\begin{align}
 \int_M\!\left[\phi(a,b+\boldsymbol\sigma)+\phi(a,b)\right]
 ={}&\int_M\left[
 B\cup_2\delta\boldsymbol\sigma+A\cup\boldsymbol\sigma
 \right],\\
 \int_M\!\left[\phi(a+\boldsymbol r,b)+\phi(a,b)\right]
 ={}&\int_M\Bigl[
 A\cup\left(
 \boldsymbol r\cup_{p-1}\delta\boldsymbol r
 +A\cup_p\delta\boldsymbol r\right)
 +\boldsymbol r\cup\left(A\cup_{p-1}A+B\right)
 \Bigr].
 \label{eq:shifted-family-elementary-variations}
\end{align}
Any remaining cell-dependent sign is included in the phase convention of the bare shift operator. Notice that the first move always involves $\cup_2$. Only the indices in the second move grow with dimension.

Place an $A$ qubit on each $p$-cell $\alpha$ and a $B$ qubit on each $(p+1)$-cell $\tau$.  The first elementary kinetic operator is
\begin{equation}
 G^B_\sigma
 :=X^B_{\delta\boldsymbol\sigma}
 \prod_{\tau'}(Z^B_{\tau'})^{
  \int_M\boldsymbol\tau'\cup_2\delta\boldsymbol\sigma}.
 \label{eq:shifted-family-G}
\end{equation}
These operators are commuting involutions.  Their mutual commutator is the integral of $\delta(\delta\boldsymbol\sigma\cup_3 \delta\boldsymbol\sigma')$, while their square is the evaluation of a Steenrod square on an exact cochain. Both vanish on closed $M$.

For a $(p-1)$-cell $r$, introduce the local incidence coefficients
\begin{align}
 d_r(\alpha)
 &:=\int_M\boldsymbol\alpha\cup
       (\boldsymbol r\cup_{p-1}\delta\boldsymbol r),
 \label{eq:shifted-family-dr}\\
 \Theta_r(\alpha_1,\alpha_2)
 &:=\int_M\left[
 \boldsymbol r\cup
  (\boldsymbol\alpha_1\cup_{p-1}\boldsymbol\alpha_2)
 +\boldsymbol\alpha_1\cup
  (\boldsymbol\alpha_2\cup_p\delta\boldsymbol r)
 \right].
 \label{eq:shifted-family-theta}
\end{align}
For any commuting involutions $z_\alpha$, set
\begin{equation}
 \mathcal D_r(z)
 :=\prod_\alpha z_\alpha^{d_r(\alpha)}
 \prod_{\alpha_1,\alpha_2}
CZ(z_{\alpha_1},z_{\alpha_2})^{
 \Theta_r(\alpha_1,\alpha_2)},
 \qquad
 \widetilde T_r^A:=X^A_{\delta\boldsymbol r}\mathcal D_r(Z^A).
 \label{eq:shifted-family-A-move}
\end{equation}
With this notation, the gauged Hamiltonian is
\begin{align}
 H_k^{\mathrm{flat}}
 ={}&-\sum_{\sigma\in M_p}G^B_\sigma
  \prod_{\alpha\in M_p}(Z^A_\alpha)^{
  \int_M\boldsymbol\alpha\cup\boldsymbol\sigma}
 -\sum_{r\in M_{p-1}}\widetilde T^A_r
  \prod_{\tau\in M_{p+1}}(Z^B_\tau)^{
  \int_M\boldsymbol r\cup\boldsymbol\tau}
 -\sum_{c_{p+1}}Z^A_{\partial c_{p+1}}
  -\sum_{c_{p+2}}Z^B_{\partial c_{p+2}}.
 \label{eq:shifted-family-hamiltonian}
\end{align}
This formula is derived for $\delta A=\delta B=0$. In particular, $\widetilde T_r^A$ need not be an involution on an arbitrary non-closed configuration. The separator construction below provides the full-space commuting completion.

For the chosen hypercubic cup diagonal, every $p$-cell $\alpha$ has a unique complementary $p$-cell $\sigma(\alpha)$, and every $(p+1)$-cell $\tau$ has a unique complementary $(p-1)$-cell $r(\tau)$:
\begin{equation}
 \int_M\boldsymbol\alpha'\cup\boldsymbol{\sigma(\alpha)}
 =\delta_{\alpha\alpha'},
 \qquad
 \int_M\boldsymbol{r(\tau)}\cup\boldsymbol\tau'
 =\delta_{\tau\tau'}.
 \label{eq:shifted-family-pairings}
\end{equation}
Define the commuting involution
\begin{equation}
 Q^B_\alpha:=G^B_{\sigma(\alpha)}.
\end{equation}
The first separator is
\begin{equation}
 \boxed{\overline Z^A_\alpha:=Z^A_\alpha Q^B_\alpha.}
 \label{eq:shifted-family-separator-A}
\end{equation}
Its $+1$ constraint identifies $Z^A_\alpha$ with $Q^B_\alpha$. We use this relation to define the substituted move and the second separator,
\begin{equation}
 \widehat T_r^A
 :=X^A_{\delta\boldsymbol r}\mathcal D_r(Q^B),
 \label{eq:shifted-family-substituted-A-move}
\end{equation}
\begin{equation}
 \boxed{\overline Z^B_\tau
 :=Z^B_\tau\widehat T^A_{r(\tau)}.}
 \label{eq:shifted-family-separator-B}
\end{equation}
The replacement $Z^A_\alpha\mapsto Q^B_\alpha$ agrees with the original kinetic term in the common $\overline Z^A=1$ sector and specifies its extension away from that sector.

The two pairings obey the cellular Stokes relation
\begin{equation}
 (\delta\boldsymbol{r(\tau)})(\alpha)
 =(\delta\boldsymbol{\sigma(\alpha)})(\tau),
 \qquad
 Z^B_\tau Q^B_\alpha Z^B_\tau
 =(-1)^{(\delta\boldsymbol{r(\tau)})(\alpha)}Q^B_\alpha.
 \label{eq:shifted-family-pairing-compatibility}
\end{equation}
We apply the endpoint identities to commuting involutions $z_\alpha$ satisfying $\prod_{\alpha\subset\partial
c}z_\alpha=1$ for every $(p+1)$-cell $c$. Their joint eigenvalues define closed $p$-cochains, and shifts by
$\delta\boldsymbol r$ preserve this condition. Writing $z^c_\alpha:=(-1)^{c(\alpha)}z_\alpha$, the
finite-variation formulas give
\begin{align}
 \mathcal D_r(z^{\delta\boldsymbol r})\mathcal D_r(z)&=1,
 \nonumber\\
 \mathcal D_r(z^{\delta\boldsymbol s})\mathcal D_s(z)
 &=\mathcal D_s(z^{\delta\boldsymbol r})\mathcal D_r(z).
\end{align}
The first relation proves $(\overline Z^B_\tau)^2=1$, and the second proves $[\overline Z^B_\tau,\overline Z^B_{\tau'}]=0$. In the mixed commutator, the sign from $Z^A_\alpha$ crossing $X^A_{\delta\boldsymbol{r(\tau)}}$ is canceled by the identical sign from $Q^B_\alpha$ crossing $Z^B_\tau$. Therefore,
\begin{equation}
 (\overline Z^A_\alpha)^2=(\overline Z^B_\tau)^2=1,
 \qquad
 [\overline Z^A_\alpha,\overline Z^A_{\alpha'}]
 = [\overline Z^A_\alpha,\overline Z^B_\tau]
 = [\overline Z^B_\tau,\overline Z^B_{\tau'}]=0.
 \label{eq:shifted-family-separator-algebra}
\end{equation}

The separator constraints also generate the two flux families in Eq.~\eqref{eq:shifted-family-hamiltonian}.  For a $(p+1)$-cell $c_{p+1}$,
\begin{equation}
 \prod_{\alpha\subset\partial c_{p+1}}\overline Z^A_\alpha
 =Z^A_{\partial c_{p+1}}.
 \label{eq:shifted-family-A-flux-generated}
\end{equation}
The accompanying $Q^B$ operators form the $B$ move with exact parameter $s=\delta q\in C^p(M;\mathbb Z_2)$ for a $(p-1)$-cochain $q$. Its $X^B$ support is $\delta s=\delta^2q=0$, and its diagonal dressing cancels by the higher-cup recursion. Hence, the first separator family imposes $\delta A=0$.

Let $q(c_{p+2})$ be the $(p-2)$-cell complementary to a $(p+2)$-cell $c_{p+2}$ and set $r=\delta\boldsymbol{q(c_{p+2})}$. Since $\delta r=0$, the exact endpoint formula yields
\begin{align}
 \left.\prod_{\tau\subset\partial c_{p+2}}
 \overline Z^B_\tau\right|_{\delta A=0}
 =(-1)^{\int_M\left[
 r\cup(A\cup_{p-1}A+B)+A\cup(r\cup_{p-2}r)
 \right]}.
 \label{eq:shifted-family-B-flux-intermediate}
\end{align}
The second term vanishes because
\begin{equation}
 r\cup_{p-2}r
 =\delta\left[
 \boldsymbol{q(c_{p+2})}\cup_{p-3}\boldsymbol{q(c_{p+2})}
 +\boldsymbol{q(c_{p+2})}\cup_{p-2}r\right],
\end{equation}
where the $\cup_{-1}$ term is absent for $p=2$. Its pairing with closed $A$ is therefore zero. The term involving $A\cup_{p-1}A$ also integrates to zero when $\delta A=0$, leaving
\begin{equation}
 \left.\prod_{\tau\subset\partial c_{p+2}}
 \overline Z^B_\tau\right|_{\delta A=0}
 =(-1)^{\int_M\boldsymbol{q(c_{p+2})}\cup\delta B}
 =Z^B_{\partial c_{p+2}}.
 \label{eq:shifted-family-B-flux-generated}
\end{equation}
Thus, no separate flux terms are needed in the full-space separator Hamiltonian
\begin{equation}
 H_k^{\mathrm{QCA}}
 :=-\sum_{\alpha\in M_p}\overline Z^A_\alpha
   -\sum_{\tau\in M_{p+1}}\overline Z^B_\tau.
 \label{eq:shifted-family-qca-hamiltonian}
\end{equation}

For $(p+1)$-cells $\tau,\tau'$, define
\begin{equation}
 m_{\tau\tau'}:=\int_M\boldsymbol\tau\cup_2\boldsymbol\tau'.
\end{equation}
The paired flippers are
\begin{equation}
\boxed{
 \overline X^A_\alpha:=X^A_\alpha,
 \qquad
 \overline X^B_\tau
 :=\left[X^B_\tau
  \prod_{\tau'}(Z^B_{\tau'})^{m_{\tau\tau'}}\right]
 \left[
  \prod_{\tau'}(\overline Z^B_{\tau'})^{m_{\tau\tau'}}
  \right].}
 \label{eq:shifted-family-flippers}
\end{equation}
The left-to-right order is part of the definition. The first bracket commutes with every $G^B_\sigma$ because its two overlap phases are both $\int_M\boldsymbol\tau\cup_2\delta\boldsymbol\sigma$. Its residual square phase is $(-1)^{m_{\tau\tau}}$, and its mutual phase with the $\tau'$ flipper is $(-1)^{m_{\tau\tau'}+m_{\tau'\tau}}$.  Moving the first brackets through the separator dressings produces the same phases once more, so both cancel. So,
\begin{equation}
 (\overline X_i)^2=(\overline Z_i)^2=1,
 \qquad
 [\overline X_i,\overline X_j]
 =[\overline Z_i,\overline Z_j]=0,
 \qquad
 \overline Z_i\overline X_j
 =(-1)^{\delta_{ij}}\overline X_j\overline Z_i,
 \label{eq:shifted-family-canonical-algebra}
\end{equation}
where $i,j$ range over both qubit species.  The flippers establish the independence of the separators, and there is one pair per physical qubit. All cup-product incidences and complementary-cell pairings have bounded
diameter for fixed $k$. Hence,
\begin{equation}
 Z^A_\alpha\mapsto\overline Z^A_\alpha,
 \quad X^A_\alpha\mapsto\overline X^A_\alpha,
 \qquad
 Z^B_\tau\mapsto\overline Z^B_\tau,
 \quad X^B_\tau\mapsto\overline X^B_\tau
 \label{eq:shifted-family-qca-map}
\end{equation}
defines a QCA in $4k$ spatial dimensions.

\subsection{General products of Wu classes}
\label{subsec:general-wu-products}

The preceding two families contain Steenrod operations tied to particular cochain degrees.  We now give a separate construction for an arbitrary ordered product of Wu classes. Let
\begin{equation}
 D=d+1,
 \qquad
 n_1+\cdots+n_L=D,
 \qquad
 \ell_i:=D-2n_i.
 \label{eq:general-wu-degrees}
\end{equation}
It is enough to consider $1\leq n_i\leq D/2$. If $n_i>D/2$, then $\nu_{n_i}$ vanishes by instability on a $D$-manifold, so the full product is zero. When $\ell_i=0$, terms containing $\cup_{-1}$ are omitted.

Introduce $L$ pairs of cocycles
\begin{equation}
    A_i\in Z^{D-n_i}(X^D;\mathbb Z_2), \qquad B_i\in Z^{n_i}(X^D;\mathbb Z_2),
\end{equation}
and fix the displayed order of all ordinary cup products. Consider
\begin{equation}
    \Phi_{\boldsymbol n}[A,B]=\frac12\int_{X^D}\left\{\sum_{i=1}^{L}\left[A_i\cup_{\ell_i}A_i+B_i\cup A_i\right]+B_1\cup B_2\cup\cdots\cup B_L\right\}.
    \label{eq:general-wu-action}
\end{equation}
The first term represents $\Sq^{n_i}A_i$. On a closed oriented manifold, the Wu formula gives
\begin{equation}
    \int_X\Sq^{n_i}A_i=\int_X\nu_{n_i}\cup A_i.
\end{equation}
Summing over $A_i$ therefore imposes $[B_i]=\nu_{n_i}$ for every $i$. The remaining response is
\begin{equation}
    \Phi_{\boldsymbol n}^{\mathrm{grav}}[X]
    =\frac12\int_X\nu_{n_1}\cup\nu_{n_2}\cup\cdots\cup\nu_{n_L}.
    \label{eq:general-wu-response}
\end{equation}

To derive the microscopic operators, let $Y^D$ have spatial boundary $M^d$ and write
\begin{equation}
 A_i=\delta a_i,
 \qquad B_i=\delta b_i,
 \qquad
 a_i\in C^{D-n_i-1}(Y;\mathbb Z_2),
 \quad
 b_i\in C^{n_i-1}(Y;\mathbb Z_2).
\end{equation}
An exact descendant of the bulk integrand is
\begin{align}
    \phi_{\boldsymbol n}(a,b) :={}&\sum_{i=1}^{L}\left[ a_i\cup_{\ell_i}\delta a_i+a_i\cup_{\ell_i-1}a_i +b_i\cup\delta a_i\right] +b_1\cup\delta b_2\cup\cdots\cup\delta b_L.
    \label{eq:general-wu-descendant}
\end{align}
Indeed,
\begin{equation}
    \delta\phi_{\boldsymbol n} =\sum_{i=1}^{L}\left[A_i\cup_{\ell_i}A_i+B_i\cup A_i\right] +B_1\cup\cdots\cup B_L.
    \label{eq:general-wu-transgression}
\end{equation}
The corresponding SPT state is
\begin{equation}
    \ket{\psi_{\boldsymbol n}}=\sum_{\{a_i,b_i\}}(-1)^{\int_M\phi_{\boldsymbol n}(a,b)}\ket{\{a_i,b_i\}}.
    \label{eq:general-wu-spt-state}
\end{equation}
For an elementary $(D-n_i-1)$-cell cochain $\boldsymbol t_i$, define the
configuration-independent self-phase
\begin{equation}
 \kappa_i(\boldsymbol t_i)
 :=\int_M\left[
 \boldsymbol t_i\cup_{\ell_i}\delta\boldsymbol t_i
 +\boldsymbol t_i\cup_{\ell_i-1}\boldsymbol t_i\right].
\end{equation}
The exact change under $a_i\mapsto a_i+\boldsymbol t_i$ is
\begin{equation}
 \frac{
 \braket{\{a_i+\boldsymbol t_i\},\{b_j\}}
 {\psi_{\boldsymbol n}}}
 {\braket{\{a_i\},\{b_i\}}{\psi_{\boldsymbol n}}}
 =(-1)^{\kappa_i(\boldsymbol t_i)+
 \int_M\left[
 \delta\boldsymbol t_i\cup_{\ell_i+1}A_i
 +B_i\cup\boldsymbol t_i\right]}.
 \label{eq:general-wu-a-variation}
\end{equation}
Similarly, let $\boldsymbol p_i$ be the indicator of an $(n_i-1)$-cell and define the ordered product
\begin{equation}
    \mathcal P_i(B;\boldsymbol p_i):=B_1\cup\cdots\cup B_{i-1}\cup\boldsymbol p_i\cup B_{i+1}\cup\cdots\cup B_L.
    \label{eq:general-wu-product-endpoint}
\end{equation}
For $i=1$, the first string of factors is absent. Integrating by parts
without reordering any factors gives
\begin{equation}
 \frac{
 \braket{\{a_j\},\{b_j+\delta_{ij}\boldsymbol p_i\}}
 {\psi_{\boldsymbol n}}}
 {\braket{\{a_j\},\{b_j\}}{\psi_{\boldsymbol n}}}
 =(-1)^{\int_M\left[
 \boldsymbol p_i\cup A_i
 +\mathcal P_i(B;\boldsymbol p_i)\right]}.
 \label{eq:general-wu-b-variation}
\end{equation}
Eqs.~\eqref{eq:general-wu-a-variation} and \eqref{eq:general-wu-b-variation} are the only endpoint identities needed for the construction.

After gauging, the $A_i$ qubits occupy $(D-n_i)$-cells $r$, and the $B_i$ qubits occupy $n_i$-cells $q$. For a $(D-n_i-1)$-cell $t$ and an $(n_i-1)$-cell $p$, define
\begin{align}
 G^{A_i}_t
 &:={}
 (-1)^{\kappa_i(\boldsymbol t)}X^{A_i}_{\delta\boldsymbol t}
 \prod_r(Z^{A_i}_r)^{
 \int_M\delta\boldsymbol t\cup_{\ell_i+1}\boldsymbol r},
 \label{eq:general-wu-GA}\\
 G^{B_i}_p
 &:=X^{B_i}_{\delta\boldsymbol p}\,\mathcal D_{i,p}(Z^B).
 \label{eq:general-wu-GB}
\end{align}
Here, the diagonal operator $\mathcal D_{i,p}$ is specified on closed configurations by
\begin{equation}
 \mathcal D_{i,p}(Z^B)\ket{\{B_j\}}
 :=(-1)^{\int_M\mathcal P_i(B;\boldsymbol p)}
 \ket{\{B_j\}}.
 \label{eq:general-wu-spectral-phase}
\end{equation}
It is a finite product of multi-qubit controlled-phase gates involving the $B_j$ species with $j\neq i$.

The gauged Hamiltonian on the common closed sector is
\begin{eqs}
 H_{\boldsymbol n}^{\mathrm{flat}}
 ={}&-\sum_{i=1}^{L}\sum_tG^{A_i}_t
 \prod_q(Z^{B_i}_q)^{\int_M\boldsymbol q\cup\boldsymbol t}
 -\sum_{i=1}^{L}\sum_pG^{B_i}_p
 \prod_r(Z^{A_i}_r)^{\int_M\boldsymbol p\cup\boldsymbol r} \\
 &-\sum_{i=1}^{L}\sum_{c\in M_{n_i+1}}Z^{B_i}_{\partial c}
 -\sum_{i=1}^{L}\sum_{c\in M_{D-n_i+1}}Z^{A_i}_{\partial c}.
 \label{eq:general-wu-flat-hamiltonian}
\end{eqs}
As before, the endpoint phases in this expression are initially determined only on closed cochain configurations.

For each $B_i$ cell $q$, let $t_i(q)$ be its complementary $(D-n_i-1)$-cell; for each $A_i$ cell $r$, let $p_i(r)$ be its
complementary $(n_i-1)$-cell. Thus,
\begin{equation}
 \int_M\boldsymbol q'\cup\boldsymbol{t_i(q)}=\delta_{qq'},
 \qquad
 \int_M\boldsymbol{p_i(r)}\cup\boldsymbol r'=\delta_{rr'}.
 \label{eq:general-wu-pairings}
\end{equation}
Define
\begin{equation}
 Q^{A_i}_q:=G^{A_i}_{t_i(q)}.
\end{equation}
The $Q^{A_i}_q$ commute and obey the same local closure relations as a closed $B_i$ configuration. We may therefore replace every spectral $Z^{B_j}_q$ in $\mathcal D_{i,p}$ by $Q^{A_j}_q$ for $j\neq i$
\begin{equation}
 \widehat G^{B_i}_p
 :=X^{B_i}_{\delta\boldsymbol p}
 \left[\mathcal D_{i,p}(Z^B)\right]_{
 Z^{B_j}_q\mapsto Q^{A_j}_q,\ j\neq i}.
 \label{eq:general-wu-substituted-GB}
\end{equation}
The full-space separators are
\begin{equation}
\boxed{
\begin{aligned}
     \overline Z^{B_i}_q
     &:=Z^{B_i}_qQ^{A_i}_q,\\
     \overline Z^{A_i}_r
     &:=Z^{A_i}_r\widehat G^{B_i}_{p_i(r)}.
    \end{aligned}}
    \label{eq:general-wu-separators}
\end{equation}
On the common $\overline Z^{B_i}=1$ sector, the substitution reduces to $Z^{B_i}_q=Q^{A_i}_q$ and reproduces the original kinetic terms in Eq.~\eqref{eq:general-wu-flat-hamiltonian}.

We next verify that Eq.~\eqref{eq:general-wu-separators} defines a commuting family on the full Hilbert space. For each species, the
complementary pairings obey
\begin{equation}
 (\delta\boldsymbol{t_i(q)})(r)
 =(\delta\boldsymbol{p_i(r)})(q).
 \label{eq:general-wu-stokes-pairing}
\end{equation}
This identity cancels the two Pauli signs in every mixed $A_i$-$B_i$ commutator. The remaining commutators are controlled by the finite differences of the single scalar phase $\int_M\phi_{\boldsymbol n}$. Since two cochain shifts commute,
\begin{equation}
 \Delta_{p_i}\Delta_{p_j}\int_M\phi_{\boldsymbol n}
 =\Delta_{p_j}\Delta_{p_i}\int_M\phi_{\boldsymbol n},
 \label{eq:general-wu-mixed-finite-difference}
\end{equation}
including when $i\neq j$. This proves that all substituted $B$ moves commute. Applying an elementary move twice gives unity, so every separator also squares to one. Hence,
\begin{equation}
 (\overline Z_I)^2=1,
 \qquad
 [\overline Z_I,\overline Z_J]=0
 \label{eq:general-wu-separator-algebra}
\end{equation}
for all species and cell labels $I,J$.

Products of these separators reproduce the flux constraints in Eq.~\eqref{eq:general-wu-flat-hamiltonian}. Multiplying $\overline Z^{B_i}_q$ around the boundary of an $(n_i+1)$-cell gives $Z^{B_i}_{\partial c}$ because the accompanying $Q^{A_i}$ operators form an exact composite move
\begin{equation}
 \prod_{q\subset\partial c_{n_i+1}}\overline Z^{B_i}_q
 =Z^{B_i}_{\partial c_{n_i+1}}.
 \label{eq:general-wu-B-flux-generated}
\end{equation}
After imposing $\delta B_j=0$ for every $j$, multiplying $\overline Z^{A_i}_r$ around a $(D-n_i+1)$-cell gives
\begin{equation}
 \left.
 \prod_{r\subset\partial c_{D-n_i+1}}\overline Z^{A_i}_r
 \right|_{\delta B_j=0}
 =Z^{A_i}_{\partial c_{D-n_i+1}}.
 \label{eq:general-wu-A-flux-generated}
\end{equation}
Here, the product phase in Eq.~\eqref{eq:general-wu-product-endpoint} is unchanged by the corresponding exact $b_i$ shift. Thus, no separate flux operators are required in the full-space separator Hamiltonian
\begin{equation}
 H_{\boldsymbol n}^{\mathrm{QCA}}
 :=-\sum_{i=1}^{L}\sum_q\overline Z^{B_i}_q
   -\sum_{i=1}^{L}\sum_r\overline Z^{A_i}_r.
 \label{eq:general-wu-qca-hamiltonian}
\end{equation}

For a fixed species $i$, define the provisional $A_i$ flipper
\begin{equation}
 V^{A_i}_r
 :=X^{A_i}_r
 \prod_{r'}(Z^{A_i}_{r'})^{
 \int_M\boldsymbol r'\cup_{\ell_i+1}\boldsymbol r}.
 \label{eq:general-wu-provisional-flippers}
\end{equation}
Its two Pauli crossings with every $G^{A_i}_t$ are equal and cancel, so $V^{A_i}_r$ commutes with all $Q^{A_i}_q$ and hence with all substituted spectral phases. It flips only its paired separator $\overline Z^{A_i}_r$. Its square and mutual-commutator phases are removed by attaching the same ordered product of $A_i$ separators. The final
flippers are
\begin{equation}
\boxed{
\begin{aligned}
 \overline X^{B_i}_q
 &:=X^{B_i}_q,\\
 \overline X^{A_i}_r
 &:=\left[
 X^{A_i}_r
 \prod_{r'}(Z^{A_i}_{r'})^{
 \int_M\boldsymbol r'\cup_{\ell_i+1}\boldsymbol r}
 \right]
 \left[
 \prod_{r'}(\overline Z^{A_i}_{r'})^{
 \int_M\boldsymbol r'\cup_{\ell_i+1}\boldsymbol r}
 \right].
\end{aligned}}
\label{eq:general-wu-commuting-flippers}
\end{equation}
The displayed order is part of the definition. Moving the first bracket through the separator bracket reproduces its square and mutual-commutator phases once more, so both cancel. The $B_i$ flippers require no additional dressing because the substitution in Eq.~\eqref{eq:general-wu-substituted-GB} removes every bare $Z^B$ from the $A$-separator family.

The complete algebra is therefore
\begin{equation}
 (\overline X_I)^2=(\overline Z_I)^2=1,
 \qquad
 [\overline X_I,\overline X_J]
 =[\overline Z_I,\overline Z_J]=0,
 \qquad
 \overline Z_I\overline X_J
 =(-1)^{\delta_{IJ}}\overline X_J\overline Z_I.
 \label{eq:general-wu-separator-flipper-algebra}
\end{equation}
The flippers exclude nontrivial local relations among the separators, and there is one pair per microscopic qubit. For fixed $\boldsymbol n=(n_1,\ldots,n_L)$, all cup-product incidences and complementary-cell pairings have bounded range. Thus,
\begin{equation}
 Z_I\longmapsto\overline Z_I,
 \qquad
 X_I\longmapsto\overline X_I
 \label{eq:general-wu-qca-map}
\end{equation}
defines a QCA realizing the response $\frac12\int\nu_{n_1}\cdots\nu_{n_L}$.

This construction establishes the microscopic QCA whenever the response is specified by a product of Wu classes. It does not imply that every such oriented characteristic number is nonzero, nor that every resulting QCA is stably nontrivial. These are subsequent classification questions, just as in the $w_2^nw_3^m$ family.

\subsection{Steenrod-decorated Wu responses}
\label{subsec:steenrod-decorated-wu}

The construction of Sec.~\ref{subsec:general-wu-products} extends from products of Wu classes to products of their Steenrod squares. The new ingredient is that a Steenrod square is nonlinear in its input cochain. So, its finite variation contains terms that are quadratic in the elementary shift. These terms make the explicit separator dressing more involved, but they do not change the separator-flipper strategy.  Once the full finite variation is retained, complementary-cell pairing and spectral substitution again produce a complete local algebra.

Let $X^D$ be a closed oriented spacetime manifold. Choose integers
\begin{equation}
 1\leq n_i\leq \frac{D}{2},
 \qquad
 0\leq r_i\leq n_i,
 \qquad
 \sum_{i=1}^L(n_i+r_i)=D,
 \qquad
 \ell_i:=D-2n_i .
 \label{eq:steenrod-wu-degrees}
\end{equation}
For a closed $n$-cochain $B$, define the cochain representative
\begin{equation}
 \mathcal S_{n,r}(B):=
 \begin{cases}
  B, & r=0,\\[2pt]
  B\cup_{n-r}B, & 1\leq r\leq n.
 \end{cases}
 \label{eq:steenrod-wu-cochain-square}
\end{equation}
Thus, $[\mathcal S_{n,r}(B)]=\Sq^r[B]$.  Introduce cocycles
\begin{equation}
 A_i\in Z^{D-n_i}(X;\mathbb Z_2),
 \qquad
 B_i\in Z^{n_i}(X;\mathbb Z_2),
 \label{eq:steenrod-wu-fields}
\end{equation}
and consider the action
\begin{equation}
 \Phi_{\boldsymbol n,\boldsymbol r}[A,B]
 =\frac12\int_{X^D}\left\{
 \sum_{i=1}^L
 \left[A_i\cup_{\ell_i}A_i+B_i\cup A_i\right]
 +\mathcal S_{n_1,r_1}(B_1)\cup\cdots\cup
  \mathcal S_{n_L,r_L}(B_L)
 \right\}.
 \label{eq:steenrod-wu-action}
\end{equation}
The displayed ordering of the ordinary cup products is fixed. Since $A_i\cup_{\ell_i}A_i$ represents $\Sq^{n_i}A_i$, the Wu formula gives
\begin{equation}
    \int_X A_i\cup_{\ell_i}A_i =\int_X\nu_{n_i}\cup A_i .
\end{equation}
Summing over $A_i$ therefore imposes $[B_i]=\nu_{n_i}$ and leaves the gravitational response
\begin{equation}
 \Phi^{\mathrm{grav}}_{\boldsymbol n,\boldsymbol r}[X]
 =\frac12\int_X
 \Sq^{r_1}\nu_{n_1}\cup\cdots\cup
 \Sq^{r_L}\nu_{n_L}.
 \label{eq:steenrod-wu-response}
\end{equation}
The product-of-Wu-classes construction of Sec.~\ref{subsec:general-wu-products} is recovered by setting every $r_i=0$.

The spatial descendant and its finite endpoint phases can be written uniformly. For $B=\delta b$, set
\begin{equation}
 \mathcal T_{n,r}(b):=
 \begin{cases}
  b, & r=0,\\[2pt]
  b\cup_{n-r}\delta b+b\cup_{n-r-1}b,
   &1\leq r\leq n,
 \end{cases}
 \label{eq:steenrod-wu-transgression}
\end{equation}
where a term containing $\cup_{-1}$ is omitted. The higher-cup recursion yields
\begin{equation}
 \delta\mathcal T_{n,r}(b)
 =\mathcal S_{n,r}(\delta b).
 \label{eq:steenrod-wu-transgression-identity}
\end{equation}
Writing $A_i=\delta a_i$ and $B_i=\delta b_i$, one convenient ordered descendant of Eq.~\eqref{eq:steenrod-wu-action} is
\begin{equation}
\begin{aligned}
 \varphi_{\boldsymbol n,\boldsymbol r}(a,b)
 :={}&\sum_{i=1}^L\left[
 a_i\cup_{\ell_i}\delta a_i
 +a_i\cup_{\ell_i-1}a_i
 +b_i\cup\delta a_i\right]\\
 &+\mathcal T_{n_1,r_1}(b_1)
 \cup\mathcal S_{n_2,r_2}(B_2)\cup\cdots\cup
 \mathcal S_{n_L,r_L}(B_L).
 \label{eq:steenrod-wu-spatial-descendant}
\end{aligned}
\end{equation}
Indeed, $\delta\varphi_{\boldsymbol n,\boldsymbol r}$ is the integrand of Eq.~\eqref{eq:steenrod-wu-action}.

We next retain the exact finite change under $b\mapsto b+p$. Let $u:=\delta p$. Define
\begin{equation}
 \Lambda_{n,r}(B;p):=
 \begin{cases}
  p, & r=0,\\[2pt]
  B\cup_{n-r}p+p\cup_{n-r}B
  +p\cup_{n-r}u+p\cup_{n-r-1}p,
  &1\leq r\leq n.
 \end{cases}
 \label{eq:steenrod-wu-finite-transgression}
\end{equation}
For closed $B$, direct application of the higher-cup recursion gives
\begin{align}
 \delta\Lambda_{n,r}(B;p)
 &={\cal S}_{n,r}(B+u)+{\cal S}_{n,r}(B),
 \label{eq:steenrod-wu-finite-transgression-identity}\\
 \mathcal T_{n,r}(b+p)+\mathcal T_{n,r}(b)
 &=\Lambda_{n,r}(\delta b;p)
 +\delta\!\left(b\cup_{n-r}p\right)
 \qquad (r>0).
 \label{eq:steenrod-wu-transgression-shift}
\end{align}
The last term in Eq.~\eqref{eq:steenrod-wu-transgression-shift} integrates to zero after multiplication by the remaining closed factors. These identities display the important nonlinear terms $p\cup_{n-r}\delta p$ and $p\cup_{n-r-1}p$.  Dropping them would generally spoil the order and mutual commutativity of the resulting separator candidates.

For a shift of the $i$th field, define the ordered endpoint polynomial
\begin{equation}
\begin{aligned}
 \mathcal P_i(B;p_i):={}&
 \mathcal S_{n_1,r_1}(B_1)\cup\cdots\cup
 \mathcal S_{n_{i-1},r_{i-1}}(B_{i-1})
 \cup\Lambda_{n_i,r_i}(B_i;p_i)\\
 &\cup\mathcal S_{n_{i+1},r_{i+1}}(B_{i+1})
 \cup\cdots\cup\mathcal S_{n_L,r_L}(B_L).
 \label{eq:steenrod-wu-endpoint-polynomial}
\end{aligned}
\end{equation}
Ordered integration by parts then gives the exact amplitude ratio
\begin{equation}
 \frac{\langle\{a_j\},\{b_j+\delta_{ij}p_i\}
 |\psi_{\boldsymbol n,\boldsymbol r}\rangle}
 {\langle\{a_j\},\{b_j\}|\psi_{\boldsymbol n,\boldsymbol r}\rangle}
 =(-1)^{\int_M[p_i\cup A_i+\mathcal P_i(B;p_i)]}.
 \label{eq:steenrod-wu-b-variation}
\end{equation}
The $a_i$ variation is unchanged from Eq.~\eqref{eq:general-wu-a-variation}:
\begin{equation}
 \frac{\langle\{a_j+\delta_{ij}s_i\},\{b_j\}
 |\psi_{\boldsymbol n,\boldsymbol r}\rangle}
 {\langle\{a_j\},\{b_j\}|\psi_{\boldsymbol n,\boldsymbol r}\rangle}
 =(-1)^{\kappa_i(s_i)+\int_M[
 \delta s_i\cup_{\ell_i+1}A_i+B_i\cup s_i]},
 \label{eq:steenrod-wu-a-variation}
\end{equation}
where
\begin{equation}
 \kappa_i(s_i)
 :=\int_M\left[s_i\cup_{\ell_i}\delta s_i
 +s_i\cup_{\ell_i-1}s_i\right].
\end{equation}

After gauging, the $A_i$ qubits occupy $(D-n_i)$-cells and the $B_i$ qubits occupy $n_i$-cells.  The $A_i$ move is the same as in Eq.~\eqref{eq:general-wu-GA}. For an $(n_i-1)$-cell $p$, define the diagonal endpoint operator on closed configurations by
\begin{equation}
 \mathcal D_{i,p}(Z^B)|\{B_j\}\rangle
 :=(-1)^{\int_M\mathcal P_i(B;p)}|\{B_j\}\rangle,
 \qquad
 G^{B_i}_p:=X^{B_i}_{\delta p}\mathcal D_{i,p}(Z^B).
 \label{eq:steenrod-wu-spectral-phase}
\end{equation}
The difference from Sec.~\ref{subsec:general-wu-products} is that $\mathcal D_{i,p}$ generally depends on $B_i$ itself. This is the direct operator-algebraic consequence of the quadratic cochain representative of $\Sq^{r_i}B_i$.

Let $t_i(q)$ and $p_i(r)$ denote the complementary cells introduced in Eq.~\eqref{eq:general-wu-pairings}, and set
\begin{equation}
 Q^{A_i}_q:=G^{A_i}_{t_i(q)}.
\end{equation}
The $Q^{A_i}_q$ commute and obey the local closure relations of a closed $B_i$ configuration. We therefore extend the endpoint phase to the full Hilbert space by substituting every spectral variable, including those of the shifted species
\begin{equation}
 \widehat G^{B_i}_p
 :=X^{B_i}_{\delta p}
 \left[\mathcal D_{i,p}(Z^B)\right]_{
 Z^{B_j}_q\mapsto Q^{A_j}_q,\;1\leq j\leq L}.
 \label{eq:steenrod-wu-substituted-move}
\end{equation}
The complete separators are
\begin{equation}
\boxed{
 \overline Z^{B_i}_q:=Z^{B_i}_qQ^{A_i}_q,
 \qquad
 \overline Z^{A_i}_r:=Z^{A_i}_r
 \widehat G^{B_i}_{p_i(r)}.}
 \label{eq:steenrod-wu-separators}
\end{equation}

The separator algebra follows without expanding the generally lengthy controlled-phase dressing. Let
\begin{equation}
    \Gamma_{i,p}(B):=\int_M\mathcal P_i(B;p),
\end{equation}
and let $B+\delta_i p$ denote a shift of only the $i$th species. Since the endpoint phases are finite differences of the single scalar phase $\int_M\varphi_{\boldsymbol n,\boldsymbol r}$, they obey
\begin{align}
 \Gamma_{i,p}(B)+\Gamma_{i,p}(B+\delta_i p)&=0,
 \label{eq:steenrod-wu-involution-coherence}\\
 \Gamma_{i,p}(B)+\Gamma_{j,p'}(B+\delta_i p)
 &=\Gamma_{j,p'}(B)+\Gamma_{i,p}(B+\delta_jp').
 \label{eq:steenrod-wu-commuting-coherence}
\end{align}
The first identity proves that every substituted move has the required order, while the second proves that two such moves commute. As in Eq.~\eqref{eq:general-wu-stokes-pairing}, complementary-cell Stokes pairing cancels the two signs in each mixed $A_i$-$B_i$ commutator. We then find
\begin{equation}
 (\overline Z_I)^2=1,
 \qquad
 [\overline Z_I,\overline Z_J]=0.
 \label{eq:steenrod-wu-commuting-separators}
\end{equation}
Products of the separators generate the two flux families exactly as in Eqs.~\eqref{eq:general-wu-B-flux-generated} and \eqref{eq:general-wu-A-flux-generated}. The interaction phase is unchanged by the corresponding exact cochain shift, so the full-space separator Hamiltonian is
\begin{equation}
    H^{\mathrm{QCA}}_{\boldsymbol n,\boldsymbol r} :=-\sum_{i=1}^L\sum_q\overline Z^{B_i}_q -\sum_{i=1}^L\sum_r\overline Z^{A_i}_r.
    \label{eq:steenrod-wu-qca-hamiltonian}
\end{equation}

The flippers require no new type of dressing.  Explicitly,
\begin{equation}
    \boxed{
    \begin{aligned}
     \overline X^{B_i}_q&:=X^{B_i}_q,\\
     \overline X^{A_i}_r&:=
     \left[X^{A_i}_r
     \prod_{r'}(Z^{A_i}_{r'})^{\int_Mr'\cup_{\ell_i+1}r}
     \right]
     \left[
     \prod_{r'}(\overline Z^{A_i}_{r'})^{
     \int_Mr'\cup_{\ell_i+1}r}
     \right].
    \end{aligned}}
    \label{eq:steenrod-wu-flippers}
\end{equation}
After the substitution in Eq.~\eqref{eq:steenrod-wu-substituted-move}, the endpoint dressing depends only on the commuting $Q^{A_j}$ operators. Therefore, the bare $B_i$ flipper commutes with every unpaired separator, while the dressed $A_i$ flipper commutes with every $Q^{A_j}$ and hence with any spectral function of them.  The operators in Eqs.~\eqref{eq:steenrod-wu-separators} and \eqref{eq:steenrod-wu-flippers} satisfy
\begin{equation}
 (\overline X_I)^2=(\overline Z_I)^2=1,
 \qquad
 [\overline X_I,\overline X_J]
 =[\overline Z_I,\overline Z_J]=0,
 \qquad
 \overline Z_I\overline X_J
 =(-1)^{\delta_{IJ}}\overline X_J\overline Z_I.
 \label{eq:steenrod-wu-canonical-algebra}
\end{equation}
Thus, the one-to-one assignment
\begin{equation}
    Z_I\longmapsto\overline Z_I, \qquad X_I\longmapsto\overline X_I
    \label{eq:steenrod-wu-qca-map}
\end{equation}
defines a QCA for every fixed set $(\boldsymbol n,\boldsymbol r)$ satisfying Eq.~\eqref{eq:steenrod-wu-degrees}.  All higher-cup incidences and complementary-cell pairings have bounded range for fixed $(\boldsymbol n,\boldsymbol r)$.

A particularly instructive new case occurs in spacetime dimension $D=10$. The number $w_4w_6$ is not a pure product of Wu classes and requires the operation $\Sq^2$. On a closed oriented 10-manifold,
\begin{equation}
 \int_Xw_4w_6
 =\int_X\nu_4\Sq^2\nu_4.
 \label{eq:w4w6-wu-characteristic-number}
\end{equation}
This is an equality of characteristic numbers, not an identity of universal cohomology classes. Indeed, $w_4=\nu_4+\nu_2^2$ and $w_6=\Sq^2\nu_4$ on a 10-manifold, while $\int_Xw_2^2w_6=0$ as shown in Appendix~\ref{app:sw-number-relations}.

To realize this response within the uniform construction, take
\begin{equation}
 (n_1,r_1)=(4,0),
 \qquad
 (n_2,r_2)=(4,2),
 \qquad
 \ell_1=\ell_2=2.
\end{equation}
Introduce $A_1,A_2\in Z^6(X^{10};\mathbb Z_2)$ and $B_1,B_2\in Z^4(X^{10};\mathbb Z_2)$. Eq.~\eqref{eq:steenrod-wu-action} becomes
\begin{equation}
    \Phi_{w_4w_6}=\frac12\int_{X^{10}}\left\{\sum_{i=1}^2\left[A_i\cup_2A_i+B_i\cup A_i\right]+B_1\cup(B_2\cup_2B_2)\right\}.
    \label{eq:w4w6-bulk-action}
\end{equation}
Summing over $A_1$ and $A_2$ imposes
$[B_1]=[B_2]=\nu_4$, giving the response in Eq.~\eqref{eq:w4w6-wu-characteristic-number}.

\section{Discussion}
\label{sec:discussion}

In this work, we have developed a workflow from the cochain-level representatives of topological response actions to quantum cellular automata. Extending results of our previous work on Clifford QCA \cite{Sun2026Clifford} to new families of non-Clifford QCA. In particular, gauging first produces a commuting Hamiltonian for the desired invertible phase. Local complementary-cell pairings then reorganize its terms into one separator for each microscopic degree of freedom. Replacing diagonal cochain variables by the corresponding commuting kinetic operators extends these separators from the closed-cochain sector to the full tensor-product Hilbert space. Finally, local dressings of the elementary shifts produce commuting flippers with the required orders and commutation relations. Once the separator-flipper algebra is complete, its identification with the on-site operator algebra defines a bounded-range QCA. Thus, a cochain-level response representative supplies not only a topological label, but also the local algebraic data needed to construct the automorphism.

The framework places the three-dimensional semion and $U(1)_4$ QCAs in a common language and extends both constructions to spatial dimensions $d=4k-1$. It also reformulates the $w_2w_3$ QCA and produces microscopic QCAs for the $w_2^n w_3^m$ and $w_2w_{4k-1}$ families, general products of Wu classes, as well as those that involve Steenrod squares. The five-dimensional $w_2^3$ and $w_3^2$ representatives illustrate an important limitation of the response-based construction: both complete automorphisms are implemented by explicit finite-depth circuits. Notably, even though the corresponding quadratic part of the action can each be implemented as Clifford circuit the local field redefinition relating the two descriptions is generally non-Clifford. Constructing a QCA, exhibiting a non-Clifford representative, detecting a topological response, and proving stable nontriviality are therefore distinct problems.

It remains to determine whether the remaining QCAs constructed above are stably nontrivial. We describe a state-based diagnostic and explain what it would imply, but we do not claim that it completes the proof for the present constructions. Let $H_0$ be an on-site commuting-projector Hamiltonian with a unique, translation-invariant product ground state $\ket{\Omega}$, and let $\alpha$ be a $d$-dimensional QCA. Then, $H_\alpha:=\alpha(H_0)$ is again a bounded-range commuting-projector Hamiltonian.  On an infinite system, its ground state is more precisely specified by the state
\begin{equation}
    \omega_\alpha:=\omega_\Omega\circ\alpha^{-1}
\end{equation}
on the quasi-local operator algebra; we denote it formally by $\ket{\psi_\alpha}$.

The inverse phase is represented by $\ket{\psi_{\alpha^{-1}}}$. Indeed, let $S$ be the product of the on-site swaps between two identical layers and set $\beta=\alpha\otimes\mathrm{id}$.  The automorphism
\begin{equation}
    \beta S\beta^{-1}S=\alpha\otimes\alpha^{-1}
\end{equation}
is implemented by a finite-depth circuit. To see this, write $S$ as a product of commuting on-site swaps.  Their images under $\beta$ remain a commuting family of uniformly bounded-support operators and can therefore be divided into a bounded number of nonoverlapping layers. So, $\ket{\psi_\alpha}\otimes\ket{\psi_{\alpha^{-1}}}$ lies in the trivial phase.

Passing to stable equivalence classes and using stacking as the group law gives a homomorphism
\begin{equation}
    \Phi_d:
    \mathrm{QCA}^{\mathrm{st}}_d
    \longrightarrow
    \mathrm{Inv}_d,
    \qquad
    [\alpha]\longmapsto[\ket{\psi_\alpha}],
    \label{eq:qca-to-invertible-phase}
\end{equation}
where $\mathrm{QCA}^{\mathrm{st}}_d$ denotes QCA classes modulo the addition of product-state ancillas, finite-depth circuits, and the allowed lattice shifts, while $\mathrm{Inv}_d$ is the group of $d$-dimensional invertible phases under stacking.  Finite-depth circuits do not change the phase of $\ket{\psi_\alpha}$, and the allowed lattice shifts preserve the chosen translation-invariant product state. The map $\Phi_d$ gives the rigorous conditional implication
\begin{equation}
    \Phi_d([\alpha])\neq 0
    \quad\Longrightarrow\quad
    [\alpha]\neq 0
    \quad\text{in }\mathrm{QCA}^{\mathrm{st}}_d.
\end{equation}
Thus, if $\ket{\psi_\alpha}$ is independently shown to be a nontrivial invertible phase, then $\alpha$ is stably nontrivial. In particular, if a $\mathbb Z_2$-valued characteristic number satisfies
\begin{equation}
    \left\langle P(w),[M_{d+1}]\right\rangle=1
\end{equation}
on some closed oriented manifold $M_{d+1}$, and if the corresponding state is proven to realize the response $(-1)^{\langle P(w),[M_{d+1}]\rangle}$, then this response obstructs trivialization of the QCA by product-state ancillas, finite-depth circuits, and lattice shifts.

More generally, the injectivity of $\Phi_d$ is unknown. A nontrivial QCA could map the reference product state into the trivial phase and therefore be invisible to this test. Thus, a vanishing closed-manifold response alone does not imply stable triviality for an arbitrary microscopic completion. Appendix~\ref{app:wu-core-reduction} proves a complementary constructive statement for the canonical circuit-compatible representatives. When an oriented Stiefel-Whitney number vanishes, a local Wu certificate reduces the response to quadratic and BF cores and hence to a finite-depth circuit. This does not establish injectivity of $\Phi_d$, because an independently chosen off-flat completion with the same response requires an additional operator-level equivalence. Surjectivity is also unknown. Phases in the image of $\Phi_d$ admit exact bounded-range commuting-projector representatives of the form $H_\alpha$, whereas quasi-adiabatic continuation generally produces quasi-local transformations with decaying tails. Whether these tails can always be removed to obtain an exact finite-range QCA remains open. The calculations in Appendix~\ref{app:w2n-w3m-so-bordism} show that the proposed characteristic-number responses are nontrivial for the $w_2^n w_3^m$ representatives with $m=0$ and $n$ even, or with $n$ and $m$ both odd. They likewise detect nontrivial candidate responses for the Wu--Bockstein family $w_2w_{4k-1}$ and for examples such as $w_4w_6$. If the corresponding microscopic states are shown to realize these bordism characters, the preceding implication would prove stable nontriviality of the associated QCAs. At present, however, this identification has not been established for the remaining constructions. Therefore, these calculations provide evidence and candidate obstructions, instead of a proof of QCA nontriviality.


The relation between the semion and $U(1)_4$ constructions presents a similar question. In a fixed dimension they have the same normalized signature response, but their quadratic data differ: $(\mathbb Z_2,q(1)=1/4)$ for the semion theory and $(\mathbb Z_4,q(a)=a^2/8)$ for $U(1)_4$. Their nonzero common response obstructs stable triviality, but cannot distinguish the two QCAs (although we believe them to correspond to different classes due to the conjectural relation between QCAs and the Witt group).  In seven spatial dimensions, their higher-form analogues have different membrane fusion groups and quadratic functions.  Determining the relation between the two microscopic QCAs, and their precise stable orders, remains open in seven and higher spatial dimensions.

Boundary physics may provide finer stable obstructions.  For the $w_2w_3$ QCA, the boundary contains a fermionic particle and a fermionic loop with mutual phase $\pi$.  This is an anomalous $(3+1)$-dimensional $\mathbb Z_2$ topological order~\cite{fidkowski2022gravitational,johnson2020topological}. Under the boundary-realizability assumption used in Refs.~\cite{haah_QCA_23,chen2023exactly}, this anomaly gives an independent obstruction to a finite-depth realization of the bulk QCA.  Constructing analogous microscopic boundary terminations for the higher-dimensional families could turn their generalized particle, loop, and membrane statistics into stable QCA invariants~\cite{kobayashi2024generalized,Feng2026AnyonicMembranes,feng2026paulistatistics,hsin2026bockstein}.

A complementary operator-algebraic goal is to construct explicitly the boundary invertible subalgebra associated with each non-Clifford QCA~\cite{Haah2023InvertibleSubalgebras}. In a related work, Ref.~\cite{chen2026invertiblesubalgebraschiralsemion}, we construct the invertible subalgebras for the semion and the $U(1)_4$ invertible subalgebras and their generalization to higher dimensions. Concrete generators for these algebras could provide a direct setting in which to compare stable classes, stacking relations, and boundary statistics, including QCAs that share the same bulk response. Extending the explicit constructions available in the Clifford case~\cite{Sun2026Clifford} is therefore a natural direction for future work.

Finally, the present construction relies on local perfect pairings on hypercubic cellulations. A cellulation-independent condition guaranteeing one independent separator per microscopic degree of freedom would make the construction intrinsic. Together with the boundary, stacking, and invertible-subalgebra problems above, such a formulation would clarify which information is fixed by the topological response and which belongs only to the full locality-preserving automorphism.

\section*{Note added}

During the completion of this manuscript, we became aware of independent works that also provide a cup product expression for the semion QCA~\cite{inamura2026noninvertiblesymmetriesmixingwitt,zhang2026quantumcellularautomatakramerswannier}.

\section*{Acknowledgments}
We thank Lukasz Fidkowski for raising the question of whether the $w_2^3$ and $w_3^2$ QCAs are stably trivial.
Z.W. and Y.-A.C. are supported by the National Natural Science Foundation of China (Grant No.~12474491) and the Central University Fundamental Research Funds (Peking University).

\appendix
\numberwithin{equation}{section}

\section{Cup-product derivation of the
\texorpdfstring{$\mathbb Z_4^{(1)}$}{Z4(1)} Hamiltonian}
\label{app:z4-cup-hamiltonian}

We derive Eq.~\eqref{eq:z4-initial-hamiltonian} directly from the four-dimensional action in Eq.~\eqref{eq:z4-semion-action}. In particular, the derivation explains the symmetrized cup-product dressing $\bface'\cup\be+\be\cup\bface'$ in the edge term. This provides a cochain counterpart to the standard categorical construction of the Walker-Wang Hamiltonian \cite{Walker2012TQFT,Shirley2022QCA,PhysRevB.90.245122}.

Let
\begin{equation}
    b=\sum_f b(f)\,\bface\in C^2(M_3,\mathbb Z_4)
    \label{eq:app-z4-spatial-field}
\end{equation}
be the spatial restriction of $B_2$, and denote the corresponding face-qudit basis state by $\ket b$. The clock and shift operators act as
\begin{equation}
    Z_f\ket b=\omega^{b(f)}\ket b,
    \qquad
    X_{\delta\be}\ket b=\ket{b+\delta\be},
    \qquad
    X_{\delta\be}:=\prod_fX_f^{(\delta\be)(f)},
    \label{eq:app-z4-basis-action}
\end{equation}
where $\omega=e^{2\pi i/4}$ and all exponents are reduced modulo four.

Let $W_4$ be a four-manifold with $\partial W_4=M_3$. If a flat spatial configuration $b$ extends to $B\in Z^2(W_4,\mathbb Z_4)$, the topological path integral assigns the boundary amplitude
\begin{equation}
    \Psi[b] =\exp\!\left(\frac{2\pi i}{4}\int_{W_4}B\cup B\right).
    \label{eq:app-z4-boundary-amplitude}
\end{equation}
Only ratios between configurations related by local exact shifts will be needed, so the following derivation applies separately in every global flux sector. In the topologically trivial sector, write $b=\delta\alpha$ and choose an extension $B=\delta A$ with $A|_{M_3}=\alpha$. Since $\delta(A\cup\delta A)=\delta A\cup\delta A$, Stokes' theorem gives the explicit transgression
\begin{equation}
    \frac14\int_{W_4}B\cup B
    =\frac14\int_{W_4}\delta(A\cup\delta A)
    =\frac14\int_{M_3}\alpha\cup\delta\alpha
    \pmod 1,
    \qquad
    \Psi[\delta\alpha]
    =\omega^{\int_{M_3}\alpha\cup\delta\alpha}.
    \label{eq:app-z4-exact-wavefunction}
\end{equation}
This expression depends only on $b=\delta\alpha$. Indeed, if $\alpha\mapsto\alpha+\lambda$ with $\delta\lambda=0$, its exponent changes by $\int_{M_3}\lambda\cup\delta\alpha=-\int_{M_3}\delta(\lambda\cup\alpha)=0$.

The elementary edge term is the finite difference of Eq.~\eqref{eq:app-z4-exact-wavefunction}. For the indicator cochain $\be$ of one edge, the one-form gauge move is $\alpha\mapsto\alpha+\be$, and hence $b\mapsto b+\delta\be$. The amplitude ratio is
\begin{equation}
    \begin{aligned}
    \frac{\Psi[\delta(\alpha+\be)]}{\Psi[\delta\alpha]}
    &=\omega^{\int_{M_3}\left(\be\cup\delta\alpha+\alpha\cup\delta\be+\be\cup\delta\be\right)}
    =\omega^{\int_{M_3}\left(\be\cup b+b\cup\be\right)}.
    \end{aligned}
    \label{eq:app-z4-finite-difference}
\end{equation}
In the second equality we used $\delta(\alpha\cup\be)=\delta\alpha\cup\be -\alpha\cup\delta\be$ and the fact that $M_3$ is closed.  We also used the elementary-cell identity $\be\cup\delta\be=0$ in the ordered cubical cup convention employed in the main text.  A convention with a nonzero, configuration-independent self-term differs only by a fixed phase redefinition of the edge shift.

Gauging the one-form symmetry sends the edge configuration and its bare shift to face variables according to
\begin{equation}
    \ket\alpha\longmapsto\ket{\delta\alpha},
    \qquad
    X_e\longmapsto X_{\delta\be}.
    \label{eq:app-z4-gauging-map}
\end{equation}
In the trivial flux sector, the resulting face-qudit state is
\begin{equation}
    \ket{\Psi}
    \propto
    \sum_{\alpha\in C^1(M_3,\mathbb Z_4)}
    \Psi[\delta\alpha]\ket{\delta\alpha}.
    \label{eq:app-z4-gauged-state}
\end{equation}
Expanding $b=\sum_{f'}b(f')\bface'$ makes the diagonal phase in Eq.~\eqref{eq:app-z4-finite-difference} explicit
\begin{equation}
    \omega^{\int_{M_3}(\be\cup b+b\cup\be)}
    =\prod_{f'}
    \left(\omega^{b(f')}\right)^{\int_{M_3}(
    \bface'\cup\be+\be\cup\bface')}.
    \label{eq:app-z4-phase-as-clocks}
\end{equation}
The gauged local generator is then
\begin{equation}
    \begin{aligned}
        \mathcal T_e
        :=&X_{\delta\be}
        \prod_{f'}Z_{f'}^{\int_{M_3}(
        \bface'\cup\be+\be\cup\bface')},
        \\
        \mathcal T_e\ket b
        =&\omega^{\int_{M_3}(b\cup\be+\be\cup b)} \ket{b+\delta\be}.
    \end{aligned}
\label{eq:app-z4-edge-generator}
\end{equation}
The ordering is important. The clock operators act first on $\ket b$, and the shift $X_{\delta\be}$ acts afterward.  Eq.~\eqref{eq:app-z4-finite-difference} gives
\begin{equation}
    \mathcal T_e\bigl(\Psi[b]\ket b\bigr)
    =\Psi[b+\delta\be]\ket{b+\delta\be},
    \label{eq:app-z4-edge-amplitude-transport}
\end{equation}
and hence $\mathcal T_e\ket{\Psi}=\ket{\Psi}$.

The action in Eq.~\eqref{eq:z4-semion-action} is defined for a cocycle $B_2$. On the enlarged face-qudit Hilbert space, this condition is imposed locally by
\begin{equation}
    Z_{\partial c}:=\prod_fZ_f^{(\delta\bface)(c)},
    \qquad
    Z_{\partial c}\ket b
    =\omega^{(\delta b)(c)}\ket b.
    \label{eq:app-z4-flux-operator}
\end{equation}
Thus, the simultaneous $+1$ eigenspace of the $Z_{\partial c}$ is precisely the flat subspace $\delta b=0\pmod 4$.

For completeness, all local constraints commute. The flux eigenvalue is unchanged under $b\mapsto b+\delta\be$ because $\delta^2=0$. For two edges $e$ and $e'$, the relative phase between the two orders of their gauge moves is
\begin{equation}
    \omega^{\int_{M_3}\left(
    \delta\be'\cup\be+\be\cup\delta\be'
    -\delta\be\cup\be'-\be'\cup\delta\be
    \right)}=1.
    \label{eq:app-z4-edge-commutation}
\end{equation}
Indeed, Stokes' theorem gives $\int\delta\be'\cup\be=\int\be'\cup\delta\be$ and $\int\delta\be\cup\be'=\int\be\cup\delta\be'$. Four successive shifts also telescope around the same finite-difference relation and return to $b+4\delta\be=b$, so $\mathcal T_e^4=1$.

The commuting parent Hamiltonian is therefore
\begin{equation}
    \begin{aligned}
        H^{\mathbb Z_4^{(1)}}= -\sum_c\left( Z_{\partial c}+Z_{\partial c}^{\dagger} \right) -\sum_e\left[X_{\delta\be} \prod_{f'}Z_{f'}^{\int_{M_3}( \bface'\cup\be+\be\cup\bface')} +\mathrm{h.c.}\right].
    \end{aligned}
    \label{eq:app-z4-derived-hamiltonian}
\end{equation}
This is Eq.~\eqref{eq:z4-initial-hamiltonian}. The first sum of terms extends the cocycle condition to the full face-qudit Hilbert space, while the second sum enforces the local finite-difference relation dictated by the four-dimensional cup-product action.

\section{Semion Hamiltonian from gauging the
\texorpdfstring{$\mathbb{Z}_2$}{Z2} one-form SPT}
\label{sec:z2-semion-construction}

In this appendix, we give an independent derivation of the semion Hamiltonian by gauging the generating $\mathbb{Z}_2$ one-form SPT. We first recall its Pontryagin-square response and its relation to the oriented signature TQFT. We then transgress the bulk action to obtain the boundary wavefunction, take a finite difference to construct its edge-flip parent term, and gauge this term to produce a closed face move. Since the gauging map lands in configurations satisfying $\delta a=0$, this procedure initially determines the kinetic operator only within the flat sector. We conclude by rewriting its phase and comparing it directly with the separator in Eq.~\eqref{eq:z4-reduced-separator}.

Let $M_4$ be an oriented four-manifold. The generating one-form SPT is characterized by the Pontryagin-square response
\begin{equation}
    S_{\mathbb{Z}_2}[B_2]
    =\frac{1}{4}\int_{M_4}
    \left(
    B_2\cup B_2
    +B_2\cup_1\delta B_2
    \right)
    \label{eq:z2-pontryagin-action}
\end{equation}
where $B_2\in Z^2(M_4,\mathbb{Z}_2)$ is a background gauge field, and all $\mathbb{Z}_2$ cochains in the integrand are understood through their canonical integer $0$--$1$ lifts.

When $M_4=X^4$ is closed, gauging the one-form symmetry amounts to summing over $B_2$. The Gauss-sum formula evaluates the invertibly normalized sum as
\begin{equation}
    Z_{\mathrm{sem}}^{\mathrm{inv}}(X^4)
    =\exp\!\left(\frac{2\pi i}{8}\sigma(X^4)\right)
    =\exp\!\left(2\pi i\int_{M_4}\frac{p_1(TX)}{24}\right).
    \label{eq:semion-signature-response-main}
\end{equation}
Thus, the normalized closed-manifold response of the gauged theory agrees with that of the invertible oriented signature TQFT. This identity does not by itself determine the stable order of the microscopic QCA; Appendix~\ref{app:semion-signature-bordism} explains the Gauss sum and this distinction in detail.

To derive the boundary Hamiltonian, we now take $\partial M_4=M_3$, where $M_3$ is the spatial slice. Locally write $B_2=[\delta\alpha]_2$, with $\alpha\in C^1(M_4,\mathbb Z_2)$. The canonical lift obeys~\cite{chen2023exactly}
\begin{equation}
    \delta\alpha \equiv [\delta\alpha]_2 + 2\left(\alpha\cup\alpha+\alpha\cup_1\delta\alpha\right) \pmod 4.
    \label{eq:z2-lift-identity}
\end{equation}
Substituting Eq.~\eqref{eq:z2-lift-identity} into Eq.~\eqref{eq:z2-pontryagin-action}, applying the higher-cup recursion, and
using Stokes' theorem gives
\begin{equation}
    S_{\mathbb Z_2}\!\left([\delta\alpha]_2\right) \equiv \int_{M_3}\left[\frac14\alpha\cup\delta\alpha +\frac12\delta\alpha\cup_1 \left(\alpha\cup\alpha+\alpha\cup_1\delta\alpha\right)\right] \pmod 1.
    \label{eq:z2-boundary-transgression}
\end{equation}
So, the fixed-point one-form SPT state is
\begin{equation}
    \begin{aligned}
        \left|\Psi_{\mathrm{SPT}}\right\rangle
        =\sum_{\alpha\in C^1(M_3,\mathbb Z_2)}
        &i^{\int_{M_3}\alpha\cup\delta\alpha}
        (-1)^{\int_{M_3}\delta\alpha\cup_1
        \left(
        \alpha\cup\alpha+\alpha\cup_1\delta\alpha
        \right)}
        |\alpha\rangle .
    \end{aligned}
    \label{eq:z2-spt-wavefunction}
\end{equation}

Set $a=[\delta\alpha]_2$.
For the indicator cochain $\be$ of an elementary edge, the canonical lift changes according to
\begin{equation}
    [\alpha+\be]_2
    =\alpha+\be-2\alpha\cup_1\be.
    \label{eq:z2-edge-lift-shift}
\end{equation}
The SPT Hamiltonian is obtained by conjugating $X_e$ with the SPT entangler. Commuting the entangler through $X$ gives the difference between the configurations at $[\alpha+\be]_2$ and $\alpha$. A direct finite difference of Eq.~\eqref{eq:z2-boundary-transgression} gives
\begin{equation}
    \boxed{
    \begin{aligned}
        T_e|\alpha\rangle
        ={}&
        i^{\int_{M_3}\left(\be\cup a+a\cup\be\right)}
        (-1)^{\int_{M_3}\left[
        \be\cup_1(a\cup_1a)
        +(a+\delta\be)\cup_1
        \left(
        a\cup_1\be+\be\cup_1a+\be\cup_1\delta\be
        \right)
        \right]}
        |[\alpha+\be]_2\rangle .
    \end{aligned}}
    \label{eq:z2-spt-parent-action}
\end{equation}
The complete cochain expansion is given in Eqs.~\eqref{eq:app-z2-transgression}--\eqref{eq:app-z2-total-variation}. The SPT Hamiltonian
\begin{equation}
    H_{\mathrm{SPT}}
    =-\sum_eT_e
    \label{eq:z2-spt-parent}
\end{equation}
has Eq.~\eqref{eq:z2-spt-wavefunction} as the ground state.

Next, we apply the gauging map for the 1-form symmetry, which replaces an edge configuration by its face coboundary and maps the bare edge Pauli to the corresponding closed face flip
\begin{equation}
    |\alpha\rangle\longmapsto|[\delta\alpha]_2\rangle,
    \qquad
    X_e\longmapsto
    X_{\delta\be}:=\prod_fX_f^{[\delta\be](f)}.
    \label{eq:z2-gauging-map}
\end{equation}
Here and below, the physical face qubit is identified with the surviving $A$ qubit of Sec.~\ref{sec:semion-qca}. We suppress the superscript $A$ in this direct construction. The gauged state is
\begin{equation}
    \begin{aligned}
        \left|\Psi_{\mathbb Z_2}\right\rangle
        \propto
        \sum_{\alpha\in C^1(M_3,\mathbb Z_2)}
        i^{\int_{M_3}\alpha\cup\delta\alpha}
        (-1)^{\int_{M_3}\delta\alpha\cup_1
        \left(
        \alpha\cup\alpha+\alpha\cup_1\delta\alpha
        \right)}
        \left|[\delta\alpha]_2\right\rangle .
    \end{aligned}
    \label{eq:z2-gauged-state}
\end{equation}
Every face configuration in this state is flat
\begin{equation}
    \delta a=0.
    \label{eq:z2-flatness}
\end{equation}
Equivalently, this can be imposed by the stabilizer $Z_{\partial c}|a\rangle=|a\rangle$ for every 3-cell $c$.

Gauging Eq.~\eqref{eq:z2-spt-parent-action} gives the non-Pauli stabilizer $T_e$ which acts as the follows
\begin{equation}
    \begin{aligned}
        T_e|a\rangle
        =&
        i^{\int_{M_3}\left(\be\cup a+a\cup\be\right)}
        (-1)^{\int_{M_3}\left[
        \be\cup_1(a\cup_1a)
        +(a+\delta\be)\cup_1
        \left(
        a\cup_1\be+\be\cup_1a+\be\cup_1\delta\be
        \right)
        \right]}
        |a+\delta\be\rangle ,
        \qquad \delta a=0.
    \end{aligned}
    \label{eq:z2-edge-move}
\end{equation}

Let us now compare it to the expression of the separator in Eq.~\eqref{eq:z4-reduced-separator}.
First, we use the exact pairing between edges and faces on the cubic lattice to write an expression of the separator on each edge. That is, we denote the corresponding edge operator by $B_e$ where %
\begin{equation}
    \overline Z_f
    =\prod_e\left( B_e\right)^{\int_{M_3}\be\cup\bface},
    \qquad e\ \text{complementary to }f.
    \label{eq:z4-face-edge-relabeling}
\end{equation}
\begin{align}
B_e = 
X_{\delta\be}^A
\prod_{f'\in\delta\be} \Bigg\{
\left(Z_{f'}^A\right)^{
\int\delta\be\cup_1\bface'}
\prod_{f''}
CZ\!\left(Z_{f'}^A,Z_{f''}^A\right)^{
\int\left(
\bface''\cup_1\bface'
+\bface'\cup_2\delta\bface''
\right)} \Bigg\}
\\
\quad\times
\prod_{f'}\left(Z_{f'}^A\right)^{
\int
\bface'\cup_1(\be\cup_1\delta\be)
+(\be\cup_1\delta\be)\cup_2\delta\bface'
}
\left(S_{f'}^A\right)^{
\int\delta\be\cup_1\bface'} (Z_{f'}^A)^{\int \be \cup \bface'}
.   
\end{align}

Let us determine the action of $B_e$ in the computational basis $\ket{a}$. Let $a\in C^2(M_3,\mathbb Z_2)$. Because $\cup_2$ is the pointwise product for degree-two cochains, $\delta\be\cup_2a$ is the restriction of $a$ to the faces in $\operatorname{supp}(\delta\be)$. Its action is therefore
\begin{equation}
    \begin{aligned}
          B_e\ket{a} &= (-1)^{\int
\delta\be\cup_1(\delta\be\cup_2a) + \left[
a\cup_1(\be\cup_1\delta\be)
+(\be\cup_1\delta\be)\cup_2\delta a
\right] + \left[
a\cup_1(\delta\be\cup_2a)
+(\delta\be\cup_2a)\cup_2\delta a
\right] } i^{\int \delta \be \cup_1 a + 2 \be \cup a}\ket{a+\delta \be}\\
&=(-1)^{\int
a\cup_1\left[
\be\cup_1\delta\be+\delta\be\cup_2a
\right]
+\delta\be\cup_1(\delta\be\cup_2a)}
(-1)^{\int\left[
\be\cup_1\delta\be+\delta\be\cup_2a
\right]\cup_2\delta a} i^{\int \delta \be \cup_1 a + 2 \be \cup a}\ket{a+\delta \be}\\
&=(-1)^{\int(a+\delta\be)\cup_1\left[
\be\cup_1\delta\be+\delta\be\cup_2a
\right]
}
(-1)^{\int\left[
\be\cup_1\delta\be+\delta\be\cup_2a
\right]\cup_2\delta a}  i^{\int \delta \be \cup_1 a + 2 \be \cup a}\ket{a+\delta \be}\\
&= 
(-1)^{\int(a+\delta\be)\cup_1
\left(\be\cup_1\delta\be+ a\cup_1\be+\be\cup_1a\right)}
(-1)^{\int
\left(\be\cup_1\delta\be+a\cup_1\be+\be\cup_1a\right)
\cup_2\delta a} i^{\int(\delta\be\cup_1a+2\be\cup a)}
\ket{a+\delta\be}.
    \end{aligned}
    \label{eq:condensedsemionstabilizeraction}
\end{equation}
where in the second line is a reorganization of terms. In the third line, we added back the cubic-lattice identity $\int\delta\be\cup_1(\be\cup_1\delta\be)=0$ and combined the first two terms by bilinearity. To get the last line, we used the higher-cup recursion to substitute where $\delta\be\cup_2a=a\cup_1\be+\be\cup_1a$ since $\be \cup_2 \delta a = \delta(\be \cup_2 a) =0$ due to the degree of $e$ being 1.

Let us show that Eq.~\eqref{eq:z2-edge-move} and Eq.~\eqref{eq:condensedsemionstabilizeraction} are equivalent up to the flatness condition $\delta a=0$. Although originally $a$ is closed modulo two, the integer coboundary of $a$ after performing the canonical lift need not vanish. Instead,
\begin{equation}
    \delta a
    \equiv2( a\cup_1 a)\pmod4.
    \label{eq:z2-bockstein-identity}
\end{equation}
Integrating the signed higher-cup recursion and using the above relation gives
\begin{equation}
    \begin{aligned}
        \int_{M_3}\left(\be\cup a+ a\cup\be\right)
        &\equiv
        \int_{M_3}\left(
        \be\cup a+ (\delta\be\cup_1 a
        +\be\cup a
        -\be\cup_1\delta a)
        \right)
        \\
        &\equiv
        \int_{M_3}\left[
        \delta\be\cup_1 a
        +2\be\cup a
        -2\be\cup_1( a\cup_1 a)
        \right]
        \pmod4.
    \end{aligned}
    \label{eq:z4-z2-phase-identity}
\end{equation}
As elsewhere, we suppress the tilde when a canonical lift appears inside an integer-valued phase exponent. So,
\begin{equation}
    i^{\int(\delta\be\cup_1a+2\be\cup a)}= i^{\int(\be\cup a+a\cup\be)} (-1)^{\int\be\cup_1(a\cup_1a)}.
    \label{eq:z2-flat-phase-cancellation}
\end{equation}
Applying Eq.~\eqref{eq:z2-flat-phase-cancellation} in Eq.~\eqref{eq:condensedsemionstabilizeraction} and then imposing $\delta a=0$ gives precisely Eq.~\eqref{eq:z2-edge-move}. Therefore, the resulting Hamiltonians $\mathbb Z_2$ construction and the condensed $\mathbb Z_4$ agree in the subsector where $\delta a =0$. This interpretation, however, does not by itself provide a transparent construction of the QCA: starting from the gauged SPT Eq.~\eqref{eq:z2-edge-move}, it is not obvious how to reorganize the terms into the locally flippable separators or to recognize that the flux terms are generated by the kinetic terms and are therefore redundant. The condensed
$\mathbb Z_4$ construction therefore supplies precisely this missing guidance and leads to the construction of the QCA.

\subsection*{Detailed \texorpdfstring{$\mathbb Z_2$}{Z2} Pontryagin-square parent-Hamiltonian derivation}
\label{app:z2-parent-derivation}

We now give the cochain calculation leading to Eqs.~\eqref{eq:z2-boundary-transgression} and~\eqref{eq:z2-spt-parent-action}. Let $M_3=\partial M_4$ be a closed spatial slice, let $\alpha\in C^1(M_4,\mathbb Z_2)$, and set $a=[\delta\alpha]_2$.  Fractional cochain expressions are evaluated using the canonical integer $0$--$1$ lifts and are understood modulo one. We use the abbreviations
\[
    K(\alpha):=\alpha\cup\alpha+\alpha\cup_1\delta\alpha,
    \qquad
    t_e(\alpha):=\alpha\cup_1\be.
\]
The lift identity in Eq.~\eqref{eq:z2-lift-identity} reads $\delta\alpha\equiv a+2K(\alpha)\pmod 4$. Here, $\be$ is the indicator of a single elementary edge. In the ordered simplicial or cubical cup convention used throughout, its local self-products obey $\be\cup\be=0$ and $\be\cup\delta\be=0$. These elementary-edge identities are used below when expanding the shifted cochains.

Substituting the lift identity into Eq.~\eqref{eq:z2-pontryagin-action}, using Eq.~\eqref{eq:higher-cup-recursion}, and applying Stokes' theorem gives
\begin{equation}
    \begin{aligned}
        S_{\mathbb Z_2}(a)
        &\equiv \int_{M_4}\Bigl[
         \frac14\delta\alpha\cup\delta\alpha
         +\frac12K(\alpha)\cup\delta\alpha
         +\frac12\delta\alpha\cup K(\alpha)
         +\frac12\delta\alpha\cup_1\delta K(\alpha)
         \Bigr] \\
        &\equiv \int_{M_4}\Bigl[
         \frac14\delta\alpha\cup\delta\alpha
         +\frac12\delta\bigl(\delta\alpha\cup_1K(\alpha)\bigr)
         \Bigr] \\
        &\equiv \int_{M_3}\Bigl[
         \frac14\alpha\cup\delta\alpha
         +\frac12\delta\alpha\cup_1K(\alpha)
         \Bigr]
        \pmod 1.
    \end{aligned}
    \label{eq:app-z2-transgression}
\end{equation}
This is the expanded form of the boundary transgression in Eq.~\eqref{eq:z2-boundary-transgression}. It gives the one-form SPT wavefunction
\begin{equation}
    \left|\Psi_{\mathrm{SPT}}\right\rangle
    =\sum_{\alpha\in C^1(M_3,\mathbb Z_2)}
    i^{\int_{M_3}\alpha\cup\delta\alpha}
    (-1)^{\int_{M_3}\delta\alpha\cup_1K(\alpha)}
    \left|\alpha\right\rangle,
    \label{eq:app-z2-spt-state}
\end{equation}
and gauging the one-form symmetry replaces the basis label by the flat face cochain
\begin{equation}
    \left|\Psi_{\mathbb Z_2}\right\rangle
    =\sum_{\alpha\in C^1(M_3,\mathbb Z_2)}
    i^{\int_{M_3}\alpha\cup\delta\alpha}
    (-1)^{\int_{M_3}\delta\alpha\cup_1K(\alpha)}
    \left|[\delta\alpha]_2\right\rangle.
    \label{eq:app-z2-gauged-state}
\end{equation}
Eqs.~\eqref{eq:app-z2-spt-state} and \eqref{eq:app-z2-gauged-state} reproduce Eqs.~\eqref{eq:z2-spt-wavefunction} and~\eqref{eq:z2-gauged-state}, respectively.

To obtain the local parent term, shift the edge cochain on one elementary edge. Its canonical integer lift transforms as
\begin{equation}
    \alpha\longmapsto\alpha':=[\alpha+\be]_2
    =\alpha+\be-2\alpha\cup_1\be
    =\alpha+\be-2t_e(\alpha).
    \label{eq:app-z2-edge-shift}
\end{equation}
We first isolate the change of the quarter-integer part of the exponent. Repeated use of Eq.~\eqref{eq:higher-cup-recursion} gives
\begin{equation}
\begin{aligned}
\Delta_i
&:=\int_{M_3}\frac14\Bigl(
 [\alpha+\be]_2\cup\delta[\alpha+\be]_2
 -\alpha\cup\delta\alpha\Bigr) \\
&\equiv\int_{M_3}\Bigl[
 \frac14\be\cup\delta\alpha
 +\frac14\alpha\cup\delta\be
 +\frac12t_e(\alpha)\cup\delta(\alpha+\be)
 +\frac12(\alpha+\be)\cup\delta t_e(\alpha)
 \Bigr] \\
&\equiv\int_{M_3}\Bigl[
 \frac14\be\cup\delta\alpha
 +\frac14\delta\alpha\cup\be
 +\frac12t_e(\alpha)\cup\delta(\alpha+\be)
 +\frac12\delta(\alpha+\be)\cup t_e(\alpha)
 \Bigr] \\
&\equiv\int_{M_3}\Bigl[
 \frac14\be\cup\delta\alpha
 +\frac14\delta\alpha\cup\be
 +\frac12\delta(\alpha+\be)\cup_1\delta t_e(\alpha)
 \Bigr] \\
&\equiv\int_{M_3}\Bigl[
 \frac14\be\cup a+\frac14a\cup\be
 +\frac12\delta(\alpha+\be)\cup_1\delta t_e(\alpha)
 +\frac12\be\cup K(\alpha)+\frac12K(\alpha)\cup\be
 \Bigr] \\
&\equiv\int_{M_3}\Bigl[
 \frac14\be\cup a+\frac14a\cup\be
 +\frac12\delta(\alpha+\be)\cup_1\delta t_e(\alpha)
 +\frac12\delta\be\cup_1K(\alpha)
 +\frac12\be\cup_1(\delta\alpha\cup_1\delta\alpha)
 \Bigr]
\pmod 1.
\end{aligned}
\label{eq:app-z2-i-variation}
\end{equation}

Next, consider the change of the half-integer cup-$1$ correction. In this part of the calculation all cochains may be reduced modulo two. Expanding $K(\alpha+\be)$ first and then applying the cup-$1$ recursion yields
\begin{equation}
\begin{aligned}
\Delta_{-1}
&:=\int_{M_3}\frac12\Bigl[
 \delta(\alpha+\be)\cup_1K(\alpha+\be)
 -\delta\alpha\cup_1K(\alpha)\Bigr] \\
&\equiv\int_{M_3}\frac12\Bigl\{
 \delta(\alpha+\be)\cup_1\bigl[
 K(\alpha)+\be\cup\alpha+\alpha\cup\be
 +\alpha\cup_1\delta\be+\be\cup_1\delta\alpha
 +\be\cup_1\delta\be\bigr]
 -\delta\alpha\cup_1K(\alpha)\Bigr\} \\
&\equiv\int_{M_3}\frac12\Bigl\{
 \delta(\alpha+\be)\cup_1\bigl[
 K(\alpha)+\be\cup_1\delta\be
 +\delta t_e(\alpha)+\delta\alpha\cup_1\be
 +\be\cup_1\delta\alpha\bigr]
 -\delta\alpha\cup_1K(\alpha)\Bigr\} \\
&\equiv\int_{M_3}\Bigl[
 \frac12\delta(\alpha+\be)\cup_1\delta t_e(\alpha)
 +\frac12\delta(\alpha+\be)\cup_1
 \bigl(\be\cup_1\delta\be+\delta\alpha\cup_1\be
 +\be\cup_1\delta\alpha\bigr) 
+\frac12\delta\be\cup_1K(\alpha)
 \Bigr]
\pmod 1.
\end{aligned}
\label{eq:app-z2-sign-variation}
\end{equation}
The first and last terms on the final line of Eq.~\eqref{eq:app-z2-sign-variation} repeat the corresponding half-integer terms in Eq.~\eqref{eq:app-z2-i-variation}. Each then appears twice in $\Delta_i+\Delta_{-1}$ and contributes an integer. After this cancellation, and after reducing $\delta\alpha$ to $a$ inside the mod-two terms, the total change of the boundary action is
\begin{equation}
    \begin{aligned}
    \Delta\Phi_e(a)
    &:=\Delta_i+\Delta_{-1} \\
    &\equiv\int_{M_3}\Bigl[
     \frac14\be\cup a+\frac14a\cup\be
     +\frac12\be\cup_1(a\cup_1a) 
     +\frac12(a+\delta\be)\cup_1
     \bigl(\be\cup_1\delta\be+a\cup_1\be+\be\cup_1a\bigr)
     \Bigr]
    \pmod 1.
    \end{aligned}
    \label{eq:app-z2-total-variation}
\end{equation}
Exponentiating Eq.~\eqref{eq:app-z2-total-variation} gives the two parent actions explicitly
\begin{equation}
    \begin{aligned}
    T_e |a\rangle
    ={}&i^{\int_{M_3}(\be\cup a+a\cup\be)} 
    (-1)^{\int_{M_3}\left[
     \be\cup_1(a\cup_1a)
     +(a+\delta\be)\cup_1
     (\be\cup_1\delta\be+a\cup_1\be+\be\cup_1a)
     \right]}
     |a+\delta\be\rangle,
     \qquad \forall\,e, \\
    Z_{\partial t}|a\rangle
    ={}&(-1)^{(\delta a)(t)}|a\rangle,
    \qquad \forall\,t.
    \end{aligned}
    \label{eq:app-z2-parent-actions}
\end{equation}
The first line is the explicit finite difference used in Eqs.~\eqref{eq:z2-spt-parent-action} and~\eqref{eq:z2-edge-move}, while the second enforces the flatness condition in Eq.~\eqref{eq:z2-flatness}. Because the calculation begins with $a=[\delta\alpha]_2$, it determines the phase in the kinetic term for a flat connection. The restriction to $\delta a=0$ of Eq.~\eqref{eq:condensedsemionstabilizeraction} from the condensed $\ZZ_4$ construction agrees with this phase factor.

\subsection*{Higher dimensions}
\label{sec:7d-gs-direct-z2}

We next derive the same flat-sector Hamiltonian in higher dimensions by gauging a $\ZZ_2$ $(2k-1)$-form SPT. Its bulk response is
\begin{equation}
    S_{\mathrm{GS}}[B_{2k}]
    =\frac14\int_{M_{4k}}\mP(B_{2k}),
    \qquad
    \mP(B_{2k})
    :=B_{2k}\cup B_{2k}+B_{2k}\cup_1\delta B_{2k}
    \quad (\mathrm{mod}\ 4),
    \label{eq:7d-gs-action}
\end{equation}
where $B_{2k}\in Z^{2k}(M_{4k},\ZZ_2)$. In $\mP(B_{2k})$, every $\ZZ_2$ cochain is replaced by its canonical integer lift.  Although $B_{2k}$ is closed modulo two, its integer lift need not be closed. The $\cup_1$ term makes $\mP(B_{2k})$ closed modulo four.

In the trivial cohomology sector, we can write $B_{2k}=[\delta\alpha]_2$ with $\alpha\in C^{2k-1}(M_{4k},\mathbb Z_2)$. We use the same symbol for its canonical integer lift. The lift obeys
\begin{equation}
    \delta\alpha
    \equiv[\delta\alpha]_2
    +2\left(\alpha\cup_{2k-2}\alpha+\alpha\cup_{2k-1}\delta\alpha\right)
    \pmod 4.
    \label{eq:7d-gs-lift}
\end{equation}
For a cocycle $\alpha$, the expression in parentheses represents $\Sq^1[\alpha]$ \cite{Steenrod1947Products,Chen2023HigherCup}. Substituting Eq.~\eqref{eq:7d-gs-lift} into Eq.~\eqref{eq:7d-gs-action} and applying Stokes' theorem gives the boundary action on the spatial manifold.
\begin{equation}
S_{\mathrm{GS}}([\delta\alpha]_2)
\equiv
\int_{M_{4k-1}}\left[
\frac14\alpha\cup\delta\alpha
+\frac12\delta\alpha\cup_1
\left(\alpha\cup_{2k-2}\alpha+\alpha\cup_{2k-1}\delta\alpha\right)
\right]
\pmod 1.
\label{eq:7d-gs-boundary-action}
\end{equation}
Thus, the fixed-point SPT wavefunction in the computational basis is
\begin{equation}
    \begin{aligned}
        \left|\Psi_{\mathrm{SPT}}\right\rangle
        =\sum_{\alpha}
        &i^{\int\alpha\cup\delta\alpha}
        (-1)^{\int\delta\alpha\cup_1
        (\alpha\cup_{2k-2}\alpha+\alpha\cup_{2k-1}\delta\alpha)}
        \left|\alpha\right\rangle .
    \end{aligned}
    \label{eq:7d-gs-spt-state}
\end{equation}

Gauging maps the configurations from $2k-1$ cells to $2k$ cells. The gauging map is
\begin{equation}
    \ket\alpha\longmapsto\ket{[\delta\alpha]_2},
    \qquad
    X_\tau\longmapsto X_{\delta\btau}.
    \label{eq:7d-gs-gauging-map}
\end{equation}
The gauged state is then
\begin{equation}
    \begin{aligned}
    \left|\Psi_{\mathrm{GS}}\right\rangle
    =\sum_{\alpha\in C^{2k-1}(M_{4k-1},\mathbb Z_2)}
    &i^{\int\alpha\cup\delta\alpha}
    (-1)^{\int\delta\alpha\cup_1
    (\alpha\cup_{2k-2}\alpha+\alpha\cup_{2k-1}\delta\alpha)}
    \left|[\delta\alpha]_2\right\rangle .
    \end{aligned}
    \label{eq:7d-gs-gauged-state}
\end{equation}
Writing $a=[\delta\alpha]_2$, every configuration in Eq.~\eqref{eq:7d-gs-gauged-state} obeys
\begin{equation}
    \delta a=0, \qquad Z_{\partial\nu}\ket a=\ket a.
    \label{eq:7d-gs-direct-flatness}
\end{equation}

The corresponding Hamiltonian term is obtained by computing the phase difference of the two wavefunction amplitudes related by $\alpha\mapsto\alpha+\btau$. Using $[\alpha+\btau]_2=\alpha+\btau-2\alpha\cup_{2k-1}\btau$, we obtain
\begin{equation}
    \boxed{
    \begin{aligned}
    T_\tau \ket a
    ={}&i^{\int(\btau\cup a+a\cup\btau)}
    (-1)^{\int\left\{
    \btau\cup_1(a\cup_{2k-1}a)
    +(a+\delta\btau)\cup_1
    \left[
    \btau\cup_{2k-1}\delta\btau
    +a\cup_{2k-1}\btau+\btau\cup_{2k-1}a
    \right]\right\}}
    \ket{a+\delta\btau},
    \quad \delta a=0.
    \end{aligned}}
    \label{eq:7d-gs-local-transition}
\end{equation}

We now compare the two constructions. For the canonical integer lift of a flat $2k$-cochain,
\begin{equation}
    \delta a \equiv 2(a\cup_{2k-1}a) \pmod4.
    \label{eq:7d-gs-flat-lift}
\end{equation}
Integrating $\delta(\btau\cup_1a)=\delta\btau\cup_1a-\btau\cup_1\delta a +\btau\cup a-a\cup\btau$ therefore gives
\begin{equation}
    \int(\btau\cup a+a\cup\btau)
    \equiv
    \int(\delta\btau\cup_1a+2\btau\cup a)
    +2\int\btau\cup_1(a\cup_{2k-1}a)
    \pmod4.
    \label{eq:7d-gs-phase-comparison}
\end{equation}
The last term in Eq.~\eqref{eq:7d-gs-phase-comparison} appears once in the power of $i$ and once in the sign in Eq.~\eqref{eq:7d-gs-local-transition}. The two factors cancel
\begin{equation}
    i^{2\int\btau\cup_1(a\cup_{2k-1}a)}
    (-1)^{\int\btau\cup_1(a\cup_{2k-1}a)}=1.
    \label{eq:7d-gs-phase-ratio}
\end{equation}
Eqs.~\eqref{eq:7d-gs-phase-comparison} and \eqref{eq:7d-gs-phase-ratio} then show that the remaining phase is exactly the amplitude ratio in Eq.~\eqref{eq:7d-gs-local-transition}. Thus, the Hamiltonians agree in the $\delta a =0$ sector.

\section{Oriented Stiefel-Whitney-number relations in dimensions four
through ten}
\label{app:sw-number-relations}

This appendix records the low-dimensional relations used to compare the gravitational responses in the main text. Let $M^D$ be a closed oriented $D$-manifold. For two degree-$D$ polynomials $P$ and $Q$ in the Stiefel-Whitney classes, we write
\begin{equation}
     P\doteq Q
     \quad\Longleftrightarrow\quad
     \left\langle P,[M^D]\right\rangle
     =\left\langle Q,[M^D]\right\rangle
     \quad\text{for every closed oriented }M^D .
    \label{eq:sw-number-equivalence}
\end{equation}
Thus, $\doteq$ denotes equality of characteristic numbers, not equality of universal classes in $H^*(BSO;\mathbb Z_2)$.  This distinction is essential. For example, $w_2w_7\doteq w_4w_5$ in dimension nine, although $w_2w_7$ and $w_4w_5$ are distinct universal cohomology classes. Here, $D=4,\ldots,10$ correspond respectively to $(3{+}1)$D through $(9+1)$D spacetime.

\subsection*{Wu formulas and the reduction procedure}

We use the Wu formula and Wu pairing \cite{MilnorStasheff1974}
\begin{align}
     \Sq^i w_j
     &=\sum_{t=0}^{i}
       \binom{j-i+t-1}{t}w_{i-t}w_{j+t},
     \qquad 0\leq i\leq j,
     \label{eq:sw-wu-formula}\\
     \left\langle\Sq^i x,[M^D]\right\rangle
     &=\left\langle\nu_i x,[M^D]\right\rangle,
     \qquad x\in H^{D-i}(M^D;\mathbb Z_2),
     \label{eq:sw-wu-pairing}
\end{align}
together with $w=\Sq(\nu)$. Here $w_0=1$, and the integer binomial coefficients are reduced modulo two. Steenrod instability says $\Sq^i x=0$ for $i>|x|$. In particular, if $2i>D$, then every $x\in H^{D-i}(M^D;\mathbb Z_2)$ has degree smaller than $i$. The nondegeneracy of the mod-two Poincar\'e pairing and Eq.~\eqref{eq:sw-wu-pairing} then imply
\begin{equation}
 \nu_i(M^D)=0,
 \qquad 2i>D.
 \label{eq:sw-high-wu-vanishing}
\end{equation}
This is the useful consequence of instability for characteristic numbers. The identity $\Sq^i(w_j)=0$ for $i>j$ by itself is generally insufficient. After expanding Eq.~\eqref{eq:sw-wu-formula}, it can reduce to a tautological cancellation.

Since $M$ is oriented, $w_1=0$. The Wu classes needed below are
\begin{align}
 \nu_1&=\nu_3=\nu_5=\nu_7=\nu_9=0,
 \nonumber\\
 \nu_2&=w_2,
 &\nu_4&=w_4+w_2^2,
 &\nu_6&=w_2w_4+w_3^2,
 \label{eq:sw-low-wu-classes}\\
 \nu_8&=w_8+w_2w_6+w_3w_5+w_4^2+w_2^4.
 \nonumber
\end{align}
Applying Eqs.~\eqref{eq:sw-wu-formula}--\eqref{eq:sw-high-wu-vanishing} to every partition of $D$ with parts at least two gives the following summary.

\begin{table}[ht!]
\centering
\footnotesize
\setlength{\tabcolsep}{4pt}
\renewcommand{\arraystretch}{1.16}
\begin{tabularx}{\textwidth}{@{}c>{\raggedright\arraybackslash}p{0.24\textwidth}Y@{}}
\toprule
$D$ & Independent mod-two numbers & Relations among the remaining degree-$D$ monomials \\
\midrule
$4$ & $w_2^2$ & $w_4\doteq w_2^2$ \\
$5$ & $w_2w_3$ & $w_5\doteq0$ \\
$6$ & $\emptyset$ & $w_6\doteq w_2w_4\doteq w_3^2\doteq w_2^3\doteq0$ \\
$7$ & $\emptyset$ & $w_7\doteq w_2w_5\doteq w_3w_4\doteq w_2^2w_3\doteq0$ \\
$8$ & $w_2^4,\ w_4^2$ &
$w_8\doteq w_2^4+w_4^2$;
$w_2w_6\doteq w_3w_5\doteq w_2^2w_4\doteq w_2w_3^2\doteq0$ \\
$9$ & $w_2^3w_3,\ w_2w_7$ &
$w_2w_7\doteq w_3w_6\doteq w_4w_5$;
$w_9\doteq w_2^2w_5\doteq w_2w_3w_4\doteq w_3^3\doteq0$ \\
$10$ & $w_4w_6$ & Every other degree-ten monomial vanishes; see
Eq.~\eqref{eq:sw-d10-vanishing-list}. \\
\bottomrule
\end{tabularx}
\caption{Independent Stiefel-Whitney numbers of closed oriented manifolds in spacetime dimensions four through ten. The equalities are understood in the sense of Eq.~\eqref{eq:sw-number-equivalence}.}
\label{tab:oriented-sw-numbers-4-10}
\end{table}

The table concerns mod-two Stiefel-Whitney numbers. In dimensions four and eight, oriented bordism has a free part, and its full integral classification also requires Pontryagin numbers. Stiefel-Whitney numbers retain only their mod-two reductions.

\subsection*{Dimensions four through seven}

In $D=4$, the only candidates are $w_4$ and $w_2^2$. Since $2\cdot4>4$, Eq.~\eqref{eq:sw-high-wu-vanishing} yields
\begin{equation}
    0\doteq\nu_4=w_4+w_2^2, \qquad\text{hence}\qquad w_4\doteq w_2^2.
    \label{eq:sw-d4-relation}
\end{equation}
Thus, $w_2^2$ is the unique independent mod-two number in $(3{+}1)$ dimensions.

In $D=5$, the candidates are $w_5$ and $w_2w_3$. The oriented form of Eq.~\eqref{eq:sw-wu-formula} gives $\Sq^1w_4=w_5$, whereas $\nu_1=w_1=0$. Therefore,
\begin{equation}
    w_5=\Sq^1w_4\doteq\nu_1w_4=0.
    \label{eq:sw-d5-relation}
\end{equation}
No Wu relation eliminates $w_2w_3$, so it is the unique independent number in $(4{+}1)$ dimensions.

In $D=6$, all four candidate numbers vanish.  Indeed,
\begin{align}
 w_3^2
 &=\Sq^1(w_2w_3)\doteq0,
 \nonumber\\
 \Sq^2(w_2^2)=w_3^2
 &\doteq\nu_2w_2^2=w_2^3,
 \nonumber\\
 0=\Sq^4w_2
 &\doteq\nu_4w_2=w_2w_4+w_2^3,
 \nonumber\\
 \Sq^2w_4=w_2w_4+w_6
 &\doteq\nu_2w_4=w_2w_4.
 \label{eq:sw-d6-relations}
\end{align}
Thus,
\begin{equation}
 w_6\doteq w_2w_4\doteq w_3^2\doteq w_2^3\doteq0.
\end{equation}

In $D=7$, the four candidates again vanish.  First,
\begin{align}
 w_2^2w_3&=\Sq^1(w_2^3)\doteq0,
 \nonumber\\
 \Sq^2(w_2w_3)=w_2w_5
 &\doteq\nu_2w_2w_3=w_2^2w_3,
 \nonumber\\
 \Sq^1(w_2w_4)&=w_3w_4+w_2w_5\doteq0,
 \nonumber\\
 w_7&=\Sq^1w_6\doteq0.
 \label{eq:sw-d7-relations}
\end{align}
It follows that $w_7\doteq w_2w_5\doteq w_3w_4\doteq w_2^2w_3\doteq0$.

\subsection*{Dimension eight}

The degree-eight candidates are
\begin{equation}
    w_8,\quad w_2w_6,\quad w_3w_5,\quad w_4^2,\quad w_2^2w_4,\quad w_2w_3^2,\quad w_2^4.
    \label{eq:sw-d8-candidates}
\end{equation}
The mixed numbers vanish by
\begin{align}
 w_3w_5&=\Sq^1(w_2w_5)\doteq0,
 \nonumber\\
 \Sq^2(w_2^3)=w_2^4+w_2w_3^2
 &\doteq\nu_2w_2^3=w_2^4,
 \nonumber\\
 \Sq^4(w_2^2)=w_2^4
 &\doteq\nu_4w_2^2=w_2^4+w_2^2w_4,
 \nonumber\\
 \Sq^2(w_2w_4)=w_2w_6+w_3w_5
 &\doteq\nu_2w_2w_4=w_2^2w_4.
 \label{eq:sw-d8-mixed-relations}
\end{align}
Finally, $2\cdot8>8$ and Eq.~\eqref{eq:sw-low-wu-classes} give
\begin{equation}
 0\doteq\nu_8
 =w_8+w_2w_6+w_3w_5+w_4^2+w_2^4,
 \end{equation}
so that
\begin{equation}
 w_8\doteq w_2^4+w_4^2.
 \label{eq:sw-d8-euler-relation}
\end{equation}
The independent degree-eight numbers are therefore $w_2^4$ and $w_4^2$.

\subsection*{Dimension nine and the
\texorpdfstring{$w_2w_7\doteq w_4w_5$}{w2w7 = w4w5} relation}

The degree-nine monomials are
\begin{equation}
    w_9,\quad w_2w_7,\quad w_3w_6,\quad w_4w_5,\quad w_2^2w_5,\quad w_2w_3w_4,\quad w_3^3,\quad w_2^3w_3.
    \label{eq:sw-d9-candidates}
\end{equation}
Four of them vanish
\begin{align}
 w_2^2w_5&=\Sq^1(w_2^2w_4)\doteq0,
 &w_3^3&=\Sq^1(w_2w_3^2)\doteq0,
 \nonumber\\
 w_9&=\Sq^1w_8\doteq0,
 &0=\Sq^6w_3&\doteq\nu_6w_3
 =w_2w_3w_4+w_3^3.
 \label{eq:sw-d9-vanishings}
\end{align}
Hence, $w_2w_3w_4\doteq0$. The remaining mixed monomials obey
\begin{align}
 0\doteq\Sq^1(w_2w_6)
 &=w_3w_6+w_2w_7,
 \label{eq:sw-d9-first-equivalence}\\
 \Sq^2(w_3w_4)
 &=w_4w_5+w_3w_6
 \doteq\nu_2w_3w_4=w_2w_3w_4\doteq0.
 \label{eq:sw-d9-second-equivalence}
\end{align}
Therefore,
\begin{equation}
 \boxed{w_2w_7\doteq w_3w_6\doteq w_4w_5.}
 \label{eq:sw-d9-main-equivalence}
\end{equation}
The two independent nine-dimensional numbers may be chosen as $w_2^3w_3$ and $w_2w_7$.

\subsection*{Dimension ten}

The complete list of degree-ten candidates is
\begin{equation}
    \begin{gathered}
     w_{10},\ w_2w_8,\ w_3w_7,\ w_4w_6,\ w_5^2,
     \ w_2^2w_6,\ w_2w_3w_5,\ w_2w_4^2,\\
     w_3^2w_4,\ w_2^3w_4,\ w_2^2w_3^2,\ w_2^5.
    \end{gathered}
     \label{eq:sw-d10-candidates}
\end{equation}
The first set of vanishings follows from the Cartan formula, instability, and Wu pairing
\begin{align}
 0=\Sq^2(w_2^4)&\doteq\nu_2w_2^4=w_2^5,
 \nonumber\\
 w_2^2w_3^2&=\Sq^1(w_2^3w_3)\doteq0,
 \nonumber\\
 w_3w_7&=\Sq^1(w_2w_7)\doteq0,
 &w_5^2&=\Sq^1(w_4w_5)\doteq0,
 \nonumber\\
 \Sq^2w_8=w_2w_8+w_{10}
 &\doteq\nu_2w_8=w_2w_8,
 \label{eq:sw-d10-basic-relations}
\end{align}
and hence $w_{10}\doteq0$. The remaining Wu relations can be organized as
\begin{align}
 0=\Sq^6(w_2^2)
 &\doteq w_2^3w_4+w_2^2w_3^2,
 \nonumber\\
 \Sq^2(w_4^2)=w_5^2
 &\doteq w_2w_4^2,
 \nonumber\\
 \Sq^2(w_2w_6)=w_3w_7
 &\doteq w_2^2w_6,
 \nonumber\\
 \Sq^4w_6=w_4w_6+w_2w_8+w_{10}
 &\doteq\nu_4w_6=w_4w_6+w_2^2w_6,
 \nonumber\\
 \Sq^2(w_3w_5)=w_5^2
 &\doteq w_2w_3w_5,
 \nonumber\\
 \Sq^1(w_2w_3w_4)
 &=w_3^2w_4+w_2w_3w_5\doteq0.
 \label{eq:sw-d10-remaining-relations}
\end{align}
Substituting Eq.~\eqref{eq:sw-d10-basic-relations} into Eq.~\eqref{eq:sw-d10-remaining-relations} gives
\begin{equation}
    \begin{gathered}
     w_{10}\doteq w_2w_8\doteq w_3w_7\doteq w_5^2
     \doteq w_2^2w_6\doteq w_2w_3w_5\doteq w_2w_4^2\doteq0,\\
     w_3^2w_4\doteq w_2^3w_4\doteq w_2^2w_3^2
     \doteq w_2^5\doteq0.
    \end{gathered}
     \label{eq:sw-d10-vanishing-list}
\end{equation}
Thus, $w_4w_6$ is the unique independent Stiefel-Whitney number in $(9+1)$ dimensions.

\subsection*{Independence and detecting manifolds}

The preceding reductions show that the displayed lists span all oriented Stiefel-Whitney numbers. Their independence can be checked on the following explicit closed oriented manifolds. In dimensions four and five,
\begin{equation}
 \int_{\mathbb{CP}^2}w_2^2=1,
 \qquad
 \int_{W^5}w_2w_3=1,
 \qquad W^5=SU(3)/SO(3).
 \label{eq:sw-low-dimensional-detectors}
\end{equation}
The second equality is the standard characteristic number of the Wu manifold~\cite{MilnorStasheff1974}. In dimension eight,
\begin{equation}
\begin{array}{c|cc}
 &\displaystyle\int w_2^4&\displaystyle\int w_4^2\\ \hline
 \mathbb{CP}^4&1&0\\
 \mathbb{CP}^2\times\mathbb{CP}^2&0&1
\end{array}
 \label{eq:sw-d8-detectors}
\end{equation}
so the two numbers in Table~\ref{tab:oriented-sw-numbers-4-10} are independent.

The Dold manifold $P(m,n)=(S^m\times\mathbb{CP}^n)/((x,[z])\sim(-x,[\overline z]))$ has generators $c$ and $d$ of degrees one and two, respectively. Its mod-two cohomology and total Stiefel-Whitney class are~\cite{Dold1956}
\begin{equation}
 H^*(P(m,n);\mathbb Z_2)
 =\mathbb Z_2[c,d]/(c^{m+1},d^{n+1}),
 \qquad
 w(P(m,n))=(1+c)^m(1+c+d)^{n+1}.
 \label{eq:sw-dold-data}
\end{equation}
Moreover,
\begin{equation}
 w_1(P(m,n))=(m+n+1)c,
 \qquad
 \left\langle c^md^n,[P(m,n)]\right\rangle=1.
 \label{eq:sw-dold-orientation-pairing}
\end{equation}
Thus, $P(1,4)$ and $P(5,2)$ are oriented.  Expanding Eq.~\eqref{eq:sw-dold-data} gives
\begin{equation}
\begin{array}{c|ccc}
 &w_2&w_3&w_7\\ \hline
 P(1,4)&d&cd&0\\
 P(5,2)&d&cd&c^5d
\end{array}
 \label{eq:sw-dold-relevant-classes}
\end{equation}
and
\begin{equation}
\begin{array}{c|cc}
 &\displaystyle\int w_2^3w_3&\displaystyle\int w_2w_7\\ \hline
 P(1,4)&1&0\\
 P(5,2)&0&1
\end{array}
 \label{eq:sw-d9-detectors}
\end{equation}
Hence, these manifolds detect the two independent nine-dimensional numbers. Finally, write $a_i=\pi_1^*w_i(W)$ and $b_i=\pi_2^*w_i(W)$ on $W^5\times W^5$.  The only bidegree-$(5,5)$ contribution to $w_4w_6$ is $(a_2b_2)(a_3b_3)=a_2a_3b_2b_3$. The Whitney product formula therefore gives
\begin{equation}
 \int_{W^5\times W^5}w_4w_6
 =\left(\int_{W^5}w_2w_3\right)^2=1,
 \label{eq:sw-d10-detector}
\end{equation}
which detects the unique ten-dimensional number. These examples establish that no further universal oriented characteristic-number relations exist among the bases displayed in Table~\ref{tab:oriented-sw-numbers-4-10}.

\section{Oriented-bordism criterion for the
\texorpdfstring{$w_2^nw_3^m$}{w2n w3m} family}
\label{app:w2n-w3m-so-bordism}

This appendix proves Eq.~\eqref{eq:general-family-so-bordism-criterion}. Let $n,m\geq0$, let $D=2n+3m\geq4$, and let $M^D$ be a closed oriented manifold. All cohomology classes and characteristic numbers below have $\mathbb Z_2$ coefficients. By ``vanishing'' we mean that the Stiefel-Whitney number vanishes on every such $M$. This is weaker than saying that the universal polynomial $w_2^nw_3^m$ vanishes in $H^*(BSO;\mathbb Z_2)$.

\paragraph{Vanishing cases.}
Write $x=w_2(TM)$ and $y=w_3(TM)$.  On an oriented manifold, the Wu and Cartan formulas give~\cite{MilnorStasheff1974}
\begin{equation}
    \Sq^1x=y,\qquad
    \Sq^1y=0,\qquad
    \Sq^2x=x^2,\qquad
    \Sq^2y=xy+w_5,\qquad
    \nu_1=0,\qquad
    \nu_2=x.
    \label{eq:w2n-w3m-basic-wu-identities}
\end{equation}
Suppose first that $m>0$ and $n$ is even.  Since $n+1$ is odd,
\begin{equation}
    x^ny^m
    =\Sq^1\!\left(x^{n+1}y^{m-1}\right).
    \label{eq:w2n-w3m-sq1-exact}
\end{equation}
Wu pairing and $\nu_1=0$ then imply
\begin{equation}
    \left\langle x^ny^m,[M]\right\rangle
    =\left\langle
    \nu_1x^{n+1}y^{m-1},[M]
    \right\rangle=0.
    \label{eq:w2n-w3m-even-n-vanishing}
\end{equation}
This includes all pure powers $w_3^m$ with $m>0$.

It remains to consider $n$ odd and $m$ even, including $m=0$. Put $z=x^{n-1}y^m$, which has degree $D-2$. Since $\nu_2=x$, Wu pairing gives
\begin{equation}
    \left\langle x^ny^m,[M]\right\rangle
    =\left\langle\nu_2z,[M]\right\rangle
    =\left\langle\Sq^2z,[M]\right\rangle.
    \label{eq:w2n-w3m-sq2-wu-pairing}
\end{equation}
When $n=1$, one has $\Sq^2z=\Sq^2(y^m)=m y^{m-1}(xy+w_5)=0$, because $m$ is even. For $n\geq3$, the Cartan formula and Eq.~\eqref{eq:w2n-w3m-basic-wu-identities} give
\begin{equation}
    \begin{aligned}
        \Sq^2z
        ={}&(n-1+m)x^ny^m
        +\binom{n-1}{2}x^{n-3}y^{m+2}
        +m x^{n-1}y^{m-1}w_5
        =\binom{n-1}{2}x^{n-3}y^{m+2}.
    \end{aligned}
    \label{eq:w2n-w3m-sq2-reduction}
\end{equation}
The first exponent in the last line is even, while $m+2>0$. Its characteristic number vanishes by Eq.~\eqref{eq:w2n-w3m-even-n-vanishing}. Eq.~\eqref{eq:w2n-w3m-sq2-wu-pairing} then shows that $\langle x^ny^m,[M]\rangle=0$. These two arguments cover every parity choice excluded by Eq.~\eqref{eq:general-family-so-bordism-criterion}.

\paragraph{Detecting manifolds.}
We now show that both remaining parity classes can occur. If $m=0$ and $n$ is even, take $M=\mathbb{CP}^n$. For the standard generator
$h\in H^2(\mathbb{CP}^n;\mathbb Z_2)$,
\begin{equation}
    w_2(T\mathbb{CP}^n)=(n+1)h=h,\qquad
    \left\langle h^n,[\mathbb{CP}^n]\right\rangle=1.
    \label{eq:w2n-w3m-cp-detector}
\end{equation}
Hence, $\langle w_2^n,[\mathbb{CP}^n]\rangle=1$.

Now suppose that $n$ and $m$ are both odd.  Consider the Dold manifold
\begin{equation}
    P(r,s)
    :=\frac{S^r\times\mathbb{CP}^s}
    {(u,[z])\sim(-u,[\overline z])},
    \qquad (r,s)=(m,n+m).
    \label{eq:w2n-w3m-dold-manifold}
\end{equation}
Its mod-two cohomology and total Stiefel-Whitney class are~\cite{Dold1956}
\begin{equation}
    H^*(P(r,s);\mathbb Z_2)
    =\frac{\mathbb Z_2[c,d]}{(c^{r+1},d^{s+1})},
    \qquad |c|=1,\quad |d|=2,\qquad
    w(P(r,s))=(1+c)^r(1+c+d)^{s+1}.
    \label{eq:w2n-w3m-dold-data}
\end{equation}
For $(r,s)=(m,n+m)$, expansion through degree three gives
\begin{equation}
    w_1=(n+2m+1)c=0,\qquad
    w_2=d+\binom{n+2m+1}{2}c^2,\qquad
    w_3=cd,
    \label{eq:w2n-w3m-dold-low-classes}
\end{equation}
where the binomial coefficient is reduced modulo two. Thus, the manifold is
oriented. Moreover,
\begin{equation}
    \begin{aligned}
        w_2^nw_3^m
        &=\left(d+\binom{n+2m+1}{2}c^2\right)^n(cd)^m
        =c^md^{n+m}.
    \end{aligned}
    \label{eq:w2n-w3m-dold-top-class}
\end{equation}
Every term in the expansion that contains at least one factor of $c^2$ contains $c^{m+2}$ and vanishes because $c^{m+1}=0$. Since $\dim P(m,n+m)=2n+3m$ and $c^md^{n+m}$ is its top generator,
\begin{equation}
    \left\langle
    w_2^nw_3^m,[P(m,n+m)]
    \right\rangle=1.
    \label{eq:w2n-w3m-dold-detector}
\end{equation}
The complex-projective and Dold detectors prove the converse to the two vanishing arguments and complete the proof of Eq.~\eqref{eq:general-family-so-bordism-criterion}. Notice that the criterion concerns only the induced $SO$-bordism invariant. The monomial $w_2^nw_3^m$ remains a nonzero universal class in $H^*(BSO;\mathbb Z_2)$ even in the parity classes whose characteristic numbers vanish identically.

\section{Vanishing Stiefel-Whitney numbers admit trivial QCA realizations}
\label{app:wu-core-reduction}

Throughout this appendix, all cochains and cohomology classes are
$\mathbb Z_2$-valued, and $\frac12$ denotes the standard embedding
$\mathbb Z_2\hookrightarrow\mathbb R/\mathbb Z$. Let $D\geq4$ be a
spacetime dimension and set $d=D-1$. For a closed oriented $D$-manifold
$X$, write $w_i=w_i(TX)$ and $\nu_i=\nu_i(TX)$. A homogeneous degree-$D$
Stiefel-Whitney polynomial specifies the gravitational response
\begin{equation}
    S_P[X]=\frac12\int_X P(w).
    \label{eq:app-sw-response}
\end{equation}
As in Appendix~\ref{app:sw-number-relations}, we write $P(w)\doteq0$ if
\begin{equation}
    \left\langle P\bigl(w(TX)\bigr),[X]\right\rangle=0
    \quad\text{for every closed oriented }X^D.
    \label{eq:app-sw-number}
\end{equation}
This condition concerns characteristic numbers, not the vanishing of
$P$ as a universal polynomial.

The result of this appendix is an existence statement. For every $P$ satisfying Eq.~\eqref{eq:app-sw-number}, we construct a local higher-form $\mathbb Z_2$ gauge-theory realization of its gravitational response and choose a complete separator--flipper algebra whose QCA is a finite-depth quantum circuit. The choice may introduce finitely many auxiliary qubit species per cell. On the hypercubic spatial lattices used in this paper, the circuit depth and gate range are independent of the system size.

More precisely, let $\mathcal C$ denote the choices of a Wu certificate, local cochain representatives, auxiliary fields, and boundary convention. We construct an action $\Phi_{P,\mathcal C}$ and a full-space completion
$\alpha_{P,\mathcal C}$ such that
\begin{equation}
 \begin{aligned}
 Z^{\mathrm{inv}}_{\Phi_{P,\mathcal C}}(X)
   &=(-1)^{\langle P(w(TX)),[X]\rangle},\\
 \alpha_{P,\mathcal C}&=\operatorname{Ad}_{U_{P,\mathcal C}},
 \qquad U_{P,\mathcal C}\text{ is an FDQC}.
 \end{aligned}
 \label{eq:app-sw-qca-existence}
\end{equation}
Here, $\operatorname{Ad}_U(O)=UOU^\dagger$, and the partition function is normalized so that the decoupled cores introduced below have value one. In particular, the right-hand side of the first line is one when $P(w)\doteq0$.

The action and the full-space completion are both part of the construction. We do not assert that $P(w)$ determines a unique QCA, or that every completion with this closed-manifold response is an FDQC. In particular, this appendix does not identify the chosen realization with an independently specified separator--flipper construction in the main text. Such an identification requires an additional operator-level
comparison.

\subsection*{Wu certificates}

The mathematical input is the Wu pairing~\cite{MilnorStasheff1974},
\begin{equation}
    \left\langle\Sq^i x,[X]\right\rangle
    =\left\langle\nu_i\cup x,[X]\right\rangle,
    \qquad x\in H^{D-i}(X;\mathbb Z_2).
    \label{eq:app-wu-pairing}
\end{equation}
Consequently, the degree-$D$ expression
\begin{equation}
    R_i(q):=\Sq^i q+\nu_i\cup q,
    \qquad |q|=D-i,
    \label{eq:app-wu-relator}
\end{equation}
has zero characteristic number. We call it a Wu relator.

We use the completeness of the Wu relations for Stiefel-Whitney numbers~\cite{Dold1956Wu,BrownPeterson1964Relations,Stong1968Cobordism}. After imposing $w_1=0$, a vanishing degree-$D$ characteristic number admits a finite Wu certificate
\begin{equation}
    P(w)=\sum_a\left[
        \Sq^{i_a}q_a(w)+\nu_{i_a}\cup q_a(w)
    \right],
    \qquad |q_a|=D-i_a,\quad 1\leq i_a\leq D.
    \label{eq:app-general-wu-certificate}
\end{equation}
This is a degree-$D$ characteristic-number relation for oriented tangent bundles. It is not an identity of cochains obtained by replacing all tangent characteristic classes by independent gauge fields. In particular, this theorem supplies no local cochain equivalence between arbitrary gauge-theory realizations of the same characteristic number.

Combining terms with equal $i_a$, and separating the possible top-index
term, gives
\begin{equation}
    P(w)=c\,\nu_D+\sum_{i\in\mathcal I}R_i(q_i),
    \qquad c\in\mathbb Z_2,\quad
    \mathcal I\subseteq\{1,\ldots,D-1\}.
    \label{eq:app-grouped-wu-certificate}
\end{equation}
Indeed, $q_D$ is constant and $\Sq^D(1)=0$ for $D>0$. Wu pairing and Steenrod instability imply
\begin{equation}
    \nu_D(TX)=0\in H^D(X;\mathbb Z_2).
    \label{eq:app-top-wu-zero}
\end{equation}
For example, pairing with the degree-zero generator on each connected component gives $\langle\nu_D,[X]\rangle=\langle\Sq^D1,[X]\rangle=0$. Mod-two Poincare duality then gives Eq.~\eqref{eq:app-top-wu-zero}. For the existence statement here, we choose the zero action and identity QCA for this summand. No $A_0$--$B_D$ spatial core is required. This is a choice of realization of a zero response, not a claim of a local primitive or a circuit trivialization for a previously fixed $\nu_D$ lattice model. If only this summand remains, the trivial realization already proves the claim. Orientation-forced zero terms are treated in the same response-level sense.

\subsection*{Quadratic cores and a chosen gauge-theory action}

For $1\leq j<D$ with $2j\leq D$, introduce cocycles $A_{D-j}\in Z^{D-j}(X;\mathbb Z_2)$ and $B_j\in Z^j(X;\mathbb Z_2)$, and define
\begin{equation}
    K_j[A_{D-j},B_j]
    :=\frac12\int_X\left(
        A_{D-j}\cup_{D-2j}A_{D-j}+B_j\cup A_{D-j}
    \right).
    \label{eq:app-quadratic-core}
\end{equation}
On a closed manifold, Wu pairing gives
\begin{equation}
    K_j[A_{D-j},B_j]
    =\frac12\int_X(\nu_j+B_j)\cup A_{D-j}.
    \label{eq:app-core-wu}
\end{equation}
Summing over $[A_{D-j}]$ is a finite Fourier transform under the nondegenerate Poincare pairing and imposes
\begin{equation}
    [B_j]=\nu_j(TX).
    \label{eq:app-B-equals-wu}
\end{equation}
For $D/2<j<D$, instability gives $\nu_j=0$ on $X$. In this range use
\begin{equation}
    K_j^{\mathrm{BF}}[A_{D-j},B_j]
    :=\frac12\int_X B_j\cup A_{D-j},
    \label{eq:app-BF-core}
\end{equation}
whose Fourier sum imposes $[B_j]=0=\nu_j$. Write $\mathcal K_j=K_j$ or $K_j^{\mathrm{BF}}$, respectively. With normalization relative to these nondegenerate Fourier pairs, each decoupled core has closed-manifold partition function one.

To express the certificate in terms of the $B$ fields, use
\begin{equation}
    w=\Sq(\nu).
    \label{eq:app-w-equals-sqnu}
\end{equation}
For closed cochains, choose local representatives
\begin{equation}
    \widehat w_k(\{B_j\})
      :=\sum_{r=0}^k\Sq^rB_{k-r},\qquad B_0=1,
    \label{eq:app-w-from-B}
\end{equation}
where $\Sq^0 B_s=B_s$, while for $1\leq r\leq s$ we use $B_s\cup_{s-r}B_s$, and terms with $r>s$ vanish. Fix the order of all ordinary cup products. Then choose
\begin{equation}
    \widehat q_i(\{B_j\}_{j\in\mathcal J})
      :=q_i(\widehat w_2,\widehat w_3,\ldots).
    \label{eq:app-q-from-B}
\end{equation}
The finite set $\mathcal J\subseteq\{1,\ldots,D-1\}$ includes $\mathcal I$ and all additional field degrees that occur in these formulas. For every additional $B_j$, introduce its paired core $\mathcal K_j$. No degree-$D$ field is needed because $|q_i|=D-i<D$. Geometric identities may reduce the field set, provided that $[\widehat q_i]=q_i(w(TX))$ whenever $[B_j]=\nu_j$. Such reductions are choices of representatives, not assumed cochain identities for arbitrary gauge configurations.

We now \emph{define} the gauge-theory realization associated with this certificate as
\begin{equation}
 \begin{aligned}
    \Phi_{P,\mathcal C}[A,B]
    := {}&\sum_{j\in\mathcal J}\mathcal K_j[A_{D-j},B_j]\\
    &+\frac12\int_X\sum_{i\in\mathcal I}
       \left[\Sq^i\widehat q_i+B_i\cup\widehat q_i\right].
 \end{aligned}
    \label{eq:app-stabilized-action}
\end{equation}
We do not obtain this definition by asserting a local coboundary relation to another realization of $P$. Instead, its response is verified directly. The $A$ sums impose $[B_j]=\nu_j$; evaluating the interaction on these classes gives the certificate Eq.~\eqref{eq:app-grouped-wu-certificate}, with its zero top-index summand omitted. Thus
\begin{equation}
    Z^{\mathrm{inv}}_{\Phi_{P,\mathcal C}}(X)
       =(-1)^{\langle P(w(TX)),[X]\rangle}=1.
    \label{eq:app-chosen-response}
\end{equation}
This proves realization of the specified closed-manifold response. The following construction separately fixes the full microscopic QCA.

\subsection*{Local field shifts and the boundary cochain}

For $i\in\mathcal I$, make the change of variables
\begin{equation}
    A_{D-i}=A'_{D-i}+\widehat q_i(B),
    \qquad B_j'=B_j.
    \label{eq:app-simultaneous-shift}
\end{equation}
It is a bijection on closed configurations: $\widehat q_i(B)$ is closed whenever every $B_j$ is closed. For a quadratic core and any closed $(D-i)$-cochain $q$, the higher-cup recursion gives the local identity
\begin{equation}
 \begin{aligned}
 K_i[A+q,B_i]-K_i[A,B_i]
  ={}&\frac12\int_X\left(\Sq^i q+B_i\cup q\right)
    +\frac12\int_X\delta\left(A\cup_{D-2i+1}q\right).
 \end{aligned}
 \label{eq:app-core-shift}
\end{equation}
For a BF core, the difference is simply $\frac12\int_X B_i\cup q$, and $\Sq^i q=0$ by instability. Therefore, the shifts cancel the entire interaction in Eq.~\eqref{eq:app-stabilized-action}. Define
\begin{equation}
 \Lambda_{\mathcal C}(A',B)
   :=\frac12\sum_{\substack{i\in\mathcal I\\ 2i\leq D}}
       A'_{D-i}\cup_{D-2i+1}\widehat q_i(B).
 \label{eq:app-explicit-boundary-cochain}
\end{equation}
Writing $F_{\mathcal C}(A',B)=(A'+\widehat q(B),B)$ and $\Phi_0=\sum_{j\in\mathcal J}\mathcal K_j$, we have, for closed bulk fields,
\begin{equation}
 \Phi_{P,\mathcal C}[F_{\mathcal C}(A',B)]
    =\Phi_0[A',B]+\int_X\delta\Lambda_{\mathcal C}(A',B).
 \label{eq:app-reduced-action}
\end{equation}
Unlike equality of characteristic numbers, this equality follows from an explicit local cochain calculation. It fixes the boundary phase needed for the circuit construction.

\subsection*{A full-space circuit realization}

First choose a complete core algebra implemented by a Clifford FDQC $U_0$. We give the general form to make clear that the circuit acts on the full gauged Hilbert space, not just on an ungauged SPT wavefunction. For a quadratic core with index $j$, set $\ell=D-2j$. Its ungauged boundary descendant is
\begin{equation}
    a\cup_\ell\delta a+a\cup_{\ell-1}a+b\cup\delta a,
    \label{eq:app-core-boundary-phase}
\end{equation}
with negative cup indices omitted. The following is the full-space version of Eq.~\eqref{eq:explicit-FDQC}.

On the spatial hypercubic lattice $M$, put $A$ qubits on $(D-j)$-cells $r$ and $B$ qubits on $j$-cells $q$. Let $t(q)$ and $p(r)$ be their complementary $(D-j-1)$- and $(j-1)$-cells:
\begin{equation}
 \int_M\boldsymbol q'\cup\boldsymbol{t(q)}=\delta_{qq'},
 \qquad
 \int_M\boldsymbol{p(r)}\cup\boldsymbol r'=\delta_{rr'}.
 \label{eq:app-core-complementary-pairing}
\end{equation}
Define the binary incidence matrices
\begin{equation}
 \begin{aligned}
 N_{rq}&=(\delta\boldsymbol{t(q)})(r)
          =(\delta\boldsymbol{p(r)})(q),\\
 M_{qr}&=\int_M\delta\boldsymbol{t(q)}\cup_{\ell+1}\boldsymbol r,\\
 S_{qq'}&=\sum_r M_{qr}N_{rq'}
      =\int_M\delta\boldsymbol{t(q)}\cup_{\ell+1}
                    \delta\boldsymbol{t(q')}.
 \end{aligned}
 \label{eq:app-core-incidence-matrices}
\end{equation}
Stokes' formula gives the equality in the first line. Higher-cup identities give $S_{qq'}=S_{q'q}$ and $S_{qq}=0$; for the diagonal, one can use
\[
 \delta\left(t\cup_{\ell+1}\delta t+t\cup_\ell t\right)
      =\delta t\cup_{\ell+1}\delta t.
\]
Integrals here are over a closed spatial lattice, or are evaluated with finite support in an infinite bulk.

Let $\mathrm{CZ}_X=(H\otimes H)\mathrm{CZ}(H\otimes H)$, and set
\begin{equation}
 \begin{aligned}
 U_j={}&\left[\prod_q(X_q^B)^{\kappa_q}\right]
 \left[\prod_{r,q}\mathrm{CZ}_X(A_r,B_q)^{N_{rq}}\right]
\left[\prod_{q,r}\mathrm{CNOT}(A_r\to B_q)^{M_{qr}}\right]
 \left[\prod_{q<q'}\mathrm{CZ}_X(B_q,B_{q'})^{S_{qq'}}\right],
 \end{aligned}
 \label{eq:app-general-core-circuit}
\end{equation}
where $\kappa_q=\int_M[t(q)\cup_\ell\delta t(q) + t(q)\cup_{\ell-1}t(q)]$ retains any configuration-independent endpoint sign. Direct conjugation gives
\begin{equation}
 \begin{aligned}
 U_j Z_r^A U_j^\dagger
   &=Z_r^A\prod_q(X_q^B)^{N_{rq}},\\
 U_j Z_q^B U_j^\dagger
   &=(-1)^{\kappa_q}Z_q^B
        \left[\prod_r(X_r^A)^{N_{rq}}\right]
        \left[\prod_r(Z_r^A)^{M_{qr}}\right],\\
 U_j X_r^A U_j^\dagger
   &=X_r^A\prod_q(X_q^B)^{M_{qr}},\\
 U_j X_q^B U_j^\dagger&=X_q^B.
 \end{aligned}
 \label{eq:app-general-core-pauli-images}
\end{equation}
The cancellation in the second line is exactly $S=MN$. The zero diagonal removes the possible ordering sign. These are the complete separator--flipper images for the quadratic core. For a BF core take $M=S=0$ and $\kappa_q=0$, leaving only the mixed $\mathrm{CZ}_X$ block. Each block has bounded-range gates and bounded local incidence, so a finite coloring gives bounded depth. The tensor product of the $U_j$ is the core circuit $U_0$.

Next implement the field redefinition on the full tensor-product Hilbert space by
\begin{equation}
 V_{\mathcal C}|A',B\rangle
       :=|A'+\widehat q(B),B\rangle.
 \label{eq:app-field-shift-circuit}
\end{equation}
Off the closed-cochain sector, $\widehat q_i$ means the same fixed local polynomial, without claiming it is a cocycle. Thus, Eq.~\eqref{eq:app-field-shift-circuit} is defined on every basis state. All controls belong to the $B$ registers and all targets to the $A$ registers, so this is an invertible controlled shift with $V_{\mathcal C}^2=1$. Locality of $\widehat q_i$ and a finite coloring make it an FDQC, generally non-Clifford.

In the primed coordinates, define the boundary phase circuit
\begin{equation}
 W_{\mathcal C}|A',B\rangle
     :=\exp\left(2\pi i\int_M\Lambda_{\mathcal C}(A',B)\right)
          |A',B\rangle.
 \label{eq:app-boundary-phase-circuit}
\end{equation}
Again, the displayed polynomial defines the operator for arbitrary cochains. It is a product of bounded-range diagonal phase gates, hence an FDQC.

We now \emph{choose} the full-space completion by transporting both members of every core separator--flipper pair. Set $T_{\mathcal C}=V_{\mathcal C}W_{\mathcal C}$ and define
\begin{equation}
 \begin{aligned}
 \overline Z^{P,\mathcal C}_s
    &:=T_{\mathcal C}(U_0Z_sU_0^\dagger)T_{\mathcal C}^\dagger,\\
 \overline X^{P,\mathcal C}_s
    &:=T_{\mathcal C}(U_0X_sU_0^\dagger)T_{\mathcal C}^\dagger.
 \end{aligned}
 \label{eq:app-transported-algebra}
\end{equation}
Here, $s$ ranges over every physical and auxiliary qubit. Unitary conjugation preserves completeness and all Pauli relations, while the finite circuit depth preserves uniformly bounded support. Therefore, these operators define a QCA on the full local operator algebra, with
\begin{equation}
 \alpha_{P,\mathcal C}(O)
       =U_{P,\mathcal C}OU_{P,\mathcal C}^\dagger,
 \qquad
 U_{P,\mathcal C}:=V_{\mathcal C}W_{\mathcal C}U_0.
 \label{eq:app-exact-qca-circuit}
\end{equation}
The order of factors follows from Eq.~\eqref{eq:app-reduced-action}. In the transported boundary convention, the flat-sector amplitudes obey
\[
 \Psi_{P,\mathcal C}(F_{\mathcal C}(A',B))
       =e^{2\pi i\int_M\Lambda_{\mathcal C}(A',B)}\Psi_0(A',B).
\]
Their finite differences give the corresponding transported gauge moves. Thus the action and the chosen microscopic completion are matched by construction. Away from the flat sector, Eq.~\eqref{eq:app-transported-algebra} is a definition, not a claim that an independently chosen completion transforms in the same way.

We have proved the existential statement
\begin{equation}
 \boxed{\begin{gathered}
 P(w)\doteq0\quad\Longrightarrow\quad
 \exists\,(\Phi_{P,\mathcal C},\alpha_{P,\mathcal C})\text{ realizing }S_P,\\
 \alpha_{P,\mathcal C}=\operatorname{Ad}_{U_{P,\mathcal C}},
 \qquad U_{P,\mathcal C}\in\mathrm{FDQC}.
 \end{gathered}}
 \label{eq:app-trivial-number-trivial-qca}
\end{equation}
In particular, $[\alpha_{P,\mathcal C}]=0$. On the Hilbert space that includes the chosen auxiliary species, the constructed automorphism is itself an FDQC; no lattice shift is required. The auxiliary species are part of this realization and are not, without a further comparison, a stabilization of an independently specified QCA.

\subsection*{Examples}

We first illustrate the construction with the degree-seven response $w_7$. Here $D=7$, so the chosen QCA realization lives in six spatial dimensions. The Wu formula gives
\begin{equation}
    \Sq^1w_6=w_1w_6+w_7,
    \qquad \nu_1=w_1,
\end{equation}
and hence
\begin{equation}
    w_7=\Sq^1w_6+\nu_1\cup w_6=R_1(w_6).
    \label{eq:app-w7-certificate}
\end{equation}
Thus, the certificate contains a single relator with $i=1$ and $q_1=w_6$.

We next express $w_6$ using the auxiliary Wu fields. In dimension seven, orientation and Steenrod instability reduce $w=\Sq(\nu)$ to
\begin{equation}
    w_6=\nu_6+\Sq^2\nu_4.
    \label{eq:app-w7-reduced-w6}
\end{equation}
A convenient local representative is therefore
\begin{equation}
    \boxed{
    \widehat q_1(B_4,B_6)
    :=
    B_6+\Sq^2B_4
    =B_6+B_4\cup_2B_4.
    }
    \label{eq:app-w7-qB}
\end{equation}
One sufficient auxiliary field set for this formula is
\begin{equation}
    \{B_1,B_4,B_6\}.
    \label{eq:app-w7-field-set}
\end{equation}
The field $B_1$ is required by the relator $R_1$, while $\widehat q_1$ itself depends only on $B_4$ and $B_6$. The unsimplified formula for $\widehat w_6$ would also contain $\Sq^1B_5$ and $\Sq^3B_3$. Their paired cores may be added, but on an oriented seven-manifold they represent $\nu_5=\nu_3=0$. They are optional stabilizations and are not needed in the reduced construction.

The three cores used in this realization are
\begin{equation}
    K_1[A_6,B_1],
    \qquad
    K_4^{\mathrm{BF}}[A_3,B_4],
    \qquad
    K_6^{\mathrm{BF}}[A_1,B_6].
    \label{eq:app-w7-cores}
\end{equation}
The first supplies the relator index $i=1$, while the two BF cores supply the fields needed to build $w_6$. For this certificate and these local representatives, choose the action
\begin{equation}
    \Phi_{w_7}:=K_1[A_6,B_1]+K_4^{\mathrm{BF}}[A_3,B_4]+K_6^{\mathrm{BF}}[A_1,B_6]+\frac12\int_{X^7}\left(\Sq^1\widehat q_1+B_1\cup\widehat q_1\right).
    \label{eq:app-w7-action}
\end{equation}
Under the single controlled shift
\begin{equation}
    A_6=A_6'+\widehat q_1(B_4,B_6),
    \label{eq:app-w7-shift}
\end{equation}
Eq.~\eqref{eq:app-core-shift} removes the Wu-relator term in Eq.~\eqref{eq:app-w7-action}. We obtain
\begin{equation}
    \Phi_{w_7}=K_1[A_6',B_1]+K_4^{\mathrm{BF}}[A_3,B_4]+K_6^{\mathrm{BF}}[A_1,B_6]+\frac12\int_{X^7}\delta\!\left(A_6'\cup_6\widehat q_1\right).
    \label{eq:app-w7-reduced}
\end{equation}
Choose the complete separator--flipper algebra by the transport in Eq.~\eqref{eq:app-transported-algebra}, using the displayed shift and boundary cochain. The resulting $w_7$ realization is implemented on the full Hilbert space by $U_{w_7,\mathcal C}=V_{\mathcal C}W_{\mathcal C}U_0$. Thus, this realization, including its auxiliary species, is an FDQC.

As a second example, consider $w_2w_8$ in $D=10$. The chosen QCA realization lives in nine spatial dimensions. Direct use of the Wu formula gives
\begin{align}
    R_3(w_7)
    &=w_3w_7,
    \label{eq:app-w2w8-R3}\\
    R_2(w_2w_6+w_8)
    &=w_2^2w_6+w_3w_7+w_{10},
    \label{eq:app-w2w8-R2}\\
    R_4(w_6)
    &=w_2^2w_6+w_2w_8+w_{10}.
    \label{eq:app-w2w8-R4}
\end{align}
Adding these three equations modulo two cancels every term except $w_2w_8$, so
\begin{equation}
    \boxed{
    w_2w_8
    =
    R_3(w_7)
    +R_2(w_2w_6+w_8)
    +R_4(w_6).
    }
    \label{eq:app-w2w8-certificate}
\end{equation}

The fields $B_2,B_3,B_4$ suffice for the following choice. Indeed, in dimension ten, orientation and instability give
\begin{equation}
    w_2=\nu_2,
    \qquad
    w_6=\Sq^2\nu_4,
    \qquad
    w_7=\Sq^3\nu_4,
    \qquad
    w_8=\Sq^4\nu_4.
    \label{eq:app-w2w8-reduced-w}
\end{equation}
We may therefore choose
\begin{equation}
    \widehat q_3(B_4):=\Sq^3B_4,\qquad \widehat q_2(B_2,B_4):=B_2\cup\Sq^2B_4+\Sq^4B_4,\qquad \widehat q_4(B_4):=\Sq^2B_4.
    \label{eq:app-w2w8-qB}
\end{equation}
Thus, the auxiliary field set for this realization is
\begin{equation}
    \{B_2,B_3,B_4\}.
    \label{eq:app-w2w8-field-set}
\end{equation}
The field $B_3$ is required because the certificate contains the relator $R_3$, even though the corresponding function $\widehat q_3$ depends only on $B_4$.

Introduce the three quadratic cores
\begin{equation}
    K_2[A_8,B_2],
    \qquad
    K_3[A_7,B_3],
    \qquad
    K_4[A_6,B_4].
    \label{eq:app-w2w8-cores}
\end{equation}
The chosen action is
\begin{equation}
    \Phi_{w_2w_8}:=K_2[A_8,B_2]+K_3[A_7,B_3]+K_4[A_6,B_4]+\frac12\int_{X^{10}}\sum_{i\in\{2,3,4\}}\left[\Sq^i\widehat q_i+B_i\cup\widehat q_i\right].
    \label{eq:app-w2w8-action}
\end{equation}
The three controlled shifts
\begin{equation}
    A_8=A_8'+\widehat q_2(B_2,B_4),
    \qquad
    A_7=A_7'+\widehat q_3(B_4),
    \qquad
    A_6=A_6'+\widehat q_4(B_4)
    \label{eq:app-w2w8-shifts}
\end{equation}
remove the entire Wu-relator interaction in Eq.~\eqref{eq:app-w2w8-action}. Up to the local boundary circuit
\begin{equation}
    \frac12\int_{X^{10}}
    \delta\!\left(
        A_8'\cup_7\widehat q_2
        +A_7'\cup_5\widehat q_3
        +A_6'\cup_3\widehat q_4
    \right),
    \label{eq:app-w2w8-boundary}
\end{equation}
the action reduces to the three decoupled cores in Eq.~\eqref{eq:app-w2w8-cores}. Transporting both the separators and the flippers as in Eq.~\eqref{eq:app-transported-algebra} defines the chosen full-space completion. Its QCA is implemented by the FDQC $U_{w_2w_8,\mathcal C}=V_{\mathcal C}W_{\mathcal C}U_0$. This proves the existence of an FDQC realization of the $w_2w_8$ response, without identifying it with a separately chosen completion.

The auxiliary fields in these examples are sufficient for the displayed local formulas. More generally, different certificates, cochain representatives, and transport conventions may give different microscopic realizations. The result is that every vanishing oriented Stiefel-Whitney number admits at least one realization of the form Eq.~\eqref{eq:app-exact-qca-circuit}. It does not establish that a vanishing response trivializes every QCA carrying that response, nor that the response determines the stable QCA class.

\section{Signature responses and quadratic data of the semion QCA families}
\label{app:semion-signature-bordism}

This appendix derives the closed-manifold responses associated with the semion and $U(1)_4$ QCA constructions in 3 spatial dimensions and with their higher-form analogues in 7 spatial dimensions. It also clarifies the information contained in these responses. A normalized partition function determines an invertible bordism invariant, whereas a quadratic function records the statistics of the boundary excitations and a QCA specifies a microscopic automorphism of the local operator algebra. These three structures are related, but the bordism invariant alone does not determine the other two. We begin by reviewing the relevant Pontryagin-square Gauss sum and then apply it in 4 and 8 spacetime dimensions.

\subsection*{The Pontryagin-square Gauss sum}

Let $X^{4r}$ be a closed oriented manifold with fundamental class $[X]\in H_{4r}(X,\mathbb Z)$.  Let $\operatorname{Tor}H^{2r}(X,\mathbb Z)$ denote the torsion subgroup, consisting of classes $x$ such that $nx=0$ for some positive integer $n$. The quotient by this subgroup is free abelian and carries the intersection form
\begin{equation}
Q_X:
\frac{H^{2r}(X,\mathbb Z)}
{\operatorname{Tor}H^{2r}(X,\mathbb Z)}
\times
\frac{H^{2r}(X,\mathbb Z)}
{\operatorname{Tor}H^{2r}(X,\mathbb Z)}
\longrightarrow \mathbb Z,
\qquad
Q_X(x,y)
=
\left\langle x\cup y,[X]\right\rangle .
\label{eq:signature-intersection-form}
\end{equation}
Here, $x\cup y\in H^{4r}(X,\mathbb Z)$, and the bracket denotes its evaluation on the fundamental class.  After extending $Q_X$ to $\mathbb R$, it becomes a nondegenerate symmetric bilinear form. Let $b^+$ and $b^-$ denote the numbers of its positive and negative eigenvalues, respectively.  The signature of $X$ is
\begin{equation}
    \sigma(X)=b^+-b^-.
    \label{eq:topological-signature-definition}
\end{equation}
It changes sign under orientation reversal and should not be confused with the signature of a spacetime metric.

The Pontryagin square is a quadratic cohomology operation
\begin{equation}
    \mP:H^{2r}(X,\mathbb Z_2)\longrightarrow H^{4r}(X,\mathbb Z_4).
\end{equation}
Its defining quadratic property is
\begin{equation}
    \mP(b+b')=\mP(b)+\mP(b')+2b\cup b'.
    \label{eq:pontryagin-square-quadratic-property}
\end{equation}
For $b\in H^{2r}(X,\mathbb Z_2)$, define
\begin{equation}
    q_X(b)=\frac14\left\langle\mP(b),[X]\right\rangle \quad\in\mathbb R/\mathbb Z.
    \label{eq:pontryagin-square-quadratic-function}
\end{equation}
The corresponding mod-two intersection pairing 
\begin{equation}
    q_X(b+b')-q_X(b)-q_X(b') =\frac12\left\langle b\cup b',[X]\right\rangle .
\label{eq:pontryagin-square-polarization}
\end{equation}
Poincar\'e duality makes this pairing nondegenerate. The Brown-Kervaire-Morita theorem then evaluates the normalized Gauss sum as \cite{Morita1971Pontrjagin,Taylor2022Gauss,Bhardwaj2020BrownKervaire}
\begin{equation}
 \boxed{
 \frac{1}{\sqrt{|H^{2r}(X,\mathbb Z_2)|}}
 \sum_{[b]\in H^{2r}(X,\mathbb Z_2)}
 \exp\!\left[
 \frac{2\pi i}{4}\left\langle\mP(b),[X]\right\rangle
 \right]
 =\exp\!\left(\frac{2\pi i}{8}\sigma(X)\right).}
 \label{eq:brown-kervaire-morita-gauss-sum}
\end{equation}
This formula remains valid in the presence of two-torsion in integral cohomology. We use the $+\mP$ convention. Replacing $\mP$ by $-\mP$, or reversing the orientation of $X$, complex conjugates the result.

A lattice gauge-theory partition function may also contain positive local normalization factors. Throughout this appendix, \emph{invertible normalization} means that such factors have been divided out, leaving the unit-modulus response in Eq.~\eqref{eq:brown-kervaire-morita-gauss-sum}. So,
\begin{equation}
 \chi_{\mathrm{sig}}:\Omega_{4r}^{SO}\longrightarrow U(1),
 \qquad
 [X]\longmapsto
 \exp\!\left(\frac{2\pi i}{8}\sigma(X)\right)
 \label{eq:signature-bordism-character}
\end{equation}
is an oriented-bordism character. Its image has order 8 because $\sigma(\mathbb{CP}^{2r})=1$.

\subsection*{The 3-dimensional semion and \texorpdfstring{$U(1)_4$}{U(1)4} QCAs}

The QCAs in three spatial dimensions are associated with responses on closed oriented 4-manifolds. The Hirzebruch signature theorem gives
\begin{equation}
    \sigma(X^4)=\frac13\int_{X^4}p_1(TX).
    \label{eq:four-dimensional-signature-theorem}
\end{equation}
Combining this identity with Eq.~\eqref{eq:brown-kervaire-morita-gauss-sum} gives the invertibly normalized semion response
\begin{equation}
    S_{\mathrm{sem}}^{\mathrm{inv}}(X^4) =\frac{\sigma(X^4)}8 =\frac1{24}\int_{X^4}p_1(TX) \pmod1.
    \label{eq:one-semion-pontryagin-response}
\end{equation}
For example, $\sigma(\mathbb{CP}^2)=1$ and $\int_{\mathbb{CP}^2}p_1=3$, so the partition function is $\exp(2\pi i/8)$. This is the signature-dependent part of the semion Crane-Yetter theory~\cite{CraneKauffmanYetter1993}.

The boundary topological order contains more information than this single phase. A pointed Abelian theory is specified by a finite Abelian group $\mathcal A$ and a quadratic function $q:\mathcal A\rightarrow\mathbb R/\mathbb Z$. Its normalized Gauss-Milgram phase is
\begin{equation}
    \xi(\mathcal A,q) =\frac1{\sqrt{|\mathcal A|}} \sum_{a\in\mathcal A}e^{2\pi i q(a)}.
    \label{eq:finite-quadratic-gauss-sum}
\end{equation}
For the semion theory $U(1)_2$,
\begin{equation}
    \mathcal A_2=\mathbb Z_2, \qquad q_2(a)=\frac{a^2}{4}, \qquad \xi(\mathcal A_2,q_2)=e^{2\pi i/8}.
    \label{eq:semion-gauss-milgram-sum}
\end{equation}
For $U(1)_4$,
\begin{equation}
 \mathcal A_4=\mathbb Z_4,
 \qquad q_4(a)=\frac{a^2}{8},
 \qquad
 \xi(\mathcal A_4,q_4)
 =\frac12\sum_{a=0}^{3}e^{2\pi i a^2/8}
 =e^{2\pi i/8}.
 \label{eq:u14-gauss-milgram-sum}
\end{equation}
Thus, the $U(1)_2$ and $U(1)_4$ theories have the same normalized signature response even though their excitation groups and quadratic functions are different. A microscopic realization of the $U(1)_4$ ququart QCA was constructed in Ref.~\cite{Shirley2022QCA}. The signature phase alone does not determine its stable QCA class or order.

The distinction is also visible in the pointed Witt group. Two pointed theories are Witt equivalent when their product with opposite chirality admits a Lagrangian subgroup, corresponding physically to a fully gapped boundary~\cite{Drinfeld2010Braided,Davydov2013Witt}. The difference between the two theories is represented by
\begin{equation}
    \left(\mathbb Z_4\oplus\mathbb Z_2,~q(x,y)=\frac{x^2}{8}-\frac{y^2}{4}\right).
    \label{eq:u12-u14-witt-difference}
\end{equation}
Its normalized Gauss sum is one, but it is not Witt trivial. Indeed, its underlying group has order 8 and therefore cannot contain a Lagrangian subgroup, whose order would have to square to 8. Hence, equality of the signature phases does not imply equality of the Witt classes.

\subsection*{The 7-dimensional higher-form constructions}

The higher-form QCAs in 7 spatial dimensions are associated with responses on closed oriented 8-manifolds. The degree-eight Hirzebruch signature theorem reads
\begin{equation}
 L_2(TX)=\frac{7p_2(TX)-p_1(TX)^2}{45},
 \qquad
 \sigma(X^8)=\int_{X^8}L_2(TX).
 \label{eq:eight-dimensional-signature-theorem}
\end{equation}
Applying Eq.~\eqref{eq:brown-kervaire-morita-gauss-sum} to a degree-four $\ZZ_2$ field gives the generalized-semion response
\begin{equation}
 S_{\mathrm{GS}}^{\mathrm{inv}}(X^8)
 =\frac{\sigma(X^8)}8
 =\int_{X^8}\frac{7p_2(TX)-p_1(TX)^2}{360}
 \pmod1.
 \label{eq:higher-semion-pontryagin-response}
\end{equation}
As a check, if $h\in H^2(\mathbb{CP}^4,\ZZ)$ is the standard generator, then
\begin{equation}
 p_1(T\mathbb{CP}^4)=5h^2,
 \qquad
 p_2(T\mathbb{CP}^4)=10h^4,
 \qquad
 \int_{\mathbb{CP}^4}\frac{7p_2-p_1^2}{360}=\frac18.
 \label{eq:cp4-higher-semion-check}
\end{equation}

The boundary interpretation is a $(6+1)$-dimensional three-form Chern-Simons theory.  For an even integral matrix $K$, we can write
\begin{equation}
 S_7[C_3]
 =\frac1{4\pi}\int K_{IJ}C_3^I\wedge dC_3^J,
 \qquad
 \mathcal A=\mathbb Z^N/K\mathbb Z^N.
 \label{eq:seven-dimensional-higher-cs}
\end{equation}
A membrane label $a\in\mathcal A$, represented by an integral vector, has quadratic function and polarization
\begin{equation}
 q(a)=\frac12a^{\mathsf T}K^{-1}a\pmod1,
 \qquad
 b(a,a')=q(a+a')-q(a)-q(a')
 =a^{\mathsf T}K^{-1}a'\pmod1.
 \label{eq:membrane-quadratic-statistics}
\end{equation}
The phase $e^{2\pi i q(a)}$ is produced by a unit framing twist of the membrane worldvolume, while $e^{2\pi i b(a,a')}$ is produced by linked worldvolumes. These are higher-dimensional analogues of topological spin and mutual braiding.

For the generalized semion, $K=(2)$, and thus
\begin{equation}
    \mathcal A=\mathbb Z_2, \qquad q(1)=\frac14, \qquad b(1,1)=\frac12.
    \label{eq:higher-semion-membrane-statistics}
\end{equation}
For the higher-form $U(1)_4$ construction, $K=(4)$, so instead
\begin{equation}
    \mathcal A=\mathbb Z_4, \qquad q(a)=\frac{a^2}{8}, \qquad b(a,a')=\frac{aa'}4.
    \label{eq:higher-u14-membrane-statistics}
\end{equation}
The corresponding higher Gauss-Milgram relation is
\begin{equation}
    \frac1{\sqrt{|\mathcal A|}}\sum_{a\in\mathcal A}e^{2\pi i q(a)}
    =e^{2\pi i c_{\mathrm{sig}}/8}, 
    \qquad
    c_{\mathrm{sig}}=\operatorname{sgn}(K) \pmod8.
    \label{eq:higher-gauss-milgram}
\end{equation}
Both $K=(2)$ and $K=(4)$ give $c_{\mathrm{sig}}=1\pmod8$ and hence the same response $e^{2\pi i\sigma(X^8)/8}$. Nevertheless, their fusion groups of membrane excitations and quadratic functions are different. Note that the coefficient $c_{\mathrm{sig}}$ is a higher-dimensional signature, or framing-anomaly, coefficient. It should not be interpreted as a two-dimensional Virasoro central charge.

The common signature phase fixes only the normalized Gauss sum. In three spatial dimensions, the $U(1)_2$ and $U(1)_4$ theories are distinguished by their pointed Witt classes.  In 7 spatial dimensions, their higher-form analogues are distinguished by their fusion groups of membrane excitations and quadratic functions. Finally, the fact that the bordism character in Eq.~\eqref{eq:signature-bordism-character} has order 8 does not prove that any microscopic QCA has stable order 8.  Establishing a stable QCA class requires the operator-algebraic construction and a separate obstruction to a finite-depth circuit.

\bibliographystyle{utphys}
\bibliography{bibliography}

@misc{yang2026threedimensionalqcastablycircuits,
      title={Three-dimensional QCA that are not stably circuits}, 
      author={Bowen Yang},
      year={2026},
      eprint={2610.40235},
      archivePrefix={arXiv},
      primaryClass={math-ph},
      url={https://arxiv.org/abs/2610.40235}, 
}

@misc{chen2026invertiblesubalgebraschiralsemion,
      title={Invertible subalgebras for chiral semion topological order}, 
      author={Yu-An Chen and Meng Sun and Nathanan Tantivasadakarn and Zongyuan Wang and Bowen Yang},
      year={2026},
      eprint={2610.18953},
      archivePrefix={arXiv},
      primaryClass={quant-ph},
      url={https://arxiv.org/abs/2610.18953}, 
}

@misc{zhang2026quantumcellularautomatakramerswannier,
      title={Quantum Cellular Automata from Kramers-Wannier Dualities and Modular Relations}, 
      author={Carolyn Zhang and Po-Shen Hsin},
      year={2026},
      eprint={2607.21728},
      archivePrefix={arXiv},
      primaryClass={quant-ph},
      url={https://arxiv.org/abs/2607.21728}, 
}

@misc{inamura2026noninvertiblesymmetriesmixingwitt,
      title={Non-Invertible Symmetries Mixing with Witt Non-Trivial Quantum Cellular Automata}, 
      author={Kansei Inamura and Oskar Wojdel and Lukasz Fidkowski and Sakura Schafer-Nameki},
      year={2026},
      eprint={2607.21698},
      archivePrefix={arXiv},
      primaryClass={quant-ph},
      url={https://arxiv.org/abs/2607.21698}, 
}

@article{ji2026quantum,
  title={Quantum Cellular Automata: The Group, the Space, and the Spectrum},
  author={Ji, Mattie and Yang, Bowen},
  journal={arXiv preprint arXiv:2602.16572},
  year={2026}
}

@article{ji2026k,
  title={K-Theoretic Obstructions to Linearizing QCA Representations},
  author={Ji, Mattie and Yang, Bowen},
  journal={arXiv preprint arXiv:2606.19657},
  year={2026}
}

@article{Shirley2022QCA,
  title = {Three-Dimensional Quantum Cellular Automata from Chiral Semion Surface Topological Order and beyond},
  author = {Shirley, Wilbur and Chen, Yu-An and Dua, Arpit and Ellison, Tyler D. and Tantivasadakarn, Nathanan and Williamson, Dominic J.},
  journal = {PRX Quantum},
  volume = {3},
  issue = {3},
  pages = {030326},
  numpages = {27},
  year = {2022},
  month = {Aug},
  publisher = {American Physical Society},
  doi = {10.1103/PRXQuantum.3.030326},
  url = {https://link.aps.org/doi/10.1103/PRXQuantum.3.030326}
}

@Article{chen2023exactly,
	title={{Exactly solvable lattice Hamiltonians and gravitational anomalies}},
	author={Yu-An Chen and Po-Shen Hsin},
	journal={SciPost Phys.},
	volume={14},
	pages={089},
	year={2023},
	publisher={SciPost},
	doi={10.21468/SciPostPhys.14.5.089},
	url={https://scipost.org/10.21468/SciPostPhys.14.5.089},
}

@article{Chen2023HigherCup,
    author = {Chen, Yu-An and Tata, Sri},
    title = "{Higher cup products on hypercubic lattices: Application to lattice models of topological phases}",
    journal = {Journal of Mathematical Physics},
    volume = {64},
    number = {9},
    pages = {091902},
    year = {2023},
    month = {09},
    issn = {0022-2488},
    doi = {10.1063/5.0095189},
    url = {https://doi.org/10.1063/5.0095189},
}

@article{haah_QCA_23,
  author = {Haah, Jeongwan and Fidkowski, Lukasz and Hastings, Matthew B.},
  title = {Nontrivial Quantum Cellular Automata in Higher Dimensions},
  journal = {Communications in Mathematical Physics},
  volume = {398},
  number = {1},
  pages = {469--540},
  year = {2023},
  doi = {10.1007/s00220-022-04528-1}
}

@article{johnson2020topological,
  title = {{(3+1)D} topological orders with only a $\mathbb{Z}_2$-charged particle},
  author = {Johnson-Freyd, Theo},
  year = {2020},
  eprint = {2011.11165},
  archivePrefix = {arXiv},
  primaryClass = {math.QA},
  url = {https://arxiv.org/abs/2011.11165}
}

@article{Chen2012cohomology,
  title = {Symmetry protected topological orders and the group cohomology of their symmetry group},
  author = {Chen, Xie and Gu, Zheng-Cheng and Liu, Zheng-Xin and Wen, Xiao-Gang},
  journal = {Physical Review B},
  volume = {87},
  number = {15},
  pages = {155114},
  year = {2013},
  doi = {10.1103/PhysRevB.87.155114}
}

@article{Walker2012TQFT,
  author = {Walker, Kevin and Wang, Zhenghan},
  title = {{(3+1)-TQFTs} and topological insulators},
  journal = {Frontiers of Physics},
  volume = {7},
  number = {2},
  pages = {150--159},
  year = {2012},
  doi = {10.1007/s11467-011-0194-z}
}

@article{Steenrod1947Products,
  author = {Steenrod, N. E.},
  title = {Products of Cocycles and Extensions of Mappings},
  journal = {Annals of Mathematics},
  volume = {48},
  number = {2},
  pages = {290--320},
  year = {1947},
  doi = {10.2307/1969172}
}

@article{fidkowski2024qca,
  title = {A quantum cellular automaton for every symmetry protected topological phase},
  author = {Fidkowski, Lukasz and Haah, Jeongwan and Hastings, Matthew B.},
  journal = {Physical Review B},
  volume = {112},
  number = {3},
  pages = {035123},
  year = {2025},
  doi = {10.1103/kw68-mkkd}
}

@article{haah2025topological,
  author = {Haah, Jeongwan},
  title = {Topological Phases of Unitary Dynamics: Classification in Clifford Category},
  journal = {Communications in Mathematical Physics},
  volume = {406},
  number = {4},
  pages = {76},
  year = {2025},
  doi = {10.1007/s00220-025-05239-z}
}

@article{Sun2026Clifford,
  author = {Sun, Meng and Yang, Bowen and Wang, Zongyuan and Tantivasadakarn, Nathanan and Chen, Yu-An},
  title = {Clifford Quantum Cellular Automata from Topological Quantum Field Theories and Invertible Subalgebras},
  journal = {PRX Quantum},
  volume = {7},
  pages = {010362},
  year = {2026},
  doi = {10.1103/4519-v15s}
}

@book{MilnorStasheff1974,
  author = {Milnor, John W. and Stasheff, James D.},
  title = {Characteristic Classes},
  series = {Annals of Mathematics Studies},
  volume = {76},
  publisher = {Princeton University Press},
  year = {1974},
  doi = {10.1515/9781400881826}
}

@article{LusztigMilnorPeterson1969,
  author = {Lusztig, George and Milnor, John and Peterson, Franklin P.},
  title = {Semi-characteristics and cobordism},
  journal = {Topology},
  volume = {8},
  number = {4},
  pages = {357--359},
  year = {1969},
  doi = {10.1016/0040-9383(69)90021-4}
}

@article{PhysRevB.90.245122,
  title = {Exactly soluble model of a three-dimensional symmetry-protected topological phase of bosons with surface topological order},
  author = {Burnell, F. J. and Chen, Xie and Fidkowski, Lukasz and Vishwanath, Ashvin},
  journal = {Physical Review B},
  volume = {90},
  number = {24},
  pages = {245122},
  year = {2014},
  doi = {10.1103/PhysRevB.90.245122}
}

@article{Dold1956,
  author = {Dold, Albrecht},
  title = {Erzeugende der Thomschen Algebra $\mathfrak N$},
  journal = {Mathematische Zeitschrift},
  volume = {65},
  number = {1},
  pages = {25--35},
  year = {1956},
  doi = {10.1007/BF01473868}
}

@article{Morita1971Pontrjagin,
  author = {Morita, Shigeyuki},
  title = {On the Pontrjagin Square and the Signature},
  journal = {Journal of the Faculty of Science, the University of Tokyo, Section IA, Mathematics},
  volume = {18},
  number = {2},
  pages = {405--414},
  year = {1971},
  doi = {10.15083/00039814}
}

@InProceedings{PedersenWeibel,
author="Pedersen, Erik K.
and Weibel, Charles A.",
editor="Ranicki, Andrew
and Levitt, Norman
and Quinn, Frank",
title="A nonconnective delooping of algebraic K-theory",
booktitle="Algebraic and Geometric Topology",
year="1985",
publisher="Springer Berlin Heidelberg",
address="Berlin, Heidelberg",
pages="166--181",
isbn="978-3-540-39413-6"
}

@article{Taylor2022Gauss,
  author = {Taylor, Laurence R.},
  title = {Gauss Sums in Algebra and Topology},
  year = {2022},
  eprint = {2208.06319},
  archivePrefix = {arXiv},
  primaryClass = {math.AT},
  url = {https://arxiv.org/abs/2208.06319}
}

@article{Bhardwaj2020BrownKervaire,
  author = {Bhardwaj, Lakshya and Lee, Yasunori and Tachikawa, Yuji},
  title = {{$SL(2,\mathbb Z)$} Action on {QFTs} with $\mathbb{Z}_2$ Symmetry and the Brown--Kervaire Invariants},
  journal = {Journal of High Energy Physics},
  volume = {2020},
  number = {11},
  pages = {141},
  year = {2020},
  doi = {10.1007/JHEP11(2020)141},
  eprint = {2009.10099},
  archivePrefix = {arXiv},
  primaryClass = {hep-th}
}

@incollection{CraneKauffmanYetter1993,
  author = {Crane, Louis and Kauffman, Louis H. and Yetter, David N.},
  title = {Evaluating the Crane--Yetter Invariant},
  booktitle = {Quantum Topology},
  editor = {Kauffman, Louis H. and Baadhio, Randy A.},
  series = {Series on Knots and Everything},
  volume = {3},
  pages = {131--138},
  publisher = {World Scientific},
  address = {Singapore},
  year = {1993},
  doi = {10.1142/9789812796387_0006},
  eprint = {hep-th/9309063},
  archivePrefix = {arXiv}
}

@article{Drinfeld2010Braided,
  author = {Drinfeld, Vladimir and Gelaki, Shlomo and Nikshych, Dmitri and Ostrik, Victor},
  title = {On Braided Fusion Categories {I}},
  journal = {Selecta Mathematica},
  volume = {16},
  number = {1},
  pages = {1--119},
  year = {2010},
  doi = {10.1007/s00029-010-0017-z},
  eprint = {0906.0620},
  archivePrefix = {arXiv},
  primaryClass = {math.QA}
}

@article{Davydov2013Witt,
  author = {Davydov, Alexei and M{\"u}ger, Michael and Nikshych, Dmitri and Ostrik, Victor},
  title = {The Witt Group of Non-Degenerate Braided Fusion Categories},
  journal = {Journal f{\"u}r die reine und angewandte Mathematik},
  volume = {677},
  pages = {135--177},
  year = {2013},
  doi = {10.1515/CRELLE.2012.014},
  eprint = {1009.2117},
  archivePrefix = {arXiv},
  primaryClass = {math.QA}
}

@article{Feng2026AnyonicMembranes,
  author = {Feng, Yitao and Xue, Hanyu and Li, Yuyang and Cheng, Meng and Kobayashi, Ryohei and Hsin, Po-Shen and Chen, Yu-An},
  title = {Anyonic Membranes and Pontryagin Statistics},
  journal = {Physical Review Letters},
  volume = {136},
  number = {8},
  pages = {086601},
  year = {2026},
  doi = {10.1103/4jww-6b6t},
  eprint = {2509.14314},
  archivePrefix = {arXiv},
  primaryClass = {quant-ph}
}

@article{Stephen2019subsystem,
  doi = {10.22331/q-2019-05-20-142},
  url = {https://doi.org/10.22331/q-2019-05-20-142},
  title = {Subsystem symmetries, quantum cellular automata, and computational phases of quantum matter},
  author = {Stephen, David T. and Nautrup, Hendrik Poulsen and Bermejo-Vega, Juani and Eisert, Jens and Raussendorf, Robert},
  journal = {{Quantum}},
  issn = {2521-327X},
  publisher = {{Verein zur F{\"{o}}rderung des Open Access Publizierens in den Quantenwissenschaften}},
  volume = {3},
  pages = {142},
  month = may,
  year = {2019}
}

@article{ma2024QCA,
  doi = {10.22331/q-2026-06-01-2123},
  url = {https://doi.org/10.22331/q-2026-06-01-2123},
  title = {Quantum {C}ellular {A}utomata on {S}ymmetric {S}ubalgebras},
  author = {Ma, Ruochen and Li, Yabo and Cheng, Meng},
  journal = {{Quantum}},
  issn = {2521-327X},
  publisher = {{Verein zur F{\"{o}}rderung des Open Access Publizierens in den Quantenwissenschaften}},
  volume = {10},
  pages = {2123},
  month = jun,
  year = {2026}
}

@article{Keyserlingk2013Threedimensional,
  title = {Three-dimensional topological lattice models with surface anyons},
  author = {von Keyserlingk, C. W. and Burnell, F. J. and Simon, S. H.},
  journal = {Phys. Rev. B},
  volume = {87},
  issue = {4},
  pages = {045107},
  numpages = {35},
  year = {2013},
  month = {Jan},
  publisher = {American Physical Society},
  doi = {10.1103/PhysRevB.87.045107}
}

@misc{Bonderson2007,
  author        = {P. H. Bonderson},
  title         = {Non\-Abelian Anyons and Interferometry},
  howpublished  = {Ph.D.\ thesis, California Institute of Technology},
  address       = {Pasadena, California},
  month         = may,
  year          = {2007},
  doi           = {10.7907/5NDZ-W890}
}

@article{fidkowski2022gravitational,
  title = {Gravitational anomaly of $(3+1)$-dimensional $\mathbb{Z}_{2}$ toric code with fermionic charges and fermionic loop self-statistics},
  author = {Fidkowski, Lukasz and Haah, Jeongwan and Hastings, Matthew B.},
  journal = {Phys. Rev. B},
  volume = {106},
  issue = {16},
  pages = {165135},
  numpages = {33},
  year = {2022},
  month = {Oct},
  publisher = {American Physical Society},
  doi = {10.1103/PhysRevB.106.165135},
  url = {https://link.aps.org/doi/10.1103/PhysRevB.106.165135}
}

@article{jones2024QCA,
	title = {Quantum {Cellular} {Automata} and {Categorical} {Dualities} of {Spin} {Chains}},
	volume = {407},
	issn = {1432-0916},
	url = {https://doi.org/10.1007/s00220-026-05571-y},
	doi = {10.1007/s00220-026-05571-y},
	number = {4},
	journal = {Communications in Mathematical Physics},
	author = {Jones, Corey and Schatz, Kylan and Williamson, Dominic J.},
	month = mar,
	year = {2026},
	pages = {66},
}

@article{tu2025anomalies,
  title = {Anomalies of Global Symmetries on the Lattice},
  author = {Tu, Yi-Ting and Long, David M. and Else, Dominic V.},
  journal = {Phys. Rev. X},
  volume = {16},
  issue = {1},
  pages = {011027},
  numpages = {66},
  year = {2026},
  month = {Feb},
  publisher = {American Physical Society},
  doi = {10.1103/m188-w1ct},
  url = {https://link.aps.org/doi/10.1103/m188-w1ct}
}

@article{Haah2023InvertibleSubalgebras,
	author = {Haah, Jeongwan},
	da = {2023/10/01},
	doi = {10.1007/s00220-023-04806-6},
	id = {Haah2023},
	isbn = {1432-0916},
	journal = {Communications in Mathematical Physics},
	number = {2},
	pages = {661--698},
	title = {Invertible Subalgebras},
	ty = {JOUR},
	url = {https://doi.org/10.1007/s00220-023-04806-6},
	volume = {403},
	year = {2023}}

@article{Zhang2023Floquet,
  title = {Bulk-Boundary Correspondence for Interacting Floquet Systems in Two Dimensions},
  author = {Zhang, Carolyn and Levin, Michael},
  journal = {Phys. Rev. X},
  volume = {13},
  issue = {3},
  pages = {031038},
  numpages = {29},
  year = {2023},
  month = {Sep},
  publisher = {American Physical Society},
  doi = {10.1103/PhysRevX.13.031038},
  url = {https://link.aps.org/doi/10.1103/PhysRevX.13.031038}
}

@article{Gross2012GNVWindex,
	author = {Gross, D. and Nesme, V. and Vogts, H. and Werner, R. F.},
	da = {2012/03/01},
	doi = {10.1007/s00220-012-1423-1},
	id = {Gross2012},
	isbn = {1432-0916},
	journal = {Communications in Mathematical Physics},
	number = {2},
	pages = {419--454},
	title = {Index Theory of One Dimensional Quantum Walks and Cellular Automata},
	ty = {JOUR},
	url = {https://doi.org/10.1007/s00220-012-1423-1},
	volume = {310},
	year = {2012}}

@article{Freedman2020ClassificationQCA,
	author = {Freedman, Michael and Hastings, Matthew B.},
	da = {2020/06/01},
	doi = {10.1007/s00220-020-03735-y},
	id = {Freedman2020},
	isbn = {1432-0916},
	journal = {Communications in Mathematical Physics},
	number = {2},
	pages = {1171--1222},
	title = {Classification of Quantum Cellular Automata},
	ty = {JOUR},
	url = {https://doi.org/10.1007/s00220-020-03735-y},
	volume = {376},
	year = {2020}}

@article{Fidkowski2020beyondcohomology,
  title = {Exactly solvable model for a $4+1\mathrm{D}$ beyond-cohomology symmetry-protected topological phase},
  author = {Fidkowski, Lukasz and Haah, Jeongwan and Hastings, Matthew B.},
  journal = {Phys. Rev. B},
  volume = {101},
  issue = {15},
  pages = {155124},
  numpages = {25},
  year = {2020},
  month = {Apr},
  publisher = {American Physical Society},
  doi = {10.1103/PhysRevB.101.155124},
  url = {https://link.aps.org/doi/10.1103/PhysRevB.101.155124}
}

@article{Haah2021CliffordQCA,
    author = {Haah, Jeongwan},
    title = {Clifford quantum cellular automata: Trivial group in 2D and Witt group in 3D},
    journal = {Journal of Mathematical Physics},
    volume = {62},
    number = {9},
    pages = {092202},
    year = {2021},
    month = {09},
    issn = {0022-2488},
    doi = {10.1063/5.0022185},
    url = {https://doi.org/10.1063/5.0022185},
}

@article{kapustin2025higher,
  title={Higher symmetries and anomalies in quantum lattice systems},
  author={Kapustin, Anton},
  journal={arXiv preprint arXiv:2505.04719},
  year={2025},
  doi={10.48550/arXiv.2505.04719}
}

@article{kapustin2014symmetry,
  title={Symmetry protected topological phases, anomalies, and cobordisms: Beyond group cohomology},
  author={Kapustin, Anton},
  journal={arXiv preprint arXiv:1403.1467},
  year={2014},
  doi={10.48550/arXiv.1403.1467}
}

@article{kobayashi2024generalized,
  title = {Generalized Statistics on Lattices},
  author = {Kobayashi, Ryohei and Li, Yuyang and Xue, Hanyu and Hsin, Po-Shen and Chen, Yu-An},
  journal = {Phys. Rev. X},
  volume = {16},
  issue = {1},
  pages = {011010},
  numpages = {51},
  year = {2026},
  month = {Jan},
  publisher = {American Physical Society},
  doi = {10.1103/6k88-w52n},
  url = {https://link.aps.org/doi/10.1103/6k88-w52n}
}

@article{feng2026paulistatistics,
      title={Pauli stabilizer formalism for topological quantum field theories and generalized statistics}, 
      author={Yitao Feng and Hanyu Xue and Ryohei Kobayashi and Po-Shen Hsin and Yu-An Chen},
      year={2026},
      eprint={2601.00064},
      archivePrefix={arXiv},
      primaryClass={quant-ph},
      url={https://arxiv.org/abs/2601.00064}, 
}

@article{QFT2,
   title={Quantum cellular automata and quantum field theory in two spatial dimensions},
   volume={102},
   ISSN={2469-9934},
   url={http://dx.doi.org/10.1103/PhysRevA.102.062222},
   DOI={10.1103/physreva.102.062222},
   number={6},
   journal={Physical Review A},
   publisher={American Physical Society (APS)},
   author={Brun, Todd A. and Mlodinow, Leonard},
   year={2020},
   month={Dec}
}

@article{PoRadical,
  title = {Radical chiral Floquet phases in a periodically driven Kitaev model and beyond},
  author = {Po, Hoi Chun and Fidkowski, Lukasz and Vishwanath, Ashvin and Potter, Andrew C.},
  journal = {Phys. Rev. B},
  volume = {96},
  issue = {24},
  pages = {245116},
  numpages = {10},
  year = {2017},
  month = {Dec},
  publisher = {American Physical Society},
  doi = {10.1103/PhysRevB.96.245116},
  url = {https://link.aps.org/doi/10.1103/PhysRevB.96.245116}
}

@article{Piroli21Fermionic,
  title={Fermionic quantum cellular automata and generalized matrix-product unitaries},
  author={Piroli, Lorenzo and Turzillo, Alex and Shukla, Sujeet K and Cirac, J Ignacio},
  journal={Journal of Statistical Mechanics: Theory and Experiment},
  volume={2021},
  number={1},
  pages={013107},
  year={2021},
  publisher={IOP Publishing},
doi={10.1088/1742-5468/abd30f}
}

@article{GongNahumPiroli21,
  title = {Coarse-Grained Entanglement and Operator Growth in Anomalous Dynamics},
  author = {Gong, Zongping and Nahum, Adam and Piroli, Lorenzo},
  journal = {Phys. Rev. Lett.},
  volume = {128},
  issue = {8},
  pages = {080602},
  numpages = {6},
  year = {2022},
  month = {Feb},
  publisher = {American Physical Society},
  doi = {10.1103/PhysRevLett.128.080602},
  url = {https://link.aps.org/doi/10.1103/PhysRevLett.128.080602}
}

@article{GongPiroliCirac21,
  title = {Topological Lower Bound on Quantum Chaos by Entanglement Growth},
  author = {Gong, Zongping and Piroli, Lorenzo and Cirac, J. Ignacio},
  journal = {Phys. Rev. Lett.},
  volume = {126},
  issue = {16},
  pages = {160601},
  numpages = {6},
  year = {2021},
  month = {Apr},
  publisher = {American Physical Society},
  doi = {10.1103/PhysRevLett.126.160601},
  url = {https://link.aps.org/doi/10.1103/PhysRevLett.126.160601}
}

@article{PoChiral,
  title = {Chiral Floquet Phases of Many-Body Localized Bosons},
  author = {Po, Hoi Chun and Fidkowski, Lukasz and Morimoto, Takahiro and Potter, Andrew C. and Vishwanath, Ashvin},
  journal = {Phys. Rev. X},
  volume = {6},
  issue = {4},
  pages = {041070},
  numpages = {31},
  year = {2016},
  month = {Dec},
  publisher = {American Physical Society},
  doi = {10.1103/PhysRevX.6.041070},
  url = {https://link.aps.org/doi/10.1103/PhysRevX.6.041070}
}

@article{PotterVishwanathFidkowski18,
  title = {Infinite family of three-dimensional Floquet topological paramagnets},
  author = {Potter, Andrew C. and Vishwanath, Ashvin and Fidkowski, Lukasz},
  journal = {Phys. Rev. B},
  volume = {97},
  issue = {24},
  pages = {245106},
  numpages = {13},
  year = {2018},
  month = {Jun},
  publisher = {American Physical Society},
  doi = {10.1103/PhysRevB.97.245106},
  url = {https://link.aps.org/doi/10.1103/PhysRevB.97.245106}
}

@article{Piroli2020,
  title = {Quantum Cellular Automata, Tensor Networks, and Area Laws},
  author = {Piroli, Lorenzo and Cirac, J. Ignacio},
  journal = {Phys. Rev. Lett.},
  volume = {125},
  issue = {19},
  pages = {190402},
  numpages = {5},
  year = {2020},
  month = {Nov},
  publisher = {American Physical Society},
  doi = {10.1103/PhysRevLett.125.190402},
  url = {https://link.aps.org/doi/10.1103/PhysRevLett.125.190402}
}

@article{Zhang2021classification,
  title = {Classification of interacting Floquet phases with {$U(1)$} symmetry in two dimensions},
  author = {Zhang, Carolyn and Levin, Michael},
  journal = {Phys. Rev. B},
  volume = {103},
  issue = {6},
  pages = {064302},
  numpages = {30},
  year = {2021},
  month = {Feb},
  publisher = {American Physical Society},
  doi = {10.1103/PhysRevB.103.064302},
  url = {https://link.aps.org/doi/10.1103/PhysRevB.103.064302}
}

@article{Sahinoglu2018,
  title = {Matrix product representation of locality preserving unitaries},
  author = {\ifmmode \mbox{\c{S}}\else \c{S}\fi{}ahino\ifmmode \breve{g}\else \u{g}\fi{}lu, M. Burak and Shukla, Sujeet K. and Bi, Feng and Chen, Xie},
  journal = {Phys. Rev. B},
  volume = {98},
  issue = {24},
  pages = {245122},
  numpages = {14},
  year = {2018},
  month = {Dec},
  publisher = {American Physical Society},
  doi = {10.1103/PhysRevB.98.245122},
  url = {https://link.aps.org/doi/10.1103/PhysRevB.98.245122}
}

@article{IgnacioCirac2017,
  title={Matrix product unitaries: structure, symmetries, and topological invariants},
  author={Cirac, J Ignacio and Perez-Garcia, David and Schuch, Norbert and Verstraete, Frank},
  journal={Journal of Statistical Mechanics: Theory and Experiment},
  volume={2017},
  number={8},
  pages={083105},
  year={2017},
  publisher={IOP Publishing},
  doi = {10.1088/1742-5468/aa7e55}
}

@article{Glorioso21,
  title = {Effective response theory for Floquet topological systems},
  author = {Glorioso, Paolo and Gromov, Andrey and Ryu, Shinsei},
  journal = {Phys. Rev. Research},
  volume = {3},
  issue = {1},
  pages = {013117},
  numpages = {23},
  year = {2021},
  month = {Feb},
  publisher = {American Physical Society},
  doi = {10.1103/PhysRevResearch.3.013117},
  url = {https://link.aps.org/doi/10.1103/PhysRevResearch.3.013117}
}

@article{Gutschow2010Clifford,
	author = {G{\"u}tschow, J. },
	da = {2010/03/01},
	doi = {10.1007/s00340-009-3840-1},
	id = {G{\"u}tschow2010},
	isbn = {1432-0649},
	journal = {Applied Physics B},
	number = {4},
	pages = {623--633},
	title = {Entanglement generation of Clifford quantum cellular automata},
	ty = {JOUR},
	url = {https://doi.org/10.1007/s00340-009-3840-1},
	volume = {98},
	year = {2010}}

@article{Yang2026CategorifyingQCA,
	author = {Yang, Bowen},
	da = {2026/03/23},
	doi = {10.1007/s00220-026-05596-3},
	id = {Yang2026},
	isbn = {1432-0916},
	journal = {Communications in Mathematical Physics},
	number = {4},
	pages = {77},
	title = {Categorifying Clifford QCA},
	ty = {JOUR},
	url = {https://doi.org/10.1007/s00220-026-05596-3},
	volume = {407},
	year = {2026}}

@article{schumacher2004reversibleQCA,
      title={Reversible quantum cellular automata}, 
      author={B. Schumacher and R. F. Werner},
      year={2004},
      eprint={quant-ph/0405174},
      archivePrefix={arXiv},
      primaryClass={quant-ph},
      url={https://arxiv.org/abs/quant-ph/0405174}, 
}

@article{hsin2026bockstein,
      title={Bockstein braiding statistics}, 
      author={Po-Shen Hsin and Yu-An Chen},
      year={2026},
      eprint={2607.02280},
      archivePrefix={arXiv},
      primaryClass={quant-ph},
      url={https://arxiv.org/abs/2607.02280}, 
}

@article{Dold1956Wu,
	author = {Dold, Albrecht},
	da = {1956/12/01},
	doi = {10.1007/BF01473879},
	id = {Dold1956},
	isbn = {1432-1823},
	journal = {Mathematische Zeitschrift},
	number = {1},
	pages = {200--206},
	title = {Vollst{\"a}ndigkeit der Wuschen Relationen zwischen den Stiefel-Whitneyschen Zahlen differenzierbarer Mannigfaltigkeiten},
	ty = {JOUR},
	url = {https://doi.org/10.1007/BF01473879},
	volume = {65},
	year = {1956}}

@article{BrownPeterson1964Relations,
title = {Relations among characteristic classes—I},
journal = {Topology},
volume = {3},
pages = {39-52},
year = {1964},
issn = {0040-9383},
doi = {https://doi.org/10.1016/0040-9383(64)90004-7},
url = {https://www.sciencedirect.com/science/article/pii/0040938364900047},
author = {Edgar H. Brown and Franklin P. Peterson}
}

@book{Stong1968Cobordism,
  author    = {Stong, Robert E.},
  title     = {Notes on Cobordism Theory},
  series    = {Mathematical Notes},
  publisher = {Princeton University Press},
  address   = {Princeton, NJ},
  year      = {1968}
}

\end{document}